\documentclass[aps,twocolumn,showpacs,showkeys]{revtex4}
\usepackage{epsfig}
\usepackage{amssymb}
\usepackage{bm}%
\usepackage{amsmath}%
\usepackage{amsfonts}%
\usepackage{graphicx}
\providecommand{\U}[1]{\protect\rule{.1in}{.1in}}

\begin{document}



\title{Superconductivity in one dimension}
\author{K.Yu. Arutyunov$^{1,2}$, D.S. Golubev$^{3,4}$, and A.D. Zaikin$^{3,4}$}
\affiliation{$^{1}$NanoScience Center, Department of Physics, University of Jyv\"askyl\"a, PB 35, FI-40014
Jyv\"askyl\"a, Finland\\
$^{2}$Nuclear Physics Institute, Moscow State University, 119992 Moscow, Russia\\
$^{3}$Forschungszentrum Karlsruhe, Institut f\"ur Nanotechnologie,
76021, Karlsruhe, Germany\\
$^{4}$I.E. Tamm Department of Theoretical Physics, P.N. Lebedev Physics Institute, 119991
Moscow, Russia}
\begin{abstract}
Superconducting properties of metallic nanowires can be entirely
different from those of bulk superconductors because of the
dominating role played by thermal and quantum fluctuations of the
order parameter.  For superconducting wires with diameters below $
\sim 50$ nm quantum phase slippage is an important process which
can yield a non-vanishing wire resistance down to very low
temperatures. Further decrease of the wire diameter, for typical
material parameters down to $\sim 10$ nm, results in proliferation
of quantum phase slips causing a sharp crossover from
superconducting to normal behavior even at $T=0$. A number of
interesting phenomena associated both with quantum phase slips and
with the parity effect occur in superconducting nanorings. We
review recent theoretical and experimental activities in the field
and demonstrate dramatic progress in understanding of the
phenomenon of superconductivity in quasi-one-dimensional
nanostructures.
\end{abstract}

\maketitle

\tableofcontents
\section{Introduction}

The phenomenon of superconductivity was discovered \cite{Kamerlingh-Onnes} as
a sudden drop of resistance to immeasurably small value. With the development
of the topic it was realized that the superconducting phase transition is
frequently not at all ''sudden'' and the measured dependence of the sample
resistance $R(T)$ in the vicinity of the critical temperature $T_{C}$ may have
a finite width. One possible reason for this behavior -- and frequently the
dominating factor -- is the sample inhomogeneity, i.e. the sample might simply
consist of regions with different local critical temperatures. However, with
improving fabrication technologies it became clear that even for highly
homogeneous samples the superconducting phase transition may remain broadened.
This effect is usually very small in bulk samples and becomes more pronounced
in systems with reduced dimensions. A fundamental physical reason behind such
smearing of the transition is \textit{superconducting fluctuations}.

An important role of fluctuations in reduced dimension is well
known. Above $T_{C}$ such fluctuations yield an enhanced
conductivity of metallic systems~\cite{almt,Maki,Thomson,LV} . For
instance, the so-called Aslamazov-Larkin fluctuation correction to
conductivity $\delta\sigma_{AL}\sim(T-T_{C})^{-(2-D/2)}$ becomes
large in the vicinity of $T_{C}$ and this effect increases with
decreasing dimensionality $D$. Below $T_{C}$ -- according to the
general theorem \cite{HMW,Mermin 1966} -- fluctuations should
destroy the long-range order in low dimensional superconductors.
Thus, it could naively be concluded that low dimensional
conductors cannot exhibit superconducting properties because of
strong phase fluctuation effects.

This conclusion, however, turns out to be somewhat premature. For
instance, 2D structures undergo Berezinskii-Kosterlitz-Thouless
(BKT) phase transition \cite{b,kt,Kosterlilz 1974} as a result of
which the decay of correlations in space changes from exponential
at high enough $T$ to power law at low temperatures. This result
implies that at low $T$ long range phase coherence essentially
survives in samples of a finite size and, hence, 2D films can well
exhibit superconducting properties.

Can superconductivity survive also in (quasi)-1D systems or do fluctuations
suppress phase coherence thus disrupting any supercurrent? The answer to this
question would clearly be of both fundamental interest and practical
importance. On one hand, investigations of this subject definitely help to
encover novel physics and shed more light on the crucial role of
superconducting fluctuations in 1D wires. On the other hand, rapidly
progressing miniaturization of nanodevices opens new horizons for applications
of superconducting nanocircuits and requires better understanding of
fundamental limitations for the phenomenon of superconductivity in reduced
dimension. A detailed review of the present status of this
field is the main purpose of this paper.

It was first pointed out by Little \cite{Little} that quasi-one-dimensional
wires made of a superconducting material can acquire a finite resistance below
$T_{C}$ of a bulk material due to the mechanism of thermally activated phase
slips (TAPS). Within the Ginzburg-Landau theory one can describe a
superconducting wire by means of a complex order parameter $\Psi
(x)=|\Psi(x)|e^{i\varphi(x)}$. Thermal fluctuations cause deviations of both
the modulus and the phase of this order parameter from their equilibrium
values. A non-trivial fluctuation corresponds to temporal suppression of
$|\Psi(x)|$ down to zero in some point (e.g., $x=0$) inside the wire, see Fig.
1. As soon as the modulus of the order parameter $|\Psi(0)|$ vanishes the
phase $\varphi(0)$ becomes unrestricted and can jump by the value $2\pi n$,
where $n$ is any integer number. After this process the modulus $|\Psi(0)|$
gets restored, the phase becomes single valued again and the system returns to
its initial state accumulating the net phase shift $2\pi n$. Provided such
phase slip events are sufficiently rare, one can restrict $n$ by $n=\pm1$ and
totally disregard fluctuations with $|n|\geq2$.

%

\begin{figure}
\begin{center}
\includegraphics[width=8cm]{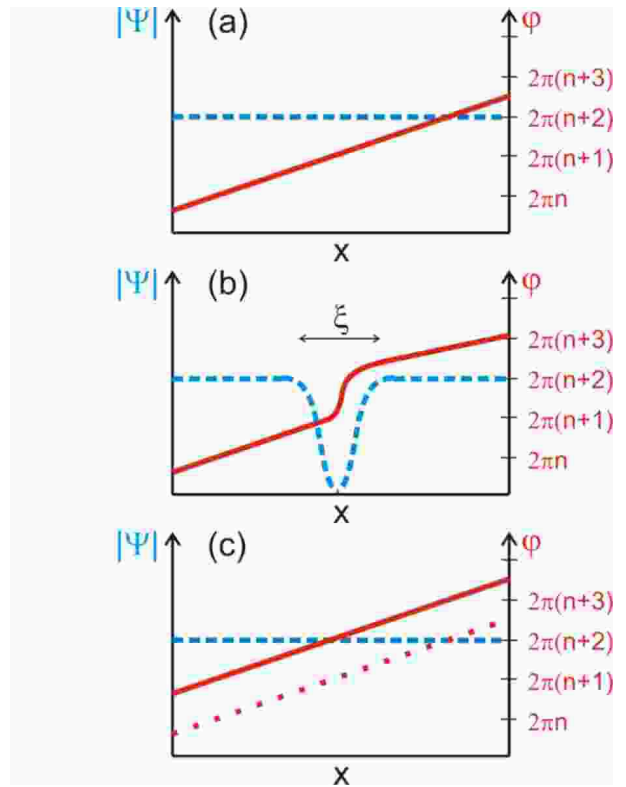}
\caption{Schematics of the phase slip process. Spatial variation
of the amplitude of the order parameter $|\Psi|$ (left axis,
dashed line) and phase $\varphi$ (right axis solid line) at
various moments of time: (a) before, (b)
during \ and (c) after the phase slippage.}%
\label{Fig PS schematics}%
\end{center}
\end{figure}

According to the Josephson relation $V=\hbar\dot{\varphi}/2e$ each
such phase slip event causes a nonzero voltage drop $V$ across the
wire. In the absence of any bias current the net average numbers
of \textquotedblright positive\textquotedblright\ ($n=+1$) and
\textquotedblright negative\textquotedblright\ ($n=-1$) phase
slips are equal, thus the net voltage drop remains zero. Applying
the current $I\propto\mid\Psi\mid ^{2}\nabla\varphi$ one creates
nonzero phase gradient along the wire and makes \textquotedblright
positive\textquotedblright\ phase slips more likely than
\textquotedblright negative\textquotedblright\ ones. Hence, the
net voltage drop $V$ due to TAPS differs from zero, i.e. thermal
fluctuations cause non-zero resistance $R=V/I$ of superconducting
wires even below $T_{C}$. We would also like to emphasize that, in
contrast to the so-called phase slip centers \cite{Meyer
IV,Tidecks,ivko,Kopnin book} produced by a large current above the
critical one $I>I_{C}$, here we are dealing with
\textit{fluctuation-induced phase slips} which can occur at
arbitrarily small values $I$.

A quantitative theory of the TAPS phenomenon was first proposed by
Langer and Ambegaokar \cite{la} and then completed by McCumber and
Halperin \cite{mh}. This LAMH theory predicts that the TAPS
creation rate and, hence, resistance of a superconducting wire $R$
below $T_{C}$ are determined by the activation exponent
\begin{equation}
R(T)\propto\exp(-U/T),\;\;\;\;\;\;\;U\sim\frac{N_{0}\Delta_{0}^{2}(T)}{2}%
s\xi(T),\label{TAPS}%
\end{equation}
where $U(T)$ is the effective potential barrier for TAPS
proportional to the superconducting condensation energy ($N_{0}$
is the metallic density of states at the Fermi energy and
$\Delta_{0}(T)$ is the BCS order parameter) for a part of the wire
of a volume $s\xi$ where superconductivity is destroyed by thermal
fluctuations ($s$ is the wire cross section and $\xi(T)$ is the
superconducting coherence length). At temperatures very close to
$T_{C}$ eq. (\ref{TAPS}) yields appreciable resistivity which was
indeed detected experimentally \cite{Webb R(T) in Sn
whiskers,Tinkham R(T) in Sn whiskers}. Close to $T_{C}$ the
experimental results fully confirm the activation behavior of
$R(T)$ expected from eq. (\ref{TAPS}). However, as the temperature
is lowered further below $T_{C}$ the number of TAPS inside the
wire decreases exponentially and no measurable wire resistance is
predicted by the LAMH theory~\cite{la,mh} except in the immediate
vicinity of the critical temperature.

Experiments \cite{Webb R(T) in Sn whiskers,Tinkham R(T) in Sn
whiskers} were done on small whiskers of typical diameters
$\sim0.5$ $\mu$m. Recent progress in nanolithographic technique
allowed to fabricate samples with much smaller diameters down to
-- and even below -- 10 nm. In such systems one can consider a
possibility for phase slips to occur not only due to thermal, but
also due to \textit{quantum} fluctuations of the superconducting
order parameter. The physical picture of quantum phase slippage is
qualitatively similar to that of TAPS (see Fig. 1) except the
order parameter $|\Psi(x)|$ gets virtually suppressed due the
process of \textit{quantum tunneling}.

Following the standard quantum mechanical arguments one can expect that the
probability of such tunneling process should be controlled by the exponent
$\sim\exp(-U/\hbar\omega_{0})$, i.e. instead of temperature in the activation
exponent (\ref{TAPS}) one should just substitute $\hbar\omega_{0}$, where
$\omega_{0}$ is an effective attempt frequency. This is because the order
parameter field $\Psi(x)$ now tunnels under the barrier $U $ rather than
overcomes it by thermal activation. Since such tunneling process should
obviously persist down to $T=0$ one arrives at a fundamentally important
conclusion that \textit{in nanowires superconductivity can be destroyed by
quantum fluctuations at any temperature including} $T=0$. Accordingly, such
nanowires should demonstrate a non-vanishing resistivity down to zero
temperature. \textit{Assuming} that $\hbar\omega_{0}\sim\Delta_{0}(T)$ one
would expect that at $\Delta_{0}(T)\lesssim T<T_{C}$ the TAPS dependence
(\ref{TAPS}) applies while at lower $T\lesssim\Delta_{0}(T)$ quantum phase
slips (QPS) take over, eventually leading to saturation of the temperature
dependence $R(T)$ to a non-zero value in the limit $T\ll\Delta_{0}$.

This behavior was indeed observed: Giordano \cite{Giordano QPS PRL 1988}
performed experiments which clearly demonstrated a notable resistivity of
ultra-thin superconducting wires far below $T_{C}$. These observations could
not be adequately interpreted within the TAPS theory and were attributed to
QPS. Later other groups also reported noticeable deviations from the LAMH
theory in thin (quasi-)1D wires. These experiments will be discussed in
Chapter 6.

It should be noted, however, that despite these developments the idea that in
realistic samples superconductivity can be destroyed by quantum fluctuations
was initially received with a large portion of scepticism. On one hand, this
was due to a number of unsuccessful attempts to experimentally observe the QPS
phenomenon. On the other hand, some early theoretical efforts have led to the
results strongly underestimating the actual QPS rate. Also, unambiguous
interpretation of the observations \cite{Giordano QPS PRL 1988} in terms of
QPS was questioned because of possible granularity of the samples used in
those experiments. If that was indeed the case, QPS could easily be created
inside weak links connecting neighboring grains. Also in this case
superconducting fluctuations play a very important role \cite{sz}
however -- in contrast to the case of uniform wires -- the superconducting
order parameter \textit{needs not to be destroyed} during a QPS event.

First attempts to theoretically analyze the QPS effects
\cite{sm,Duan,Chang} -- as well as a number of later studies --
were based on the so-called time-dependent Ginzburg-Landau (TDGL)
equations. Unfortunately the TDGL approach is by far insufficient
for the problem in question for a number of reasons: (i) A trivial
reason is that the Ginzburg-Landau (GL) expansion applies only at
temperatures close to $T_{C}$ whereas in order to describe QPS one
usually needs to go to lower temperatures down to $T\to 0$. (ii)
More importantly, also at $T_{C}-T\ll T_{C}$ TDGL equation remains
applicable only in a special limit of gapless superconductors,
while it fails in a general situation considered here. (iii) TDGL
approach does not account for dissipation effects due to
quasiparticles inside the QPS core (in certain cases also outside
this core) which are expected to reduce the probability of QPS
events similarly to the standard problem of quantum tunneling with
dissipation \cite{cl,cl1,weiss}. (iv) TDGL approach is not fully
adequate to properly describe excitation of electromagnetic modes
around the wire during a QPS event (this effect turns out to be
particularly important for sufficiently long wires). Thus,
TDGL-based description of QPS effects simply cannot be trusted,
and a much more elaborate theory is highly desirable in this
situation.

A microscopic theory of QPS processes in superconducting nanowires was
developed \cite{ZGOZ,ZGOZ2,GZ01} with the aid of the imaginary time effective
action technique \cite{GZ01,ogzb}. This theory remains applicable down to
$T=0$ and properly accounts for non-equilibrium, dissipative and
electromagnetic effects during a QPS event. One of the main conclusions of
this theory is that in sufficiently dirty superconducting nanowires with
diameters in the 10 nm range QPS probability can already be large enough to
yield experimentally observable phenomena. Also, further interesting effects
including quantum phase transitions caused by interactions between quantum
phase slips were predicted \cite{ZGOZ,ZGOZ2}.

An important parameter of this theory is the QPS fugacity
\[
y\sim S_{\mathrm{core}}\exp(-S_{\mathrm{core}}),\quad S_{\mathrm{core}}\sim
g_{\xi},
\]
where $g_{\xi}$ is the dimensionless conductance of the wire
segment of length $\xi$. Provided $g_{\xi}$ is very large,
typically $g_{\xi}\gtrsim100$, the fugacity $y$ remains
vanishingly small, QPS events are very rare and in many cases can
be totally neglected. In such case the standard BCS mean field
description should apply and a finite (though possibly
sufficiently long) wire remains essentially superconducting
outside an immediate vicinity of $T_{C}$. For smaller
$g_{\xi}\lesssim 10\div 20$ QPS effects already become important
down to $T\to 0$. Finally, at even smaller $g_{\xi}\sim1$ strong
fluctuations should wipe out superconductivity everywhere in the
wire. We also point out that in the case of nanowires considered
here the parameter $g_{\xi}$ is related to the well known Ginzburg
number as $Gi_{\mathrm{1D}}\sim 1/g_{\xi}^{2/3}$, i.e. the
condition $g_{\xi}\sim 1$ also implies that the fluctuation region
becomes of order $T_{C}$.

Another important parameter is the ratio between the (\textquotedblright
superconducting\textquotedblright) quantum resistance unit $R_{q}=\pi
\hbar/2e^{2}=6.453$ k$\Omega$ and the wire impedance $Z_{w}=\sqrt{\tilde{L}%
/C}$:
\[
\mu=R_{q}/2Z_{w},
\]
where $C$ is the wire capacitance per unit length, $\tilde{L}=4\pi\lambda
_{L}^{2}/s$ is the wire kinetic inductance and $\lambda_{L}$ is the London
penetration depth. Provided this parameter becomes of order one, $\mu\sim1$,
superconductivity in sufficiently long wires gets fully suppressed due to
intensive fluctuations of the phase $\varphi$ of the superconducting order
parameter. We note that both $g_{\xi}$ and $\mu$ scale with the wire cross
section $s$ respectively as $g_{\xi}\propto s$ and $\mu\propto\sqrt{s}$. It
follows immediately that \textit{with decreasing the cross section below a
certain value the wire inevitably looses intrinsic superconducting properties
due to strong fluctuation effects}. For generic parameters both conditions
$g_{\xi}\sim1\div10$ and $\mu\sim1$ are typically met for wire diameters in
the range $\sqrt{s}\lesssim10$ nm.

A number of recent experimental observations are clearly
consistent with the above theoretical conclusions. Perhaps the
first unambiguous evidence for QPS effects in quasi-1D wires was
reported by Bezryadin, Lau and Tinkham \cite{BT} who fabricated
sufficiently uniform superconducting wires with thicknesses down
to $3\div5$ nm and observed that several samples showed no signs
of superconductivity even at temperatures well below the bulk
critical temperature. Those results were later confirmed and
substantially extended by different experimental groups. At
present there exists an overwhelming experimental evidence for QPS
effects in superconducting nanowires fabricated to be sufficiently
uniform and homogeneous. Below we will analyze the main
experimental results and compare them with theoretical
predictions.

Yet another interesting issue is related to persistent currents
(PC) in superconducting nanorings. It was demonstrated \cite{MLG}
that QPS effects can significantly modify PC in such systems and
even lead to exponential suppression of supercurrent for
sufficiently large ring perimeters. Another important factor that
can substantially affect PC in isolated superconducting nanorings
at low $T$ is the electron parity number. Of particular interest
is the behavior of rings with odd number of electrons which can
develop \textit{spontaneous} supercurrent in the ground state
without any externally applied magnetic flux \cite{SZ}.

The structure of our Review is as follows. A theory of superconducting
fluctuations in nanowires will be addressed in Chapters 2-5. In Chapter 2 we
discuss a general derivation of the real time effective action of a
superconductor suitable for further investigations of fluctuation effects at
temperatures below $T_{C}$. We also formulate the Langevin equations and
analyze their relation to TDGL-type of equations frequently used in the
literature. In Chapter 3 we adopt our general formalism to the case of
superconducting nanowires and demonstrate the importance of superconducting
fluctuations in such structures. In Chapter 4 we will briefly review LAMH
theory of thermally activated phase slips. Quantum phase slip effects will be
analyzed in details in Chapter 5. Chapter 6 is devoted to an elaborate
discussion of key experiments in the field and their interpretation in terms
of existing theories. In Chapter 7 we will analyze persistent currents in
superconducting nanorings. Chapter 8 contains a brief summary of our main
observations and conclusions. Some technical details are presented in Appendix.

\section{Effective action, Langevin and Ginzburg-Landau equations}

\subsection{General formulation}

The starting point of our analysis is the formal expression for the quantum
evolution operator on the Keldysh contour or the so-called Keldysh ``partition
function''. As usually \cite{sz}, the kernel of the evolution operator
${\mathcal{J}}$ can be expressed in the form of a path integral over quantum
fields defined both at the forward (below denoted by the subscript $_{F}$) and
the backward (denoted by the subscript $_{B}$) branches of the Keldysh
contour. We have
\begin{align}
{\mathcal{J}} & = \int{\mathcal{D}}\psi_{F/B}^{\alpha}\,{\mathcal{D}}\bar
\psi_{F/B}^{\alpha}{\mathcal{D}}\Delta_{F/B}\,{\mathcal{D}}V_{F/B}%
\,{\mathcal{D}}{\bm A}_{F/B}\nonumber\\
& \times\, e^{iS[\psi_{F}^{\alpha},\Delta_{F},V_{F},\bm{A}_{F}]-iS[\psi
_{B}^{\alpha},\Delta_{B},V_{B},\bm{A}_{B}]},
\end{align}
Here $\psi_{F/B}^{\alpha},\bar\psi_{F/B}^{\alpha}$ are electron Grasmann
fields, $\alpha=\uparrow,\downarrow$ is the spin index, $\Delta_{F/B}$ are
superconducting complex order parameter fields (which emerge as a result of
the standard Hubbard-Stratonovich decoupling of the BCS coupling term
\cite{sz}), $V_{F/B}$ and ${\bm A}_{F/B}$ are respectively the scalar and
vector potentials. The action $S$ is defined as follows
\begin{align}
iS & =\sum_{\alpha}\int d^{4}X\bigg[ \bar\psi^{\alpha}\bigg( i\frac{\partial
}{\partial t}+\frac{1}{2m} \left( \nabla+i\frac{e}{c}\bm{A}\right) ^{2}
\nonumber\\ &
+\,\epsilon_{F} -U(\bm{r})+eV \bigg)\psi^{\alpha}  -\Delta\bar\psi^{\uparrow}\bar\psi^{\downarrow}- \Delta^{*}
\psi^{\downarrow}\psi^{\uparrow} \bigg]
\nonumber\\ &
+\,\int d^{4}X\left( \frac{{\bm E}%
^{2}-{\bm H}^{2}}{8\pi}-\frac{N_{0}|\Delta|^{2}}{\lambda}\right) .
\end{align}
Here $\epsilon_{F}$ is the Fermi energy, $U({\bm r})$ is the disorder
potential, $N_{0}$ is the density of states per unit spin at the Fermi level,
and $\lambda$ is the BCS coupling constant. In our notations the electron
charge is $e^{*}=-e$, i.e. we define $e>0$. Here and below we employ the
notation $d^{4}X=dt\,d^{3}{\bm
r}$.

As the above action is quadratic in the electron fields one can integrate them
out exactly. After that one arrives at the path integral
\begin{align}
{\mathcal{J}}=\int{\mathcal{D}}\Delta_{F/B}\,{\mathcal{D}}V_{F/B}%
\,{\mathcal{D}}{\bm A}_{F/B}\; e^{iS_{\mathrm{eff}}[\Delta_{F/B}%
,V_{F/B},{\bm A}_{F/B}]},
\end{align}
where the effective action reads
\begin{align}
iS_{\mathrm{eff}}=i\int d^{4}X\left( \frac{{\bm E}^{2}-{\bm H}^{2}}{8\pi
}-\frac{N_{0}|\Delta|^{2}}{\lambda}\right)  +\,\mathrm{Tr}\,\ln\check
{\mathcal{G}}^{-1}.\label{Seff}%
\end{align}
The inverse $4\times4$ matrix Green-Keldysh function $\check{\mathcal{G}}%
^{-1}$ can be split into $2\times2$ sub-blocks (indicated by a hat):
\begin{align}
\check{\mathcal{G}}^{-1}=\left(
\begin{array}
[c]{cc}%
\hat{\mathcal{G}}^{-1}_{11} & -\hat\sigma_{z}\hat\Delta\\
-\hat\sigma_{z}\hat\Delta & \hat{\mathcal{G}}^{-1}_{22}%
\end{array}
\right) \label{G-1}%
\end{align}
where
\begin{align}
\hat{\mathcal{G}}^{-1}_{11} & = \hat\sigma_{z}\left( i\frac{\partial}{\partial
t}-H_{0} +\frac{i}{2}\{\nabla,\hat{\bm v}_{S}\}-\frac{m\hat{\bm v}_{S}^{2}}%
{2}+e\hat\Phi\right) \nonumber\\
\hat{\mathcal{G}}^{-1}_{22} & = \hat\sigma_{z}\left( i\frac{\partial}{\partial
t}+H_{0} +\frac{i}{2}\{\nabla,\hat{\bm v}_{S}\}+\frac{m\hat{\bm v}_{S}^{2}}%
{2}-e\hat\Phi\right) .
\end{align}
Here $H_{0}=-\nabla^{2}/2m +U(\bm{r})-\epsilon_{F}$ is the single electron
Hamiltonian,
\begin{align}
\hat{\bm v}_{S}=\frac{1}{2m}\left(
\begin{array}
[c]{cc}%
\nabla\varphi_{F} +\frac{2e}{c}{\bm A}_{F} & 0\\
0 & \nabla\varphi_{B} +\frac{2e}{c}{\bm A}_{B}%
\end{array}
\right) ,
\end{align}
\begin{align}
\hat\Phi=\left(
\begin{array}
[c]{cc}%
V_{F} -\frac{\dot\varphi_{F}}{2e} & 0\\
0 & V_{B} -\frac{\dot\varphi_{B}}{2e}%
\end{array}
\right) ,
\end{align}
define the gauge invariant combiantions of the superconducting order parameter
phase $\varphi$ and the electromagnetic potentials. The matrix $\hat\Delta$ is
constructed analogously. It reads
\begin{align}
\hat\Delta=\left(
\begin{array}
[c]{cc}%
\Delta_{F} & 0\\
0 & \Delta_{B}%
\end{array}
\right) .
\end{align}
Note that both fields $\Delta_{F}$ and $\Delta_{B}$ are now real since we have
already decoupled the phase factors $\varphi_{F/B}$ by the gauge transformation.

\subsection{Perturbation theory}

Although the above expression for the effective action $S_{\mathrm{eff}}$
(\ref{Seff}) is exact it remains too complicated for practical calculations.
In order to proceed let us -- analogously to the derivation in the Matsubara
technique \cite{ogzb,GZ01} -- restrict our analysis to quadratic fluctuations.
For this purpose we split the order parameter into the BCS mean field term
$\Delta_{0}$ and the fluctuating part. Performing a shift $\Delta_{F/B}%
\to\Delta_{0}+\Delta_{F/B}$ we re-define the order parameter field in a way to
describe fluctuations by the fields $\Delta_{F/B}$. Expanding the effective
action (\ref{Seff}) in powers of $\hat\Delta$, $\hat{\bm v}_{S}$ and $\hat
\Phi$ to the second order we obtain the action describing the quadratic
fluctuations in the system
\begin{align}
iS_{\mathrm{eff}}^{(2)} =iS_{1}+ iS_{\Delta}+i S_{\mathrm{em}} +
iS_{\varphi\Delta},
\end{align}
where
\begin{align}
iS_{1} & = \,\mathrm{Tr}\,\bigg[  \hat\sigma_{z} \frac{i}{2}\{\nabla,\hat
{v}_{S}\}(\hat G+\hat{\bar G})
\nonumber\\ &
+\, \hat\sigma_{z} e\hat\Phi(\hat G-\hat{\bar G})
-\hat\sigma_{z}\hat\Delta(\hat F+\hat{\bar F}) \bigg]
\end{align}
defines the first order (in $\hat\Phi,\hat{\bm v}_{S},\hat\Delta$)
contribution to the action,
\begin{align}
&  iS_{\Delta}= -i\frac{N_{0}}{\lambda}\int d^{4}X\, \big(|\Delta_{0}%
+\Delta_{F}|^{2}-|\Delta_{0}+\Delta_{B}|^{2}\big)\nonumber\\
&  -\,\frac{1}{2}\,\mathrm{Tr}\,\left\{  (\hat F\hat\sigma_{z}\hat\Delta)^{2}
+ (\hat{\bar F}\hat\sigma_{z}\hat\Delta)^{2} + 2\hat{\bar G}\hat\sigma_{z}%
\hat\Delta\hat G\hat\sigma_{z}\hat\Delta\right\} ,\;\;\;\;\label{SDelta}%
\end{align}
accounts for fluctuations of the absolute value of the order parameter field,
\begin{align}
&  iS_{\mathrm{em}} = i\int d^{4}X\;\frac{\bm{E}_{F}^{2} - \bm{E}_{B}^{2}
-\bm{H}_{F}^{2}+\bm{H}_{B}^{2}}{8\pi}
\nonumber\\ &
-\,\frac{1}{2}\,\mathrm{Tr}%
\,\bigg\{ m\hat{\bm v}_{S}^{2}(\hat G-\hat{\bar G})
 +\left[ \hat\sigma_{z}\left( \frac{i}{2}\{\nabla,\hat{\bm v}_{S}%
\}+e\hat\Phi\right) \hat G\right] ^{2}
\nonumber\\ &
+ \hat\sigma_{z}\left( \frac{i}%
{2}\{\nabla,\hat{\bm v}_{S}\}+e\hat\Phi\right) \hat F \hat\sigma_{z}\left(
\frac{i}{2}\{\nabla,\hat{\bm v}_{S}\}-e\hat\Phi\right) \hat{\bar F}\nonumber\\
&  +\,\left[ \hat\sigma_{z}\left( \frac{i}{2}\{\nabla,\hat{\bm v}_{S}%
\}-e\hat\Phi\right) \hat{\bar G}\right] ^{2} \bigg\}\label{sem}%
\end{align}
describes electromagnetic fields and their coupling to the phase of the order
parameter field and
\begin{align}
&  iS_{\varphi\Delta}=\,\mathrm{Tr}\,\bigg\{  \hat\sigma_{z}\left( \frac{i}%
{2}\{\nabla,\hat{\bm v}_{S}\}+e\hat\Phi\right) \hat F \hat\sigma_{z}\hat
\Delta\hat G
\nonumber\\ &
+\,\hat\sigma_{z}\left( \frac{i}{2}\{\nabla,\hat{\bm v}_{S}%
\}+e\hat\Phi\right) \hat G \hat\sigma_{z}\hat\Delta\hat{\bar F}
\nonumber\\
&    +\, \hat\sigma_{z}\left( \frac{i}{2}\{\nabla,\hat{\bm v}%
_{S}\}-e\hat\Phi\right) \hat{\bar G} \hat\sigma_{z}\hat\Delta\hat F
\nonumber\\ &
+\,\hat\sigma_{z}\left( \frac{i}{2}\{\nabla,\hat{\bm v}_{S}\}-e\hat\Phi\right)
\hat{\bar F} \hat\sigma_{z}\hat\Delta\hat{\bar G} \bigg\} .
\end{align}
is responsible for coupling of electromagnetic and phase fluctuations to the
absolute value of the order parameter field.

All the above contributions to the action are expressed in terms of
equilibrium non-perturbed normal and anomalous $2\times2$ Green-Keldysh
matrices which enter as sub-blocks $\hat G,\hat F,\hat{\bar G},\hat{\bar F}$
into the $4\times4$ matrix
\begin{align}
\check{\mathcal{G}}_{0}=\left(
\begin{array}
[c]{cc}%
\hat G & \hat F\\
\hat{\bar F} & \hat{\bar G}%
\end{array}
\right) ,\label{GGGG}%
\end{align}
obtained by inverting the expression (\ref{G-1}) at $\hat\Phi=0$, $\hat
{\bm A}=0$ and $\Delta_{F}=\Delta_{B}=\Delta_{0}$. The formal expressions for
the sub-blocks $\hat G,\hat F,\hat{\bar G},\hat{\bar F}$ are defined in
Appendix A.1.

One can demonstrate that the matrix Green-Keldysh functions satisfy the Ward
identities (\ref{Ward2}) and (\ref{Ward1}) specified in Appendix A.2. Making
use of eq. (\ref{Ward2}) we can rewrite the first order contribution to the action
in the form
\begin{align}
iS_{1}  & = -\,\mathrm{Tr}\,\left[  \hat\sigma_{z}\hat\Delta(\hat F+\hat{\bar
F}) \right]  +i\Delta_{0} \,\mathrm{Tr}\,\left[  \hat\sigma_{z}\hat
\varphi(\hat F-\hat{\bar F}) \right] \nonumber\\
&  +\,\frac{ie}{2mc}\,\mathrm{Tr}\,\left[ \hat\sigma_{z}\{\nabla,\hat
{\bm A}\}(\hat G+\hat{\bar G})\right]  +e\,\mathrm{Tr}\,\left[ \hat\sigma
_{z}\hat V(\hat G-\hat{\bar G})\right] .\nonumber
\end{align}
Observing that $\hat F=\hat{\bar F}$ and introducing the current $\bm{j}_{0}$
and the charge density $\rho_{0}$ in the non-perturbed state we further
transform the action $iS_{1}$ and get
\begin{eqnarray}
iS_{1} &=& -2\,\mathrm{Tr}\,\left[  \hat\sigma_{z}\hat\Delta\hat F \right]
-\frac{i}{c}\int d^{4}X\,{\bm j}_{0}({\bm A}_{F}-{\bm A}_{B})
\nonumber\\ &&
-\,i\int d^{4}X\,
\rho_{0}(V_{F}-V_{B}).\label{S11}%
\end{eqnarray}
In addition, assuming the non-perturbed system state to be in thermodynamic
equlibrium, we set both the current and the charge density equal zero. Then we
obtain
\begin{align}
iS_{1}= -2\,\mathrm{Tr}\,\left[  \hat\sigma_{z}\hat\Delta\hat F \right]
.\label{S1}%
\end{align}
Finally, we assume that the equilibrium value of the order parameter
$\Delta_{0}(T)$ satisfies the standard BCS gap equation
\begin{align}
\frac{1}{\lambda}=\int_{0}^{\omega_{D}}d\xi_{p}\frac{\tanh\frac{\sqrt{\xi
_{p}^{2}+\Delta_{0}^{2}}}{2T}}{\sqrt{\xi_{p}^{2}+\Delta_{0}^{2}}},\label{BCS}%
\end{align}
where $\omega_{D}$ is the Debye frequency. In this case $iS_{1}$ is cancelled by
the first order contribution coming from $iS_{\Delta}$ and the action does not
contain the first order terms in $\delta\hat\Delta$ any more.

The Ward identities (\ref{Ward2}) and (\ref{Ward1}) also allow one to
transform the contribution $iS_{\mathrm{em}}$ (\ref{sem}) and cast it to the
form
\[
iS_{\mathrm{em}}=iS_{J}+iS_{L}+iS_{D},
\]
where the terms $iS_{J}$, $iS_{L}$ and $iS_{D}$ define the terms of a different
physical origin which we will identify respectively as Josephson, London and
Drude contributions to the effective action. They read
\begin{align}
iS_{J}= 2e^{2}\,\mathrm{Tr}\,\left[ \hat\sigma_{z}\hat\Phi\hat F\hat\sigma
_{z}\hat\Phi\hat{\bar F}\right] ,\label{SJ}%
\end{align}
\begin{eqnarray}
iS_{L} &=& -i\int d^{4}X\;\frac{\bm{H}_{F}^{2}-\bm{H}_{B}^{2}}{8\pi}
\nonumber\\ &&
+\,2\,\mathrm{Tr}\,\left[ \hat\sigma_{z}\{\nabla,\hat{\bm v}_{S}\}\hat F
\hat\sigma_{z}\{\nabla,\hat{\bm v}_{S}\}\hat{\bar F} \right] ,\label{SL}%
\end{eqnarray}
\begin{eqnarray}
&& iS_{D}   = i\int d^{4}X\;\frac{\bm{E}_{F}^{2}-\bm{E}_{B}^{2}}{8\pi}
\nonumber\\ &&
+\,\frac{1}{2}\,\mathrm{Tr}\,\bigg[\left(  \hat{\sigma}_{z}\{\nabla,\hat{\bm u}%
\}\hat{G}\right)  ^{2}+\left(  \hat{\sigma}_{z}\{\nabla,\hat{\bm u}\}\hat
{\bar{G}}\right)  ^{2}
\nonumber\\ &&
 -\,2\hat{\sigma}_{z}\{\nabla,\hat{\bm u}\}\hat{F}\hat{\sigma}_{z}%
\{\nabla,\hat{\bm u}\}\hat{\bar{F}}-4m\hat{\sigma}_{z}\hat{\bm u}^{2}(\hat
{G}-\hat{\bar{G}})\bigg],\;\;
\end{eqnarray}
where
\begin{align}
\hat{\bm u}=\frac{e}{2m}\left( \int_{t_{0}}^{t} dt^{\prime}\,\nabla\hat V
+\frac{1}{c}\hat{\bm A}\right) .\label{u}%
\end{align}
At low frequencies and temperatures the Josephson contribution $iS_{J}$ can be
large, thus suppressing fluctuations of the gauge invariant potential
$\hat{\Phi}$. In this case one can set $\hat{\Phi}=0$ and get $\hat{\dot{\phi
}}=2e\hat{V}$, which is just the well known Josephson relation between the
phase and the electric potential. Note that for ultra-thin superconducting
wires the Josephson relation can be violated, as it will be demonstrated below.

The London contribution $iS_{L}$ is responsible for the screening of the
magnetic field penetrating inside the superconductor. Finally, the Drude
contribution $iS_{D}$ remains non-zero in the normal state where it accounts
for Ohmic dissipation due to flowing electric currents. Since the correction
to the magnetic susceptibility in normal metals is usually small, one can
ignore the vector potential in the expression for $\hat{\bm u}$ (\ref{u}).
Afterwards one can again apply the Ward identities (\ref{Ward1}) and rewrite
$iS_{D}$ in the form
\begin{align}
iS_{D} & = i\int d^{4}X\;\frac{\bm{E}_{F}^{2} - \bm{E}_{B}^{2}}{8\pi}
\nonumber\\ &
-\,\frac{e^{2}}{2}\,\mathrm{Tr}\,\bigg[ \left( \hat\sigma_{z}\hat V\hat G\right)
^{2} + \left( \hat\sigma_{z}\hat V\hat{\bar G}\right) ^{2} + 2\hat\sigma
_{z}\hat V\hat F \hat\sigma_{z}\hat V\hat{\bar F} \bigg].\;\;\;\;\;\label{SD}%
\end{align}

At last, let us consider the cross term $iS_{\varphi\Delta}$. Again applying
the identity (\ref{Ward2}) we cast this term to the form similar to eq.
(\ref{S11}):
\begin{align}
iS_{\varphi\Delta} & = -\,\mathrm{Tr}\,\left[  \hat\sigma_{z}\hat\Delta\left(
\delta\hat F-\delta\hat{\bar F}\right)  \right] \nonumber\\
&  -\frac{i}{c}\int d^{4}X\,\delta{\bm j}_{0}({\bm A}_{F}-{\bm A}_{B}) -i\int
d^{4}X\, \delta\rho_{0}(V_{F}-V_{B}),\nonumber
\end{align}
where $\delta{\bm j},\delta\rho_{0}\propto\Delta_{F},\Delta_{B}$ define the
first order corrections to the current and the charge density due to
fluctuating order parameter fields. One can verify that in the absence of both
particle-hole asymmetry and charge imbalance these corrections vanish.
Likewise, in this case we have $\delta\hat F=\delta\hat{\bar F}$. Thus,
we conclude that
\begin{align}
iS_{\varphi\Delta}=0.
\end{align}

For clarity, we now summarize again the results derived this section. The
complete expression for the effective action describing quadratic fluctuations
in a superconductor reads
\begin{align}
iS_{\mathrm{eff}}^{(2)} = iS_{1}+iS_{\Delta}+iS_{J}+iS_{L}+iS_{D},
\end{align}
where the terms $iS_{1}$, $iS_{\Delta}$, $iS_{J}$, $iS_{L}$ and $iS_{D}$ are
defined respectively in eqs. (\ref{S1}), (\ref{SDelta}), (\ref{SJ}),
(\ref{SL}) and (\ref{SD}).

\subsection{Gaussian fluctuations in dirty superconductors}

Below we will mainly be interested in the limit of so-called dirty
superconductors, i.e. we assume that the concentration of defects in the
system is sufficiently high and the electron motion is diffusive. In the case
of ultra-thin superconducting channels only this limit appears to be of
practical interest, since usually the electron elastic mean free path $l$ does
not exceed the diameter of the wire. Since we will mainly focus our attention
on wires with diameters in the 10 nm range, realistic values of $l$ should be
typically in the same range, i.e. we have $l\ll\xi_{0}\sim\hbar v_{F}%
/\Delta_{0}\sim1$ $\mu$m and $l\ll\lambda_{L}$.

In order to account for processes with characteristic length scales exceeding
the electron mean free path $l$ it will be convenient for us to perform
disorder averaging directly in the effective action. To this end we substitute
explicit expressions for the Green functions (\ref{G}) into the effective
action derived in the previous section and then apply the standard rules of
averaging for the electron wave functions. In the diffusion approximation we
have
\begin{align}
& \sum_{mn} \left\langle \delta(\xi_{n}-\xi_{1})\delta(\xi_{m}-\xi_{2}) \chi
_{n}(\bm{r})\chi_{n}(\bm{r}^{\prime})\chi_{m}(\bm{r}^{\prime})\chi_{m}(\bm{r})
\right\rangle
\nonumber\\ &
=\frac{N_{0}}{\pi}\,\mathrm{Re}\, {\mathcal{D}}(\xi_{1}-\xi
_{2},\bm{r},\bm{r}^{\prime}),\nonumber
\end{align}
\begin{align}
&  \sum_{mn}\left\langle \delta(\xi_{n}-\xi_{1})\delta(\xi_{m}-\xi_{2}) \left(
\nabla^{\alpha}_{\bm{r}_{1}}-\nabla^{\alpha}_{\bm{r}_{4}}\right)  \left(
\nabla^{\beta}_{\bm{r}_{3}}-\nabla^{\beta}_{\bm{r}_{2}}\right)  \right.
\nonumber\\
&  \left.  \chi_{n}(\bm{r}_{1})\chi_{n}(\bm{r}_{2})\chi_{m}(\bm{r}_{3}%
)\chi_{m}(\bm{r}_{4}) \right\rangle \big|_{\bm{r}_{4}=\bm{r}_{1}%
=\bm{r},\;\bm{r}_{3}=\bm{r}_{4}=\bm{r}^{\prime}}\nonumber\\
&  =\,\frac{4m^{2}D}{\pi}\delta_{\alpha\beta}\, \mathrm{Re}\,\big[ i(\xi
_{1}-\xi_{2}){\mathcal{D}}(\xi_{1}-\xi_{2},\bm{r},\bm{r}^{\prime
})\big].\label{av}%
\end{align}
Here $D=v_{F}l/3$ is the diffusion constant and ${\mathcal{D}}(\omega
,\bm{r},\bm{r}^{\prime})$ is the diffuson defined as a solution of the
diffusion equation
\begin{align}
\left( -i\omega-D\nabla^{2}_{\bm r}\right) {\mathcal{D}}(\omega
,\bm{r},\bm{r}^{\prime})=\delta(\bm{r}-\bm{r}^{\prime}).
\end{align}
In the following we will mainly consider spatially extended systems
in which case one has
\begin{align}
{\mathcal{D}}(\omega,\bm{r},\bm{r}^{\prime})=\int\frac{d^{3}\bm{q}}{(2\pi
)^{3}}\, \frac{e^{i\bm{q}(\bm{r}-\bm{r}^{\prime})}}{-i\omega+Dq^{2}}.
\end{align}

Employing eqs. (\ref{av}) we arrive at the following expression for the
effective action
\begin{align}
  iS_{\mathrm{eff}}^{(2)}&= i\int d^{4}X\, \frac{\bm{E}^{-}\bm{E}^{+}%
-\bm{H}^{-}\bm{H}^{+}}{4\pi}
\nonumber\\ &
+\, i\int d^{4}Xd^{4}X^{\prime}\bigg\{ \Delta
^{-}(X) K_{\Delta}^{X,X^{\prime}} \Delta^{+}(X^{\prime})
\nonumber\\
&  +\, \Phi^{-}(X)K_{J}^{X,X^{\prime}}\Phi^{+}(X^{\prime}) - {\bm{v}}_{S}%
^{-}(X) K_{L}^{X,X^{\prime}}{\bm{ v}}_{S}^{+}(X^{\prime})
\nonumber\\ &
+\,{\bm{E}}%
^{-}(X)K_{D}^{X,X^{\prime}}{\bm{ E}}^{+}(X^{\prime}) \bigg\}
\nonumber\\
&  -\,\int d^{4}Xd^{4}X^{\prime}\bigg\{ \Delta^{-}(X) \tilde K_{\Delta
}^{X,X^{\prime}} \Delta^{-}(X^{\prime})
\nonumber\\ &
+\,\Phi^-(X)\tilde K_{J}^{X,X^{\prime}}\Phi
^{-}(X^{\prime})
- {\bm{v}}_{S}^{-}(X) \tilde K_{L}^{X,X^{\prime}}{\bm{ v}}_{S}%
^{-}(X^{\prime})
\nonumber\\ &
+\,{\bm{E}}^{-}(X)\tilde K_{D}^{X,X^{\prime}}{\bm{ E}}%
^{-}(X^{\prime}) \bigg\},\label{action}%
\end{align}
where we introduced "classical" $\Delta^{+}=(\Delta_{F}+\Delta_{B})/2$ and
"quantum" $\Delta^{-}=\Delta_{F}-\Delta_{B}$ components of the order parameter
field and used analogous definitions for other fluctuating variables ${\bm{
v}}_{S}^{\pm}$, $\Phi^{\pm}$ and ${\bm{E}}^{\pm}$. The four kernels
$K_{j}^{X,X^{\prime}}$ ($j=\Delta,L,J,D$) are defined as follows
\begin{align}
K_{j}^{X,X^{\prime}} & =\int\frac{d^{4}Q}{(2\pi)^{4}}\,e^{-i\omega
(t-t^{\prime})+i\bm{q}(\bm{r}-\bm{r}^{\prime})} \chi_{j}(Q),\nonumber\\
\tilde K_{j}^{X,X^{\prime}} & =\int\frac{d^{4}Q}{(2\pi)^{4}}\,e^{-i\omega
(t-t^{\prime})+i\bm{q}(\bm{r}-\bm{r}^{\prime})} \,\frac{\mathrm{Im}\,[\chi
_{j}(Q)]}{2}\coth\frac{\omega}{2T},
\end{align}
where we denote $d^{4}Q=d\omega d^{3}\bm{q}$.

Explicit expressions for the functions $\chi_{\Delta}$, $\chi_{J}$, $\chi_{L}$
and $\chi_{D}$ are rather cumbersome. They are presented in Appendix A.3
respectively in eqs. (\ref{chiDelta}), (\ref{chiJ}), (\ref{chiL}) and
(\ref{chiD}). Here we provide simple analytical expressions valid in some
limiting cases.

Let us first concentrate on the low temperature limit $T\ll\Delta_{0}$. In
this limit at small frequencies and wave vectors $\omega,Dq^{2}\ll\Delta_{0}$
we find
\begin{eqnarray}
\chi_{\Delta} &=&-2N_{0}\left(  1-\frac{\omega^{2}}{12\Delta_{0}^{2}}+\frac{\pi
Dq^{2}}{8\Delta_{0}}\right)  ,
\nonumber\\
\chi_{J}&=& 2e^{2}N_{0}\left(  1+\frac
{\omega^{2}}{6\Delta_{0}^{2}}-\frac{\pi Dq^{2}}{8\Delta_{0}}\right),
\nonumber\\
\chi_{L}&=&2\pi m^{2}N_{0}D\Delta_{0},
\nonumber\\
\chi_{D}&=&\frac{e^{2}N_{0}D}{4\Delta_{0}},
\label{chiT0}%
\end{eqnarray}
while in the limit $|\omega|,Dq^{2}\gg\Delta_{0}(0)\equiv\pi e^{-\gamma}T_{C}$
($\gamma\simeq0.577$ is the Euler constant) one finds
\begin{align}
\chi_{\Delta}  & =-2N_{0}\ln\frac{-i\omega+Dq^{2}}{\Delta_{0}(0)},
\nonumber\\
\chi_{J}  & =-\frac{8e^{2}N_{0}\Delta_{0}^{2}(T)}{\omega(\omega^{2}+D^{2}%
q^{4})}
\nonumber\\ &\times\,
\left[  \omega\ln\frac{-i\omega+Dq^{2}}{\Delta_{0}(0)}-iDq^{2}\ln
\frac{-i\omega}{2\Delta_{0}(0)}\right]  ,
\nonumber\\
\chi_{L}  & =\frac{8m^{2}N_{0}D\Delta_{0}^{2}(T)}{\omega^{2}+D^{2}q^{4}}
\nonumber\\ & \times\,
\left[  Dq^{2}\ln\frac{-i\omega+Dq^{2}}{\Delta_{0}(0)}+i\omega\ln
\frac{-i\omega}{2\Delta_{0}(0)}\right]  ,
\nonumber\\
\chi_{D}  & =\frac{\sigma}{-i\omega+Dq^{2}},\label{chiD1}%
\end{align}
where $\sigma=2e^{2}N_{0}D$ is the normal state Drude conductivity. Here we
explicitly indicated the temperature dependence of the superconducting
gap $\Delta_{0}(T)$ in order to emphasize that these asymptotic expressions are
valid at all temperatures rather than only in the limit $T\ll\Delta_{0} $.

Let us now consider higher temperatures $|T-T_{C}|\ll T_{C}$. At $T> T_{C}$
our general expression for $\chi_{\Delta}$ reduces to the standard result
\begin{align}
\chi_{\Delta} & = -2N_{0}\left[ \ln\frac{T}{T_{C}}+\Psi\left( \frac{1}{2}+
\frac{-i\omega+Dq^{2}}{4\pi T}\right)  -\Psi\left( \frac{1}{2}\right) \right]
,\label{Psi}%
\end{align}
$\chi_{D}$ is again defined by the Drude formula (\ref{chiD1}), while two
other $\chi$-functions vanish identically in this temperature interval,
$\chi_{J}=\chi_{L}=0$. The latter observation implies that phase fluctuations
remain unrestricted in this case. Hence, no Taylor expansion of the action in
the phase $\varphi$ can be performed. In this case it is more convenient to
undo the gauge
transformation restoring the initial dependence of the action on the complex
order parameter field and then to expand the action
in this field. For simplicity ignoring electromagnetic fields and expanding
the action to the second order in $\Delta^{\pm}$ we find
\begin{align}
&  iS_{\mathrm{eff}}^{(2)}= \frac{i}{2}\int d^{4}Xd^{4}X^{\prime}\left\{
\left( \Delta^{-}(X)\right) ^{*} K_{\Delta}^{X,X^{\prime}} \Delta
^{+}(X^{\prime}) \right. \nonumber\\
&  \left.  +\, \Delta^{-}(X) K_{\Delta}^{X,X^{\prime}} \left( \Delta
^{+}(X^{\prime})\right) ^{*}\right\}
\nonumber\\ &
-\,\int d^{4}Xd^{4}X^{\prime}\left(
\Delta^{-}(X)\right) ^{*} \tilde K_{\Delta}^{X,X^{\prime}} \Delta
^{-}(X^{\prime}).\label{action1}%
\end{align}
In the limit of small frequencies and wave vectors $|\omega|,Dq^{2}\ll2\pi
T_{C} $ one recovers the standard expression
\begin{align}
\chi_{\Delta}=-2N_{0}\ln\frac{T}{T_{C}}-\frac{\pi N_{0}}{4T}(-i\omega
+Dq^{2}),\label{limt}%
\end{align}
which usually serves as a starting point for the derivation of the so-called
time dependent Ginzburg-Landau equation (TDGL) which we will address shortly below.

Turning to temperatures below the critical one, $T<T_{C}$, and expanding the
$\chi$-functions (\ref{chiDelta})-(\ref{chiD}) in powers of $\Delta_{0}$ we
obtain
\begin{eqnarray}
\chi_{\Delta} & =& - \frac{7\zeta(3)}{2\pi^{2}}\frac{N_{0}\Delta_{0}^{2}}%
{T^{2}}
\nonumber\\ &&
-\,\frac{\pi N_{0}}{4T}(-i\omega+Dq^{2})-\frac{\pi N_{0}}{2T}%
\frac{\Delta_{0}^{2}}{-i\omega+Dq^{2}},
\nonumber\\
\chi_{J} & = &\frac{7\zeta(3)}{2\pi^{2}}\frac{e^{2}N_{0}\Delta_{0}^{2}}{T^{2}}
\nonumber\\ &&
+\,\frac{\pi e^{2}N_{0}\Delta_{0}^{2}}{T}\left( \frac{1}{-i\omega+Dq^{2}}
+\frac{\Delta_{0}^{2}}{\omega^{2}}\frac{-2i\omega+Dq^{2}}{\left(
-i\omega+Dq^{2}\right) ^{2}}\right) ,
\nonumber\\
\chi_{L} & = &\frac{\pi m^{2}N_{0}D\Delta_{0}^{2}}{T} \left( \frac
{-2i\omega+Dq^{2}}{-i\omega+Dq^{2}}-\frac{\Delta_{0}^{2}}{\left(
-i\omega+Dq^{2}\right) ^{2}}\right) ,\nonumber\\
\chi_{D} & =& \frac{2 e^{2}N_{0}D}{-i\omega+Dq^{2}}\left( 1-\frac{7\zeta
(3)}{2\pi^{2}}\frac{\Delta_{0}^{2}}{T^{2}}\right)
\nonumber\\ &&
-\,\frac{\pi e^{2}
N_{0}D\Delta_{0}^{2}}{2T}\frac{1}{(-i\omega+Dq^{2})^{2}}.\label{chis}%
\end{eqnarray}
These results apply for $\Delta_{0}\ll|\omega|,Dq^{2}$ and in the limit of
small wave vectors and frequencies, $|\omega|,Dq^{2}\ll2\pi T_{C}$. Here
$\Delta_{0}(T)$ obeys the standard BCS self-consistency gap equation at $T\sim
T_{C}$:
\begin{align}
\ln\frac{T_{C}}{T}=\frac{7\zeta(3)}{8\pi^{2}}\frac{\Delta_{0}^{2}}{T^{2}}+
\mathcal{O}\left(  \frac{\Delta_{0}^{4}}{T^{4}}\right) .
\end{align}

For $|\omega|,Dq^{2}\ll\Delta_{0}$ we obtain non-analytic expressions. For
example, $\chi_{J}$ in this limit reads
\begin{align}
& \chi_{J} = \frac{e^{2}N_{0}\Delta_{0}}{2T}\bigg\{  \frac{7\zeta(3)}{\pi^{2}%
}\frac{\Delta_{0}}{T}
\nonumber\\ &
+\,\frac{-i\omega}{\sqrt{\omega^{2}+D^{2}q^{4}}} \bigg[
\ln\frac{\sqrt{\omega^{2}+D^{2}q^{4}}+Dq^{2}}{\sqrt{\omega^{2}+D^{2}q^{4}%
}-Dq^{2}} +\pi i\frac{\omega}{|\omega|}\bigg] \bigg\}.
\end{align}

More accurate expressions for the kernels $\chi_{\Delta}$, $\chi_{J}$ and
$\chi_{L} $ valid at temeperatures close to $T_{C}$ and at any $|\omega|,
Dq^{2}\ll2\pi T_{C}$ are given in Appendix, see Eqs. (\ref{chiDeltaap1}%
-\ref{chiLap}). Only in the limit $|\omega|\ll Dq^{2}\ll\Delta_{0}$
those expressions match with the well known results for the
coefficients of the linearized (time-independent) Ginzburg-Landau equation:
\begin{align}
& \chi_{\Delta}= - \frac{7\zeta(3)}{2\pi^{2}}\frac{N_{0}\Delta_{0}^{2}}{T^{2}%
},\;\; \chi_{J}= \frac{7\zeta(3)}{2\pi^{2}}\frac{e^{2}N_{0}\Delta_{0}^{2}%
}{T^{2}},\;\;
\nonumber\\ &
\chi_{L} = \frac{\pi m^{2}N_{0}D\Delta_{0}^{2}}{T}.\label{GL}%
\end{align}
At frequencies $Dq^{2}\ll|\omega|\ll\Delta_{0}$ the functions
$\chi_{\Delta}$ and $\chi_{J}$ turn out to be parametrically different
taking much higher values:
\begin{align}
\chi_{\Delta}= -\frac{\pi N_{0}\Delta_{0}}{4T},\;\; \chi_{J}= \frac{\pi
e^{2}N_{0}\Delta_{0}}{4T},\label{GL1}%
\end{align}
while $\chi_{L}$ is still given by Eq. (\ref{GL}). The Drude susceptibility
$\chi_{D}$ may be taken in the usual form (\ref{chiD1}) in both cases. Thus,
already at small frequencies (well below the gap $\Delta_{0} \ll T_{C}$)
microscopic results can strongly deviate from those frequently used within
semi-phenomenological TDGL approach. At higher frequencies and/or wave vectors
$|\omega|+Dq^{2}\gg\Delta_{0}$ this difference becomes even more pronounced,
cf. eqs. (\ref{chis}).

One can demonstrate that these kernels are not independent and obey the
following exact identity
\begin{align}
\chi_{\Delta}= -\frac{\chi_{J}}{e^{2}} +\frac{\omega^{2}}{4e^{2}\Delta_{0}%
^{2}}\chi_{J} -\frac{q^{2}}{4m^{2}\Delta_{0}^{2}}\chi_{L},
\end{align}
which directly follows from the Ward identities (\ref{Ward1}). In addition, in
the diffusive limit the kernels $\chi_{J}$ and $\chi_{L}$ are related to each
other as
\begin{align}
\chi_{L}(\omega,\bm{q}) = \chi_{L}(\omega, 0) - \frac{m^{2}D^{2}q^{2}}{e^{2}%
}\chi_{J}(\omega,\bm{q}).
\end{align}
This latter relation applies only for dirty superconductors.

For clarity, it is worthwhile to display the relation between the $\chi
$-kernels derived here and some other quantities analyzed in the literature.
For example, one can introduce the complex conductivity $\sigma(Q)$ of a
superconductor \cite{MB}
\begin{align}
\bm{j}_{Q}=\sigma(Q)\bm{E}_{Q},
\end{align}
where $\bm{j}_{Q}$ and $\bm{E}_{Q}$ are the Fourier components of respectively
the current density and the electric field. In Ref. \cite{AG} the function
${\mathcal{Q}}(Q)$ was analyzed which expresses the current via the vector
potential:
\begin{align}
\bm{j}_{Q}=-{\mathcal{Q}}(Q)\bm{A}_{Q}.
\end{align}
Both ${\mathcal{Q}}(Q)$ and $\sigma(Q)$ are related to the kernels $\chi_{L}$
and $\chi_{D}$ as follows
\begin{align}
{\mathcal{Q}}(Q)=\frac{-i\omega}{c}\sigma(Q) =\frac{1}{c}\left[ \frac{e^{2}%
}{m^{2}}\chi_{L}(Q)-\omega^{2}\chi_{D}(Q)\right] .\label{Qsigma}%
\end{align}

\subsection{Langevin equations}

Let us now rewrite our results in a slightly different manner. The effective
action $S^{(2)}_{\mathrm{eff}}$ can be equivalently defined by means of the
following formula
\begin{align}
e^{iS^{(2)}_{\mathrm{eff}}}=\left\langle e^{iS_{\xi}[\Delta^{\pm},{\bm v}%
_{S}^{\pm},\Phi^{\pm}, {\bm E}^{\pm},{\bm H}^{\pm},\xi_{\Delta},\xi_{J}%
,{\bm \xi}_{L}]} \right\rangle _{\xi_{j}},\label{stoch}%
\end{align}
where
\begin{align}
  iS_{\xi} &= i\int d^{4}X\, \frac{\bm{E}^{-}\bm{E}^{+}-\bm{H}^{-}\bm{H}^{+}%
}{4\pi}
\nonumber\\ &
+\,i\int d^{4}Xd^{4}X^{\prime}\bigg\{ \Delta^{-}(X) K_{\Delta
}^{X,X^{\prime}} \Delta^{+}(X^{\prime})\nonumber\\
&  +\,\Phi^{-}(X)K_{J}^{X,X^{\prime}}\Phi^{+}(X^{\prime})+{\bm{E}}^{-}%
(X)K_{D}^{X,X^{\prime}}{\bm{ E}}^{+}(X^{\prime})
\nonumber\\ &
-\,\ {\bm{v}}_{S}^{-}(X)
K_{L}^{X,X^{\prime}}{\bm{ v}}_{S}^{+}(X^{\prime}) \bigg\}\nonumber\\
&  -\,i\int d^{4}X\,\big[ \xi_{\Delta}\Delta^{-} +\xi_{J}\Phi^{-} + {\bm\xi
}_{L}{\bm v}_{S}^{-} \big],\label{Sxi}%
\end{align}
and averaging is performed over three different stochastic variables
$\xi_{\Delta}$, $\xi_{J}$, ${\bm\xi}_{L}$ defined by the pair correlators
\begin{align}
\langle\xi_{\Delta}(t,{\bm r})\xi_{\Delta}(0,0)\rangle & = \int\frac{d^{4}%
Q}{(2\pi)^{4}}\,e^{-i\omega t+i{\bm qr}}\, \coth\frac{\omega}{2T}\,
\nonumber\\ &\times\,
\mathrm{Im}\,\big[\chi_{\Delta}(Q)\big],\nonumber\\
\langle\xi_{J}(t,{\bm r})\xi_{J}(0,0)\rangle & = \int\frac{d^{4}Q}{(2\pi)^{4}%
}\,e^{-i\omega t+i{\bm qr}}\, \coth\frac{\omega}{2T}\,
\nonumber\\ &\times\,
\mathrm{Im}%
\,\big[\chi_{J}(Q)+q^{2}\chi_{D}(Q)\big],
\nonumber\\
\langle\xi_{L}^{\alpha}(t,{\bm r})\xi_{L}^{\beta}(0,0)\rangle & =
\delta_{\alpha\beta}\int\frac{d^{4}Q}{(2\pi)^{4}}\,e^{-i\omega t+i{\bm qr}}\,
\coth\frac{\omega}{2T}\,\nonumber\\
& \times\, \mathrm{Im}\,\left[ -\chi_{L}(Q)+\frac{m^{2}\omega^{2}}{e^{2}}%
\chi_{D}(Q)\right] ,
\nonumber\\
\langle{\bm\xi}_{L}(t,{\bm r})\xi_{J}(0,0)\rangle & = -\int\frac{d^{4}Q}%
{(2\pi)^{4}}\,e^{-i\omega t+i{\bm qr}}\, \coth\frac{\omega}{2T}\,
\nonumber\\ &\times\,
\frac{m\omega{\bm q}}{e}
\, \mathrm{Im}\,\left[ \chi_{D}(Q)\right] .\label{correl}%
\end{align}
All other cross correlators of the above stochastic variables are equal to zero.

The representation (\ref{stoch}) is just the result of the standard
Hubbard-Stratonovich decoupling transformation in the effective action
(\ref{action}). We have also used the identity
\begin{align}
{\bm E}^{-}=-\nabla V^{-}-\frac{1}{c}\frac{\partial{\bm A}}{\partial t}
=-\nabla\Phi^{-} -\frac{m}{e} \frac{\partial{\bm v}_{S}}{\partial t}.
\end{align}

Let us now find the least action paths for $S_{\xi}$. Setting the variational
derivatives of the action (\ref{Sxi}) with respect to quantum fields
$\Delta^{-}$, $\varphi^{-}$, $V^{-}$ and ${\bm A}^{-}$ equal to zero we arrive
at four different equations for the fields $\Delta^{+}$, $\varphi^{+}$,
$V^{+}$ and ${\bm A}^{+}$ which provide the minimum for the action $S_{\xi}$
(\ref{Sxi}). The first equation describes fluctuations of the absolute value
of the order parameter. It reads
\begin{align}
\int d^{4}X^{\prime}\,K_{\Delta}^{X,X^{\prime}}\;\Delta^{+}(X^{\prime}%
)=\xi_{\Delta}(X).\label{EqDelta}%
\end{align}
The second one is the continuity equation for the supercurrent.
We obtain
\begin{align}
\frac{\partial\rho_{S}}{\partial t}+\nabla\bm{j}_{S} = -\frac{\partial\xi_{J}%
}{\partial t}+\frac{e}{m}\nabla{\bm \xi}_{L},\label{Eqjs}%
\end{align}
where we introduced the superconducting density $\rho_{S}$ and the
superconducting current density $\bm{j}_{S}$
\begin{align}
\rho_{S} &=-\int d^{4}X^{\prime}\,K_{J}^{X,X^{\prime}}\Phi^{+}(X^{\prime}),
\nonumber\\
{\bm j}_{S} &=-\frac{e}{m}\int d^{4}X^{\prime}\,K_{L}^{X,X^{\prime}%
}{\bm v}_{S}^{+}(X^{\prime}).\label{rhos_js}%
\end{align}
The remaining two saddle point equations take the form
\begin{align}
\frac{\nabla{\bm E}^{+}(X)}{4\pi} +\int d^{4}X^{\prime}\,K_{D}^{X,X^{\prime}%
}\;\nabla{\bm E}^{+}(X^{\prime}) -\rho_{S}(X)=\xi_{J}(X)\label{3rd}%
\end{align}
and
\begin{align}
\frac{\nabla\times{\bm H}^{+}}{4\pi}=\frac{1}{4\pi c}\frac{\partial{\bm E}%
^{+}}{\partial t} +\frac{\bm{j}_{S}}{c}+\frac{\bm{j}_{N}}{c}-\frac{e}%
{m}{\bm \xi}_{L}.\label{maks}%
\end{align}
Here
\begin{align}
{\bm j}_{N}=\int dX^{\prime}\,K_{D}^{X,X^{\prime}}\frac{\partial{\bm E}%
^{+}(X^{\prime})}{\partial t^{\prime}}\label{jn}%
\end{align}
is the normal quasiparticle current.

Eqs. (\ref{EqDelta}), (\ref{Eqjs}), (\ref{3rd}) and (\ref{maks}) together with
noise correlators (\ref{correl}) represent the set of \textit{Langevin
equations} fully describing quantum dynamics of the order parameter and
electromagnetic fields for dirty superconductors within the Gaussian
approximation. As it is clear from our derivation, these equations remain
valid provided the electron distribution function is not driven far from
equilibrium. Generalization of our approach to non-equilibrium situations is
also possible but will not be discussed here.

\subsection{Time dependent Ginzburg-Landau equation}

Now let us establish the relation between our results and the approach based
on the so-called time dependent Ginzburg-Landau equation (TDGL) which is
widely used to model various non-stationary effects in superconductors at
temperatures close to $T_{C}$. For example, above the critical temperature
this TDGL approach allows to correctly evaluate the so-called Aslamazov-Larkin
fluctuation correction to the conductivity of the system. Below $T_{C}$ it
enables one to describe formation of phase slip centers and the resistive
state of current biased superconducting wires. Relative simplicity of the TDGL
approach makes it possible to apply powerful numerical methods thus making
this technique particularly appealing. In many cases the TDGL-based analysis
was employed even far beyond its applicability range, e.g., in order to
describe quantum phase slips in superconducting nanowires at $T \to0$.

The TDGL equation is usually written in the following simple form \cite{LV}
\begin{align}
& \bigg[ \frac{\partial}{\partial t}-2ieV+\frac{1}{\tau_{GL}}-D\left(
\nabla+i\frac{2e}{c}\bm{A}\right) ^{2}
\nonumber\\ &
+\frac{7\zeta(3)}{\pi^{3}T}|\Delta
|^{2}\bigg] \Delta=\tilde\xi_{\Delta},\label{TDGL1}%
\end{align}
where
\begin{equation}
\tau_{GL}=\frac{\pi}{8|T-T_{C}|}\label{tauGL}%
\end{equation}
is the so-called Ginzburg-Landau time and
\begin{align}
\langle\tilde\xi_{\Delta}^{*}(t,\bm{r})\tilde\xi_{\Delta}(t^{\prime
},\bm{r}^{\prime})\rangle= \frac{16 T^{2}}{\pi N_{0}}\delta(t-t^{\prime}%
)\delta(\bm{r}-\bm{r}^{\prime}).
\end{align}
Although this form can be justified for gapless
superconductors at high concentration of magnetic impurities, in general no
consistent microscopic derivation of eq. (\ref{TDGL1}) was ever performed.
Nevertheless, it is usually believed that eq. (\ref{TDGL1}) is microscopically
justified at temperatures above $T_{C}$ where the average value of the BCS
order parameter is zero, $\Delta_{0}=0$. Unfortunately, our present
microscopic derivation (as well as earlier imaginary time analysis
\cite{GZ01,ogzb}) does not fully support this statement. An independent
analysis based on the real-time non-linear $\sigma$-model was recently
performed by Levchenko and Kamenev \cite{LK} who also noticed that even at
$T>T_{C}$ eq. (\ref{TDGL1}) is not quite correct. These authors formulated a
more accurate real-time TDGL equation in the form convenient for a comparison
with eq. (\ref{TDGL1}):
\begin{align}
&  \bigg[ \frac{\partial}{\partial t}-2ie\frac{\partial{\mathcal{K}}%
(X)}{\partial t} +\frac{1}{\tau_{GL}}-D\left( \nabla+i\frac{2e}{c}%
\bm{A}(X)\right) ^{2}
\nonumber\\ &
+\,\frac{7\zeta(3)}{\pi^{3}T}|\Delta(X)|^{2}\bigg]
\Delta(X)  - \frac{7\zeta(3)}{\pi^{3}T} \Delta(X) \int d^{4}X^{\prime}{\mathcal{D}%
}(X,X^{\prime})
\nonumber\\ & \times\,
\left(  \Delta^{*}(X^{\prime})\frac{\partial\Delta(X^{\prime}%
)}{\partial t^{\prime}} -2ei |\Delta(X^{\prime2 }\frac{\partial{\mathcal{K}%
}(X^{\prime})}{\partial t^{\prime}}\right)
  =\tilde\xi_{\Delta}(X),\label{TDGL2}%
\end{align}
where ${\mathcal{K}}(X)=\int dX^{\prime}{\mathcal{D}(X,X^{\prime}%
)}\big[ V(X^{\prime}) - D\nabla{\bm A}(X^{\prime})\big]$.

It is obvious that eq. (\ref{TDGL2}) does not in general coincide with eq.
(\ref{TDGL1}). At the same time, it is satisfactory to observe that eq.
(\ref{TDGL2}) agrees with our results up to terms $\sim\Delta^{+},\varphi^{+}%
$. In order to demonstrate this fact it is necessary to identify
$\Delta=(\Delta_{0}+\Delta^{+})e^{i\varphi^{+}}$ and consider
terms linear in $\Delta^{+}$ and $\varphi^{+}$. In this way one
arrives at our Langevin eqs. (\ref{EqDelta})-(\ref{Eqjs}) with
$\Delta_{0}=$const, where the functions $\chi_{\Delta}$,
$\chi_{J}$, $\chi_{L}$ coincide with those given by Eq.
(\ref{chis}) in the leading order in $\Delta_{0}$. Thus, we
conclude that in the limit $\Delta_{0} \to0$ our Langevin
equations are equivalent to TDGL-type of equation \cite{LK} within
the order $\sim\Delta_{0}^{2}$. Some differences, however, arise
for higher order terms, namely for terms $\sim\Delta_{0}^{4}$
originating from the functions $\chi_{J}$ and $\chi_{L}$ as well
as for terms $\sim\Delta_{0}^{2}$ emerging from $\chi_{\Delta}$.
This observation indicates that eq. (\ref{TDGL2}) is still not
fully justified at temperatures $T<T_{C}$.

In fact, it is not quite clear to us whether it would be of any practical
importance to pursue the GL expansion up to terms $\sim\Delta^{3}$ in the
TDGL-type equations. Of course, a regular expansion in powers of $\Delta^{\pm
}$ can be performed in the initial effective action (\ref{Seff}). In the order
$\sim\Delta^{4}$ this expansion generates many complicated non-local (both in
space and in time) terms containing the quantum field $(\Delta^{-})^{n}$ with
$n$ ranging from 1 to 4. In order to recover terms $\sim\Delta^{3}$ in the
TDGL equation one should disregard all terms in the action $\sim(\Delta
^{-})^{n}$ with $n\geq2$. In certain situations this approximation might be
difficult to justify. In addition, the remaining terms $\sim\Delta^{3}$ are
hardly tractable except in the zero frequency limit. Finally, the whole
approach remains restricted to temperatures $T\sim T_{C}$. In view of all
these problems it appears more appealing to perform the expansion of the
effective action in superconducting fluctuations around the mean field value
$\Delta_{0}$. This strategy was pursued in the bulk of this chapter.
Restricting these expansion to second order terms in $\Delta^{\pm}$ we arrive
at the Langevin equations (\ref{EqDelta})-(\ref{Eqjs}) with $\chi_{\Delta}$,
$\chi_{J}$ and $\chi_{L}$ defined in Eqs. (\ref{chiDelta}-\ref{chiL}). This
approach remains applicable at all temperatures down to $T=0$ and is
sufficient for practical calculations in a large number of situations.

\section{Thin metallic wires}

We now turn to the specific case of sufficiently long and very thin
superconducting wires which will be of particular interest for us here. For
such systems the terms describing the action of free electro-magnetic field
can be rewritten in the form
\begin{eqnarray}
&& i\int d^{4}X\,\frac{\bm{E}^{-}\bm{E}^{+}-\bm{H}^{-}\bm{H}^{+}}{4\pi}
\nonumber\\ &&
\rightarrow\, i\int dtdx\,\left(  CV^{+}V^{-}-\frac{A^{+}A^{-}}{L}\right)  .
\end{eqnarray}
Here we have defined the coordinate along the wire $x$, the capacitance per
unit length of the wire $C$ and the inductance times unit length $L$. $A$
stands for the component of the vector potential parallel to the wire. For a
cylindric wire with radius $r_{0}$ embedded in a dielectric environment with
susceptibility $\epsilon$, the capacitance $C$ and inductance $L$ are
\begin{equation}
C\approx\frac{\epsilon}{2\ln(R_{0}/r_{0})},\;\;\;L\approx2\ln(R_{0}/r_{0}),
\end{equation}
where $R_{0}$ is the distance from the center of the wire and the bulk
metallic electrode.

In order to transform other terms one should apply a simple rule
\begin{align}
\int d^{4}X \to s\int dt\,dx,\;\;\; \int\frac{d^{4}Q}{(2\pi)^{4}}\to\frac
{1}{s}\int\frac{d\omega dq}{(2\pi)^{2}},
\end{align}
where $s$ is the wire cross section.

\subsection{Propagating modes}

In the low temperature limit $T\ll\Delta_{0}$ all $\chi$-functions
(\ref{chiT0}) are real and, hence, the noise terms in all four Langevin
equations (\ref{EqDelta}-\ref{maks}) vanish. This enables propagation of
electromagnetic modes along a quasi-1D superconducting wire. The equations of
motion for such a wire take the form
\begin{align}
\frac{1}{12\Delta_{0}^{2}}\frac{\partial^{2}\Delta^{+}}{\partial t^{2}}
-\frac{\pi D}{8\Delta_{0}}\frac{\partial^{2}\Delta^{+}}{\partial x^{2}}
+\Delta^{+} & =0
\nonumber\\
\frac{\partial^{2}\varphi^{+}}{\partial t^{2}}-2e\frac{\partial V^{+}%
}{\partial t} -\pi D\Delta_{0}\left( \frac{\partial^{2}\varphi^{+}}{\partial
x^{2}}+ \frac{2e}{c}\frac{\partial A^{+}}{\partial x}\right)  & =0
\nonumber\\
CV^{+} -\frac{\sigma s}{8\Delta_{0}}\frac{\partial^{2} V^{+}}{\partial x^{2}}
+2e^{2}N_{0}s\left( V^{+} -\frac{1}{2e}\frac{\partial\varphi^{+}}{\partial
t}\right)  & =0
\nonumber\\
\frac{A^{+}}{L}+\frac{\sigma s}{8\Delta_{0}c} \left( \frac{\partial^{2}
V}{\partial t\partial x} +\frac{1}{c}\frac{\partial^{2} A^{+}}{\partial t^{2}
}\right)  &
\nonumber\\
+\,\frac{\pi\sigma\Delta_{0} s}{2ec}\left(  \frac{\partial\varphi^{+}
}{\partial x}+\frac{2e}{c}A^{+} \right)  & =0.\nonumber
\end{align}
For dirty metallic wires with the diameter of the order of superconducting
coherence length $\xi=\sqrt{D/\Delta_{0}}$ one finds $1/L\gg\pi\sigma
\Delta_{0}s/c^{2}$ and $C\ll2e^{2}N_{0}s$. In this case the last equation
gives $A^{+}\to0$, while the second and third equations describe the
propagation of the plasmon Mooij-Sch\"on mode \cite{ms} with dispersion
\begin{align}
\omega=c_{0}q,
\end{align}
where the velocity of this mode $c_{0}$ is
\begin{align}
c_{0}\simeq\frac{1}{\sqrt{\tilde LC}}=\sqrt{\frac{\pi\sigma\Delta_{0}s}{C}%
}\label{c0}%
\end{align}
and $\tilde L= 4\pi\lambda_{L}^{2}/s=1/2\pi e^{2}N_{0}\Delta_{0}Ds$ is the
kinetic inductance of a superconducting wire.

\subsection{Gaussian fluctuations of the order parameter}

The effective action (\ref{action}) fully accounts for Gaussian fluctuations
in diffusive superconducting structures. For instance, from Eqs.
(\ref{EqDelta}-\ref{maks}) one readily establishes the correlation functions
for all fluctuating variables in our problem. For the order
parameter fields $\Delta^{\pm}$ in a quasi-1D wire we have
\begin{align}
\langle\Delta^{+}(t,r)\Delta^{+}(0,0)\rangle & = -\frac{1}{s}\int\frac{d\omega\,
dq}{(2\pi)^{2}}\,e^{-i\omega t+iqr}
\nonumber\\ &\times\,
\mathrm{Im}\,\left[ \frac{1}%
{\chi_{\Delta}(Q)}\right] \coth\frac{\omega}{2T}.\label{corr}%
\end{align}
Correlation functions ${\bm{ v}}_{S}^{\pm}$, $\Phi^{\pm}$ and ${\bm{E}}^{\pm}$
are defined analogously via the corresponding kernels $\chi_{J}$, $\chi_{L}$
and $\chi_{D}$.

Consider Gaussian fluctuations of the order parameter in thin one-dimensional
wires. The simplest possible average $\langle\Delta^{+}(0,0)\Delta
^{+}(0,0)\rangle/\Delta_{0}^{2}$ is divergent since the function $\chi
_{\Delta}$ grows very slowly at large $\omega$ and $\bm{q}$. Let us define and
analyze a slightly different object
\begin{align}
{\mathcal{R}}= \frac{\left\langle \Delta^{+}\left( 1/\Delta_{0},\xi\right)
\;\Delta^{+}(0,0)\right\rangle } {\Delta_{0}^{2}}.
\end{align}
One can verify that, for example, the non-local kernel $K_{L}^{X,X^{\prime}}$
significantly decays as long as $|t|$ exceeds $1/\Delta_{0}$ and $|\bm{r}|$
becomes bigger than the coherence length $\xi$. Therefore, the parameter
$\mathcal{R}$ provides a qualitative measure of the ratio between the
fluctuation correction to the current and its mean field value.

In order to estimate the parameter $\mathcal{R}$ we note that at low
temperatures $T\ll\Delta_{0}$ the kernel $\chi_{\Delta}$ (\ref{chiDelta}) can
be expressed in the form
\[
\chi_{\Delta}=-N_{0}F_{0}\left(  \omega/\Delta_{0},\xi q\right)  ,
\]
where $F(x,y)$ is a certain dimensionless function. Then we obtain
\[
\mathcal{R}\approx\frac{\alpha_{0}}{sN_{0}\sqrt{D\Delta_{0}}},
\]
where
\[
\alpha_{0}=\frac{1}{2\pi^{2}}\int_{0}^{1}dx\int_{-1}^{1}dy\,\mathrm{Im}%
\,\left(  \frac{1}{F_{0}(x,y)}\right)
\]
is a numerical prefactor. Below we will demonstrate that the probability for
quantum tunneling of the order parameter field $\Delta$ in superconducting
nanowires is proportional to $\sim\exp[-S_{QPS}]$, where $S_{QPS}%
\sim1/\mathcal{R}$ is the action of single quantum phase slips
(QPS). Hence, for $\mathcal{R}\ll1$ Gaussian fluctuations are
small and QPS events are rare, which are important pre-conditions
for the BCS mean field theory. On the other hand, at
$\mathcal{R}\gtrsim1$ one enters the regime of strong non-Gaussian
fluctuations which fully suppress the mean field order parameter
thus driving the wire to a normal state. The concept of QPS also
becomes ill-defined in this regime of strong quantum fluctuations.

The same conclusions can be extracted from the so-called Ginzburg-Levanyuk
criterion. Let us consider the Ginzburg number $Gi$ defined as the value
$(T_{C}-T)/T_{C}$ at which the fluctuation correction to the specific heat
becomes equal to the specific heat jump at the phase transition point. In the
case of quasi-1D wires this number reads \cite{LV}:
\begin{align}
Gi_{\mathrm{1D}}=\frac{1.3}{(p_{F}^{2}s)^{2/3}(T_{C}\tau_{e})^{1/3}%
},\label{Gi}%
\end{align}
where $\tau_{e}=l/v_{F}$ is the elastic mean free time. Typically in thick
wires one finds $Gi_{1D}\ll1$ and fluctuations become strong only in a very
narrow region close to $T_{C}$, i.e. at $|T_{C}-T|/T_{C} < Gi_{\mathrm{1D}}$.
One can also rewrite eq. (\ref{Gi}) as
\begin{align}
Gi_{\mathrm{1D}}=\frac{0.15}{(sN_{0}\sqrt{D\Delta_{0}})^{2/3}} \sim
\mathcal{R}^{2/3}\sim\frac{1}{S_{QPS}^{2/3}},\label{Gi2}%
\end{align}
or simply $Gi_{\mathrm{1D}} \sim1/g_{\xi}^{2/3}$, where $g_{\xi}$ is the
dimensionless conductance of the wire segment of length $\xi$. Thus, for
$g_{\xi} \sim1$ the width of the fluctuation region $\delta T$
is comparable to $T_{C}$
and the BCS mean field approach becomes obsolete down to $T=0$.


\subsection{Matsubara effective action}

To complete our analysis we will briefly address the imaginary time
(Matsubara) version of the effective action. Technically it is more convenient
to deal with this form of the action provided one needs to account for quantum
tunneling processes. This is precisely what we will do below when we describe
quantum phase slips in superconducting nanowires. The calculation is described
in details in Refs. \cite{ogzb,GZ01} and is completely analogous to one
carried out above in real time.

Our starting point is the path integral representation of the grand partition
function
\begin{align}
{\mathcal{Z}}=\int{\mathcal{D}}\Delta\,{\mathcal{D}}V\,{\mathcal{D}}{\bm A}\;
e^{-S_{E}},
\end{align}
where $S_{E}$ is the Euclidean version of the effective action, the
fluctuating order parameter field $\Delta$ as well as scalar and vector
potentials $V$ and ${\bm A}$ depend on coordinate $x$ along the wire and
imaginary time $\tau$ restricted to the interval $0\leq\tau\leq\beta\equiv
1/T$. As before, assuming that deviations of the amplitude of the order
parameter field from its equilibrium value $\Delta_{0}$ are relatively small
we expand the effective action in powers of $\delta\Delta(x,\tau
)=\Delta(x,\tau)-\Delta_{0}$ and in the electromagnetic fields up to the
second order terms. The next step is to average over the random potential of
impurities. After that the effective action becomes translationally invariant
both in space and in time. Performing the Fourier transformation we obtain
\cite{ogzb,GZ01}
\begin{align}
S_{E} & =\frac{s}{2}\int\frac{d\omega dq}{(2\pi)^{2}}\left\{  \frac{|A|^{2}%
}{Ls}+\frac{C|V|^{2}}{s}+ \tilde\chi_{D}\left|  qV+\frac{\omega}{c}A\right|
^{2} \right.
\nonumber\\
&  \left. +\,\tilde\chi_{J}\left|  V +\frac{i\omega}{2e}\varphi\right| ^{2}
+\frac{\tilde\chi_{L}}{4m^{2}}\left| iq\varphi+\frac{2e}{c} A
\right| ^{2} + \tilde\chi_{\Delta}|\delta\Delta|^{2} \right\} .\label{a105}%
\end{align}
The functions $\tilde\chi_{\Delta}$, $\tilde\chi_{J}$, $\tilde\chi_{L}$ and
$\tilde\chi_{D}$, which depend both on the frequencies and the wave vectors,
are expressed in terms of the averaged products of the Matsubara Green
functions \cite{ogzb,GZ01}. These functions represent the imaginary time
version of the analogous real time functions $\chi_{j}$ already encountered
above. In order to recover the expressions for $\tilde\chi_{j} (\omega, k)$
one just needs to substitute $-i\omega\to|\omega|$ in eqs. (\ref{chiDelta}%
)-(\ref{chiD}) for $\chi_{j} (\omega, k)$. The action $S_{E}$ (\ref{a105})
represents the imaginary time analogue of the real time effective action
(\ref{action}).

Note that the action $S_{E}$ (\ref{a105}) is quadratic both in the voltage $V$
and the vector potential $A$. Hence, these variables can be integrated out
exactly. Performing this integration one arrives at the effective action which
only depends on $\varphi$ and $\delta\Delta$. We obtain
\begin{equation}
S=\frac{s}{2}\int\frac{d\omega dq}{(2\pi)^{2}}\left\{ {\mathcal{F}}%
(\omega,q)|\varphi|^{2}+ \tilde\chi_{\Delta}|\delta\Delta|^{2}\right\}
.\label{a116}%
\end{equation}
The general expression for ${\mathcal{F}}(\omega,q)$ and the saddle point
relations between the electromagnetic potentials and the fluctuating phase
$\varphi$ are presented in Appendix A.4 (eqs. (\ref{calF})-(\ref{Aphi})) for completeness.

As we already discussed, usually the wire geometric inductance remains
unimportant. Therefore here and below we put $L=0$. Then eqs. (\ref{calF}%
)-(\ref{Aphi}) get simplified and read
\begin{equation}
\mathcal{F}(\omega,q)= \frac{\left( \frac{\tilde\chi_{J}}{4e^{2}}\omega
^{2}+\frac{\tilde\chi_{L}}{4m^{2}}q^{2} \right)  \left( \frac{C}{s}+\tilde
\chi_{D} q^{2} \right) + \frac{\tilde\chi_{J}\tilde\chi_{L}}{4m^{2}}q^{2}}
{\frac{C}{s}+\tilde\chi_{J}+\tilde\chi_{D} q^{2}} ,\label{a106}%
\end{equation}
and
\begin{equation}
V=\frac{\tilde\chi_{J}} {\frac{C}{s}+\tilde\chi_{J}+\tilde\chi_{E} q^{2} }
\left( \frac{-i\omega}{2e}\varphi\right) , \quad A=0.\label{noJ}%
\end{equation}
Note that according to eq. (\ref{noJ}) the Josephson relation $V=\dot
\varphi/2e$ is in general \textit{not} satisfied. This relation may
approximately hold only in the limit $\tilde\chi_{J}\gg C/s+\tilde\chi
_{D}q^{2}$. Making use of the results presented in Appendix one easily
observes that in the important limit of small elastic mean free paths $l$ the
latter condition is obeyed only at low frequencies and wave vectors
$\omega/\Delta_{0}\ll1$ and $Dq^{2}/\Delta_{0}\ll1$.

Let us now perform yet one more approximation and expand the action in powers
of $\omega$ and $q^{2}$. Keeping the terms of the order $q^{4}$ and
$\omega^{2}q^{2}$ we find
\begin{align}
& S_{E}=\frac{s}{2}\int\frac{d\omega dq}{(2\pi)^{2}}
\bigg\{
\bigg( \frac{C}
{s}\omega^{2} +\pi\sigma\Delta_{0} q^{2}+ \frac{\pi^{2}}{8}\sigma Dq^{4}
\nonumber\\ &
+\,\frac{\pi\sigma}{8\Delta_{0}}\omega^{2} q^{2} \bigg)  \left| \frac{\varphi}{2e}\right| ^{2}
+2N_{0}\bigg( 1+\frac{\omega^{2}}{12\Delta^{2}_{0}}+\frac{\pi Dq^{2}%
}{8\Delta_{0}} \bigg) |\delta\Delta|^{2} \bigg\} .
\label{action11}%
\end{align}
The term $\propto\omega^{4}$ turns out to be equal to zero. At even smaller
wave vectors, $Dq^{2}/2\Delta\ll2C/\pi e^{2}N_{0}s \ll1,$ we get
\begin{align}
S_{E} & =\frac{1}{2}\int\frac{d\omega dq}{(2\pi)^{2}} \left\{  \left(
{C}\omega^{2}+\pi\sigma\Delta_{0} s q^{2}\right)  \left| \frac{\varphi}%
{2e}\right| ^{2} + s\tilde\chi_{A} |\delta\Delta|^{2}\right\}
.\label{longwave}%
\end{align}
Here, as before, we have assumed $C/2e^{2}N_{0}s\ll1$. The first two terms in
this action correspond to the effective Hamiltonian of the form
\[
\int dx [(\partial_{\tau}\varphi)^{2}/2+(\partial_{x}\varphi)^{2}/2]
\]
which again defines the Mooij-Sch\"on plasma modes propagating along the wire
with the velocity $c_{0}$ (\ref{c0}).

The effective action (\ref{a116}) allows to directly evaluate the fluctuation
correction to the order parameter in superconducting nanowires. Performing
Gaussian integration over both $\varphi$ and $\delta\Delta$ we arrive at the
wire free energy
\begin{align}
F=F_{BCS} -\frac{T}{2}\sum_{\omega,q}\left[  \ln\frac{\lambda{\mathcal{F}%
}(\omega,q)}{2N_{0}\Delta_{0}^{2}} + \ln\frac{\lambda\tilde\chi_{\Delta
}(\omega,q)}{2N_{0}} \right] ,
\end{align}
where $F_{BCS}$ is the standard BCS free energy. The order
parameter is defined by the saddle point equation $\partial
F/\partial\Delta=0$ and can be written in the form
$\Delta=\Delta_{0}-\delta\Delta_{0}$, where $\Delta_{0}$
is the solution of the BCS self-consistency equation $\partial F_{BCS}%
/\partial\Delta_{0}=0$ (\ref{BCS}) and the fluctuation correction
$\delta\Delta_{0}$ has the form
\begin{align}
\delta\Delta_{0}&=-\frac{T}{2}\left(
\frac{\partial^{2}F_{BCS}}{\partial \Delta_{0}^{2}}\right) ^{-1}
\nonumber\\ &\times\,
\frac{\partial}{\partial\Delta_{0}}\sum _{\omega,q} \left[
\ln\frac{\lambda{\mathcal{F}}(\omega,q)}{2N_{0}\Delta _{0}^{2}} +
\ln\frac{\lambda\tilde\chi_{\Delta}(\omega,q)}{2N_{0}} \right]
.\label{dDelta}%
\end{align}
Making use of the above expressions for the functions ${\mathcal{F}}%
(\omega,q)$ and $\tilde\chi_{\Delta}(\omega,q)$ and having in mind that for a
wire of length $X$ one has $\partial F/\partial\Delta_{0}=2N_{0}sX$, at $T
\to0$ we obtain
\begin{align}
\frac{\delta\Delta_{0}}{\Delta_{0}}\sim \frac{1}{g_{\xi}} \sim
 Gi_{\mathrm{1D}}^{3/2}.\label{estim22}%
\end{align}
In eq. (\ref{estim22}) fluctuations of both the phase and the
absolute value of the order parameter give contributions of the
same order. The estimate (\ref{estim22}) again demonstrates that
at low temperatures suppression of the order parameter in
superconducting nanowires due to Gaussian fluctuations remains
weak as long as $g_{\xi} \gg1$ and it becomes important only for
extremely thin wires with $Gi_{\mathrm{1D}} \sim1$.

Finally let us return to the action (\ref{a105}) which we will use to
illustrate a deficiency of the TDGL approach in the imaginary time.
Considering the superconducting part of the action only and assuming that
temperature is close to $T_{C}$ we can identify $\delta\Delta$ with $\Delta$
and set $\Delta_{0}=0$ in all the $\tilde\chi$-kernels. Exactly as in the
real-time approach one then has $\tilde\chi_{J}=\tilde\chi_{L}=0$ and the
phase fluctuations become unrestricted. For this reason one should again undo
the gauge transformation and return to the complex order parameter field.

For simplicity let us ignore both the scalar and the vector potentials. The
TDGL action for the wire is then usually written in the form
\begin{eqnarray}
&& S_{\mathrm{TDGL}}  =  N_{0}Ts\sum_{\omega_{n}}
\int dx \frac{\pi|\omega_{n}|}{8T}|\Delta|^{2}
+ N_{0}s\int d\tau dx
\nonumber\\&&\times\,
\left( \frac{\pi D}{8T}
\left| \frac{\partial\Delta}{\partial x}\right| ^{2}
+ \frac{T-T_{C}}{T_{C}} |\Delta|^{2}+\mathcal{O}(|\Delta|^{4})\right).
\label{Ssgl}
\end{eqnarray}
This form can be obtained from the action (\ref{a105}) by formally expanding
the kernel $\tilde\chi_{\Delta}$ in Matsubara frequencies and wave vectors
$\omega_{\mu},Dq^{2} \ll4\pi T$, cf. eqs. (\ref{Psi}) and (\ref{limt}). Note,
however, that since the validity of the GL expansion is restricted to
temperatures $T\sim T_{C}$, the Matsubara frequencies $\omega_{n}=2\pi nT$ are
never really smaller than $4\pi T$. Hence, the expansion $\Psi(1/2+|\omega
_{n}|/4\pi T)-\Psi(1/2)\to\pi\omega_{n}/8T$ which yields (\ref{Ssgl}) is never
correct except in the stationary case $\omega_{n}=0$. Already these simple
arguments illustrate the failure of the TDGL action (\ref{Ssgl}) in the
Matsubara technique. Further problems with this TDGL approach arise in the
presence of the electromagnetic potentials $V$ and $A$. We refer the reader to
the paper \cite{ogzb} for the corresponding analysis.

\section{Thermally activated phase slips}

As we already discussed, sufficiently thin superconducting wires can acquire
non-zero resistance even below $T_{C}$ due to fluctuations of the
superconducting order parameter. In this section we will address thermal
fluctuations which are particularly important in the immediate vicinity of
$T_{C}$.

The theory of thermally activated phase slips (TAPS) was developed by Langer
and Ambegaokar \cite{la} and then completed by McCumber and Halperin
\cite{mh}. Here we will briefly review this LAMH theory with minor
modifications related to the fact that the TDGL-based approach is not
sufficiently accurate to correctly determine the pre-exponent in the
expression for the TAPS rate.

This rate $\gamma_{TAPS}$ is defined by the standard activation dependence
\begin{align}
\gamma_{TAPS}=Be^{-\delta F/T}\label{Gamma111}%
\end{align}
where $\delta F$ is the free energy difference which determines an effective
potential barrier which the system should overcome in order to create a phase slip.

\subsection{Activation exponent}

In order to evaluate $\delta F$ we make use of the general expression for the
Ginzburg-Landau free energy functional
\begin{align}
F[\Delta(x)]&= sN_{0}\int dx \bigg(  \frac{\pi D}{8T}\left| \frac
{\partial\Delta}{\partial x}\right| ^{2}
\nonumber\\ &
+\, \frac{T-T_{C}}{T_{C}} |\Delta|^{2}
+ \frac{7\zeta(3)}{16\pi^{2}T^{2}}|\Delta|^{4} \bigg) .\label{FGL}%
\end{align}
The saddle point paths for this functional are determined by the standard GL
equation
\begin{align}
-\frac{\pi D}{8T}\frac{\partial^{2}\Delta}{\partial x^{2}} + \frac{T-T_{C}%
}{T_{C}} \Delta+ \frac{7\zeta(3)}{8\pi^{2}T^{2}}|\Delta|^{2}\Delta=0.
\end{align}
In the case of a long quasi-1D wire this equation has two solutions, a trivial
one
\begin{align}
\Delta= \Delta_{0}\equiv\sqrt{\frac{8\pi^{2}T(T_{C}-T)}{7\zeta(3)}%
},\label{Delta0}%
\end{align}
providing the minimum for the free energy and a metastable one
\begin{align}
\Delta_{M}(x)=\Delta_{0}\tanh\left(  \sqrt{\frac{4(T_{C}-T)}{\pi D}} \,
x\right) .\label{DeltaM}%
\end{align}
The potential barrier $\delta F$ in eq. (\ref{Gamma111}) is set by the
difference
\begin{align}
\delta F=F[\Delta_{M}(x)]-F[\Delta_{0}]=\frac{16\pi^{2}}{21\zeta(3)}
sN_{0}\sqrt{\pi D}(T_{C}-T)^{3/2}.\label{dF}%
\end{align}
Note that this result applies as long as the transport current $I$ across the
wire is sufficiently small. With increasing $I$ the height of the potential
barrier $\delta F$ decreases and finally vanishes as $I$ approaches the
critical (depairing) current of the wire. The corresponding expression for
$\delta F(I)$ can be found in Refs. \cite{la,mh}. Recently a microscopic
calculation of $\delta F$ in the case of a clean single channel
superconducting wire was reported in Ref. \cite{ZhLKV}.

\subsection{Pre-exponent}

Now let us turn to the pre-exponent $B$ in the expression for the TAPS rate
(\ref{Gamma111}). In order to evaluate $B$ one should go beyond the stationary
free energy functional (\ref{FGL}) and include time-dependent fluctuations of
the order parameter field $\Delta(x,\tau)$. In Ref. \cite{mh} this task was
accomplished within the framework of a TDGL-based analysis. Employing TDGL
equation it is possible to re-formulate the problem in terms of the
corresponding Fokker-Planck equation \cite{Langer} which can be conveniently
solved for the problem in question. Since the important time scale within the
TDGL approach is the Ginzburg-Landau time $\tau_{GL}$ (\ref{tauGL}), this time
also enters the expression for the pre-exponent $B$ derived in \cite{mh}.

Unfortunately, as it was demonstrated, e.g., in Chapters 2 and 3,
the TDGL approach fails below $T_{C}$. Hence, one should employ a
more accurate effective action analysis. The microscopic effective
action for superconducting wires is rather complicated and it
cannot be easily reduced to any Fokker-Planck-type of equation.
For this reason, below we will take a somewhat different route
\cite{GZ08} and combine our effective action approach with the
well known general formula for the decay rate of a metastable
state (see, e.g., \cite{sz,weiss})
\begin{equation}
\gamma(T)=2\mathrm{Im}F(T).\label{ImF}%
\end{equation}
As our effective action does not contain the parameter $\tau_{GL}$ we expect
that our final result for the pre-exponent will not contain this parameter
either.

Following the standard procedure \cite{ABC} we expand the general expression
for the effective action around both saddle point solutions (\ref{Delta0}) and
(\ref{DeltaM}) up to quadratic terms in both the phase $\varphi$ and
$\delta\Delta$. Neglecting the contributions from fluctuating electromagnetic
fields we obtain
\begin{equation}
S_{0/M}=F[\Delta_{0/M}]+ \delta^{2} S_{0/M},
\end{equation}
where
\begin{align}
\delta^{2} S_{0} & = \frac{sT}{2} \sum_{\omega_{n}}\int dxdx^{\prime
}\;
\nonumber\\ &\times\,
\big[\delta\Delta(\omega_{n},x)\tilde\chi_{\Delta}^{(0)}(|\omega
_{n}|;x-x^{\prime}) \delta\Delta(\omega_{n},x^{\prime})\nonumber\\
&  +\, \varphi(\omega_{n},x)k_{\varphi}^{(0)}(|\omega_{n}|;x-x^{\prime
})\varphi(\omega_{n},x^{\prime}) \big],\nonumber\\
\delta^{2} S_{M} & = \frac{sT}{2} \sum_{\omega_{n}}\int dxdx^{\prime
}\;
\nonumber\\ &\times\,
\big[\delta\Delta(\omega_{n},x)\tilde\chi_{\Delta}^{(M)}(|\omega
_{n}|;x,x^{\prime}) \delta\Delta(\omega_{n},x^{\prime})\nonumber\\
&  +\, \varphi(\omega_{n},x)k_{\varphi}^{(M)}(|\omega_{n}|;x,x^{\prime
})\varphi(\omega_{n},x^{\prime}) \big].
\end{align}
Here $\omega_{n}=2\pi Tn$ are Bose Matsubara frequencies. The functions $\tilde
\chi_{\Delta}^{(0)}$ and $k_{\varphi}^{(0)}$ are expressed in terms of the
kernels $\tilde\chi_{\Delta}$, $\tilde\chi_{J}$ and $\tilde\chi_{L}$ as
follows:
\begin{align}
\tilde\chi_{\Delta}^{(0)}(|\omega_{n}|;x-x^{\prime}) & =\int\frac{dq}{2\pi}\,
e^{iq(x-x^{\prime})}\,\tilde\chi_{\Delta}(\omega_{n},q),\nonumber\\
k_{\varphi}^{(0)}(|\omega_{n}|;x-x^{\prime})  & = \int\frac{dq}{2\pi}\,
e^{iq(x-x^{\prime})}\,
\nonumber\\ &\times\,
\left(  \frac{\omega_{n}^{2}}{4e^{2}}\tilde\chi
_{J}(\omega_{n},q) +\frac{q^{2}}{4m^{2}}\tilde\chi_{L}(\omega_{n},q) \right) .
\end{align}
The functions $\tilde\chi_{\Delta}^{(M)}$ and $k_{\varphi}^{(M)}$ describe the
fluctuations around the coordinate dependent metastable state $\Delta_{M}(x)$,
and, therefore, cannot be straitforwardly related to $\tilde\chi_{\Delta}$,
$\tilde\chi_{J}$ and $\tilde\chi_{L}$. Fortunately, the explicit form of
$\tilde\chi_{\Delta}^{(M)}$ and $k_{\varphi}^{(M)}$ is not important for us here.

The pre-exponent $B$ in eq. (\ref{Gamma111}) is obtained by integrating over
fluctuations $\delta\Delta$ in the expression for the grand partition. One
arrives at a formally diverging expression which signals decay of a
metastable state. After a proper analytic continuation one arrives at the
decay rate in the form (\ref{Gamma111}) with
\begin{align}
B=2T\,\mathrm{Im}\, \prod_{\omega_{n}} \sqrt{\frac{\det\tilde\chi_{\Delta
}^{(0)}(\omega_{n})\, \det k_{\varphi}^{(0)}(\omega_{n})} {\det\tilde
\chi_{\Delta}^{(M)}(\omega_{n})\, \det k_{\varphi}^{(M)}(\omega_{n})}%
}\label{B1}%
\end{align}
Here it is necessary to take an imaginary part since one of the eigenvalues of
the operator $k_{M}(0)$ is negative.

The key point is to observe that at $T \sim T_{C}$ all Matsubara frequencies
$|\omega_{n}|=2\pi T|n|$ -- except for one with $n=0$ --
strongly exceed the order
parameter, $|\omega_{n}|\gg\Delta_{0}(T)$. Hence, for all such values the
function $\chi_{\Delta}(i|\omega_{n}|)$ approaches the asymptotic form
(\ref{Psi}) which is not sensitive to superconductivity and we have
$\det\tilde\chi_{\Delta}^{(0)}(\omega_{n}) \simeq\det\tilde\chi_{\Delta}%
^{(M)}(\omega_{n}) $ and $\det k_{\varphi}^{(0)}(\omega_{n}) \simeq\det
k_{\varphi}^{(M)}(\omega_{n})$. The corresponding determinants in eq.
(\ref{B1}) cancel out and only the contribution from $\omega_{n}=0$ remains.
It yields
\begin{align}
B\simeq2T\,\mathrm{Im}\, \sqrt{\frac{\det\tilde\chi_{\Delta}^{(0)}(0)\, \det
k_{\varphi}^{(0)}(0)} {\det\tilde\chi_{\Delta}^{(M)}(0)\, \det k_{\varphi
}^{(M)}(0)}}.\label{B2}%
\end{align}
The ratio of these determinants can be evaluated with the aid of the GL free
energy functional (\ref{FGL}) with the result \cite{mh}
\begin{align}
\mathrm{Im}\, \sqrt{\frac{\det\tilde\chi_{\Delta}^{(0)}(0)\, \det k_{\varphi
}^{(0)}(0)} {\det\tilde\chi_{\Delta}^{(M)}(0)\, \det k_{\varphi}^{(M)}(0)}}
=\frac{2\sqrt{3}}{\sqrt{\pi}}\frac{X}{\xi(T)}\sqrt{\frac{\delta F}{T}},
\end{align}
where $\delta F$ is the free energy barrier (\ref{dF}), $X$ is the wire length
and $\xi(T)=\sqrt{\pi D/4(T_{C}-T)}$ is the superconducting coherence length
in the vicinity of $T_{C}$.

Combining all the above results we arrive at the TAPS rate
\begin{align}
\gamma_{TAPS} =\frac{4\sqrt{3}}{\sqrt{\pi}} T\frac{X}{\xi(T)}\sqrt
{\frac{\delta F}{T}} \exp\left[ -\frac{\delta F}{T}\right] .\label{Gamma222}%
\end{align}
As we expected, this result (\ref{Gamma222}) does not contain the
Ginzburg-Landau time $\tau_{GL}$ and exceeds the corresponding expression
\cite{mh} by the factor $\sim T\tau_{GL}$.

Eq. (\ref{Gamma222}) is applicable at $T_{C}-T \ll T_{C}$ and as long as
$\delta F\gg T$. Combining these two inequalities with eq. (\ref{dF}) we
arrive at the condition
\begin{equation}
Gi_{\mathrm{1D}}\ll\frac{T_{C}-T}{T_{C}} \ll1,\label{abe}%
\end{equation}
where the Ginzburg number $Gi_{\mathrm{1D}}$ is defined in eq. (\ref{Gi}). The
double inequality (\ref{abe}) is standard for the GL theory. Obviously, it
also restricts the applicability range of the LAMH theory.

\subsection{Temperature-dependent resistance and noise}

Every phase slip event implies changing of the superconducting phase in time
in such a way that the total phase difference values along the wire before and
after this event differ by $\pm2\pi$. Since the average voltage is
linked to the
time derivative of the phase by means of the Josephson relation, $\langle
V\rangle=\langle\dot{\varphi}/2e\rangle$, for the net voltage drop across the
wire we obtain
\begin{equation}
V=\frac{\pi}{e}\left[  \Gamma_{2\pi}(I)-\Gamma_{-2\pi}(I)\right]  ,
\end{equation}
where $\Gamma_{\pm2\pi}$ are the TAPS rates corresponding to the phase changes
by $\pm2\pi$. In the absence of any bias current $I\rightarrow0$ both rates
are equal $\Gamma_{\pm2\pi}=\gamma_{TAPS}$ and the net voltage drop $V$
vanishes. In the presence of a non-zero bias current the symmetry between
these two rates is lifted since -- in complete analogy to the case of
Josephson junctions (cf., e.g., \cite{sz}) -- the potential barrier for these
two processes differ. As long as the bias current $I$ is sufficiently small,
we obtain
\begin{equation}
\Gamma_{\pm2\pi}(I)=\gamma_{TAPS}e^{\pm\pi I/2eT}.
\end{equation}
Thus, at such values of $I$ and at temperatures slightly below $T_{C}$ the
$I-V$ curve for quasi-1D superconducting wires takes the form
\begin{equation}
V=\frac{\pi}{e}\gamma_{TAPS}\,\sinh\frac{\pi I}{2eT},
\label{IVTAPS}
\end{equation}
with $\gamma_{TAPS}$ defined in eq. (\ref{Gamma222}). This important result
\cite{la,mh} implies that thermal fluctuations effectively destroy long range
phase coherence in the system and the wire acquires non-vanishing resistance
$R=V/I$ even below $T_{C}$. This resistance demonstrates strong (exponential)
dependence on temperature and the wire cross section
\begin{equation}
R(T)\propto\exp(-\delta F(T)/T),\label{TAPS R(T)}%
\end{equation}
leading to effective fluctuation-induced broadening of the superconducting
phase transition which can be detected experimentally. The corresponding
discussion is presented in Chapter 6.

To complete our description of thermal fluctuations in superconducting wires
we point out that in addition to non-zero resistance (\ref{TAPS R(T)}) TAPS
also cause the voltage noise below $T_{C}$. Treating TAPS as independent
events one immediately concludes that they should obey Poissonian statistics.
Hence, the voltage noise power $S_{V} = \langle VV \rangle$ is proportional to
the TAPS rate $\gamma_{TAPS}$. The contributions from TAPS changing the phase
by $\pm2\pi$ add up and we obtain
\begin{align}
S_{V}=\frac{4\pi^{2}}{e^{2}}\;\gamma_{TAPS}\;\cosh\frac{\pi I}{2eT}.
\end{align}
Similarly to the wire resistance the voltage noise rapidly decreases as one
lowers the temperature away from $T_{C}$. Only in the vicinity of the critical
temperature this noise remains appreciable and can be detected in experiments.

\section{Theory of quantum phase slips in superconducting nanowires}

As temperature goes down thermal fluctuations decrease and, hence,
TAPS become progressively less important and eventually die out in
the limit $T \to0$. At low enough temperatures \textit{quantum
fluctuations} of the order parameter field $\Delta$ take over and
essentially determine the behavior of ultra-thin superconducting
wires. As we have already discussed, the most important quantum
fluctuations in such wires are \textit{Quantum Phase Slips} (QPS).
Each QPS event involves suppression of the order parameter in the
phase slip core and a winding of the superconducting phase around
this core. This configuration describes quantum tunneling of the
order parameter field through an effective potential barrier and
can be conveniently described within the imaginary time formalism.
Below we will elaborate on the microscopic theory of quantum phase
slips in superconducting nanowires. In doing so, to a large extent
we will follow the papers \cite{ZGOZ,ZGOZ2,GZ01}.

\subsection{QPS action}

Let us denote the typical size of the QPS core as $x_{0}$ and the typical
(imaginary time) duration of the QPS event as $\tau_{0}$. At this stage both
these parameters are not yet known and remain to be determined from our
subsequent analysis. It is instructive to separate the total action of a
single QPS $S_{QPS}$ into a core part $S_{\mathrm{core}}$ around the phase
slip center for which the condensation energy and dissipation by normal
currents are important (scales $x\leq x_{0}$, $\tau\leq\tau_{0}$), and a
hydrodynamic part outside the core $S_{\mathrm{out}}$ which depends on the
hydrodynamics of the electromagnetic fields, i.e.
\begin{equation}
S_{QPS}=S_{\mathrm{core}}+S_{\mathrm{out}}.\label{QPSac}%
\end{equation}
Let us first evaluate the hydrodynamic part $S_{\mathrm{out}}$. This task is
simplified by the fact that outside the core the absolute value of the order
parameter field remains equal to its mean field value $\Delta_{0}$, and only
its phase $\varphi(x,\tau)$ changes in space and time. Without loss of
generality we can assume that the absolute value of the order parameter is
equal to zero at $\tau=0$ and $x=0$. For sufficiently long wires and outside
the QPS core the saddle point solution corresponding to a single QPS event
should satisfy the identity
\begin{equation}
\partial_{x}\partial_{\tau}\tilde{\varphi}-\partial_{\tau}\partial_{x}%
\tilde{\varphi}=2\pi\delta(\tau,x)
\end{equation}
which follows from the fact that after a wind around the QPS center the phase
should change by $2\pi$. In a way QPS is just a vortex in space-time with the
phase distribution $\varphi(x,\tau)$ described by the saddle point solution
\begin{equation}
\tilde{\varphi}(x,\tau)=-\arctan(x/c_{0}\tau).\label{QPSphi}%
\end{equation}
Substitutung the solution (\ref{QPSphi}) into the action (\ref{action11}) we
obtain
\begin{equation}
S_{\mathrm{out}}=\mu\ln[\min(c_{0}\beta,X)/\max(c_{0}\tau_{0},x_{0})]\;,
\label{hydr}
\end{equation}
where the parameter
\begin{equation}
\mu=\frac{\pi}{4e^{2}c_{0}(L+\tilde{L})}\simeq\frac{\pi}{4\alpha}\sqrt
{\frac{sC}{4\pi\lambda_{L}^{2}}}%
\end{equation}
sets the scale for the the hydrodynamic contribution to the QPS action. Here
and below $\alpha=e^{2}/\hbar c\simeq1/137$ is the fine structure constant. We
also note that at $T\rightarrow0$ the contribution $S_{\mathrm{out}}$
(\ref{hydr}) diverges logarithmically for infinitely long wires thus making
\textit{single} QPS events unlikely in this limit.

Let us now turn to the core contribution to the action of
a single QPS. In order to exactly evaluate this contribution it is necessary
to explicitly find the QPS saddle point of the full non-linear effective
action. This is a formidable task which can hardly be accomplished in
practice. On the other hand, this task is greatly simplified if one is aiming
at estimating the term $S_{\mathrm{core}}$ up to a numerical prefactor of
order one. Below we will recover the full microscopic expression for the core
contribution $S_{\mathrm{core}}$ leaving only this numerical prefactor
undetermined. In this way we fully capture all essential physics of QPS. The
dimensionless prefactor can be regarded as a
fit parameter which can be extracted, e.g., from the comparison with available
experimental data.

The above strategy allows us to approximate the complex order parameter field
inside the QPS core by two simple functions which should satisfy several
requirements. The absolute value of the order parameter $|\Delta(x,\tau)|$
should vanish at $x=0$ and $\tau=0$ and coincide with the mean field value
$\Delta_{0}$ outside the QPS core. The phase $\varphi(x,\tau)$ should flip at
$x=0$ and $\tau=0$ in a way to provide the change of the net phase difference
across the wire by $2\pi$. On top of that, in a short wire and outside the QPS
core the phase $\varphi$ should not depend on the spatial coordinate in the
zero bias limit. All sufficiently smooth functions obeying these
requirements can be used to estimate $S_{\rm core}$. For concreteness, in
what follows we will choose
\begin{equation}
|\delta\Delta(x,\tau)|=\Delta_{0} \exp(-x^{2}/2x_{0}^{2}-\tau^{2}/2\tau
_{0}^{2}) .\label{Deltaqps}%
\end{equation}
for the amplitude of the order parameter field and
\begin{equation}
\varphi(x,\tau)=-\frac{\pi}{2}\tanh\left( \frac{x\tau_{0}}{x_{0}\tau}\right)
.\label{phiqps2}%
\end{equation}
for its phase. Rewriting the action (\ref{action11}) in the space-time domain
\begin{align}
S & =\frac{s}{2}\int dx\, d\tau\bigg\{  \frac{C}{4e^{2}s} \left(
\frac{\partial\varphi}{\partial\tau} \right) ^{2}
+\frac{\pi N_{0}D\Delta_{0}}{2} \left( \frac{\partial\varphi}{\partial x}\right) ^{2}
\nonumber\\ &
+\, \frac{\pi\sigma}{32e^{2}\Delta_{0}}\left( \frac{\partial^{2} \varphi}{\partial x\partial\tau}
\right) ^{2} \bigg\}
+ sN_{0}\int dx\, d\tau
\nonumber\\ &\times\,
\bigg\{  \delta\Delta^{2}+\frac{1}{12\Delta_{0}
^{2}}\left( \frac{\partial\delta\Delta}{\partial\tau} \right) ^{2} + \frac{\pi
D}{8\Delta_{0}} \left( \frac{\partial\delta\Delta}{\partial x} \right) ^{2}
\bigg\} .\label{action21}%
\end{align}
(where we dropped unimportant terms $\propto(Dq^{2})^{2}$) and substituting
the trial functions (\ref{Deltaqps}), (\ref{phiqps2}) into the action
(\ref{action21}) one finds
\begin{align}
S(x_{0},\tau_{0}) & =\left[  a_{1}\frac{C}{e^{2}} + a_{2} sN_{0}\right]
\frac{x_{0}}{\tau_{0}}+ a_{3} sN_{0}D\Delta_{0}\frac{\tau_{0}}{x_{0}%
}\nonumber\\
& +\, a_{4}\frac{s\sigma}{2e^{2}\Delta_{0}}\frac{1}{x_{0}\tau_{0}} + a_{5}
sN_{0}\Delta^{2}_{0} x_{0}\tau_{0}+ a_{6} \frac{\tilde C}{e^{2}\tau_{0}%
},\label{action1111}%
\end{align}
where $a_{j}$ are numerical factors of order one which depend on the precise
form of the trial functions, $\tilde C=CX$ is the total capacitance of the
wire and $X$ is the wire length. Note that fictitious divergencies emerging
from a singular behavior of the function (\ref{phiqps2}) at $x=x_{0}$ and
$\tau=\tau_{0}$ are eliminated since the order parameter vanishes in this
space-time point.

Let us first disregard capacitive effects neglecting the last term in eq.
(\ref{action1111}). Minimizing the remaining action with respect to the core
parameters $x_{0}$, $\tau_{0}$ and making use of the inequality
$C/e^{2}N_{0}s\ll1$, we obtain
\begin{equation}
x_{0}^{4}=\frac{a_{3}a_{4}}{a_{2}a_{5}}\frac{\sigma D}{2e^{2}N_{0}\Delta
_{0}^{2}},\;\;\;\;\;\;\tau_{0}^{4}=\frac{a_{2}a_{4}}{a_{3}a_{5}}\frac{\sigma
}{2e^{2}N_{0}D\Delta_{0}^{4}}.\label{xtau0}%
\end{equation}
These values provide the minimum for the QPS action, and we find
\begin{equation}
S_{\mathrm{core}}=2sN_{0}(\sqrt{a_{2}a_{3}D\Delta_{0}}+\sqrt{a_{4}a_{5}%
\sigma\Delta_{0}/2e^{2}N_{0}}).\label{S321}%
\end{equation}
Substituting the Drude expression for the normal conductance of our wire
$\sigma=2e^{2}N_{0}D$ into eqs. (\ref{xtau0}) and (\ref{S321}) we arrive at
the final results for the core parameters
\begin{equation}
x_{0}=\left(  \frac{a_{3}a_{4}}{a_{2}a_{5}}\right)  ^{1/4}\sqrt{\frac
{D}{\Delta_{0}}},\;\;\;\;\;\;\tau_{0}=\left(  \frac{a_{2}a_{4}}{a_{3}a_{5}%
}\right)  ^{1/4}\frac{1}{\Delta_{0}}\label{xtau}%
\end{equation}
and for the core action
\begin{equation}
S_{\mathrm{core}}=\pi AN_{0}s\sqrt{D\Delta_{0}}=A\frac{R_{q}}{R_{N}}\frac
{X}{\xi}=\frac{A}{4}g_{\xi}.\label{otvet}%
\end{equation}
Here $A=2(\sqrt{a_{2}a_{3}}+\sqrt{a_{4}a_{5}})/\pi$ is the numerical prefactor
$R_{N}$ is the total normal state wire resistance, $R_{q}=\pi\hbar
/2e^{2}=6.453$ k$\Omega$ is the ``superconducting'' resistance quantum,
$\xi=\sqrt{D/\Delta_{0}}$ is the superconducting coherence length and
$g_{\xi}=4(R_{q}/R_{N})(X/\xi)$
is the dimensionless normal conductance of a wire segment of length $\xi$.

As it was already pointed out, the results (\ref{xtau}) and (\ref{otvet}) hold
provided the capacitive effects are small. This is the case for relatively
short wires
\begin{equation}
X \ll\xi\frac{e^{2}N_{0}s}{C}.\label{short}%
\end{equation}
In the opposite limit the same minimization procedure of the action
(\ref{action1111}) yields
\begin{equation}
x_{0} \sim\xi,\;\;\;\;\; \Delta\tau_{0} \sim\sqrt{XC/\xi e^{2}N_{0}s}
\gg1.\label{cap}%
\end{equation}
The QPS core action then takes a somewhat more complicated form
\begin{equation}
S_{\mathrm{core}}=A^{\prime}\frac{R_{q}}{R_{N}}\left( \frac{X}{\xi}\right)
^{3/2}\sqrt{\frac{C}{e^{2}N_{0}s}},\label{otvet2}%
\end{equation}
where $A^{\prime}$ is again a numerical prefactor. Eqs. (\ref{QPSac}),
(\ref{hydr}), (\ref{otvet}) and (\ref{otvet2}) provide complete information
about the action for single QPS in diffusive superconducting nanowires.

Let us analyze the above expressions. Introducing the number of conducting
channels in the wire ${\mathcal{N}}=p_{F}^{2}s/4\pi$, setting $C\sim1$ and
making use of the condition $e^{2}/\hbar v_{F}\sim1$ satisfied for typical
metals, one can rewrite the inequality (\ref{short}) in a very simple form
\begin{equation}
X\ll\xi{\mathcal{N}}.\label{short2}%
\end{equation}
To give an idea about the relevant length scales, for typical values $\xi
\sim10$ nm and $N\sim10^{2}\div10^{3}$ -- according to eq. (\ref{short2}) --
the wire can be considered short provided its length does not exceed $1\div10$
$\mu$m. This condition is satisfied in a number of experiments, e.g., in Refs.
\cite{BT,Lau MoGe PRL,Bezryadin MoGe review JPCM 2008,Zgirski NanoLett 2005,Zgirski QPS PRB 2008}. On the other hand,
in experiments \cite{Giordano QPS PRL 1989,Giordano QPS PRB 1991,Giordano Physica B 1994,Altomare Al nanowire PRL 2006} much
longer wires with lengths up to $X\sim100$ $\mu$m were studied. Apparently
such samples are effectively in the long wire regime $X\gg\xi{\mathcal{N}}$.

In the case of short wires we observe a clear separation between different
fluctuation effects contributing to the QPS action: Fluctuations of the order
parameter field and dissipative currents determine the core part (\ref{otvet})
while electromagnetic fluctuations are responsible for the hydrodynamic term
(\ref{hydr}). In the case of longer wires capacitive effects also contribute
to the core part (\ref{otvet2}). We also observe that in the short wire limit
different contributions to the QPS action depend differently on the wire
thickness (or the number of conducting channels ${\mathcal{N}}$): The core
part (\ref{otvet}) decreases linearly with the wire cross section,
$S_{\mathrm{core}}\propto{\mathcal{N}}$, whereas the hydrodynamic contribution
shows a weaker dependence $\mu\propto\sqrt{{\mathcal{N}}}$. In the long wire
limit the dependence of the core part (\ref{otvet2}) on ${\mathcal{N}}$ also
becomes weaker, $S_{\mathrm{core}}\propto\sqrt{{\mathcal{N}}}$, due to
capacitive effects.

Yet another important observation is that in the interesting range of wires
thicknesses $r_{0}\equiv\sqrt{s} \gtrsim5\div10$ nm the core part
$S_{\mathrm{core}}$ usually exceeds the hydrodynamic term $\sim\mu$. E.g. for
$C \sim1$ we obtain $\mu\approx30 (r_{0}/\lambda_{L})$. Setting $r_{0}
\sim5\div10$ nm and estimating $\mu$ and $S_{\mathrm{core}}$ for typical
system parameters $p^{-1}_{F} \sim0.2$ nm, $l\sim7$ nm, $\xi\sim10$ nm,
$\lambda_{L}\sim100$ nm, $v_{F}=10^{6}$ m/s and $\Delta_{0} \sim1\div10$ K we
find $\mu\sim1 \div3$ and $S_{\mathrm{core}} \gg\mu$. The latter inequality
becomes even stronger for thicker wires. Note that the condition
$S_{\mathrm{core}} \gg\mu$ allowed us to ignore the hydrodynamic part of the
QPS action while minimizing the core part with respect to $x_{0}$ and
$\tau_{0}$.

At the first sight the result (\ref{otvet}) derived in the short
wire limit could create an illusion that our microscopic
description would not be needed in order to recover the correct
form of the core part $S_{\mathrm{core}}$. Indeed, the same form
could be guessed, e.g., from an oversimplified TDGL-based approach
or just from the ``condensation energy'' term (proportional to
$\chi_{A}$) without taking into account dissipative effects. For
instance, minimization of the contribution
$\sim|\delta\Delta|^{2}$ (the last three terms in eq.
(\ref{action21})) is formally sufficient to arrive at the correct
estimate $S_{\mathrm{core}} \sim N_{0}s\sqrt{D\Delta_{0}}$. The
same equation demonstrates, however, that not only the amplitude
but also the phase fluctuations of the order parameter field
provide important contributions to the QPS action. If the latter
fluctuations were taken into account \textit{without} including
dissipative effects (this would correspond to formally setting
$\sigma\to0$ in eq. (\ref{action21})) minimization of the core
action would immediately yield the meaningless result $x_{0}
\to0$, $\tau_{0} \to0$ (cf., eq. (\ref{xtau0})) implying that the
hydrodynamic contribution $S_{\mathrm{out}}$ could not be
neglected in that case. Minimization of the total action $S_{QPS}$
would then yield the estimate for the action parametrically
different from that of eq. (\ref{otvet}). On the other hand, in
the strong damping limit (approached by formally setting
$\sigma\to\infty$ in eq. (\ref{action21})) the core size would
become very large and the action $S_{\mathrm{core}}$ (\ref{S321})
would diverge meaning that no QPS would be possible at all. These
observations clearly illustrate crucial importance of dissipative
effects. Under the condition $C/e^{2}N_{0}s \ll1$ (usually well
satisfied in metallic nanowires) dissipation plays a dominant role
during the phase slip event, and the correct QPS core action
\textit{cannot} be obtained without an adequate microscopic
description of dissipative currents flowing inside the wire. Only
employing the Drude formula for the wire conductivity
$\sigma=2e^{2}N_{0}D$ enables one to recover the correct result
(\ref{otvet}) whereas for some other models of dissipation
different results for the core action would follow, cf., e.g.,
Ref. \cite{ZGOZ}.

To complete our discussion of the QPS action let us recall that in
the course of our derivation we employed two approximations: (i)
we expanded the action up to the second order in
$\delta\Delta(x,\tau)=\Delta(x,\tau)-\Delta_{0}$ and (ii) in eq.
(\ref{action11}) we expanded the action (\ref{a116}) in powers of
$\omega /\Delta_{0}$ and $Dq^{2}/\Delta_{0}$. The approximation
(i) is sufficient everywhere except inside the QPS core where
$\Delta(x,\tau)$ is small. In these space- and time-restricted
regions one can expand already in $\Delta(x,\tau)$ and again
arrive at eq. (\ref{a116}) with $\delta\Delta
(x,\tau)\to\Delta(x,\tau)$ and with all the $\chi$-functions
defined in Appendix A3 with $\Delta_{0} \equiv0$. Both expansions
match smoothly at the scale of the core size $x_{0} \sim\xi$,
$\tau_{0} \sim1/\Delta_{0}$. Hence, the approximation (i) is
sufficient to derive the correct QPS action up to a numerical
prefactor of order one.

The approximation (ii) is sufficient within the same accuracy. One can
actually avoid this approximation and substitute the trial functions
(\ref{Deltaqps}), (\ref{phiqps2}) directly into the action (\ref{a116}) .
Neglecting capacitive effects in the limit (\ref{short}) one can rewrite the
QPS action as a function of the dimensionless parameters $x_{0}/\xi$ and
$\Delta_{0} \tau_{0}$ only. Making use of the general results for the $\chi
$-functions collected in Appendix A3 and minimizing the QPS action with respect
to $x_{0}$ and $\tau_{0}$ one again arrives at the result (\ref{otvet}) with
$A \sim1$. If the inequality (\ref{short}) is violated, the accuracy of our
expansion in powers of $\omega/\Delta_{0}$ can only become better (cf. eq.
(\ref{cap})).

\subsection{QPS rate}

We now proceed further and evaluate the QPS rate $\gamma_{QPS}$. Provided the
QPS action is sufficiently large $S_{QPS} \gg1$ this rate can be expressed in
the form
\begin{equation}
\gamma_{QPS}=B\exp(-S_{QPS}).\label{gamma}%
\end{equation}
The results for the QPS action derived above allow to determine the rate
$\gamma_{QPS}$ with the exponential accuracy. Here we evaluate the
pre-exponential factor $B$ in eq. (\ref{gamma}). For this purpose we will make
use of the standard instanton technique \cite{ABC}.

Consider the grand partition function of the wire $Z$. As we already discussed
this function can be expressed via the path integral
\begin{equation}
Z=\int{\mathcal{D}} \Delta{\mathcal{D}}\varphi\exp( -S ),\label{Z}%
\end{equation}
which will be evaluated the saddle point approximation. The least action
paths
\begin{equation}
\delta S /|\delta\Delta| =0, \;\;\;\;\;\;\delta S /\delta\varphi
=0\label{instantons}%
\end{equation}
determine all possible QPS configurations. Integrating over small fluctuations
around all QPS trajectories one represents the grand partition function $Z$ in
terms of infinite series where each term describes the contribution of one
particular QPS saddle point. Provided interaction between different quantum
phase slips is sufficiently weak one can perform a summation of these series
in a straightforward manner with the result
\begin{equation}
Z=\exp(-F/T),\label{ZF}%
\end{equation}
where $F$ defines the wire free energy
\begin{eqnarray}
F&=&F_{0}-T \frac{\int{\mathcal{D}}\delta Y\exp(-\delta^{2}S_{1}[\delta
Y])}{\int{\mathcal{D}}\delta Y\exp(-\delta^{2}S_{0}[\delta Y])} \exp
(-S_{QPS})
\nonumber\\
&\equiv& F_{0} -\frac{\gamma_{QPS}}{2}.\label{freeinst}%
\end{eqnarray}
Here $F_{0}$ is the free energy without quantum phase slips, $\delta
Y=(\delta\Delta,\delta\varphi)$ describe fluctuations of relevant
coordinates (fields), $\delta^{2}S_{0,1}[\delta Y]$ are the quadratic in
$\delta Y$ parts of the action, and the subscripts "0" and "1" denote the
action respectively without and with one QPS.

The integrals over fluctuations in eq. (\ref{freeinst}) can be evaluated
exactly only in simple cases. Technically such a calculation can be quite
complicated even if explicit analytical expressions for
the saddle point trajectories are avalilable. In our case such
expressions for the QPS trajectories are not
even known. Hence, an exact evaluation of the path integrals in eq.
(\ref{freeinst}) is not possible. Furthermore, any attempt to find an explicit
value for such a prefactor would make little sense simply because the
numerical value of $A$ in eq. (\ref{otvet}) is not known exactly.

What to do in this situation? Below we will present a simple approach which
allows to establish the correct general expression for the pre-exponent $B$ up
to an unimportant numerical prefactor. Our approach may be useful not only in
the case of superconducting wires but for various other situations since
numerical prefactors in the pre-exponent are usually of little importance.

In order to evaluate the ratio of the path integrals in eq. (\ref{freeinst})
let us introduce the basis in the functional space $\Psi_{k}(z)$ in which the
second variation of the action around the instanton $\delta^{2}S_{1}[\delta
Y]$ is diagonal. Here the basis functions depend on a general vector
coordinate $z$ which is simply $z=(\tau,x)$ in our case. The first $N$
functions $\Psi_{k}$ are the so-called ``zero modes'' reflecting the instanton
action invariance under shifts in certain directions in the functional space.
In our case the problem has two zero modes corresponding to shifts of the QPS
position along the wire and in imaginary time, i.e. $N=2$. Obviously such
shifts do not cause any changes in the instanton energy. The eigenfunctions
corresponding to these zero modes are: $\Psi_{k}(X)=\partial\tilde Y/\partial
z_{k}$, where $\tilde Y(z)$ the instanton (or QPS) trajectory, $k \leq N$ and
the number of zero modes $N$ coincides with the dimension of the vector $z$.
An arbitrary fluctuation $\delta Y(z)$ can be represented in terms of the
Fourier expansion
\begin{equation}
\delta Y(z)=\sum\limits_{k=1}^{N} \delta z_{k} \frac{\partial\tilde
Y(z)}{\partial z_{k}}+ \sum\limits_{k=N+1}^{\infty} u_{k}\Psi_{k}%
(z).\label{expan}%
\end{equation}
Then we get
\begin{eqnarray}
\delta^{2}S_{0}[\delta Y]&=&\frac{1}{2}\sum\limits_{k,n=1}^{\infty} A_{kn}%
u_{k}u_{n}^{*},
\nonumber\\
\delta^{2}S_{1}[\delta Y]&=&\frac{1}{2}\sum
\limits_{k=N+1}^{\infty} \lambda_{k} |u_{k}|^{2},\label{deS}
\end{eqnarray}
where for $k\leq N$ the Fourier coefficients $u_{k}\equiv\delta z_{k}$ are
just the shifts of the instanton position along the $k-$th axis and
$\lambda_{k}$ are the eigenvalues of $\delta^{2}S_{1}[\delta Y]$. Integrating
over the Fourier coefficients one arrives at the standard formula for the
ratio of determinants with excluded zero modes \cite{ABC}
\begin{eqnarray}
&& \frac{\int{\mathcal{D}}\delta Y\exp(-\delta^{2}S_{1}[\delta Y])}
{\int{\mathcal{D}}\delta Y\exp(-\delta^{2}S_{0}[\delta Y])} =
\int\limits_{0}^{L_{1}} d\delta x_{1} .. \int\limits_{0}^{L_{N}} d\delta x_{N}
\nonumber\\ &&\times\,
\sqrt{\frac{\det A_{kn}}{(2\pi)^{N}\prod\limits_{k=N+1}^{\infty}\lambda_{k}}
},\label{detex}
\end{eqnarray}
where $L_{k}$ is the system size in the $k-$th dimension. Now we will argue
that with a sufficient accuracy in the latter formula one can keep the
contribution of only first $N$ eigenvalues. Indeed, the contribution of the
``fast'' eigenmodes (corresponding to frequencies and wave vectors much larger
than the inverse instanton size in the corresponding dimension) is insensitive
to the presence of an instanton. Hence, the corresponding eigenvalues are the
same for both $\delta^{2}S_{0}$ and $\delta^{2}S_{1}$ and just cancel out from
eq. (\ref{detex}). In addition to the fast modes there are several eigenmodes
with frequencies (wave vectors) of order of the inverse instanton size. The
ratio between the product of all such modes for $\delta^{2}S_{1}$ and the
product of eigenvalues for $\delta^{2}S_{0}$ with the same numbers is
dimensionless and may only affect a numerical prefactor which is not
interesting for us here. Dropping the contribution of all such eigenvalues one
gets
\begin{eqnarray}
&& \frac{\int{\mathcal{D}}\delta Y\exp(-\delta^{2}S_{1}[\delta Y])}
{\int{\mathcal{D}}\delta Y\exp(-\delta^{2}S_{0}[\delta Y])} \approx
\int\limits_{0}^{L_{1}} d\delta x_{1} .. \int\limits_{0}^{L_{N}} d\delta x_{N}
\nonumber\\ &&\times\,
\sqrt{\frac{\det A_{kn}|_{k,n\leq N}}{(2\pi)^{N}}}.\label{appz}
\end{eqnarray}
What remains is to estimate the parameters $A_{kk}$ for $k \leq N$. For this
purpose let us observe that the second variation of the action becomes
approximately equal to the instanton action, $\delta^{2}S_{1}=\frac{1}%
{2}A_{kk}z_{0k}^{2}\approx S_{QPS}$, when the shift in the $k-$th direction
becomes equal to the instanton size in the same direction $\delta z_{k}%
=z_{0k}$. Then we find $A_{kk}\approx2S_{QPS}/z^{2}_{0k}$ and
\begin{equation}
\det A_{kn}|_{k,n<N}\approx\prod\limits_{k=1}^{N} A_{kk}\approx\frac
{(2S_{QPS})^{N}}{\prod\limits_{k=1}^{N} z_{0k}^{2}}.\label{prod22}%
\end{equation}
Finally, combining eqs. (\ref{gamma}), (\ref{freeinst}), (\ref{appz}) and
(\ref{prod22}) we obtain
\begin{equation}
B=bT\left( \prod\limits_{k=1}^{N} \frac{L_{k}}{z_{0k}}\right)  \left(
\frac{S_{QPS}}{\pi} \right) ^{N/2}.\label{Ffinal}%
\end{equation}
Here $b$ is an unimportant numerical prefactor. This formula demonstrates that
the functional dependence of the pre-exponent can be figured out practically
without any calculation. It is sufficient to know just the instanton action,
the number of the zero modes $N$ and the instanton effective size $z_{0k}$ for
each of these modes.

In fact, a similar observation has already been made \cite{ABC} in the case of
local Lagrangians equal to the sum of kinetic and potential energies. Here we
demonstrated that the result (\ref{Ffinal}) can be applied to even more
general effective actions, including nonlocal ones.

Turning to the interesting for us case of QPS in superconducting nanowires we
set $L_{1}\equiv1/T$, $L_{2}\equiv X$. Then eq. (\ref{Ffinal}) yields
\begin{equation}
B=b\frac{S_{QPS}X}{\tau_{0}x_{0}}.\label{B}%
\end{equation}
This equation provides an accurate expression for the pre-exponent $B$ up to a
numerical factor $b$ of order one. This result is parametrically different
from the resuls derived within the TDGL-type of analysis \cite{Chang} or
suggested phenomenologically in Ref. \cite{Giordano QPS PRL 1988}. The
inequality $S_{\mathrm{core}}\gg\mu$ allows to substitute $S_{\mathrm{core}}$
(\ref{otvet}) instead of $S_{QPS}$ in eq. (\ref{B}). Substituting also
$x_{0}\sim\xi$ and $\tau_{0}\sim1/\Delta_{0}$, for the QPS rate we finally
obtain
\begin{equation}
\gamma_{QPS}\sim\Delta_{0}\frac{R_{q}}{R_{N}}\frac{X^{2}}{\xi^{2}}%
\exp(-S_{QPS})\sim g_{\xi}\Delta_{0}\frac{X}{\xi}\exp(-Ag_{\xi}%
/4).\label{gammaf}%
\end{equation}
This result concludes our calculation of the tunneling rate for single quantum
phase slips.

Finally we would like to emphasize that the method employed here works
successfully in various other problems described both by local and nonlocal in
time Lagrangians. Several examples of such problems are discussed in Ref.
\cite{GZ01}. Here we mention only one such example. It is the well known
problem quantum tunneling with dissipation \cite{cl}. In the limit of strong
dissipation quantum decay of a metastable state was treated by Larkin and
Ovchinnikov \cite{LO} who found the exact eigenvalues and, evaluating the
ratio of the determinants, obtained the prefactor in expression for the decay
rate $B \propto\eta^{7/2}/m^{2}$, where $\eta$ is an effective friction
constant and $m$ is the particle mass. This result would imply that the
pre-exponential factor in the decay rate \cite{LO} should be very large and
may even diverge if one formally sets $m \to0$. Later it was realized
\cite{ZP} that this divergence should be regularized by means of proper
renormalization of the bare parameters in the effective action. After that the
high frequency contribution to the pre-exponent is eliminated and one arrives
at the result \cite{ZP} $B \propto1/\sqrt{\eta}$ which does not contain the
particle mass $m$ at all. This result allowed to fully resolve a discrepancy
between theory and experiments \cite{Lukens}. Note that the result \cite{ZP}
is trivially reproduced from eq. (\ref{Ffinal}): The pre-exponent $B$
can also be expressed in the form $B \sim\sqrt{S_{b}}/\tau_{0}$, where $S_{b}
\propto\eta$ is the instanton (bounce) action and $\tau_{0} \propto\eta$ is
its typical size. It is remarkable that in this particular case our approach
correctly reproduces even the numerical prefactor.

\subsection{QPS interactions and quantum phase transitions}

Although typically the hydrodynamic part of the QPS action (\ref{hydr}) can be
smaller than its core part (\ref{otvet}) the former also plays an important
role since it determines interactions between different quantum phase slips.
Consider two such phase slips (two vortices in space-time) with the
corresponding core coordinates $(x_{1},\tau_{1})$ and $(x_{2},\tau_{2})$.
Provided the cores do not overlap, i.e. provided $|x_{2}-x_{1}|>x_{0}$ and
$|\tau_{2}-\tau_{1}|>\tau_{0}$, the core contributions are independent and
simply add up. In order to evaluate the hydrodynamic part we substitute the
superposition of two solutions $\tilde{\varphi}(x-x_{1},\tau-\tau_{1}%
)+\tilde{\varphi}(x-x_{2},\tau-\tau_{2})$ satisfying the identities
\begin{equation}
\partial_{x}\partial_{\tau}\tilde\varphi-\partial_{\tau}\partial_{x}%
\tilde\varphi=2\pi\nu_{1,2}\delta(\tau-\tau_{1,2} ,x-x_{1,2})
\end{equation}
(where $\nu_{1,2}=\pm1$ are topological charges of two QPS fixing the phase
change after a wind around the QPS center to be $\pm2\pi$) into the action
(\ref{action11}) and obtain
\begin{equation}
S^{(2)}_{QPS}=2S_{\mathrm{core}}-\mu\nu_{1}\nu_{2}\ln\left[
\frac{(x_{1}-x_{2})^{2}+c_{0}^{2}(\tau_{1}-\tau_{2})^{2}}{\xi^{2}}\right]
,\label{Sint}%
\end{equation}
i.e. different quantum phase slips interact logarithmically in
space-time. QPSs with opposite (equal) topological charges attract
(repel) each other.

The next step is to consider a gas of $n$ quantum phase slips. Again assuming
that the QPS cores do not overlap we can substitute a simple superposition of
the saddle point solutions for $n$ quantum phase slips $\varphi= \sum^{n}%
_{i}\tilde{\varphi}(x-x_{i},\tau-\tau_{i})$ into the action and find
\begin{equation}
S^{(n)}_{QPS}=nS_{\mathrm{core}}+S^{(n)}_{\mathrm{int}},\label{mul0}%
\end{equation}
where
\begin{equation}
S^{(n)}_{\mathrm{int}}=-\mu\sum\limits_{i\neq j}\nu_{i}\nu_{j} \ln
\biggl{(}\frac{\rho_{ij}}{x_{0}}\biggr{)}+ \frac{\Phi_{0}}{c}I\sum
\limits_{i}\nu_{i}\tau_{i}\;.\label{mul}%
\end{equation}
Here $\rho_{ij}=(c^{2}_{0}(\tau_{i}-\tau_{j})^{2}+(x_{i}-x_{j})^{2})^{1/2}$
defines the distance between the i-th and j-th QPS in the $(x,\tau)$ plane,
$\nu_{i}=\pm1$ are the QPS topological charges and $\Phi_{0}=hc/2e$ is the
flux quantum. In eq. (\ref{mul}) we also included an additional term which
keeps track of the applied current $I$ flowing through the wire. This term
trivially follows from the standard contribution to the action \cite{sz}
\[
\int d\tau\int dx(I/2e)\partial_{x} \varphi.
\]
The grand partition function of the wire is represented as a sum over all
possible configurations of quantum phase slips (topological charges):
\begin{eqnarray}
Z &=& \sum_{n=0}^{\infty}\frac{1}{2n!}\left( \frac{y}{2}\right) ^{2n} \int
_{x_{0}}^{X}\frac{dx_{1}}{x_{0}}...\int_{x_{0}}^{X}\frac{dx_{2n}}{x_{0}}
\nonumber\\ &&\times\,
\int_{\tau_{0}}^{\beta}\frac{d\tau_{1}}{\tau_{0}}...
\int_{\tau_{0}}^{\beta
}\frac{d\tau_{2n}}{\tau_{0}} \sum_{\nu_{i}=\pm1}\exp(- S^{(2n)}_{\mathrm{int}%
})\label{partit}%
\end{eqnarray}
where an effective fugacity $y$ of these charges is related to the QPS
rate as
\begin{equation}
y=\frac{x_{0}\tau_{0}B}{X}\exp(-S_{\mathrm{core}}) \sim S_{\mathrm{core}}
\exp(-S_{\mathrm{core}}).\label{fug}%
\end{equation}
We also note that only neutral QPS configurations with
\begin{equation}
\sum_{i}^{n} \nu_{i}=0\label{nutot}%
\end{equation}
(and hence $n$ even) contribute to the partition function (\ref{partit}). This
fact is a direct consequence of the boundary condition $\varphi(x,\tau
)=\varphi(x,\tau+\beta)$ in the path integral for the partition
function~\cite{sz}.

It is easy to observe that for $I=0$ Eqs.~(\ref{mul}), (\ref{partit}) define
the standard model for a 2D gas of logarithmically interacting charges
$\nu_{i}$. The only specific feature of our present model as compared to the
standard situation is that here the space and time coordinates are not
equivalent and one can consider different limiting cases of ``long'' and
``short'' wires.

Let us first consider the limit of very long wires and assume that
$T \to 0$. Following the standard analysis of logarithmically
interacting 2D Coulomb gas \cite{b,kt,Kosterlilz 1974} which is
based on the renormalization group (RG) equations both for the
interaction parameter $\mu$ and the charge fugacity $y$. Defining
the scaling parameter $\ell=\ln(\rho/\xi)$ we have
\cite{b,kt,Kosterlilz 1974}
\begin{equation}
\partial_{\ell}\mu=-4\pi^{2} \mu^{2}y^{2}, \quad\partial_{\ell}y=(2-\mu
)y.\label{btkrg}%
\end{equation}
Following the standard line of reasoning we immediately conclude that a
quantum phase transition for phase slips occurs in a long superconducting wire
at $T \to0$ and
\begin{equation}
\mu=\mu^{*}\equiv2+4\pi y \approx2.\label{RGbtk}%
\end{equation}
This is essentially a Berezinskii-Kosterlitz-Thouless (BKT) phase
transtion~\cite{b,kt,Kosterlilz 1974} for charges $\nu_{i}$ in
space-time. The difference from the standard BKT transition in 2D
superconducting films is only that in our case the transition is
driven by the wire thickness $\sqrt{s}$ (which enters into $\mu$)
and not by temperature. In other words, for thicker wires with
$\mu>\mu^{*}$ quantum phase slips with opposite topological
charges are bound in pairs (dipols) and the $linear$ resistance of
a superconducting wire is strongly suppressed and $T$-dependent.
This resistance tends to vanish in the limit $T \to0$. Thus, we
arrive at an important conclusion: at $T=0$ a long quasi-1D
superconducting wire remains in a superconducting state, with
vanishing linear resistance, provided its thickness is
sufficiently large and, hence, the electromagnetic interaction
between phase slips is sufficiently strong, i.e. $\mu>\mu^{*}$.

On the other hand, for $\mu<\mu^{*}$ the density of free (unbound) quantum
phase slips in the wire always remains finite, such fluctuations destroy the
phase coherence (and, hence, superconductivity) and bring the wire into the
normal state with non-vanishing resistance even at $T=0$. Thus, another
important conclusion is that superconductivity in sufficiently thin wires is
\textit{always} destroyed by quantum fluctuations.

The above analysis is valid for sufficiently long wires. For typical
experimental parameters, however, $X<c_{0}/T$ (or even $X\ll c_{0}/T$), and
the finite wire size needs to be accounted for. For this purpose we modify the
above RG treatment in the following manner. Starting from small scales
$\rho\sim x_{0}$ we increase the scaling parameter $\ell$ and renormalize both
$\mu$ and $y$ according to RG equations (\ref{btkrg}). Solving these equations
up to $\ell=\ell_{X}=\ln(X/\xi)$ we obtain the renormalized fugacity
$\tilde{y}=y(\ell_{X})$. For larger scales $\ell>\ell_{X}$ only the time
coordinate matters and we arrive at the partition function fully equivalent to
one for a (0D) superconducting weak link (Josephson junction) in the presence
of quantum fluctuations of the phase. These systems are described in details
in Ref. \cite{sz}, therefore an extended discussion of this issue can be
avoided here. We only point out that our renormalized fugacity $\tilde{y}$ is
equivalent to the tunneling amplitude of the Josephson phase (normalized by
the Josephson plasma frequency), i.e.
\begin{equation}
\tilde{y}\rightarrow\sqrt{S_{0}}\exp(-S_{0})\label{equival}%
\end{equation}
where $S_{0}=\sqrt{8E_{J}/E_{C}}$ is the action of an instanton (kink) in a
Josephson junction, $E_{J}$ and $E_{C}$ are respectively the Josephson and the
charging energies.

The subsequent analysis essentially depends on the presence (or absence) of
additional dissipation in our system. In the absence of dissipation at $T
\to0$ the Josephson junction \textit{always} remains in the normal (i.e.
non-superconducting) state since superconductivity is suppressed by quantum
fluctuations of the phase and tunneling of Cooper pairs is prohibited
\cite{sz}. Due to eq. (\ref{equival}) exactly the same conclusion
applies for sufficiently short superconducting wires studied here. In the
presence of additional dissipation, however, quantum fluctuations of the phase
can be suppressed and, hence, superconductivity can be restored \cite{sz}.

Of particular importance is the case of Ohmic dissipation which is
realized either provided the wire is shunted by a normal resistor
$R_{S}$, or the wire itself has a non-vanishing normal
conductivity $\sigma=\sigma_{qp} $ even far from the QPS core. The
latter situation can occur, e.g., at finite (and not too low)
temperatures due to the presence of a sufficient number of
quasiparticles above the gap or possibly also due to
nonequilibrium effects. Bearing in mind that in a number of
experiments with ultra-thin wires (to be discussed below) QPS
effects were observed already at sufficiently high temperatures
$T\gtrsim\Delta_{0}(T)$ it is worth to briefly address the model
with Ohmic dissipation here.

In this case our consideration should be modified employing a two
stage scaling procedure \cite{ZGOZ2}. At it was already explained,
we first proceed with 2D RG equations (\ref{btkrg}) up to the
scale $\ell=\ell_{X}=\ln(X/\xi)$. For simplicity we assume that
Ohmic dissipation does not significantly affect RG eqs.
(\ref{btkrg}) at such small scales. Eventually we arrive at the
renormalized fugacity $\tilde{y}=y(\ell_{X})$. At larger scales
the space coordinate is irrelevant and the problem reduces to that
of a 1D Coulomb gas with logarithmic interaction. Therefore, (for
$\tilde{y}\ll1$) further scaling is defined by
\cite{sz,s,s1,s2,s3}
\begin{equation}
\partial_{\ell}\tilde{y}=(1-\gamma)\tilde{y},\quad\partial_{\ell}%
\gamma=0,\label{RG2}%
\end{equation}
where the interaction parameter $\gamma$ depends on the dissipation strength
being equal to either (i) $\gamma=R_{q}/R_{S}$ or to (ii) $\gamma=\pi
s\sigma_{qp}/2e^{2}X$. For $\gamma>1$ the fugacity scales down to zero, which
again corresponds to a superconducting phase, whereas for $\gamma<1$ it
increases indicating a resistive phase in complete analogy to a single
Josephson junction with Ohmic dissipation. In the case (ii) the phase
transition point again depends on the wire cross section $s$ as well as on its
total length $X$ and the value $\sigma_{qp}$.

Thus, two different quantum phase transitions can occur in
superconducting nanowires. One of them is the BKT-like phase
transition which is controlled by the strength $\mu$ of inter-QPS
electromagnetic interactions and eventually by the wire thickness.
Another one is the Schmid phase transition occurring in the
presence of Ohmic dissipation \cite{ZGOZ2,Buchler} provided, e.g.,
by an external shunt resistance $R_{S}$. Note that this situation
is somewhat reminiscent of that occurring in chains of resistively
shunted Josephson junctions or granular arrays
\cite{many,FvdZ,Chbm,Chbm1,PLA1,Fi87,Z88,Ch88,PZ89,Kor,Kor1,Zw,Bobbert,PLA2}
where, however, Schmid-like QPT is driven by local (rather than
global) shunt resistance.

\subsection{Wire resistance at low temperatures}

Let us now turn to the calculation of the wire resistance in the
presence of quantum phase slips. We first consider the limit of
long wires. At any nonzero $T$ such wires have a nonzero
resistance $R(T,I)$ even in the \textquotedblleft
ordered\textquotedblright\ phase $\mu>\mu^{\ast}$. In order to
evaluate $R(T)$ in this regime we proceed perturbatively in the
QPS fugacity $y$. Since for $\mu>\mu^{\ast}$ quantum phase slips
form closed pairs (dipols) and, hence, interactions between
different dipols can be neglected. For this reason it suffices to
evaluate the correction $\delta F$ to the wire free energy due to
one bound pair of quantum phase slips with opposite topological
charges. This procedure is completely analogous to that described
in details in Ref.~\cite{sz} (see Chapter 5.3 of that paper for an
extended discussion). Taking into account only logarithmic
interactions within bound pairs of quantum phase slips we can
easily sum up the series in eq. (\ref{partit}) and arrive at the
result
\begin{equation}
\delta F=\frac{Xy^{2}}{x_{0}\tau_{0}}\int_{\tau_{0}}^{\beta}\frac{d\tau}%
{\tau_{0}}\int_{x_{0}}^{X}\frac{dx}{x_{0}}e^{(\Phi_{0}I\tau/c)-2\mu\ln
[\rho(\tau,x)/x_{0}]}\;,\label{fren}%
\end{equation}
where $\rho=(c_{0}^{2}\tau^{2}+x^{2})^{1/2}$. It is convenient to
first integrate over the spatial coordinate $x$ and take the wire
length $X\rightarrow\infty$. For nonzero $I$ the expression in
Eq.~(\ref{fren}) is formally divergent for
$\beta\rightarrow\infty$ and acquires an imaginary part Im $\delta
F$ after analytic continuation of the integral over the temporal
coordinate $\tau$~\cite{sz}. This indicates a QPS-induced
instability of the superconducting state of the wire: the state
with a zero phase difference
$\delta\varphi(X)=\varphi(X)-\varphi(0)=0$ decays into a lower
energy state with $\delta\varphi(X)=2\pi$. The corresponding decay
rate $\Gamma_{2\pi}$ is defined by eq. (\ref{ImF}). The rate for
the opposite transition from $2\pi$ to 0 (or from 0 to $-2\pi$) --
which is nonzero as long as $T>0$ -- is defined analogously with
$I\rightarrow-I$. The average voltage drop across the wire is then
given by the difference between these rates
\begin{equation}
V=(\Phi_{0}/c)[\Gamma_{2\pi}(I)-\Gamma_{2\pi}(-I)]\label{vvv}%
\end{equation}
Evaluating the imaginary part of the free energy $\mbox{Im}\delta F$ from
(\ref{ImF}), (\ref{vvv}) we finally obtain
\begin{eqnarray}
V&=&\frac{\Phi_{0}Xy^{2}}{c\tau_{0}x_{0}}\frac{\sqrt{\pi}\Gamma(\mu-\frac{1}%
{2})}{\Gamma(\mu)\Gamma(2\mu-1)}\sinh\left(  \frac{\Phi_{0}I}{2cT}\right)
\nonumber\\ &&\times\,
\left|\Gamma\left(  \mu-\frac{1}{2}+\frac{i}{\pi}\frac{\Phi_{0}I}{2cT}\right)
\right|^{2}[2\pi\tau_{0}T]^{2\mu-2}\;,\label{volt}%
\end{eqnarray}
where $\Gamma(x)$ is the Euler gamma-function. For the wire resistance
$R(T,I)=V/I$ this expression yields
\begin{equation}
R\propto y^{2}T^{2\mu-3}\label{2mu-3}%
\end{equation}
for $T\gg\Phi_{0}I$ and
\begin{equation}
R\propto y^{2}I^{2\mu-3}\label{2mui}%
\end{equation}
for $T\ll\Phi_{0}I$. Thus, for sufficiently thick wires with $\mu>\mu^{\ast}$,
we expect a strong temperature dependence of the linear resistance which
eventually vanishes in the limit $T\rightarrow0$ indicating the
superconducting behavior of the wire.

Unfortunately for thinner wires with $\mu<\mu^{*}$ and at low
temperatures the above simple analysis becomes insufficient
because of the presence of unbound QPS and the necessity to
account for many-body effects in the gas of quantum phase slips.
We expect that the temperature dependence of the wire resistivity
should become linear at the transition to the disordered (i.e.
non-superconducting) phase. At $T\ll\Phi_{0}I/c$ we expect a
strongly nonlinear $I-V$ characteristics $V\sim I^{a}$ in thick
wires, and the universal power $a(\mu^{*})=2$ in thin wires at the
transition into the resistive state with $V\sim I$. Note that in
contrast to the BKT transition in 2D superconducting films, in the
case of wires the jump is expected to be from $a=2$ to 1 rather
than from $a=3$ to 1.

In the case of short wires $X\ll c_{0}/T$ we again proceed in two steps. A 2D
scaling analysis yields the ``global'' fugacity $\tilde{y}$. In analogy with
resistively shunted Josephson junctions~\cite{sz}, the voltage drop from the
imaginary part of the free energy reads
\begin{equation}
V=\frac{2\Phi_{0}\tilde{y}^{2}}{\Gamma(2\gamma)c\tilde{\tau}_{0}} \sinh\left(
\frac{\Phi_{0}I}{2cT}\right)  \left|\Gamma(\gamma+\frac{i\Phi_{0}I}{2\pi
cT})\right|^{2} [2\pi\tilde{\tau}_{0} T]^{2\gamma-1}\;,\label{volt2}%
\end{equation}
where $\tilde\tau_{0} \sim XC/e^{2}\gamma$. This equation gives $R\propto
y^{2}T^{2\gamma-2}$ and $R\propto y^{2}I^{2\gamma-2}$ respectively at high and
low $T$. The above result is valid for $\gamma>1$ and also for smaller
$\gamma$ at not very small $T$~\cite{sz}. In the low temperature limit in the
metallic phase the linear resistance becomes~\cite{sz}
\begin{equation}
R=S\sigma_{qp}/X,\label{Rshort0}%
\end{equation}
or $R=R_{S}$, i.e. $R$ is just equal either to the quasiparticle resistance of
the wire or to the shunt resistance. The physics of this result is exactly the
same as in the case of resistively shunted Josephson junctions: the
superconducting channel turns out to be completely blocked by quantum
fluctuations and the current can only flow through the normal resistance.

\subsection{Discussion}

It is worthwhile to emphasize that the above results were derived for
sufficiently \textit{uniform} nanowires with (almost) constant cross-section
$s$. As we already mentioned above, the very existence of quantum phase slips
and/or the possibility to experimentally observe QPS effects in such wires
were debated at an early stage of the research. At present there already
exist a large
number of independent experiments by different groups providing an
overwhelming evidence that QPS effects essentially
determine low temperature properties of ultra-thin superconducting wires. A
detailed comparison between these experimental results and the above
theoretical predictions will be carried out in the next chapter.

It is quite obvious that our QPS theory does not require strict
uniformity of the system, it will also apply provided there exist
relatively small diameter variations along the wire which are
practically unavoidable in any realistic experimental situation.
The probability for QPS to occur will be higher in thinner parts
of the wire than in thicker ones, otherwise all the physics
remains essentially the same. This situation should be contrasted
to that of chains of Josephson junctions and 1D granular wires in
which case quantum phase slips occur inside the junctions (tunnel
barriers) and do not require any suppression of the
superconducting order parameter. In other words, such quantum
phase slips have no core at all. Accordingly, no condensation
energy is lost during such phase slip events and also dissipative
effects remain insignificant (unless Josephson junctions are
shunted by external normal resistors).

Note that, although the QPS core physics for uniform and granular quasi-1D
wires is entirely different, the behavior of these systems at larger scales
can be qualitatively similar. For instance, it is well known that the BKT-like
quantum phase transition (driven by the ratio $E_{J}/E_{C}$) occurs in 1D
superconducting granular arrays \cite{BD,many,FvdZ} similarly to the case of
uniform wires \cite{ZGOZ}. Also, Schmid-like quantum dissipative phase
transitions in arrays and chains of resistively shunted Josephson junctions
were studied in a great detail, see
\cite{many,FvdZ,PLA1,Fi87,Z88,Ch88,PZ89,Kor,Zw,Bobbert,PLA2} and further
references therein. One of the main conclusions reached there was that
dissipative phase transitions yield \textit{local} phase coherence in a long
chain of junctions provided each of such junctions is shunted by an effective
Ohmic resistance below some value of order the quantum resistance unit $R_{q}%
$. In contrast, at $T\to 0$ BKT-like quantum phase
transition yields global phase coherence in Josephson chains
provided $E_{J}\gtrsim E_{C}$. As a result, one arrives at a
non-trivial phase diagram with all possible combinations of local
and global order \cite{many,FvdZ}. Let us also mention that more
recently the models of resistively shunted superconducting
granular chains were studied \cite{Chak,Dem,Refnew} aiming to interpret
some recent experimental data \cite{BT,Bezryadin MoGe review JPCM 2008}
seemingly indicating that
the superconducting phase transition can be driven by the global
wire resistance. Within this approach dissipation is introduced
phenomenologically employing an effective two-fluid model.
It remains unclear, however,
what could be the physical origin of Ohmic dissipation in sufficiently long
and nominally uniform superconducting wires at temperatures well
below the gap $\Delta_{0}$. An alternative proposal aimed to explain
recent experimental observations in relatively short nanowires
\cite{BT,Bezryadin MoGe review JPCM 2008} is discussed below in Sec. 6.3

We should also stress that even if the diameter remains strictly
constant the wire can be considered uniform only at length scales
exceeding the elastic electron mean free path $l$ whereas at the
scale $\sim l$ this uniformity is naturally broken by the presence
of impurities and defects. Scattering of electrons at such
impurities and defects plays a very important role since it breaks
Galilei invariance thus preventing from momentum conservation for
electrons propagating along the wire \cite{ogzb}. Momentum
non-conservation is in turn necessary for QPS to occur since
quantum phase slips ``unwind'' the supercurrent thus generating
momentum transfer. At $T=0$ (i.e. in the absence of phonons) this
extra momentum cannot be absorbed in perfectly clean and uniform
wires in which case QPS would be strictly prohibited. On a formal
level, this implies that the QPS action $S_{QPS}$ should become
large in the clean limit \cite{Khl}. In fact, this effect can be
observed within our analysis too, one just needs to formally take
the limit $\sigma\to\infty$ in eq. (\ref{S321}).

For completeness, let us mention about several proposals concerning the nature
of QPT in long superconducting nanowires with impurities. Khlebnikov and
Pryadko \cite{KhlPr} argued that this transition should occur at $\mu=1$ as a
dissipative QPT \cite{s} rather than the BKT-like QPT discussed here.
Accordingly, for the wire resistance they suggested
\begin{equation}
R \propto T^{2\mu-2}\label{2mu-2}
\end{equation}
instead of $R \propto T^{2\mu-3}$, cf. eq. (\ref{2mu-3}). Sachdev \textit{et
al.} \cite{Sachdev}, on the contrary, found QPT of the BKT universality class
but claimed it to be superconductor-insulator transition (SIT) rather than
superconductor-metal transition (SMT) as originally suggested in Ref.
\cite{ZGOZ}.

In their analysis Khlebnikov and Pryadko \cite{KhlPr} used a simplified
Gross-Pitaevski action which is in many respects different from the
microscopic effective action for superconducting wires derived in Chapter 2.
Within this phenomenological approach the authors \cite{KhlPr} first evaluated
the QPS rate for a given disorder configuration and only then performed
averaging over disorder. In contrast, within our analysis averaging over
disorder is performed in the effective action before introducing QPS. This
latter way is appropriate for diffusive superconducting wires, e.g., because
the QPS core size $\sim\xi$ greatly exceeds the electron mean free path $l$.
The condition $\xi\gg l$ allows to treat the wire as effectively homogeneous
for QPS and the problem acquires an extra zero mode corresponding arbitrary
shifts of the QPS core along the wire. This zero mode is lacking within the
analysis \cite{KhlPr}.%

\begin{figure}
\begin{center}
\includegraphics[width=6cm]{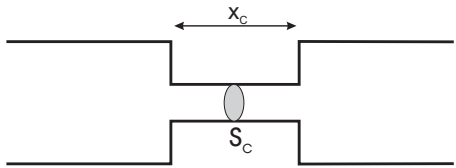}
\caption{Superconducting nanowire containing a thinner part (constriction) of
length $x_{c}$ and cross section $s_{c}$.}%
\label{F2 constriction}%
\end{center}
\end{figure}

Although for homogeneous (at scales $\gtrsim l$) and uniform wires QPT is of
the BKT universality class and the temperature dependence of the resistance in
the ordered (superconducting) phase is given by eq. (\ref{2mu-3}), there exist
other situations in which the dependence (\ref{2mu-2}) may apply at not very
low $T$. Consider, e.g., a long superconducting wire with cross-section $s$
and length $X$ which is uniform everywhere except a small part (constriction)
of length $x_{c}\ll X$ and cross-section $s_{c}<s$, as shown in Fig. 2. QPS
would then occur easier inside the constriction than in the rest of the wire.
Assuming for simplicity that $x_{c}$ is of order of (or just slightly larger
than) $\xi$, for the corresponding QPS fugacity inside the constriction area
we obtain
\begin{equation}
y_{c}\sim S_{\mathrm{core}}(s_{c}/s)\exp(-S_{\mathrm{core}}(s_{c}%
/s)),\label{fug2}%
\end{equation}
where $S_{\mathrm{core}}$ is the QPS core action (\ref{otvet}) for the wire.
As before, we can express the partition function $Z$ in terms of series in
powers of the QPS fugacity (cf., eq. (\ref{Z})) which is now equal to $y_{c}$
inside the constriction and to $y$ otherwise. In other words, we should now
deal with two different types of logarithmically interacting QPS. Quantum
phase slips in the wire are treated exactly as before, and we again arrive at
the BKT-like QPT at $\mu=\mu^{\ast}\approx2$. QPS inside the constriction
interact logarithmically only in time, thus giving rise to the Schmid-like QPT
at $\mu=1$. In this case the quasi-1D superconducting wire with
Mooij-Sch\"{o}n plasmons as elementary excitations plays the role of an
effective \textit{dissipative} environment for the constriction. Such QPT was
discussed by Hekking and Glazman \cite{HG} in the case of a narrow
superconducting ring interrupted by the Josephson junction. In the situation
considered here this QPT appears to be of little importance since $\mu^{\ast
}>1$, i.e. at $T\rightarrow0$ BKT-like QPT occurs when quantum phase slips
inside the constriction are still bound in pairs.

On the other hand, at not very low temperatures and provided $y_{c}\gg y$
quantum phase slips inside the constriction may give an additional
contribution to the wire resistance with the temperature dependence
(\ref{2mu-2}). In order to demonstrate that we modify eq. (\ref{fren}) by
taking into account the difference between QPS fugacities in the wire and in
the constriction. This is trivially handled by splitting the space integrals
into those over the wire and constriction areas. As a result, the linear
resistance of the system becomes
\[
R(T)=R_{w}(T)+R_{c}(T),
\]
where $R_{w}(T)$ is determined by QPS inside the wire (\ref{2mu-3}), i.e.
$R_{w}(T)\propto y^2T^{2\mu-3}$, while the additional contribution $R_{c}(T)$
comes from QPS inside the constriction. This contribution has the form
$R_{c}(T)\propto y^2_cT^{2\mu-2}$. Hence, at sufficiently high temperatures
the constriction contribution $R_{c}(T)$ can
dominate over $R_{w}(T)$, in which case the dependence (\ref{2mu-2}) applies.
However, at lower temperatures $R_{c}(T)$ becomes
irrelevant and $R(T)$ again is determined by eq. (\ref{2mu-3}). Obviously, the
same arguments hold in the case of several constrictions or, more
generally, for some wires with strongly fluctuating cross-section. A similar
scenario was recently discussed in Ref. \cite{Pai} where Gaussian fluctuations
of the wire thickness were considered.

Finally, let us briefly address the non-superconducting phase $\mu< \mu*$. In
our opinion, frequently used oversimplified approaches, like TDGL or effective
Luttinger liquid models, are by far insufficient in order to judge whether
this disordered phase is actually metallic or insulating, i.e. whether we are
dealing with SMT or SIT at $T=0$ and $\mu= \mu^{*}$. Also RG equations
(\ref{btkrg}) \textit{cannot} unambiguously resolve this issue. To illustrate
this point let us compare the behavior of Josephson chains and homogeneous
wires. The QPS fugacities $y$ of both systems obey RG equations (\ref{btkrg})
and, hence, for $\mu< \mu^{*}$ these fugacities grow upon renormalization in
both cases. In the case of Josephson chains it only implies strong
fluctuations of the phases across barriers with \textit{no suppression} of
superconductivity inside metallic grains. In this case inter-grain tunneling
of Cooper pairs is prohibited by electron-electron interactions and the system
is in the insulating (Coulomb blockade) phase.

On the contrary, large renormalized QPS fugacity $y$ in the case of
homogeneous wires implies that the QPS gas becomes dense, the QPS cores
overlap and, hence, for $\mu< \mu^{*}$ and $T \to0$ superconductivity gets
eventually destroyed everywhere in the wire. In other words, in this regime
quantum fluctuations drive homogeneous ultra-thin superconducting wires
\textit{normal}. This important difference between homogeneous and granular
wires was emphasized in Ref. \cite{ZGOZ}. What remains is to figure out
whether such normal diffusive wires stay metallic or turn insulating at low
enough $T$.

Few years after the work \cite{ZGOZ} it was realized that,
similarly to highly conducting tunnel barriers
\cite{PZ91,GZ94,GZ94-1} disordered metallic wires with large
conductances also demonstrate the feature of \textit{weak Coulomb
blockade} \cite{Naz,GZ00,GZ04,GZ04-1,BN} caused by non-trivial
interplay between scattering and electron-electron interactions.
The presence of an (exponentially small) Coulomb gap implies that
disordered wires most likely turn insulating at $T=0$. However, in
generic metallic wires with many conducting channels such a state
can only be observed at exponentially low $T $, while for any
experimentally attainable temperature one expects a metallic
behavior with a weak trend for the wire resistance to grow with
decreasing $T$ due to Coulomb blockade corrections \cite{GZ00}.
Interestingly, such a trend was indeed observed in recent
experiments \cite{Bezryadin MoGe TAPS EPL-2006,Bezryadin
condmat07,Bezryadin MoGe review JPCM 2008} and the measured
dependence $R(T)$ was found to closely follow theoretical
predictions \cite{GZ00}. Further theoretical analysis of the role
of electron-electron interactions in quasi-1D structures
\cite{Bezryadin MoGe TAPS EPL-2006,Bezryadin condmat07,Bezryadin
MoGe review JPCM 2008} would be highly desirable \cite{GZ06}.


\section{Experiments on superconducting nanowires}

\subsection{General considerations}

It follows from our theoretical analysis that both TAPS and -- in particular--
QPS can be observed in sufficiently thin superconducting wires, i.e.
experiments aimed to observe QPS-related phenomena should be performed on
samples with smallest possible cross section values $s$. On top of that, it
is highly desirable to deal with structurally (and chemically) homogeneous
samples in order to eliminate spatial variations of $T_{C}$ along the
wire and rule out the effect of (possibly existing) constrictions and tunnel
barriers. It has been shown that structural imperfections of real quasi-1D
structures (non-uniform cross section, existence of probes and finite length
effects) might effectively mask the phenomena related to thermal or quantum
fluctuations \cite{Zgirski inhomogeneity PRB}.

In addition to the above requirements, in order to provide optimal conditions
for observation of fluctuation effects it is crucial to properly choose the
sample material. This is obvious from the fact that, for instance, the QPS
core action (\ref{otvet}) at low enough $T$ can be represented in the form
\[
S_{\mathrm{core}}\sim T_{C}^{1/2}s/\rho_{N},
\]
where $\rho_{N}\equiv1/\sigma$ is the wire resistivity in the
normal state. Hence, it is desirable to select superconducting
materials with smaller values of $T_{C}$ (or $\Delta_{0}(0)$) and
perform experiments on sufficiently dirty nanowires with higher
values of $\rho_{N}$. The latter requirement is in line with the
well known general observation that fluctuation effects are more
pronounced in dirtier systems. This requirement also implies that
the electron elastic mean free path should obey the
\textquotedblleft dirty limit\textquotedblright\ condition
$l<\xi\sim\sqrt{l\xi_{0}}$ where $\xi _{0}\sim\hbar
v_{F}/\Delta_{0}$ is the BCS \textquotedblleft
clean\textquotedblright\ coherence length. The parameters for
various conventional superconductors are listed in Table I.
Judging from these numbers the most suitable materials for
experimental investigations of QPS effects are those with higher
resistivity values, i.e. $MoGe$ and $\alpha$:InO. On the other
hand, a certain disadvantage of these materials is that theu can
be strongly inhomogeneous. In this respect, Zr or Ti can be
advantageous for experimental studies of quantum fluctuations.

Though measuring the temperature dependence of the system resistance $R(T)$ in
the vicinity of the transition point might seem a routine experimental task,
in the case of nanostructures more care is required. Typically, $R(T)$
dependencies are measured in the current-biased regime. A standard requirement
is to keep the bias current $I$ much smaller than the critical (depairing)
current $I_{C}$ in order to avoid hysteresis effects due to overheating.
Additionally, in order to stay within the linear response regime, the
measuring current $I$ should remain smaller than the charactristic scale
$I_{0}=k_{B}T_{C}/\Phi_{0}$ \cite{la,mh} equal to few tens of nA for the
majority of materials (see Table I). The characteristic normal state
resistance of a typical metallic nanowire with diameter $\sqrt{s}\simeq$ 10 nm
and length $X$ $\simeq$ 1 $\mu$m is of order few k$\Omega$. Hence, in the
normal state the expected voltage is at least few $\mu$\textit{V}, which is
certainly not a problem to measure. However, with decreasing temperature only
slightly below $T_{C}$ the resistance $R(T)$ drops exponentially and the
measured signal quickly reduces to the \textit{nV} range. Employing
room-temperature electronics, it is preferable to use ac lock-in technique in
order to increase the signal-to-noise ratio. This can be associated with some
hidden problems. One of them is that even a tiny fraction of dc component
(e.g. from the ground loop) adds a parasitic signal $\sim dV/dI(T)$ to the
\textquotedblright valuable\textquotedblright\ one $R(T)\equiv V(T)/I$. For
this reason it is advisable to decouple the ac current source from the sample
using a low-noise transformer.
\vspace{0.3cm}

{\scriptsize
\begin{tabular}
[c]{|c|c|c|c|c|c|c|c|}\hline
Material & $T_{c}$ & $\lambda_{0}$ & $\xi_{0}$ & $B_{c}(0)$ & $\Delta(0)$ &
$v_{F}$ & $\rho_{N}$\\\hline
& $K$ & $nm$ & $\mu m$ & $mT$ & $\mathit{meV}$ & $10^{6}$ $m/s$ & $\mu
\Omega\times cm$\\\hline
W & 0.015 & - & 165* & 0.1 & 0.002* & 1.8* & 5.7\\\hline
Ir & 0.11 & - & 22.5* & 1.6 & 0.017* & 1.8* & 5.1\\\hline
Hf & 0.13 & - & 19.0* & 1.3 & 0.020* & 1.8* & 35\\\hline
Ti & 0.40 & - & 6.2* & 5.5 & 0.061* & 1.8* & 42.0\\\hline
Zr & 0.61 & - & 4.0* & 4.7 & 0.093* & 1.8* & 44.0\\\hline
Zn & 0.86 & 28 & 1.8 & 5.5 & 0.120 & 1.82 & 5.9\\\hline
Al & 1.19 & 16 & 1.6 & 10 & 0.175 & 2.02 & 2.7\\\hline
$\alpha$:InO & 2.8 & - & 0.330 & - & 0.430* & - & 3000\\\hline
In & 3.41 & 65 & 0.360 & 28 & 0.520 & 1.74 & 8.4\\\hline
Sn & 3.73 & 50 & 0.230 & 30.5 & 0.557 & 1.88 & 11.0\\\hline
MoGe (film) & 5.5 & 720 & 0.005 & 66 & 1.1 & - & 200\\\hline
Pb & 7.2 & 40 & 0.090 & 80.3 & 1.365 & 1.82 & 20.7\\\hline
Nb & 9.25 & 85 & 0.030 & 206 & 1.520 & 1.37 & 12.5\\\hline
\end{tabular}
}
\vspace{0.3cm}

{\small
Table-I. Material parameters of several superconductors. The data
are taken from Refs. \cite{Tinkham superconductivity
book,Superconductor Material,Kittel solid state,Ashcroft -
Mermin}; for $MoGe$ from Refs. \cite{Lau MoGe PRL,Bezryadin
private,Lau privat} and for $\alpha$:InO from Refs.
\cite{Johansson InO nanowire,Johansson privat}. The numbers marked
with * are estimated values.
}

Another problem might originate from the presence of rf filters which are
mandatory to protect nano-sized samples from noisy electromagnetic
environments. Very often such filters are just \textit{RLC} circuits shunting
the rf component to the ground through a capacitor as high as few $\mu F$.
Such configuration might provide good results with high-Ohmic systems (e.g.
tunnel structures). In a superconducting nanowire measured in 4-probe
configuration each electrode contacting the ''body'' of the sample is
typically made of the same material. Hence, the resistance of these
probes also varies from few
k$\Omega$ down to zero over the same temperature range
within the transition. Depending on particular configuration, at a
certain $T$ the ac sample impedance might become comparable to that of the
current leads through the ground, causing the current re-arrangement
throughout the sample. Even at rather low measuring frequencies $\sim$ 10
\textit{Hz} the parasitic effect might manifest itself as a non-monotonous
$R(T)$ dependence with unusual ''bumps'' or ''foot'' at the bottom of the
transition. Additional complication might arize from the non-negligible
dependence of the gain of the nanovolt pre-amplifier on the total impedance of
the load.

Concluding this part, an experimentalist should be extremely
cautious in designing the measuring set-up for unambiguous
interpretation of the data on superconducting nanowires.

Structural and geometrical homogeneity of quasi-1D samples is the
central question in interpretation of experimental data related to
contribution of superconducting fluctuations. Theoretically one
usually assumes that (i) the critical temperature $T_{C}$ of
quasi-1D wires under consideration remains spatially constant,
i.e. it does not vary along the wire, (ii) the cross section $s$
does not vary along the wire either and (iii) the measuring probes
are non-invasive. Unfortunately, in realistic nanowires none of
these conditions is usually well satisfied. To which extent can
these imperfections be neglected while interpreting the
experimental data using models developed under the assumptions
(i)-(iii)?

If \textit{structural} imperfections (e.g. non-uniform chemical composition)
can alter the critical temperature, then the shape of the experimentally
observed $R(T)$ dependence is determined by the sequence of transitions of
various parts of the wire with different local critical temperature $T_{C}(x)
$ . If the degree of such inhomogeneity is not too strong, a step-like $R(T)$
transition may not be observed as the variations of the critical temperature
are averaged on the scale of the coherence length $\xi$ resulting in a
relatively wide ''smooth-looking'' $R(T)$ dependence.

One might naively expect that working with superconducting samples fabricated
of initially pure material can help to eliminate the problem. Unfortunately
this is not the case. First, properties of low-dimensional superconductors are
known depend on the fabrication process details, such as thin film deposition
rate, residual pressure in the vacuum chamber, material of the substrate etc.
Second, even if to make an effort to keep the fabrication parameters constant,
it is hard to get rid of size-dependent effects.

\begin{figure}
\begin{center}
\includegraphics[width=8cm]{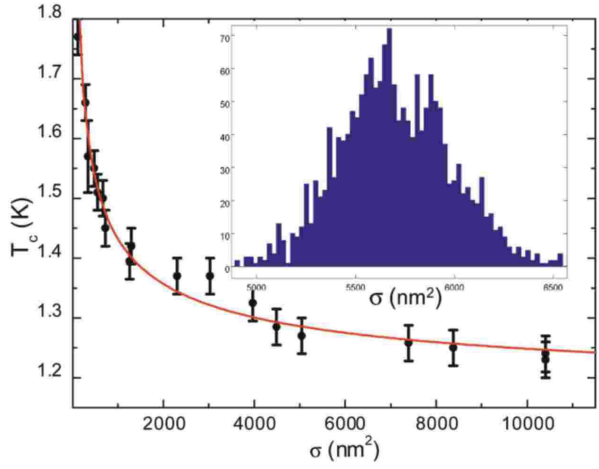}
\caption{Empirical dependence of the
critical temperature $T_{C}$ on cross section
$\sigma\equiv s$ for aluminum nanowires. Line is a guide for the eye. Inset:
SPM measured distribution of cross sections for a typical lift-off fabricated
nanowire: length $X$ =10 $\mu$m and effective diameter $\sqrt{s}\approx$ 75 nm
\cite{Zgirski inhomogeneity PRB}. }%
\label{F3 Tc(size) Al}%
\end{center}
\end{figure}

It is a well-known experimental fact that the critical temperature
$T_{C}$ of thin superconducting films frequently differs from one
for bulk samples. The same tendency is observed in metallic
nanostructures. In indium, aluminum and zink $T_{C}$ increases
with decreasing characteristic dimension \cite{Giordano QPS PRL
1988,Zgirski NanoLett 2005,Zgirski QPS PRB 2008,Altomare Al
nanowire PRL 2006,Tian Zn nanowire PRL 2005}. On the contrary, in
lead, niobium and MoGe an opposite tendency is observed
\cite{BT,Bezryadin MoGe TAPS EPL-2006,Dynes QPS PRL 1993,Bezryadin
Nb nanowires,Bezryadin MoGe and Nb wires}. No noticeable
variations of $T_{C}$ were detected in tin nanowires \cite{Tian Sn
nanowire APL 2003,Tian Sn nanowire PRB 2005,Piraux Sn nanowire APl
2004,Piraux Sn nanowire NanoSci NanoTech 2005}. The origin of this
phenomenon is not clear. There exist models predicting both
suppression \cite{Fin} and enhancement \cite{Shanenko Tc(size) PRB
2006} of the mean field critical temperature for low dimensional
superconductors. One can take the empirical fact of variation of
$T_{C}$ with effective diameter of a superconducting nanowire as
granted. Using $Al$ lift-off fabricated nanowires as a
representative example (Figs. \ref{F3 Tc(size) Al} and \ref{F4
R(T) Al inhjomogeneous}) it has been shown that size-dependent
effects are extremely important for interpretation of fluctuation
phenomena in quasi-1D systems \cite{Zgirski inhomogeneity PRB}.
The conclusion of a quantitative analysis \cite{Zgirski
inhomogeneity PRB} is rather disappointing: even chemically pure
nanostructures cannot be considered as sufficiently homogeneous as
soon as the size dependence of the critical temperature $T_{C}$
comes into play. One should study atomically homogeneous systems
as single crystalline whiskers \cite{Webb R(T) in Sn
whiskers,Tinkham R(T) in Sn whiskers}. Unfortunately, modern
nanotechnology does not enable growth of high-quality quasi-1D
single crystalls of arbitrary diameter made of all materials of
interest. The lithographic processes results in much lower quality
samples. The only exception are materials, such as $Al$, $Zn$ and
$In$, where the critical temperature decreases with reduction of
the effective dimension (e.g., nanowire diameter). Under certain
conditions, experiments on lift-off fabricated nanowires made of
these materials can be interpreted using fluctuation models
developed for perfect 1D channels. Provided broadening of the
$R(T)$ dependencies is detected below the bulk critical
temperature $T_{C}^{bulk}$, the size effects cannot account for
the phenomenon, and presumably \textquotedblright
real\textquotedblright\ physics is observed.

\begin{figure}
\begin{center}
\includegraphics[width=8cm]{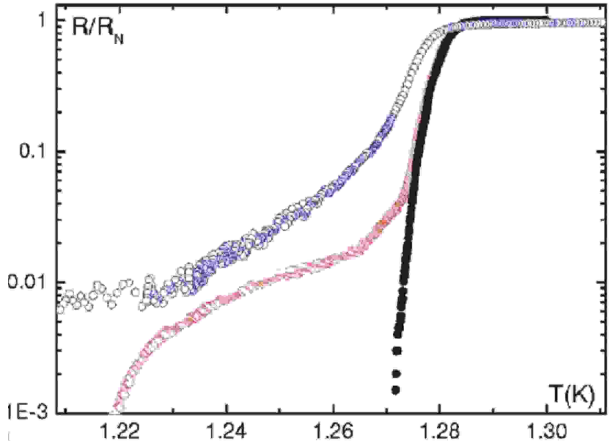}
\caption{Bullets ($\bullet$) correspond to simulated $R(T)$ transition without
taking into consideration the nodes. Diamonds ($\diamond$) represent the
results of
similar calculations with contribution of the node regions. Open circles
($\circ$) denote the experimentally measured $R(T)$ dependence. Simulations and
experimental data are for the same wire as in the inset from the previous
figure \cite{Zgirski inhomogeneity PRB}.}%
\label{F4 R(T) Al inhjomogeneous}%
\end{center}
\end{figure}

\subsection{Experiments on TAPS}
As we have already discussed in Chapter 4 the TAPS mechanism
provides non-zero voltage drop across the superconducting wire at
temperatures below $T_C$. For reference purposes let us rewrite
again the expression for the $I-V$ curve derived in the presence
of TAPS at sufficiently low bias currents $I$. Combining eqs.
(\ref{Gamma222}) and (\ref{IVTAPS}) and restoring some fundamental
constants (set equal to unity in our theory analysis) we obtain
\[
V=\frac{4\sqrt{3\pi}}{e}k_{B}T\frac{X}{\xi(T)}\sqrt{\frac{\delta F}{k_{B}T}%
}\exp\left[  -\frac{\delta F}{k_{B}T}\right] \sinh\left[\frac{\pi
\hbar I}{2ek_BT}\right]  ,
\]
where $\xi(T)$ is the temperature-dependent coherence length of a
dirty wire, $k_{B}$ is the Boltzmann constant,
\[
\delta F=\frac{16\pi^{2}k_{B}^{3/2}}{21\zeta(3)}sN_{0}\sqrt{\pi \hbar D}%
(T_{C}-T)^{3/2}
\]
is the the potential barrier for TAPS and $\zeta(3)\simeq 1.202$.

\begin{figure}
\begin{center}
\includegraphics[width=8cm]{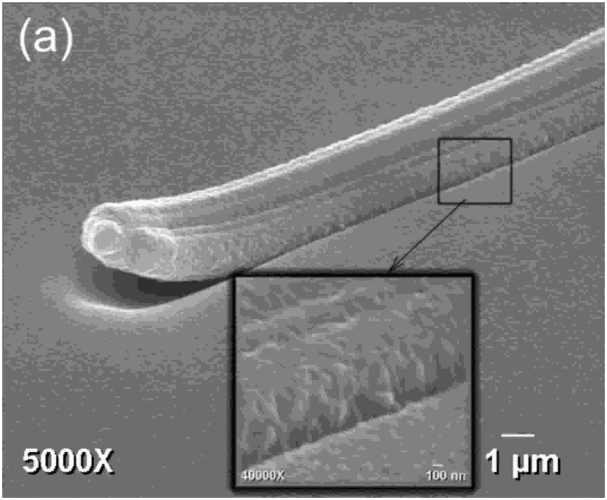}%
\caption{SEM image of a typical tin whisker on the surface of spin-on-glass
\cite{Arutyunov SOG whiskers}. Inset shows the magnified view of the crystal
surface \cite{Arutyunov patents in nanotechnology}. }%
\label{F 5 whisker}%
\end{center}
\end{figure}

Very quickly after development of the LAMH theory \cite{la,mh} two
experimental groups reported experiments aimed at verification of
the TAPS model. At 70-s microfabrication technique was not much
developed. In these early studies of 1D superconductivity
\cite{Webb R(T) in Sn whiskers,Tinkham R(T) in Sn whiskers}
metallic whiskers (Fig. \ref{F 5 whisker}) with characteristic
diameter $\sim$ 1 $\mu$m and lengths up to 1 mm were utilized.
Growth and basic properties of these highly anisotropic objects
have been widely described in the literature \cite{Givargizov
whisker growth}. In the particular case of tin whiskers
\textquotedblright squeeze\textquotedblright\ method was typically
applied for their growing \cite{Fisher whisker growth}. Then the
crystals were literally hand picked from the ingots and positioned
on substrates. The electrodes were made either by conducting paste
or epoxy \cite{Lutes Sn filaments}, soldered by Wood's metal
\cite{Meyer IV,Webb R(T) in Sn whiskers} or Sn-Pb alloy
\cite{Tinkham R(T) in Sn whiskers}, or squeezed by a soft metal
(e.g. indium) \cite{Gaidukov whiskers}. An obvious disadvantage of
the procedure is inevitable damage of the crystal within the locus
of the electric probes. An attempt to overcome these difficulties
was made by combining planarization technique with electron beam
lithography \cite{Arutyunov SOG whiskers,Arutyunov patents in
nanotechnology} (Figs. \ref{F 5 whisker}\ and \ref{F6 SOG
structure}). The technique appears promising, but extremely time
consuming and resulting in a rather low yield of
\textquotedblright working\textquotedblright\ samples.%

\begin{figure}
\begin{center}
\includegraphics[width=8cm]{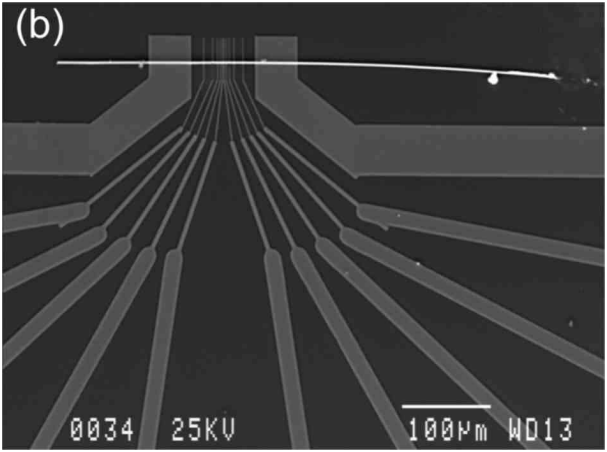}%
\caption{SEM image of a whisker-based microstructure on \textit{Si/SiO
}substrate covered with \textit{Spin-On-Glass}. Slightly bent bright
horizontal line is the tin whisker. Electrodes are made on top of the crystal
by e-beam lithography followed by evaporation of copper contacts
\cite{Arutyunov SOG whiskers}.}%
\label{F6 SOG structure}%
\end{center}
\end{figure}
\begin{figure}
\begin{center}
\includegraphics[width=9cm]{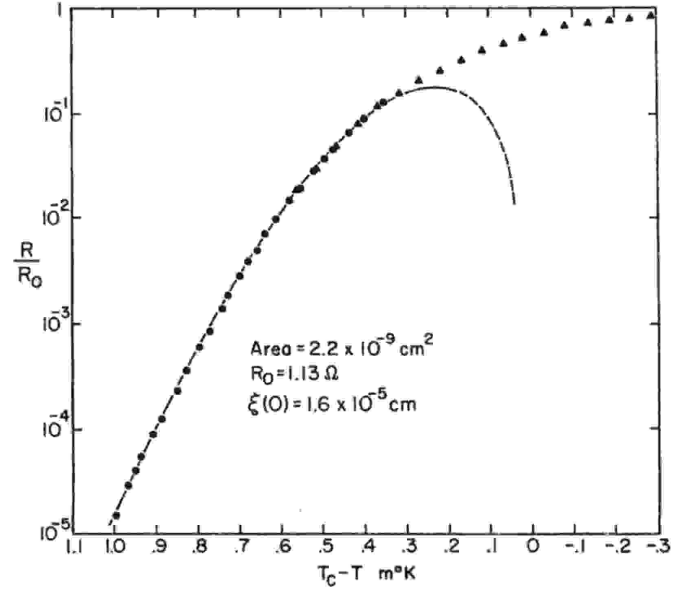}%
\caption{$R(T)$ dependence for a tin whisker \cite{Webb R(T) in Sn whiskers}.
 Solid symbols are the experimental data, dashed line represents the
fit to the TAPS model \cite{la,mh}.}%
\label{fig7 R(T) in whisker Tinkham}%
\end{center}
\end{figure}

An example of experimental $R(T)$ dependence measured on tin whisker is shown
in Fig. \ref{fig7 R(T) in whisker Tinkham}. One observes a very good
quantitative agreement between the experiment \cite{Webb R(T) in Sn
whiskers,Tinkham R(T) in Sn whiskers} and the TAPS model \cite{la,mh}. The
superconducting transition $R(T)$ is very steep: resistance of the sample
drops five orders of magnitude within the temperature range $\delta T$ $\sim$
1 mK below $T_{C}$. This is a consequence of rather large effective diameter
values $\sqrt{s}\simeq0.5$ $\mu$m and extremely high homogeneity of single
crystals. In the case of thinner single-crystalline structures
the width of the $R(T)$-curve would be larger, cf. eq. (\ref{TAPS R(T)}).
Unfortunately, the dimensions of whiskers made of
superconducting materials do not vary much. Manual manipulation of sub-1 $\mu
m$ is extremely time consuming and results in a low yield of suitable samples,
while the dimensions of whiskers are still too large to use scanning probe
(SPM) technique well developed for manipulation of nano-sized objects, such as
carbon nanotubes.

An alternative to whiskers is superconducting microcylinders encapsulated in a
dielectric substrate (e.g. glass) fabricated using Taylor-Ulitovski method
\cite{Taylor micricylinder method,Taylor microcylinder patent,Ulitovski
  microcylinder method,Ulitovski microcylinder patent1,Ulitovski microcylinder
  patent2}.
The quality of these wires is not as high as that of
atomically-pure whiskers. However, an important advantage of this
approach is the possibility to fabricate very long samples.
Fabrication of a wire with metal core of about few $\mu$m and
length up to a km (!) is a routine work, while fine adjustment of
parameters enables fabrication of sub-100 nm filaments with length
up to few cm \cite{Nikolaeva sub 1 mkm Bi wires,Nikolaeva Bi-Te
nanofilaments,Nikolaeva confinement in Bi nanowires}. A remarkable
feature of wires obtained by this method is that typically they
are single crystals. A number of superconducting microcylinders
(Sn and In) fabricated using Taylor-Ulitovski method were studied
\cite{Arutyunov Sn and In filaments JAP,Arutyunov Sn and In
filaments Physica C}. Bigger length values of the filaments
($X\sim$ 1 cm ) as compared to those for previously studied
whiskers ($X\sim$ 0.5 mm ) enabled observation of interesting
features related to interaction of current-induced phase slips.
The cross section values for microcylinders did not differ much
from those for whiskers, while the homogeneity was clearly worse:
No qualitatively new phenomena related to fluctuation-governed 1D
superconductivity has been observed.

There were several reports on experimental studies of 1D
superconductivity in lift-off-fabricated superconducting
nanostructures (wires and loops) with the line-width much smaller
than of whiskers or microcylinders \cite{Santhanam nMR
anomaly,Santhanam R(T) anomaly,Moschalkov Little-Parks
anomaly,Moshchalkov R(T) anomaly,Arutyunov non-locality}. The
advantage of the method is the ability to fabricate samples with a
complicated shape in a reproducible way. Typically, the metal is
deposited through PMMA mask using thermal or e-beam evaporation,
or sputtering. The quality of the structures is far from perfect:
the majority of superconducting thin-film structures fabricated at
room temperatures are polycrystalline with the grain size of about
few tens of nm. Deposition of metal (in particular low melting
ones, such as tin or indium) on a cryogenically cooled substrate
might reduce the grain size. In contrast to whiskers or
microcylinders, the lift-off-fabricated superconducting
nanostructures studied so far were all in the dirty limit
$l\ll\xi$. Attempts to \textit{quantitatively} describe the shape
of superconducting transition of these quasi-1D systems using the
TAPS model failed: Experimental curves for $R(T)$ were always
significantly broader than theoretical predictions. Nevertheless,
it was believed that with a certain adjustment of fit parameters
(e.g., reduction of the effective size of the phase slip center
\cite{Moshchalkov R(T) anomaly}) a reasonable agreement between
the experiments and the TAPS model predictions could be achieved.
Later it was shown that inevitable inhomogeneity of
lift-off-fabricated nanostructures, the presence of node regions
and finite size effects can dramatically broaden the
experimentally observed dependencies $R(T)$ making any comparison
with the TAPS theory inconclusive or even impossible (Fig. \ref{F4
R(T) Al inhjomogeneous}) \cite{Zgirski inhomogeneity PRB}.

\begin{figure}
\begin{center}
\includegraphics[width=8cm]{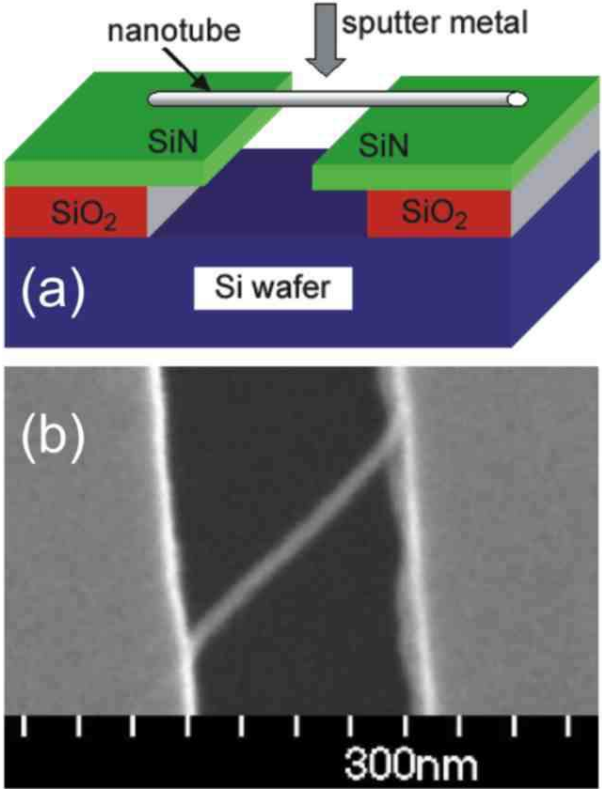}%
\caption{Molecular template decoration method. (a) Schematics of the
fabrication technique. (b) SEM image of a carbon-nanotube-based structure
\cite{Bezryadin molecular decoration 1}, \cite{Bezryadin molecular decoration
3}.}
\label{F8 Bezryadin method}%
\end{center}
\end{figure}

Very thin superconducting nanowires have been studied using the
template decoration technique: Deposition of a metal film on top
of a suspended insulating molecule used as a template
\cite{Bezryadin molecular decoration 1,Bezryadin molecular
decoration 2,Bezryadin molecular decoration 3,Johansson InO
nanowire} (Fig. \ref{F8 Bezryadin method}). Only pseudo-four probe
configuration is available with the deposited metal film from both
sides of the trench serving as electrodes. The approach enabled
fabrication of superconducting nanowires (MoGe, Nb and
$\alpha$:InO) with effective diameters down to $\sim$3 nm and
lengths up to $\sim$1 $\mu$m.
TEM study revealed amorphous structure for MoGe and $\alpha$:InO.%

\begin{figure}
\begin{center}
\includegraphics[width=8cm]{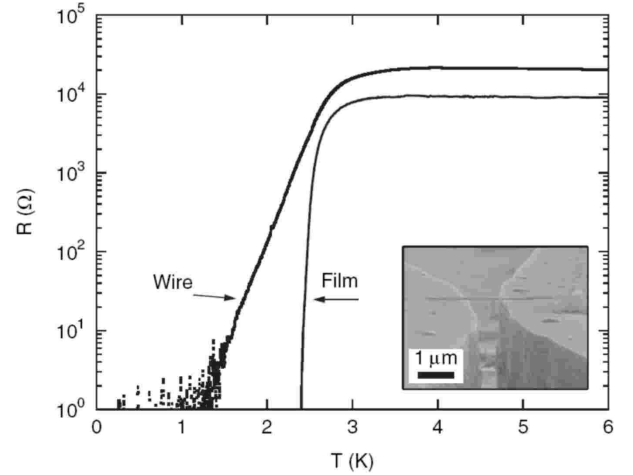}%
\caption{$R(T)$ obtained from $\alpha$:InO nanowire with width 100 nm together
with the data from a similarly prepared 500 $\mu$m wide film. Inset: A
scanning electron micrograph (SEM) of a typical device \cite{Johansson InO
nanowire}.}%
\label{F9 Johansson InO}%
\end{center}
\end{figure}

The $R(T)$ dependencies in $\alpha$:InO appeared to be significantly wider
than it is predicted by the TAPS model, though no claims about the impact of
the QPS mechanism were made \cite{Johansson InO nanowire}. Possibly,
broadening of experimental curves $R(T)$ in $\alpha$:InO can be associated
with inevitable inhomogeneity of the samples introduced during their
fabrication. It
is well known that physical properties of $\alpha$:InO low dimensional systems
strongly vary by annealing \cite{InO film Shahar PRB 1992}.

Interpretation of some experiments on \textit{MoGe} nanowires
remains under debates. In earlier reports \cite{BT,Lau MoGe PRL}
very broad $R(T)$ dependencies were observed and associated with
quantum phase slips. However, in later works \cite{Bezryadin MoGe
and Nb wires,Bezryadin MoGe TAPS EPL-2006,Bezryadin MoGe review
JPCM 2008,Bezryadin condmat07} broadening of a superconducting
transition in similar \textit{MoGe} structures was interpreted
within the TAPS model with no QPS contribution (Fig. \ref{F10 MoGe
short with TAPS}). On the other hand, later it was argued
\cite{Meidan,Meidan2} that fits were actually produced outside the
applicability range of the LAMH theory (\ref{abe}). Moreover, for
the experimental parameters \cite{Bezryadin MoGe and Nb
wires,Bezryadin MoGe TAPS EPL-2006,Bezryadin MoGe review JPCM
2008,Bezryadin condmat07} the strong inequalities (\ref{abe}) are
not satisfied, i.e. this theory has almost no applicability range
at all, see also Fig. \ref{F11 TAPS range MoGe}. Thus, employing
the LAMH model in order to fit the resistance curves $R(T)$ for
these samples is problematic. In addition, even if one performs
such fits one is bound to use fit values for the electron mean
free path much larger than the wire diameter \cite{Bezryadin MoGe
TAPS EPL-2006} which is rather unrealistic for such structures
(Fig. \ref{F10 MoGe short with TAPS}). At this point let us recall
that estimates for the coherence length
$\xi(0)\sim(\xi_{0}l)^{1/2}$ both in \textit{MoGe} thin films
\cite{Graybeal QPS in MoGe PRL 1987,Graybeal 2D MoGe films PRB
1984} and wires \cite{BT} typically yield values $\sim5$ to 7 nm
which translates into mean free path values $l$ of only few nm,
i.e. much shorter than used in the fits \cite{Bezryadin MoGe TAPS
EPL-2006}. We will return to possible interpretation of these
experiments below in connection with QPS effects.

Transmission electron microscopy (TEM) study of \textit{Nb}
nanowires, fabricated using template decoration method, revealed
polycrystalline structure with an average grain size about few nm.
In order to fit the data obtained in these \textit{Nb} structures
it has been proposed \cite{Lau MoGe PRL} to represent the measured
wire conductance $1/R(T)$ as a sum of contributions from normal
electrons above the gap and from TAPS,
$1/R(T)=1/R_{N}+1/R_{TAPS}$. The experimental curves $R(T)$ could
be reasonably well fitted \cite{Bezryadin Nb nanowires,Bezryadin
MoGe and Nb wires} to this phenomenological model (Fig. \ref{F12
R(T) Nb wires}). Still, the ability to detect TAPS contribution in
\textit{Nb} is rather unexpected since the energy gap $\Delta_{0}$
is very large in this material (see Table I). It is also quite
surprising that the inhomogeneities in granular \textit{Nb}
nanowires do not seem to broaden the $R(T)$ curves, which can be
fitted by the TAPS model developed for homogeneous quasi-1D
systems.

\begin{center}%
\begin{figure}
\begin{center}
\includegraphics[width=8cm]{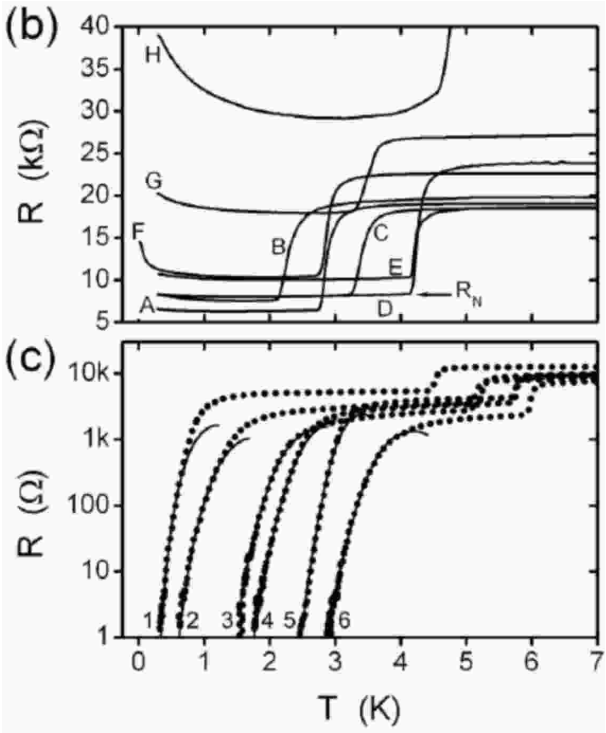}%
\caption{Resistance versus temperature plots for "short" MoGe nanowires:
insulating (top) and superconducting (bottom) \cite{Bezryadin MoGe TAPS
EPL-2006}. Solid lines are fits to the TAPS model. The best fit values of the
coherence lengths $\xi(0)$ are 70.0, 19.0, 11.5, 9.4, 5.6, and 6.7 nm, and
lengths $L$ are 177, 43, 63, 93, 187, 99 nm for samples 1--6, respectively.
Double-step shape of $R(T)$ transitions in the top and bottom plots comes from
the superconducting transition of the contact regions contributing to the
2-probe measurement configuration. }%
\label{F10 MoGe short with TAPS}%
\end{center}
\end{figure}
\begin{figure}
\begin{center}
\includegraphics[width=8cm]{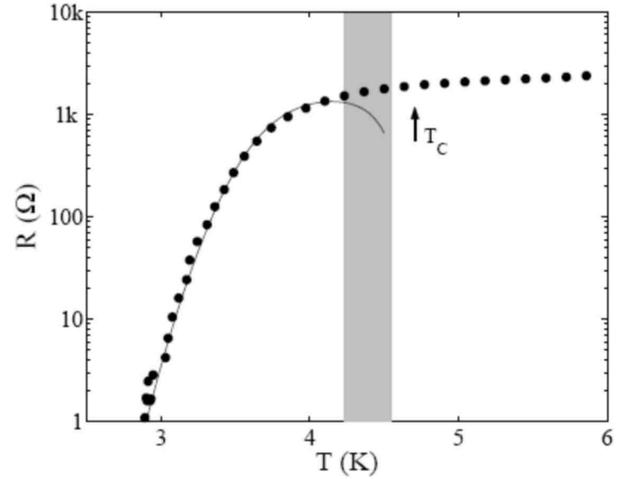}%
\caption{Fits of the data points for the sample No. 6 of Ref.
\cite{Bezryadin MoGe TAPS EPL-2006} to the LAMH theory. The fit
parameters are the coherence length, $\xi$ = 6.7 nm, and the
critical temperature, $T_{C}$ = 4.7 K, marked by an arrow. The
LAMH theory can only be applied within the strip 4.23 K $<T<4.55$
K \cite{Meidan,Meidan2} (shown in grey) which is clearly outside
the region where $R$ strongly depends on
temperature.}%
\label{F11 TAPS range MoGe}%
\end{center}
\end{figure}

\end{center}%

\begin{figure}
\begin{center}
\includegraphics[width=8cm]{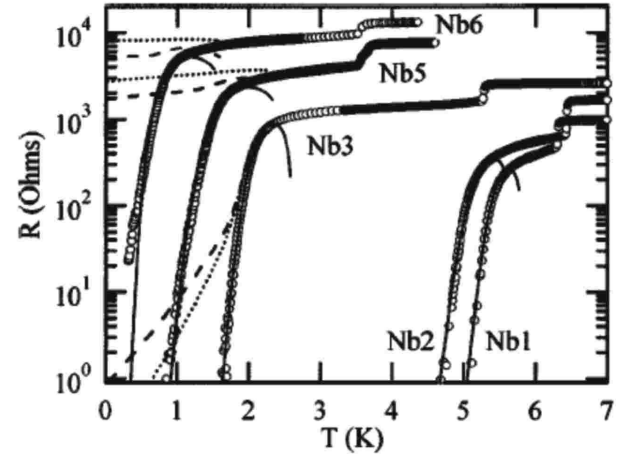}%
\caption{Temperature dependence of the resistance of
superconducting Nb nanowires obtained by template decoration
method using carbon nanotube as a substrate. The following fit
parameters for the samples Nb1, Nb2, Nb3, Nb5, and Nb6 are used:
Transition temperatures (K) are $T_{C}^{wire}$ = 5.8, 5.6, 2.7,
2.5, and 1.9 K; lengths (nm) are $X$ = 137, 120, 172, 177 and 113;
normal state resistance ($\Omega$) are $R_{N}$ = 470, 650, 1600,
4250 and 9500; coherence lengths (nm) are $\xi(0) $= 8.5, 8.1, 18,
16, and 16.5, respectively. Solid lines show fits to the
phenomenological TAPS model \cite{Lau MoGe PRL}. The dashed and
dotted lines are some theoretical curves
that include QPS contributions.}%
\label{F12 R(T) Nb wires}%
\end{center}
\end{figure}

\subsection{Experimental observations of QPS effects in superconducting
nanowires}

Let us now turn to the experiments detecting QPS effects in superconducting
nanowires. Mooij and coworkers \cite{Mooji} discussed the possibility to
observe quantum phase slips experimentally and attempted to do so as early as
in 1987. According to our theory, the crossover between TAPS and QPS regimes
can be expected at temperatures $T \sim\Delta_{0}(T)$. For superconductors
with typical material parameters and $T_{C} \gtrsim$ 1K (see Table I) this
condition implies that QPS effects may become important already at $T_{C}-T
\gtrsim$ 100 mK. As we already discussed, in order to be able to observe QPS
it is necessary to fabricate nanowires with effective diameters in the 10 nm
range, $\sqrt{s}\sim$ 10 nm. The wires should be sufficiently uniform and
homogeneous in order not to override fluctuation effects by trivial broadening
of $R(T)$ dependencies due to wire imperfections and inhomogeneities
\cite{Zgirski inhomogeneity PRB}. Thus, proper fabrication technology is
vitally important for experimental studies of QPS in quasi-1D superconductors.

\begin{figure}
\begin{center}
\includegraphics[width=8cm]{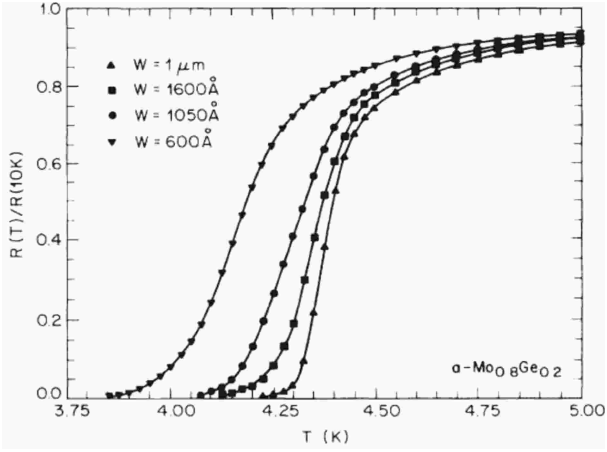}%
\caption{Resistive transitions for a set of MoGe wires with
different width $w$ upon a single film of thickness $t=5$ nm
\cite{Graybeal QPS in MoGe PRL
1987}.}%
\label{F13 R(T) MoGe Graybeal}%
\end{center}
\end{figure}

Perhaps the first experimental indication of the effect of quantum
fluctuations was obtained in amorphous MoGe nanowires with effective diameter
down to $\sqrt{s}\simeq$ 30 nm \cite{Graybeal QPS in MoGe PRL 1987}. The
samples in 4-probe configuration with length $X$ up to 20 $\mu$m were
fabricated using e-beam lithography. Though the paper \cite{Graybeal QPS in
MoGe PRL 1987} was mainly focused on the effect of disorder in low-dimensional
superconductors, it was clearly stated that for the narrowest samples
significantly broader curves $R(T)$ were observed than it is predicted by the
TAPS model (see Fig. \ref{F13 R(T) MoGe Graybeal}). The effect of quantum
fluctuations was pointed out as a possible reason for this disagreement.%

\begin{figure}
\begin{center}
\includegraphics[width=8cm]{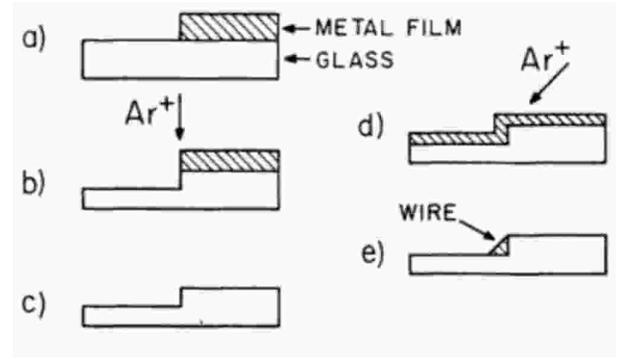}%
\caption{Step decoration technique \cite{Girdano step technique PRL 1979}.}%
\label{F14 Giordano step technique}%
\end{center}
\end{figure}

Detailed experimental studies of transport properties of
superconducting nanowires have been carried out by Giordano and
co-workers \cite{Giordano QPS PRL 1988,Giordano QPS PRL
1989,Giordano QPS PRB 1991,Giordano Physica B 1994,Giordano QPS
PRB 1990}. In these experiments \textit{In} and \textit{In-Pb
}wires with triangular cross-section were fabricated using
step-edge lithographic technique \cite{Giordano step technique APL
1980,Girdano step technique PRL 1979}. This method utilizes the
shadow effect produced by extremely shallow steps in substrates
and the corresponding tilted ion beam milling of the deposited
material (Fig. \ref{F14 Giordano step technique}). Polycrystalline
wires were fabricated with effective diameters $\sqrt{s}$ in the
range 40 - 100 nm with the grain size from 10 to 20 nm. The
uniformity of the samples was controlled with SEM and
with the accuracy of $\sim 10$ nm.%

\begin{figure}
\begin{center}
\includegraphics[width=8cm]{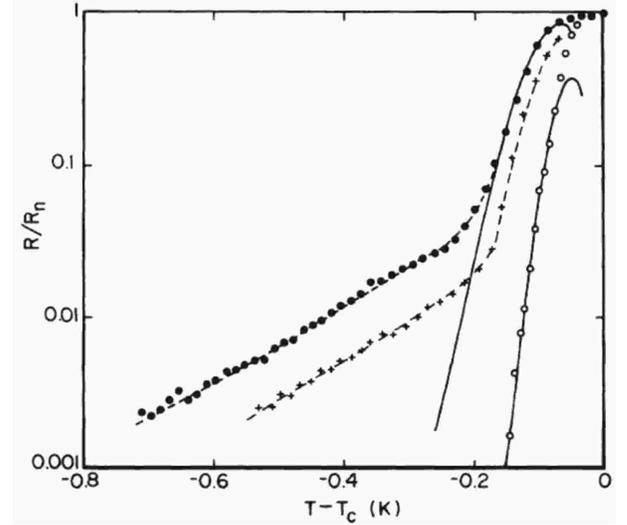}%
\caption{Resistance (normalized by its normal state value) as a function of
temperature for three \textit{In} wires; the sample diameters were 41 nm
($\bullet$), 50.5 nm ($+$) and 72 nm ($\circ$). The solid curves are fits to
the TAPS model, while the dashed curves indicate fits to the phenomenological
model \cite{Giordano QPS PRL 1988}. The sample lengths are 80, 150 and 150
$\mu$m, and the normal state resistance values are 5.7, 7.1 and 1.2 k$\Omega$,
respectively \cite{Giordano QPS PRL 1988}.}%
\label{F15 Giordano R(T) In}%
\end{center}
\end{figure}

For wider structures a reasonable agreement with the TAPS model
was observed. On the other hand,  for thinner wires with
$\sqrt{s}\lesssim$ 50 nm clear deviations from TAPS predictions
were demonstrated (see Fig. \ref{F15 Giordano R(T) In}). The
discrepancy was interpreted as a manifestation of quantum phase
slippage. In order to explain their observations Giordano and
co-workers proposed phenomenological description based on the
Caldeira-Leggett model for macroscopic quantum tunneling with
dissipation \cite{cl}. Though qualitative agreement with this
simple phenomenological model has been obtained (Fig. \ref{F15
Giordano R(T) In}), quantitative interpretation of the data
\cite{Giordano QPS PRL 1988} is problematic due to poor uniformity
of the samples: $\pm$ 10 nm for wires with $\sqrt{s}\lesssim$ 50
nm \cite{Giordano QPS PRL 1988}.

Recently, Pai \textit{et al.} \cite{Pai} performed a fit of the same
experimental data \cite{Giordano QPS PRL 1988} to the power-law
dependence $R(T)\propto T^{2\mu-2}$ derived
from the result (\ref{2mu-3}) assuming Gaussian fluctuations of the wire
thickness. Unfortunately the temperature dependence of the QPS fugacity
(\ref{fug}) was not taken into account in Ref. \cite{Pai}. This temperature
dependence (predominantly determined by that of the gap, $S_{\mathrm{core}}%
\propto\Delta_{0}^{1/2}(T)$) should, however, still be significant
at temperatures not very far from $T_{C}$ and will alter the fits.

It appears that strong granularity of the wires was most likely a very
important factor in the experiments by Giordano and co-workers. For instance,
estimating $S_{\mathrm{core}}$ for the thinnest wires \cite{Giordano QPS PRB
1991} with diameters 16 and 25 nm, for the experimental parameters we obtain
respectively $S_{\mathrm{core}} \sim280 A$ and $\sim700 A$. For uniform wires
such huge values of the QPS action would totally prohibit any signature of
quantum phase slips. Since QPS effects were very clearly observed, it appears
inevitable that these samples contained constrictions with diameters $\sim3
\div10$ times smaller than the average thickness values presented in
\cite{Giordano QPS PRB 1991}.

\begin{figure}
\begin{center}
\includegraphics[width=8cm]{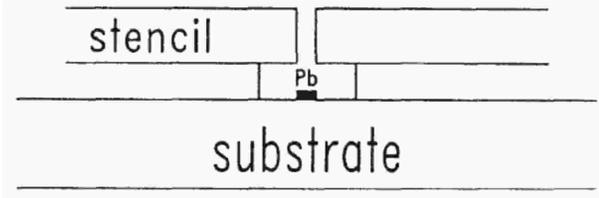}%
\caption{Cross-sectional view of the stencil structure. Metallic films were
evaporated through the shadow mask structure fabricated on the substrate
\cite{Dynes QPS PRL 1993}.}%
\label{F16 Dynes stencil technique}%
\end{center}
\end{figure}

Another set of experiments was performed by Dynes and co-workers
\cite{Dynes QPS PRL 1993,Dynes QPS and nMR PRL 1997,Dynes QPS PRL
1996}. Suspended stencil technique has been developed (Fig.
\ref{F16 Dynes stencil technique} ) enabling fabrication of
quench-condensed granular nanowires with cross-section area $s$
down to 15 nm$^{2}$ and length $X$ ranging from 1 to 2 $\mu$m from
various materials including several superconductors, such as
\textit{Pb, Sn, Pb-Bi}. The samples edge roughness was claimed to
be about 3 nm. A remarkable feature of the method is the ability
to vary the wire thickness $t$ and, hence, its cross section $s$
at a constant width $w$ \textit{in situ} inside the cryostat
in-between the sessions of truly 4-probe $R(T)$ measurements.

\begin{figure}
\begin{center}
\includegraphics[width=8cm]{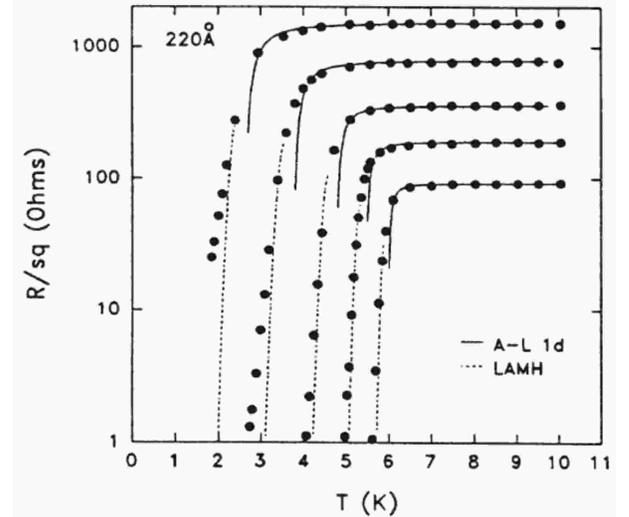}%
\caption{$R(T)$ dependencies for the same \textit{Pb} nanowire with width $w=$
22 nm and various thicknesses. Solid lines at the top of the transitions are
fits to the Aslamazov - Larkin theory. Dashed lines are fits to the TAPS
model, which clearly yields steeper $R(T)$ dependencies compared to the
experimental data for the thinnest wires \cite{Dynes QPS PRL 1993}.}%
\label{F17 Dynes R(T) Pb}%
\end{center}
\end{figure}
\begin{figure}
\begin{center}
\includegraphics[width=8cm]{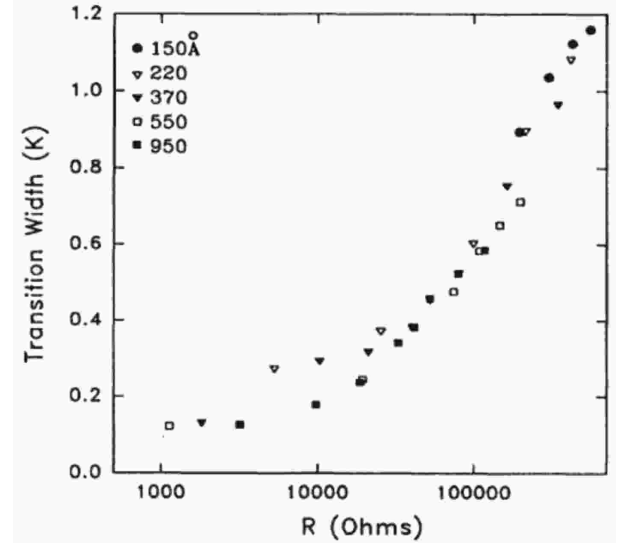}%
\caption{Width of the transition (defined as a temperature interval between
20\% and 80\% of the normal state resistance \cite{Dynes QPS PRL 1993}) as a
function of the normal state resistance $R_{N}$.}%
\label{F18 Dynes R(T)width Pb}%
\end{center}
\end{figure}

The experiments \cite{Dynes QPS PRL 1993} clearly indicated
systematic deviations of the experimental data points $R(T)$ from
the TAPS model predictions (Fig. \ref{F17 Dynes R(T) Pb}). This
discrepancy increases as the wires become narrower. The width of
the superconducting transition was found to scale with the normal
state resistance $R_{N}$ (Fig. \ref{F18 Dynes R(T)width Pb}). It
should be noted that in lead nanowires as narrow as 15 nm and as
thin as 10 nm no low temperature resistance tails were observed.
Instead, a less dramatic but systematic broadening of the
superconducting transition beyond the TAPS limit was noted
\cite{Dynes QPS PRL 1993,Dynes QPS and nMR PRL 1997}. On the other
hand, long resistance tails were always present in tin structures
fabricated and measured using similar technique \cite{Dynes QPS
PRL 1996}. Very probably, the discrepancy comes from the
difference in the heights of the potential barrier $\delta F$
between these two materials (see Table I): Lead wires should be
significantly narrower than tin ones in order to obtain a similar
magnitude of fluctuation effects.

In addition, we point out that the experiments \cite{Dynes QPS PRL
1993,Dynes QPS and nMR PRL 1997,Dynes QPS PRL 1996} demonstrated
clear superonducting transitions reaching experimental "zero" in
nanowires with normal state resistance $R_{N}\gg$ 10 k$\Omega$.
This observation is not in line with the conjecture
\cite{BT,Bezryadin MoGe review JPCM 2008} that superconductivity
in \textit{MoGe} nanowires could only be possible for wire
resistances below the quantum resistance unit $R_{q}\simeq $ 6.45
k$\Omega$.

A clear manifestation of a crucial role of QPS effects in superconducting
nanowires was provided in the experiments by Tinkham and co-workers
\cite{BT,Lau MoGe PRL}. These authors developed a novel technique
which allowed them to fabricate sufficiently uniform superconducting wires
considerably thinner than 10 nm with lengths ranging between $\sim100$ nm and
1 $\mu$m. This was achieved by sputtering a superconducting alloy of amorphous
$Mo_{79}Ge_{21}$ over a free-standing carbon nanotube or bundle of tubes laid
down over a narrow and deep slit etched in the substrate (Fig.
\ref{F8 Bezryadin method}). Three out of eight samples studied in the
experiments \cite{BT} demonstrated no sign of superconductivity even well
below the bulk critical temperature $T_{C}$. Furthermore, in the low
temperature limit the resistance of these samples was found to show a slight
upturn with decreasing $T$. In view of that one can conjecture that these
samples may actually become insulating at $T\rightarrow0$. The resistance of
other five samples \cite{BT} decreased with decreasing $T$. Also for these
five samples no sharp superconducting phase transition was observed.

All three non-superconducting wires (i1,i2 and i3) \cite{BT} had
the normal state resistance above the quantum unit $R_{q}$, while
$R_N$ for the remaining five ``superconducting'' samples was lower
than $R_{q}$. This observation allowed the authors \cite{BT} to
suggest that a dramatic difference in the behavior of these two
groups of samples (otherwise having similar parameters) can be due
to the dissipative quantum phase transition (QPT)
\cite{ZGOZ2,Buchler} (see also Chapter 5) analogous to Schmid
phase transition \cite{sz,weiss} observed earlier in Josephson
junctions \cite{Mikko}. Already at this stage we would like to
emphasize that an important pre-requisite for this QPT is the
existence of a source for linear Ohmic dissipation \textit{at low
energies} which typically requires the presence of some normal
shunt resistance. It appears that no such condition was fulfilled
in the experiments \cite{BT}. Hence, the interpretation of the
data \cite{BT} in terms of a dissipative QPT is problematic.

\begin{figure}
\begin{center}
\includegraphics[width=8cm]{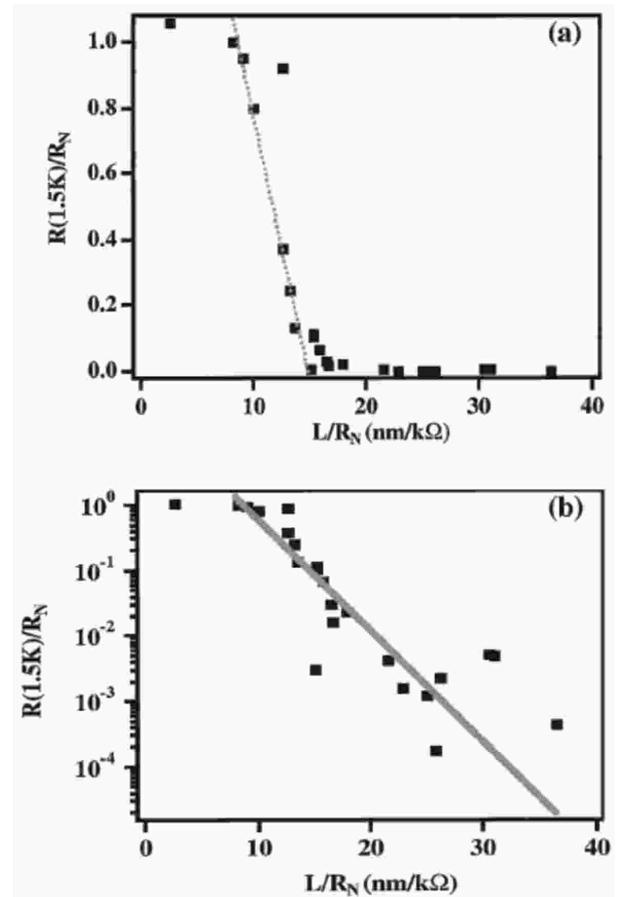}%
\caption{Resistance at 1.5 K normalized to normal state resistance as a
function of $L/R_{N}$. (a) Linear plot. The dotted line is a guide to the eye.
(b) Semilog plot with an exponential fit. Slope of the fitted line is 0.39
k$\Omega$/ nm \cite{Lau MoGe PRL}.}%
\label{F19 Lau R(L)}%
\end{center}
\end{figure}

And indeed, this interpretation was not confirmed in the later experiments of
the same group \cite{Lau MoGe PRL} who observed superconducting behavior in
samples with normal resistances as high as 40 k$\Omega$ $\gg R_{q}$. The
authors \cite{Lau MoGe PRL} concluded that \textquotedblright the relevant
parameter controlling the superconducting transition is not the ratio of
$R_{q}/R_{N}$, but appears to be resistance per unit length, or equivalently,
the cross-sectional area of a wire\textquotedblright\ (Fig.\ref{F19 Lau R(L)}).%

This conclusion clearly favors interpretation of the data \cite{BT,Lau MoGe
PRL} either in terms of a BKT-like QPT \cite{ZGOZ,ZGOZ2} at $\mu\approx2$ or
just as a sharp crossover between the regimes of vanishingly small and
sufficiently high QPS rates $\gamma_{QPS}$ which will correspond respectively
to superconducting and normal behavior of the nanowires. In both cases the
crucial parameter is the wire cross section $s$. Both these QPT and crossover
are expected to occur for wire diameters $\sim10$ nm.

Let us estimate the QPS core action $S_{\mathrm{core}}$ defined by
eq. (\ref{otvet}) for the eight samples studied in the experiments
\cite{BT}. With the density of states $N_{0}=1.86\times10^{13}$
s/m$^{3}$, the superconducting critical temperature
$T_{C}\simeq5.5$ K and the measured resistivity $\rho=1.8$
$\mu\Omega/$m we obtain the coherence length $\xi \simeq7$ nm in
agreement with the estimate \cite{BT}. Our estimates for the
action $S_{\mathrm{core}}$ are summarized in the following Table
II:

\begin{center}%
\begin{tabular}
[c]{|c|c|c|}\hline
sample & $R/X$, k$\Omega$/nm & $S_{\mathrm{core}}$\\\hline
i1 & 0.122 & 7.8 $A$\\\hline
i2 & 0.110 & 8.7 $A$\\\hline
i3 & 0.079 & 12.7 $A$\\\hline
s1 & 0.038 & 25.1 $A$\\\hline
s2 & 0.028 & 33.7 $A$\\\hline
s3 & 0.039 & 22.6 $A$\\\hline
ss1 & 0.054 & 15.4 $A$\\\hline
ss2 & 0.044 & 19.6 $A$\\\hline
\end{tabular}

\end{center}

We observe that for $A \sim1$ the QPS rate $\gamma_{QPS} \propto
\exp(-S_{\mathrm{core}})$ is expected to be much higher in the normal samples
i1, i2 and i3 than in five remaining wires which demonstrated the
superconducting behavior in the low temperature limit. This observation is
consistent with the interpretation in terms of the crossover
between the regimes with low and high QPS rates.

Further support for this interpretation comes from the data \cite{Lau MoGe
PRL} obtained for more than 20 different nanowires. The resistance of these
wires measured at $T=1.5$ K is presented in Fig. \ref{F19 Lau R(L)} versus the
inverse normal resistance per unit length $X/R_{N}\propto s$. In Fig.
\ref{F19 Lau R(L)} (a) one observes a sharp crossover between normal and
superconducting behavior at wire diameters $\sqrt{s}\sim10$ nm. This crossover
could be interpreted as an indication to the QPS-binding-unbinding QPT
\cite{ZGOZ,ZGOZ2} controlled by the parameter $\mu\propto\sqrt{s}$. Note,
however, that all wires studied in \cite{Lau MoGe PRL} were quite short.
Hence, this QPT should inevitably be significantly broadened by finite size effects.

Re-plotting the same data on a semilog scale (Fig. \ref{F19 Lau R(L)} (b)) we
indeed observe a rather broad distribution of measured resistances which --
despite some scatter -- can be fitted to the linear dependence on the wire
cross section $s$. This fit is highly suggestive of the
crossover scenario although it cannot yet rule out the (broadened by size
effects) QPT either.

\begin{figure}
\begin{center}
\includegraphics[width=8cm]{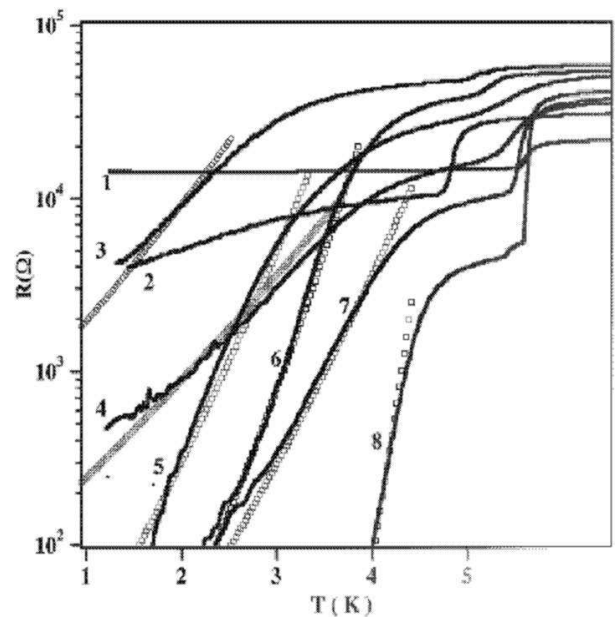}%
\caption{Superconducting transitions of "long" \textit{MoGe} nanowires on top
of insulating carbon nanotube used as the substrate \cite{BT}, \cite{Lau MoGe
PRL}. Double-step shape of the $R(T)$ dependences comes from the
superconducting transition of the contact regions contributing to the 2-probe
measurement configuration. The samples' normal state resistances and lengths
are 1: 14.8 k$\Omega$, 135 nm; 2: 10.7 k$\Omega$, 135 nm; 3: 47 k$\Omega$, 745
nm; 4: 17.3 k$\Omega$, 310 nm; 5: 32 k$\Omega$, 730 nm; 6: 40 k$\Omega$, 1050
nm; 7: 10 k$\Omega$, 310 nm; 8: 4.5 k$\Omega$, 165 nm.\ Symbols stand for
calculations using eq. \ref{RTT} with the single numerical coefficient
$A$=0.7. The critical temperature $T_{C}$ and the dirty-limit coherence length
$\xi(0)$ used as fitting parameters for samples 3-8 are 3: 5.0 K, 8 nm; 4: 6.4
K, 8.5 nm; 5: 4.6 K, 8.9 nm; 6: 4.8 K, 8.9 nm; 7: 5.6 K, 11.9 nm; 8: 4.8 K,
8.5 nm. Data for samples 1 and 2 cannot be fitted by any reasonable set of
parameters.}%
\label{F20 Lau MoGe our fits}%
\end{center}
\end{figure}

In order to finally discriminate between these two options it is necessary to
analyze the temperature dependence of the resistance $R(T)$. According to our
theory, in the linear regime the (caused by QPS) wire resistance is defined by
eq. (\ref{2mu-3}). The dependence $R\propto y^{2}$ comes from pairs of QPS
events where the fugacity (\ref{fug}) depends on temperature via
$S_{\mathrm{core}}\propto\Delta_{0}^{1/2}(T)$. An additional (weak) power-law
dependence $\propto T^{2\mu-3}$ enters because of interactions between quantum
phase slips in space-time. In order to fit the data \cite{Lau MoGe PRL} we
will ignore this power-law dependence and use the simplified formula
\begin{equation}
R(T)=b_{1}\frac{\Delta_{0}(T)S_{\mathrm{core}}^{2}X}{\xi(T)}\exp
(-2S_{\mathrm{core}}),\label{RTT}%
\end{equation}
with $S_{\mathrm{core}}$ defined in (\ref{otvet}) and $b_{1}$ being an
unimportant constant factor. As this formula can only be applied to samples
with vanishing low $T$ resistances $R(T\rightarrow0)\rightarrow0$, we select
the data \cite{Lau MoGe PRL} for the samples 3 to 8 seemingly demonstrating
such a behavior. The results of the fits to eq. (\ref{RTT}) are presented in
Fig. \ref{F20 Lau MoGe our fits}. We observe a very good agreement between our
QPS theory and experiment which was obtained using the same value $A\simeq0.7$
for all six samples.

We also note that the fits remain essentially unchanged if we take into
account an additional power law dependence $\propto T^{2\mu-3}$ (with $\mu$
estimated from the wire parameters). This observation proves that inter-QPS
interaction is indeed insignificant for the interpretation of the data
\cite{Lau MoGe PRL}. Thus, similarly to Lau \textit{et al.} \cite{Lau MoGe
PRL}, we conclude that the temperature dependence of the resistance of their
ultra-thin $MoGe$ wires is determined by QPS effects and the observed sharp
transition between normal and superconducting behavior is most likely a
thickness-governed crossover between the regimes of respectively large and
small QPS rates $\gamma_{_{QPS}}$.%

A bulk of recent experimental data
\cite{Bezryadin MoGe and Nb wires,Bezryadin MoGe TAPS EPL-2006,Bezryadin condmat07}
accumulated on
\textit{MoGe} nanowires fabricated using the molecular template method was
recently reviewed in detail by Bezryadin \cite{Bezryadin MoGe review JPCM
2008}. The main observations can be summarized as follows: ($i$) "shorter"
nanowires ($X\lesssim200$ nm) demonstrate either \textquotedblright weakly
insulating\textquotedblright\ behavior with clear features of weak Coulomb
blockade \cite{Naz,GZ00,GZ04,BN}, or relatively steep superconducting
transition $R(T)$ with virtually no samples showing an intermediate regime,
($ii$) "longer" samples (200 nm $\lesssim X\lesssim$ 1 $\mu$m) typically
showed the behavior which -- similarly to the data \cite{Lau MoGe PRL} -- can
reasonably well be interpreted in terms of a crossover between the regimes of
small and large QPS rates (corresponding to respectively superconducting and
normal behavior).

The $R(T)$ curves of longer wires in the regime ($ii$) showed a decrease of
the resistance with cooling no matter whether their normal state resistance
$R_{N}$ was smaller or bigger than $R_{q}$. The crossover between normal and
superconducting behavior of the samples was controlled by the wire cross
section $s$ or, equivalently, by the ratio $R_{N}/X$. As in Ref. \cite{Lau
MoGe PRL}, the overall picture is consistent with the QPS scenario \cite{GZ01}.

Turning to the regime of shorter wires ($i$), it was argued
\cite{Bezryadin condmat07,Bezryadin MoGe review JPCM 2008} that
the transition between superconducting and normal behavior for
these samples is most likely controlled by the \textit{total}
normal wire resistance $R_{N}$ rather than by the resistance per
unit length $R_{N}/X$, see Fig. 3 of Ref. \cite{Bezryadin
condmat07}. This observation implies that a significant fraction
of short wires with lengths $X\lesssim200$ nm showed
superconducting-like behavior even though the corresponding
estimated values of the QPS core action could be as high as
$S_{\mathrm{core}}\sim1$ (in which case for longer wires one
expects strong proliferation of QPS leading to total destruction
of superconductivity). One should, therefore, explain why  these
samples stay superconducting instead of being normal or even
insulating in the limit of low enough $T$.

An empirical fact that short samples \cite{Bezryadin
condmat07,Bezryadin MoGe review JPCM 2008} stayed normal for
$R_{N}\gtrsim R_{q}$ and turned superconducting as soon as
$R_{N}\lesssim R_{q}$ allowed the authors to assume the presence
of some kind of QPT that could, for instance, be similar to Schmid
dissipative phase transition in Josephson junctions
\cite{s,Chbm,sz,weiss}. Thus, in some sense the authors
\cite{Bezryadin condmat07} revived an earlier idea \cite{BT}, now
in application to short wires only. Supporting this idea Meidan
\textit{et al.} \cite{Meidan,Meidan2} suggested a phenomenological
scheme essentially employing the well known RG analysis developed
for resistively shunted Josephson junctions \cite{s,sz,weiss}.
Unfortunately, as was also noticed in Refs. \cite{Meidan,Meidan2},
the experimental parameters correspond to large fugacity values
$y\sim1$, i.e. the RG approach \cite{s} was employed in Ref.
\cite{Meidan,Meidan2} well beyond its applicability range. Another
-- even more significant problem -- is that in the structures
\cite{Bezryadin condmat07,Bezryadin MoGe review JPCM 2008} no
source for linear Ohmic dissipation seems to exist at low
energies. Meidan \textit{et al.} phenomenologically argued that
such dissipation could occur due to proliferation of QPS inside
the wire (which would make the wire to a large extent normal).
However, even \textit{intrinsically normal} wires in contact with
superconducting electrodes develop a \textit{proximity-induced
gap} in its energy spectrum \cite{GK,BBS2,Been,Pilgram,Zhou,KKZ}
of order Thouless energy $\epsilon_{\mathrm{Th}}=D/X^{2}$. Hence,
Ohmic dissipation can hardly occur in such SNS structures at
energies $\lesssim\epsilon_{\mathrm{Th}}$. Since for short samples
\cite{Bezryadin condmat07,Bezryadin MoGe review JPCM 2008} with
$X\sim40\div200$ nm the energy $\epsilon_{\mathrm{Th}}$ can easily
reach few K, it is hard to expect that a dissipative QPT can occur
in such systems.

The same arguments suggest that the proximity effect may actually play an
important role in the interpretation of the experimental data \cite{Bezryadin
condmat07,Bezryadin MoGe review JPCM 2008} for short samples. Even if
the wire is intrinsically normal, supercurrent can flow through such a weak
link due to the proximity effect. Accordingly, superconductivity in the short
samples with $R_{N}\lesssim R_{q}$ can simply be associated with the onset of
dc Josephson current in an SNS junction. The corresponding scenario is as follows.

In the interesting limit $X\gg\xi$ the Josephson energy $E_{J}(T)$ of an SNS
junction \cite{FNT,WZK,Dubos} is exponentially small at high
temperatures $T\gg\epsilon_{\mathrm{Th}}$. Hence, at such $T$ we have
$E_{J}(T)\ll T$ and the weak Josephson current is fully suppressed by thermal
fluctuations. Upon decreasing temperature one eventually reaches the regime
$T\lesssim\epsilon_{\mathrm{Th}}$, in which case $E_{J}$ becomes \cite{Dubos}
\begin{equation}
E_{J}\sim\epsilon_{\mathrm{Th}}R_{q}/R_{N}.
\end{equation}
Thus, for $R_{N}\lesssim R_{q}$ at $T\ll\epsilon_{\mathrm{Th}}$ we have
$E_{J}\gg T$, i.e. thermal fluctuations are negligible and the SNS junction
becomes superconducting. In the vicinity of $T\sim\epsilon_{\mathrm{Th}}$ this
argument also allows to predict exponential decrease of the wire resistance
$R(T)\propto\exp(\epsilon_{\mathrm{Th}}R_{q}/R_{N}T)$ in a qualitative
agreement with exponential dependencies observed in \cite{Bezryadin MoGe TAPS
EPL-2006,Bezryadin MoGe review JPCM 2008}. On the other hand, for
$R_{N}\gtrsim R_{q}$ at $T\sim\epsilon_{\mathrm{Th}}$ we still have $E_{J}<T$,
i.e. supercurrent is disrupted by thermal fluctuations down to lower $T$ and
significantly broader $R(T)$ curves are expected.

At lower $T$ thermal fluctuations become unimportant but, on the other hand,
quantum fluctuations take over destroying the Josephson current for
$E_{C}\gtrsim E_{J}$, where $E_{C}=e^{2}/2C$ is an effective charging energy
of an SNS junction. Thus, in the low temperature limit one would expect to see
the crossover at $E_{J}\sim E_{C}$ or, equivalently, at
\begin{equation}
R_{q}/R_{N}\sim E_{C}/\epsilon_{\mathrm{Th}}.\label{crov}%
\end{equation}
Now, similarly to the case of Josephson junctions \cite{sz}, one can
demonstrate that the capacitance $C$ is given by the expression $C=C_{g}%
+\delta C$, where $C_{g}$ is the junction geometric capacitance and $\delta C$
is the renormalization term $\delta C\sim\hbar/R_{N}\epsilon_{\mathrm{Th}}$.
Since in our case geometric capacitance is most likely very small $C_{g}%
\ll\delta C$, one has $E_{C}\sim R_{N}\epsilon_{\mathrm{Th}}/R_{q}$.
Substituting this estimate into eq. (\ref{crov}), one immediately obtains the
crossover condition
\[
R_{N}\sim R_{q}
\]
in agreement with experimental findings. We believe that these
simple arguments are sufficient to understand a
superconducting-to-normal crossover observed in short wires
\cite{Bezryadin MoGe TAPS EPL-2006,Bezryadin condmat07,Bezryadin
MoGe review JPCM 2008}. These arguments also demonstrate that for
such samples Josephson physics (both classical and quantum)
appears to be more important than that of TAPS and QPS. Hence,
although short wires can also demonstrate interesting phenomena,
longer wires appear to be more suitable for experimental
investigations of QPS effects.

Now let us turn to other experiments where QPS in superconducting
nanowires have been observed.

The original method of forcing the molten alloy into a porous
media \cite{Bogomolov capillary wire Sov Phys Sol St 1971} or
capillary \cite{Michael In microcylinders JLTP 1974} enables
fabrication of metallic wires with diameters $\gtrsim$1 $\mu$m. At
these early stages of research no studies of the shape of
superconducting transition were performed in such structures.
Chemical or electrochemical methods developed later allowed
filling of very narrow pores with a wide variety of materials
including superconductors: $Pb$ \cite{Yi Pb electrodeposited
nanowires APL 1999,Piraux IV in Pb and Sn nanowire PRL 2003,Piraux
Pb nanowire PRB 2004,Piraux Pb nanowire IV APL 2008}, $Sn$
\cite{Piraux Sn nanowire APl 2004,Piraux Sn nanowire NanoSci
NanoTech 2005,Piraux IV in Pb and Sn nanowire PRL 2003,Piraux
multicontact Sn nanowire APL 2007,Piraux Pb and Sn granular
nanowires APL 2003} and $Zn$ \cite{Tian Zn nanowire PRL 2005,Tian
Zn nanowire PRB 2006,Tian specific heat Zn nanowire PRL 2007}. The
most widely used media are tracketched membranes and alumina
nanoporous films \cite{Piraux Sn nanowire NanoSci NanoTech 2005}.
The diameter of pores in these materials can be rather uniform
ranging from $\sim$15 nm to few $\mu$m. One side of the membrane
is coated with a thick metal film serving as a cathode. The
``sandwich'' is immersed into a corresponding electrolyte.
Standard three-terminal reduction of ions in the solution leads to
growth of nanowires from the cathode. Typically the nanowires are
polycrystalline. However, with properly adjusted electrodeposition
conditions nanowires can be made crystalline \cite{Tian Sn
nanowire PRB 2005}. The method enables fabrication of large
amounts of almost vertically aligned nanowires of desired
morphology and diameters. Electric contacts to a bundle of
nanowires can be made easily connecting the thick bottom cathode
(from which the filaments start to grow) and the top metal plate
on top of the nanopore membrane. The number of nanowires
contacting metallic electrodes depends on squeezing pressure of
the top electrode and the filling factor of the membrane
\cite{Tian Zn nanowire PRB 2006}. For unambiguous interpretation
of the data it is desirable to perform electric measurements on a
single nanowire. If the density of the nanopores is sufficiently
high, a possible solution is to dissolve the hosting membrane,
isolate a single nanowire and fabricate contacts using
conventional lithographic technique \cite{Piraux single nanowire
probing APL 2002,Piraux Sn nanowire NanoSci NanoTech 2005}. A
nanoindentation-based method has been proposed enabling electric
measurements on a single nanowire inside the hosting membrane
\cite{Piraux single wire contact Nanotechnology 2005}.

\begin{figure}
\begin{center}
\includegraphics[width=9cm]{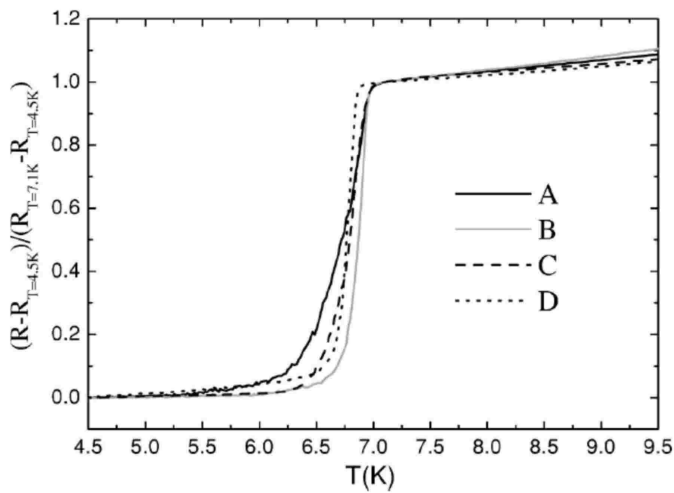}%
\caption{$R(T)$ dependence for four Pb nanowires (A) - (D) with diameters 40,
55, 55 and 70 nm, correspondingly. The transition is broader for narrower
wires \cite{Piraux Pb nanowire PRB 2004}.}%
\label{F21 Piraux R(T) Pb}%
\end{center}
\end{figure}
\begin{figure}
\begin{center}
\includegraphics[width=8cm]{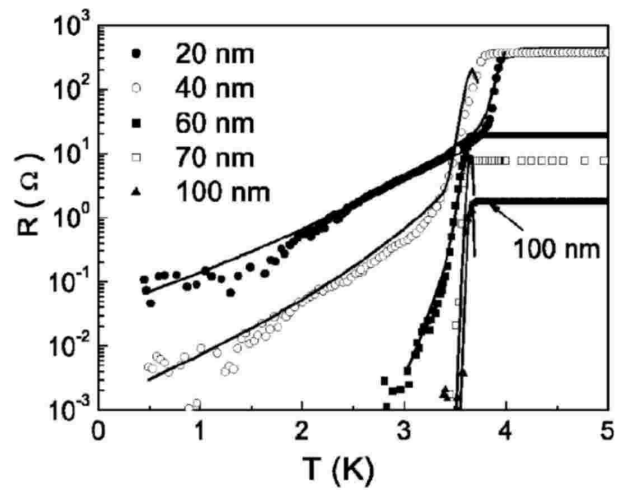}%
\caption{$R(T)$ curves for 20, 40, 60, 70, and 100 nm wide and 6 $\mu$m long
Sn nanowire arrays
containing respectively 18, 1, 8, 15 and 53 wires in the bundle. The solid
lines for 20, 40 and 60 nm wires are the results based on the TAPS model near
$T_{C}$ and QPS model below $T_{C}$ \cite{Tian Sn nanowire PRB 2005}.}%
\label{F22 Tian R(T) Sn}
\end{center}
\end{figure}

Experiments with \textit{Zn} nanowires \cite{Tian Zn nanowire PRL
2005,Tian Zn nanowire PRB 2006,Tian specific heat Zn nanowire PRL 2007} lead
to observations of unexpected ``antiproximity'' effect. The shape of the
supeconducting transition $R(T)$ of a bundle of \textit{Zn} nanowires with
length up to few $\mu$m depends on the material of the electrodes. It was
found that superconductivity is completely or partially suppressed in samples
with superconducting electrodes, while clear superconducting transition has
been observed in systems with normal electrodes. The origin of the phenomenon
is not yet clear, while a model has been proposed \cite{Antiproximity theory
Clarke PRL 2006}.

Experiments with individual \textit{Pb} nanowires grown using nanopore
electrochemical method show clear superconducting transitions, which get
broadened with reduction of the sample diameter (Fig. \ref{F21 Piraux R(T) Pb}%
). Unfortunately no theory fits was provided by the authors, while
the linear scale of the reported $R(T)$ dependencies complicates an
independent quantitative comparison \cite{Piraux Pb nanowire PRB 2004}.

Experiments on bundles of several \textit{Sn} nanowires show clear broadening
of the $R(T)$ transitions in the narrowest samples (Fig.
\ref{F22 Tian R(T) Sn}), which was associated with QPS effects \cite{Tian Sn
nanowire PRB 2005}. As the number of wires in the measured bundle is not known
precisely, quantitative comparison with theoretical predictions appears
complicated. In addition, the experiments were presumably performed outside
the linear regime: Only rather high dc excitation current densities
$j_{bias}\sim$ 10$^{4}$ A/cm$^{2}$ were used being just slightly below the
experimentally measured critical current density at low $T$, $j_{C}(T\ll
T_{C})\sim$ 10$^{5}$ A/cm$^{2}$.

\begin{widetext}

\begin{figure}
\begin{center}
\includegraphics[width=17cm]{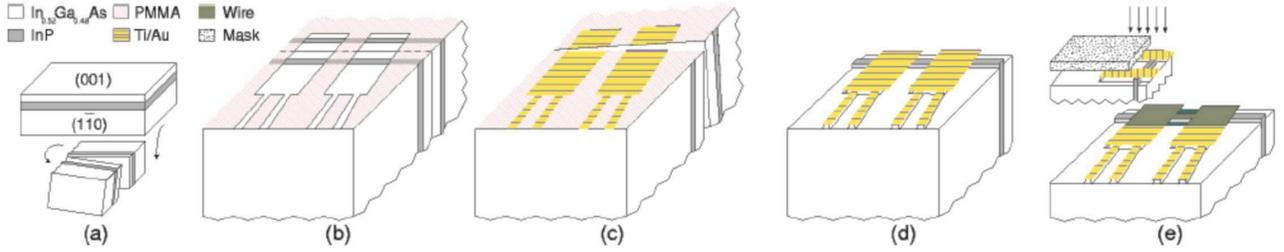}%
\caption{Schematics of the sample fabrication. (a) The sample is cleaved in
small strips which are cut in half and glued together with the two (001) plane
facing each other, (b) PMMA is spun on the (1$\overline{1}$0) crystallographic
plane of the two pieces and a pattern is written using standard EBL and then
developed, (c) thermal evaporation is used to deposit a film of Ti/Au (the
portion of the film deposited on the PMMA is not shown) and then the two
halves are separated, (d) after lift-off and oxygen plasma etching, wet
etching is used to define the InP ridge, (e) the wire is formed
through the final evaporation
appropriately masking the substrate of the sample
(top: side view of the evaporation arrangement, bottom: final result)
\cite{Altomare AuPd and Al nanowire fabrication APL 2005}.}%
\label{F23 Altomare method}%
\end{center}
\end{figure}

\end{widetext}

An advanced method of nanowire fabrication uses MBE grown \textit{InP} layer
on a cleaved \textit{In}$_{\mathit{x}}$\textit{Ga}$_{\mathit{1-x}}%
$\textit{As/InP} substrate as a support for thermally evaporated
metal (Fig. \ref{F23 Altomare method}). The applicability of the
approach was demonstrated on \textit{AuPd} and \textit{Al}
\cite{Altomare AuPd and Al nanowire fabrication APL 2005,Altomare
AuPd and Al nanowire fabrication cond-mat 2007}. The method allows
fabrication of very long (up to 100 $\mu$m) superconducting
(\textit{Al}) nanowires with effective diameters down to 7 nm. The
approach enables pseudo-4-terminal measurements (Fig. \ref{F24
Altomare R(T) Al}). The observed broad $R(T)$ dependencies were
associated with QPS. Though rather scarce experimental data make
quantitative conclusions on the QPS mechanism difficult, the
experiments on long \textit{Al} nanowires \cite{Altomare Al
nanowire PRL 2006,Altomare Al nanowire cond-mat-2005} have a clear
message: No correlations between the total normal state wire
resistance $R_{N}$ (compared to the quantum resistance unit $R_{q}
\simeq 6.45$ k$\Omega$) and superconductivity
in such wires was found.%

\begin{figure}
\begin{center}
\includegraphics[width=8cm]{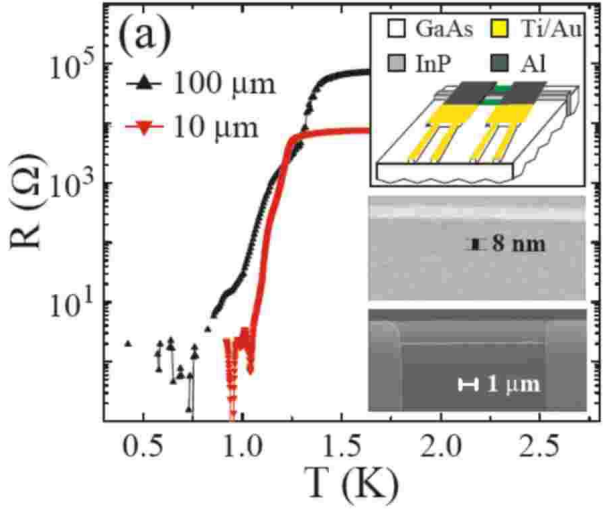}%
\caption{$R(T)$ dependencies for Al nanowires. At temperatures $\lesssim
$1 K, samples s2 100 $\mu  $m long, $R_{N}$ =86 k$\Omega$ ($\blacktriangle$)
and s1: 10 $\mu  $m long, $R_{N}$ = 8.3 k$\Omega$ ($\blacktriangledown$)
became superconducting. The insets show SEM images and schematic of a typical
device containing a 10 $\mu  $m long, nominally 8 nm wide Al nanowire (similar
to samples s1 and s2): the Al layer does not entirely cover the Au/Ti pads
which were measured in series with the superconducting wire (in the main panel
this series resistance has been subtracted) \cite{Altomare Al nanowire
cond-mat-2005}.}%
\label{F24 Altomare R(T) Al}%
\end{center}
\end{figure}

A convenient material to study the phenomena associated with phase slips is
aluminum. Its bulk critical temperature $T_{C}^{bulk}\sim1.2$ K is relatively
low (see Table I), hence, the QPS rate should be comparatively high enabling
pronounced manifestation of fluctuation effects. An additional useful feature
of aluminum is its peculiar size dependence of $T_{C}$. Although the origin of
this effect remains unclear, an increase of $T_{C}$ with reduction of the
characteristic dimension of aluminum structures (wire diameter or film
thickness) is a well-known experimental fact and can be taken as granted
\cite{Shanenko Tc(size) PRB 2006}. This effect does not allow to interpret
broadening of $R(T)$ dependencies at temperatures $T<T_{C}^{bulk}$ in terms of
sample inhomogeneities, such as constrictions.

\begin{figure}
\begin{center}
\includegraphics[width=9cm]{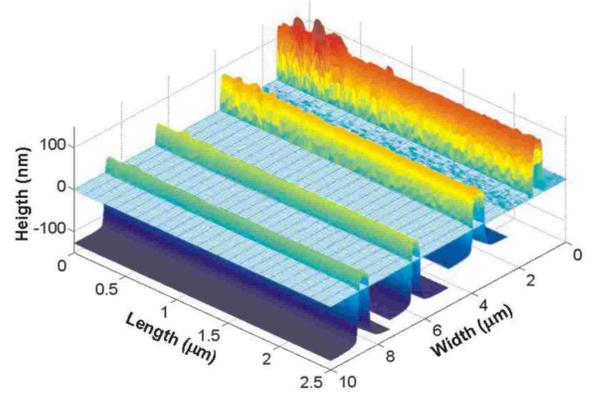}%
\caption{SPM images showing evolution of the shape of the same aluminium
nanowire after several sessions of ion beam sputtering. Bright color above the
horizonatal plane (initial level of the substrate) corresponds to metal, dark
color below indicates sputtered \textit{Si} substrate. Note the reduction of the
initial surface roughness of the nanowire \cite{Zgirski QPS PRB 2008}.}%
\label{F25 Al nanowire sputtering}%
\end{center}
\end{figure}

Aluminium was chosen for investigations of 1D superconductivity in Refs.
\cite{Zgirski NanoLett 2005,Zgirski QPS PRB 2008}. It was demonstrated that
low energy $Ar^{+}$ ion sputtering can progressively and non-destructively
reduce dimensions of various nanostructures including nanowires
\cite{Savolainen IBE method APA 2004,Zgirski IBE method Nanotechnology
2008}. The penetration depth of $Ar^{+}$ ions into \textit{Al} matrix at
acceleration voltages of $\sim$ 500 eV is about 2 nm and is comparable to the
thickness of naturally formed oxide. The accuracy of the effective diameter
determination from the normal state resistance by SEM and SPM measurements is
about $\pm$ 2 nm. Only those samples which showed no obvious geometrical
imperfections were used for further experiments. To a large extend the method
allows one to study the evolution of the size phenomenon, eliminating
artifacts related to uniqueness of samples fabricated in independent
processing runs. The ion beam treatment polishes the surface of the samples
removing inevitable roughness just after fabrication \cite{Zgirski IBE method
Nanotechnology 2008} (Fig. \ref{F25 Al nanowire sputtering}). If there were no
detectable geometrical imperfections in the original (thick) wires, they
could not be introduced in the course of diameter reduction by low energy ion
sputtering.%

\begin{figure}
\begin{center}
\includegraphics[width=8cm]{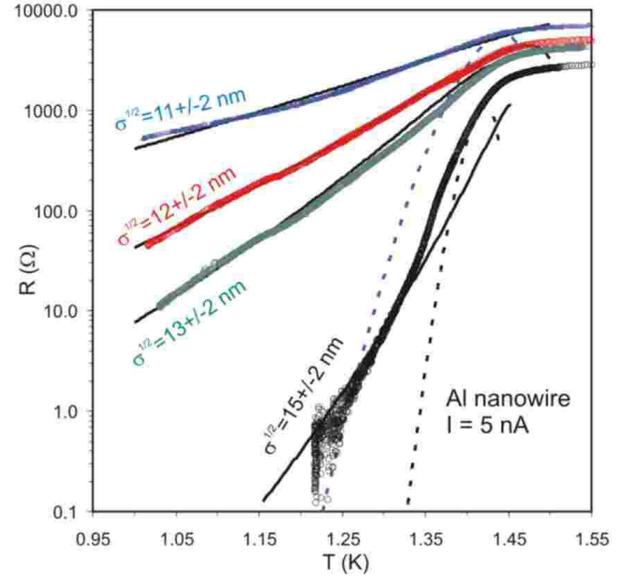}%
\caption{$R(T)$ curves for the thinnest samples obtained by
progressive diameter reduction for the same aluminium nanowire
with length $X$ = 10 $\mu m$. The TAPS model fitting is shown with
dashed lines for 11 and 15 nm samples with the best fit mean free
path $l=$ 3 and 10 nm, correspondingly, $T_{C}=$ 1.46 K and
critical magnetic field $B_{c}(0)=$ 10 mT. Fits to eq. (\ref{RTT})
are shown by solid lines. For 11, 12, 13 and 15 nm wires the fit
parameters are: $A\simeq0.1$; $T_{c}$: 1.64 K, 1.52 K, 1.47 K, and
1.47 K; mean free path $l$ : 7.5 nm, 8.2 nm, 9.5 nm and 9.5 nm;
normal state resistance $R_{N}$ : 7200 k$\Omega$, 5300 k$\Omega$,
4200
k$\Omega$ and 2700 k$\Omega$ \cite{Zgirski QPS PRB 2008}.}%
\label{F26 R(T) Al Zgirski PRB 2008}%
\end{center}
\end{figure}

After a sequence of sputterings (alternated with $R(T)$ measurements) the wire
diameter was reduced from $\sqrt{s}\sim100$ nm down to $\sqrt{s}$ $\lesssim$
10 nm. Experiments were performed on several sets of aluminum nanowires with
length $X$ equal to 1, 5 and 10 $\mu$m. For larger diameters $\sqrt{s}\gtrsim$
20 nm the shape of the $R(T)$ dependence is rather "sharp" and can be
qualitatively described by the TAPS mechanism. Note that the abovementioned
size dependent variation of $T_{C}$ in aluminum nanowires results in
broadening of the $R(T)$ transition and significantly reduces applicability of
the TAPS model \cite{Zgirski inhomogeneity PRB} (Fig.
\ref{F4 R(T) Al inhjomogeneous}). When the wire diameter is further
reduced, deviations from the TAPS behavior become obvious (Fig.
\ref{F26 R(T) Al Zgirski PRB 2008}). Fits to the TAPS model fail to provide
any reasonable quantitative agreement with experiment for diameter values
below $s^{1/2}\leq$ 20 nm even if one hypothetically assumes the existence of
unrealistically narrow constrictions not observed by SPM. And, as we already
discussed, broadening of the $R(T)$ dependencies in aluminum nanowires at
$T<T_{C}^{bulk}$ can hardly be ascribed to geometrical imperfections, such as
constrictions. We conclude that the most natural interpretation of our
observations is associated with quantum fluctuations.

Let us now perform a detailed comparison of the data with theoretical
predictions \cite{ZGOZ,ZGOZ2,GZ01} discussed in Chapter 5. To begin with, we
should select a proper formula for the current-voltage characteristics of the
wire which can be applied to the samples \cite{Zgirski QPS PRB 2008}. One can
employ the general formula (\ref{volt}) derived for sufficiently long wires.
It remains not completely clear if this assumption is well satisfied for the
wires used in experiments \cite{Zgirski QPS PRB 2008}, though it appears that
at least the longest wires with $X=10$ $\mu$m are already in this regime
($X/\xi\gtrsim100$ in this case). In addition, one can recall that in the
experimental temperature range not far from $T_{C}$ there can still be
sufficiently many quasiparticles above the gap which can produce additional
dissipation. Although it would not be fully justified to use eq. (\ref{volt2})
in this case, still dissipation can result in additional inter-QPS interaction
which can slightly modify the power-law dependencies (\ref{2mu-3}),
(\ref{2mui}).

Fortunately, this effect, even if exists, does not change our fitting
procedure. In order to describe the QPS contribution to the wire resistance we
will use the formula
\begin{equation}
R(T)=b_{2}\frac{\Delta_{0}(T)S_{\mathrm{core}}^{2}X}{\xi(T)}\exp
(-2S_{\mathrm{core}})T^{\kappa},\label{b1}%
\end{equation}
while for the non-linear (current-dependent) wire resistance we will use
\begin{equation}
R(I)\propto I^{\kappa}.\label{b2}%
\end{equation}
Here $b_{2}$ is an unimportant constant which remains the same for all
samples. Let us again remind the reader that the dependence $R \propto y^{2}$
comes from pairs of QPS events. As we already discussed, the fugacity
(\ref{fug}) depends on temperature via $S_{\mathrm{core}} \propto\Delta
_{0}^{1/2}(T)$. An additional power-law dependence enters because of inter-QPS
interaction. The parameter $\kappa$ can be equal to $\kappa=2\mu-3$ (as in
eqs. (\ref{2mu-3}), (\ref{2mui})) or take somewhat different values due to
dissipative contribution. Below we will just use it as a fit parameter instead of
the parameter $\mu$.

\begin{figure}
\begin{center}
\includegraphics[width=8cm]{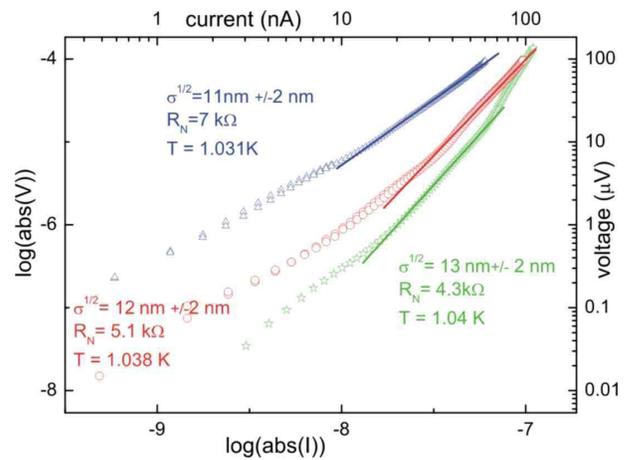}%
\caption{$V(I)$ dependencies of the same samples as in Fig. 26
taken at close temperatures stabilized with accuracy $\pm$ 0.1 mK.
Solid lines correspond to the dependence $R \sim I^{\kappa}$ with
$\kappa \simeq $ 0.58, 1.31 and 1.49 (from top to bottom)
\cite{Zgirski QPS PRB 2008}.}%
\label{F27 I-V Al PRB 2008}%
\end{center}
\end{figure}

We first determine this parameter by fitting the $I-V$ curves at a given
temperature to the dependence (\ref{b2}), see Fig. \ref{F27 I-V Al PRB 2008}.
From these fits we obtain the values for parameter $\kappa$ (and,
correspondingly, for $\mu$) for each sample. Now we perform the fits of the
resistance data $R(T)$ taken in the linear regime $I<I_{0}=k_{B}T_{C}/\Phi
_{0}\ll I_{c}$ to the dependence (\ref{b1}). There are four parameters to
specify for our fits: critical temperature $T_{C}$, normal state resistance
$R_{N}$, electron elastic mean free path $l$ (used to re-calculate the dirty
limit coherence length $\xi=0.85(\xi_{0}l)^{1/2}$ ), and the numerical factor
$A$ of order unity. The critical temperature and the normal state resistance
can be trivially deduced from the experimental data for $R(T)$. The electron
mean free path $l$ can be roughly estimated from the normal state resistivity
$\rho_{N} $ as the product $\rho_{N} l=5\times10^{-16}\Omega$m$^{2} $ is a
well-tabulated value for dirty aluminum. Since the cross section for
ultra-narrow nanowires is known from SPM measurements within $\pm$ 2 nm
accuracy, there remains a certain room to choose particular values of the mean
free path. As a rule of thumb, for all \textit{Al} nanowires with effective
diameter $\sqrt{s}\lesssim$ 20 nm the best-fitted mean free path (at low
temperatures) was found to be roughly equal to one-half of the diameter (Fig.
\ref{F26 R(T) Al Zgirski PRB 2008}, caption). This estimate appears quite
reasonable taking into consideration that at these scales and temperatures
electron scattering occurs merely at the sample boundaries.

Our experimental data and the fits to the theoretical dependence
are presented in Fig. \ref{F26 R(T) Al Zgirski PRB 2008}. We
observe that as the sample diameter decreases the $R(T)$ curves
become progressively broader, exactly as it is predicted by the
QPS theory. Our fits demonstrate a good agreement between theory
and experiment, thus confirming the important role of QPS effects
in our thinnest wires. For all aluminum nanowires the best-fit
value for the parameter $A$ was determined to be $A\simeq0.1$.
This value is smaller than that extracted from the fits to the
experimental data by Lau \textit{et al.} \cite{Lau MoGe PRL}. We
believe that this difference can be attributed to different
geometry and degree of inhomogeneity of samples used in these two
experiments.

We also point out that the temperature dependence of all fitted curves $R(T)$
is merely determined by that of the QPS action $S_{\mathrm{core}}\propto
\Delta_{0}^{1/2}(T)$ which enters the exponent in the expression for the QPS
fugacity (\ref{fug}) whereas an additional power-law dependence $\propto
T^{\kappa}$ turns out to be insignificant: The fits remain essentially
unchanged if we set $\kappa\rightarrow0$. This is quite natural since
$S_{\mathrm{core}}\gg\kappa$ and, on top of that, the temperature dependence
of $S_{\mathrm{core}}(T)$ dominates over that emerging from QPS interaction in
the temperature interval under consideration.%

To complete our discussion of QPS experiments we point out that very
recently an additional experimental evidence for quantum phase slip effects
in short MoGe wires was reported \cite{Bezr08}. In this experiment the
superconducting wires were biased by a large current (just slightly below
the critical one) in order to decrease an effective potential barrier for QPS.
Similarly to macroscopic quantum tunneling experiments (see,
e.g., Ref. \cite{Lukens} and further references therein) the contribution of the QPS
mechanism was identified in Ref. \cite{Bezr08} by observing
low temperature broadening of the distribution for switching currents
at which the nanowire is driven to the
normal state by quantum fluctuations of the order parameter. Further
quantitative investigations of this effect as well as the analysis of
the role of possibly existing
external noise and some other extrinsic factors would be highly desirable.

Summarizing our analysis of the experimental data on
superconducting nanowires, we can conclude that homogeneity of
these wires is one of the central issues which can influence the
data interpretation. The methods enabling experiments on the same
nanowire while increasing \cite{Dynes QPS PRL 1993,Dynes QPS and
nMR PRL 1997,Dynes QPS PRL 1996} or decreasing \cite{Zgirski
NanoLett 2005,Zgirski QPS PRB 2008} its characteristic dimension
to a large extent eliminate artifacts related to the uniqueness of
samples fabricated in independent processing runs. On the other
hand, series of experiments on $MoGe$ wires \cite{BT,Lau MoGe
PRL,Bezryadin MoGe review JPCM 2008} provide large statistics of
independent samples. Another conclusion is that longer nanowires
appear to be more suitable for experimental investigations of QPS
effects while the behavior of very short wires could be to a large
extent determined by the Josephson physics (classical or quantum)
which can essentially mask the QPS one.

The whole scope of the analyzed experimental data obtained in different groups
within different fabrication methods employing different materials clearly
demonstrates that at sufficiently low temperatures quantum phase slips provide
an important mechanism which may cause non-zero resistance of superconducting
nanowires. A vast majority of the data on wires with diameters in the 10 nm
range confirms that the TAPS scenario cannot explain the broad $R(T)$
dependencies observed in these structures and favors QPS
\cite{ZGOZ,ZGOZ2,GZ01} as the only possible scenario. Although in general this
scenario can include a number of rather sophisticated physical effects, such
as quantum phase transitions (see Chapter 5), it appears that in many cases
the rapid change from superconducting to normal behavior can simply be
interpreted as a crossover between the regimes of low and high QPS rates
\cite{GZ01} controlled by the wire cross section $s$ or, equivalently, by the
the wire normal state resistance per unit length $R_{N}/X$. Also the
dominating contribution to the temperature dependence observed, e.g., in
sub-15-nm $MoGe$ \cite{BT,Lau MoGe PRL} and aluminium \cite{Zgirski QPS PRB
2008} nanowires is well described by that of the QPS rate
\cite{ZGOZ,ZGOZ2,GZ01}, see Figs. \ref{F20 Lau MoGe our fits} and
\ref{F26 R(T) Al Zgirski PRB 2008}. Although more experiments would be highly
desirable, already at this stage we can conclude that the existing
experimental data unambiguously confirm our understanding of basic features of
QPS physics in superconducting nanowires.

\subsection{Open questions and related topics}

\subsubsection{Negative magneotresistance}

Application of a weak perpendicular magnetic field to thin
superconducting strips governed by fluctuations revealed an
unusual effect: negative magnetoresistance (nMR) (Figs. \ref{F28
nMR Dynes} and \ref{F29 nMR Zgirski}). The phenomenon has been
observed in lead \cite{Dynes QPS and nMR PRL 1997} and aluminium
\cite{Zgirski QPS PRB 2008} nanowires. The effect is seen only in
very narrow quasi-1D superconducting channels in the resistive
state sufficiently far from the critical temperature $T_{C}$. A
trivial explanation related to the Kondo mechanism is problematic,
as the presence of magnetic impurities in those experiments is
believed to be negligible and -- in any case -- has not been
independently proven. Even if we were to assume the presence of
Kondo impurities, several features of nMR would still remain
unclear. First, the corresponding magnetic fields are too small to
polarize magnetic moments of any impurities at such values of $T$.
In addition, it is known that aluminum is immune to creation of
localized magnetic moments with concentration up to few at \%
\cite{Magnetic ions in Al Jalkanen SolStComm 2007}. Second, there
is a pronounced diameter dependence, making nMR observable only in
thinnest wires. Third, the onset of superconductivity is not
affected by weak magnetic fields: nMR is observed only at the
bottom part of the $R(T)$ transition. Perhaps we should also
mention another effect -- enhancement of the critical current by
magnetic field -- which was observed in \textit{MoGe} and
\textit{Nb} nanowires \cite{Bezryadin nMR MoGe and Nb PRL 2006}
and attributed to interplay between spin-exchange scattering from
residual magnetic impurities and the orbital and Zeeman effects
\cite{Bezryadin nMR theory EPL 2006}. However, it remains unclear
if this effect is related to nMR in any way. On top of that,
presently there exists no evidence for the presence of magnetic
impurities in experiments with aluminium
\cite{Zgirski QPS PRB 2008} and lead \cite{Dynes QPS and nMR PRL 1997}.%

\begin{figure}
\begin{center}
\includegraphics[width=8cm]{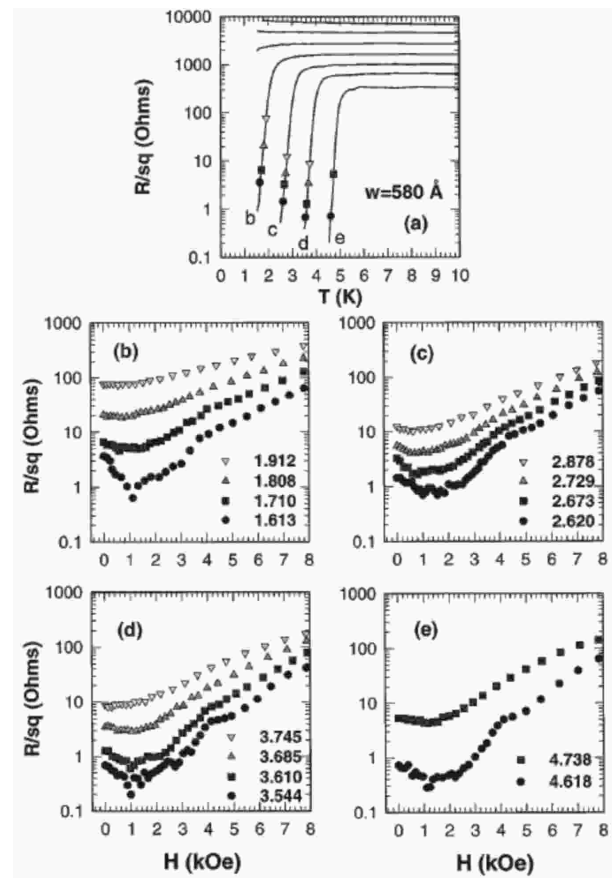}%
\caption{(a) Resistive transitions in zero magnetic field for 58 nm wide lead
wires of different thicknesses. Superconductivity is suppressed with reduction
of the wire thickness. The symbols represent the points at which the
magnetoresistance was measured. (b)--(e) Magnetoresistance for the wire at
different thicknesses. The numbers indicate the temperature in Kelvin
\cite{Dynes QPS and nMR PRL 1997}.}%
\label{F28 nMR Dynes}%
\end{center}
\end{figure}
\begin{figure}
\begin{center}
\includegraphics[width=8cm]{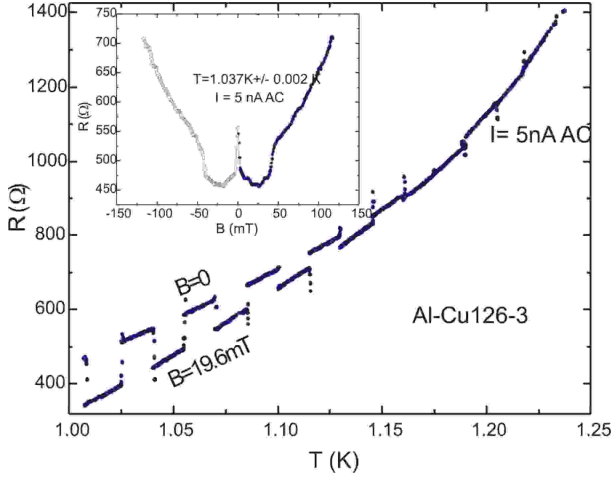}%
\caption{Slowly ($\sim$1 h) recorded dependence $R(T,B_{ext}=const)$
for the 11-nm sample from Fig. \ref{F26 R(T) Al Zgirski PRB 2008}.
The perpendicular magnetic field
$B_{ext}=$19.6 mT was switched on and off several times while sweeping
the temperature. The top branch corresponds to zero
field, while the lower one to the field ``on''.
Inset: resistance versus perpendicular magnetic field
$R(B_{ext},T=const)$ for the same sample measured at constant temperature and
small ac current \cite{Zgirski QPS PRB 2008}.}%
\label{F29 nMR Zgirski}%
\end{center}
\end{figure}

NMR has been observed in tin microbridges \cite{Kadin pair breaking JLTP 1978}
and aluminium nanostructures demonstrating the resistive transition anomaly
(see below) \cite{Santhanam nMR anomaly}. However, the dimensions of those
structures were significantly larger as compared to ultrathin nanowires
\cite{Dynes QPS and nMR PRL 1997,Zgirski QPS PRB 2008}. Presumably the origin
of nMR in these earlier reports can be ascribed to charge imbalance
\cite{Kadin pair breaking JLTP 1978} or to sample inhomogeneity
\cite{Arutyunov R(T) anomaly PRB 1999}. NMR has been predicted in disordered
superconducting wires \cite{PA}. However, the contribution responsible for nMR
within this model is exponentially small as compared to the effective
resistance produced by TAPS. This is clearly not the case in experiments
\cite{Dynes QPS and nMR PRL 1997,Zgirski QPS PRB 2008}.

A plausible explanation of the nMR effect could be related to the reduction of
the energy gap $\Delta_{0}$ in the magnetic field. It was argued \cite{ZGOZ2}
that this reduction leads to the following trade-off. On one hand, the core
action $S_{\mathrm{core}} \propto\Delta_{0}^{1/2}$ decreases with increasing
magnetic field and, hence, the QPS rate becomes bigger. On the other hand, the
quasiparticle resistance $R_{qp}\propto\exp(\Delta_{0}/T)$ decreases too
implying increasing dissipation which, in turn, suppresses the QPS rate.
Provided the latter trend dominates over the former one, the contribution of
quantum fluctuations to the wire resistance gets reduced with increasing
magnetic field, thus leading to nMR \cite{ZGOZ2}. Yet another idea employs
possible formation of a charge imbalance region accompanying each phase slip
event \cite{Arutyunov nMR Physica C 2008}. This non-equilibrium region, if it
exists, would provide dissipation outside the core of a phase slip. The
corresponding Ohmic contribution can be effectively suppressed by the magnetic
field, resulting in nMR. However, so far the validity of the charge imbalance
concept was only demonstrated at temperatures sufficiently close to $T_{C}$
and its applicability to QPS is by no means obvious. This mechanism still
requires a solid theoretical justification.

\bigskip

\subsubsection{Step-like R(T) dependencies in crystalline 1D structures}

\begin{figure}
\begin{center}
\includegraphics[width=8cm]{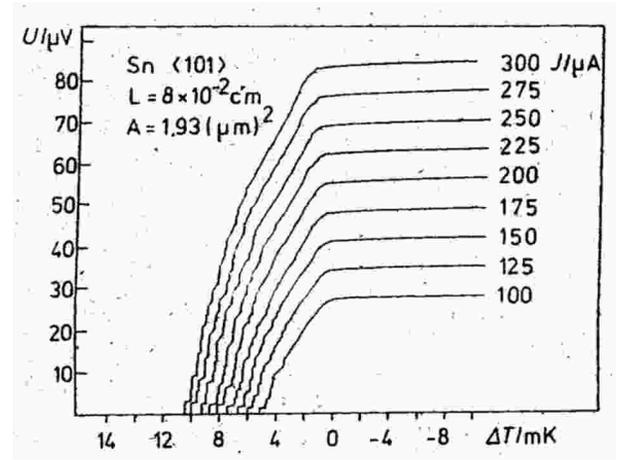}%
\caption{Step-like $V(T)$ dependence in tin whisker at various excitation
currents \cite{Meyer RT steps}.}%
\label{F30 Meyer R(T) steps}%
\end{center}
\end{figure}

\begin{figure}
\begin{center}
\includegraphics[width=8cm]{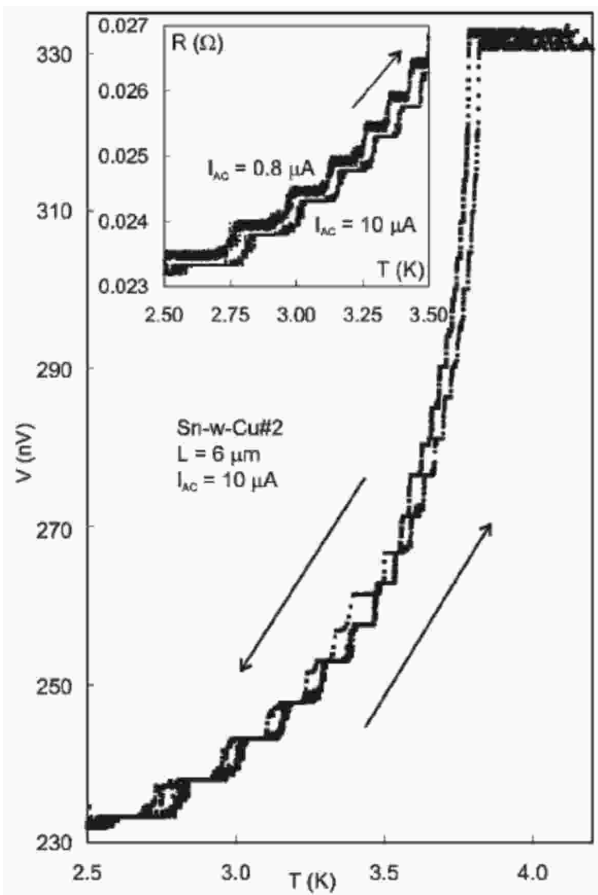}%
\caption{An example of a pronounced step-like $V(T)$ dependence in
a hybrid whisker-based structure similar to that of Fig. \ref{F6
SOG structure}. Arrows indicate the direction of the temperature
sweep. Small hysteresis is a consequence of the temperature
measurement finite response time. The inset shows an enlarged view
of several $R(T)$ steps measured at two different currents. The
curves are shifted slightly due to zero offset drift of the
front-end battery powered preamplifier \cite{Arutyunov SOG whiskers}.}%
\label{F31 step R(T) SOG whisker}%
\end{center}
\end{figure}

Rather unusual step-like shape of $R(T)$ transition has been reported in tin
whiskers \cite{R(T) steps Webb-Warburton PRL 1968}. In relatively long
structures $X\sim0.8$ mm with "manually" fixed electrodes using conducting
epoxy and/or Wood metal \cite{Meyer RT steps} the steps were observed at
relatively high measuring current densities $j\sim10^{4}$ A/cm$^{2}$ (Fig.
\ref{F30 Meyer R(T) steps})\textbf{. } In hybrid whisker-based structures with
lithographically-fabricated electrodes \cite{Arutyunov SOG whiskers} (Fig.
\ref{F6 SOG structure}) the steps on $R(T)$ dependencies were observed between
closely located probes ($X<$ 10 $\mu$m) and at current densities as low as
$j\sim10^{2}$ A/cm$^{2}$\textbf{\ }(Fig. \ref{F31 step R(T) SOG whisker}%
)\textbf{. } For comparison, the experimental critical current
density in pure tin whiskers at sufficiently small temperatures
reaches its theoretical value $j_{c}(0)\sim10^{7}$ A/cm$^{2}$. The
origin of the phenomenon is not clear. The step-like $R(T)$
dependencies were not reported in similar experiments using tin
whiskers \cite{Webb R(T) in Sn whiskers,Tinkham R(T) in Sn
whiskers}. Note that the steps on $R(T)$ transition should not be
confused with step-like $I-V$ characteristics observed in 1D
superconducting wires very close to the critical temperature and
at bias currents exceeding the temperature-dependent critical
value $I>I_{C}(T)$. The origin of the $I-V$ steps is usually
associated with current-induced non-equilibrium effects
\cite{Tidecks,Kopnin book}.

\subsubsection{Resistive transition anomaly}

\begin{figure}
\begin{center}
\includegraphics[width=8cm]{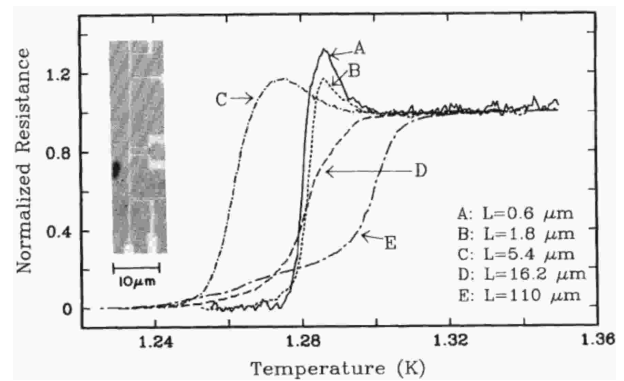}%
\caption{Normalized resistance as a function of temperature for five
aluminium samples A-E showing the resistance anomaly. Longer wires display a
smaller and broader peak. Inset: The sample configuration for three shortest
wires \cite{Santhanam R(T) anomaly}.}%
\label{F32 Santhanam RT anomaly}%
\end{center}
\end{figure}

Intensive studies of various lift-off fabricated quasi-1D nanostructures
revealed an unusual "resistive transition anomaly": Resistance increase above
the normal state value $R_{N}$ at the top of the superconducting transition
\ (Fig. \ref{F32 Santhanam RT anomaly})
\cite{Santhanam R(T) anomaly,Moschalkov Little-Parks anomaly,Moshchalkov R(T)
  anomaly,Arutyunov R(T) anomaly PRB 1999,Kim R(T) anomaly Physica B 1994,Kwong R(T) anomaly PRB 1991,Park
  R(T) anomaly PRL 1995,Strunk R(T) anomaly PRB
  1996,Strunk R(T) anomaly Superlat 1996,Strunk R(T) anomaly JAP 1998,Strunk R(T) anomaly PRB 1998}. No
special correlation has been observed between the absolute value
of the sample length $X$ (distance between the probes) and the
magnitude of the anomaly. Pronounced "bumps" were observed only in
relatively short structures with width $w$ $\gtrsim X/10$. No
dependence upon the film thickness within the range 30--90 nm has
been noticed either. The same statement holds for the sample
topology: both single-connected and non-single-connected
structures displayed the bumps. For the same multiterminal
structure, the anomaly could be clearly observed for a pair of
voltage contacts, while the neighboring segment showed no signs of
the effect. For a given set of contacts, the magnitude of the
resistive anomaly sometimes depends on the cooling history.
Heating up to $\sim$50 K could eliminate the effect. Variation of
the measuring current or application of the magnetic field can
modify the magnitude of the bump (Fig. \ref{F33 R(T) anomaly
current-field}).

\begin{figure}
\begin{center}
\includegraphics[width=8cm]{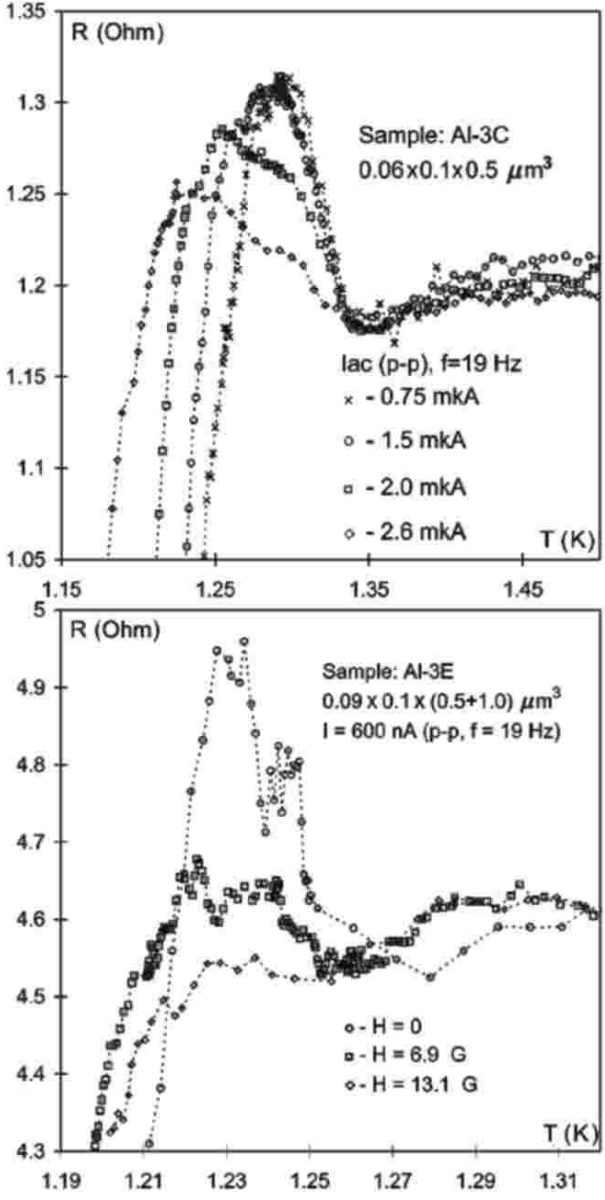}%
\caption{Top part of the resistive transition $R(T)$ for aluminium
multiterminal nanostructure at different measuring ac currents (top) and at
various external magnetic fields normal to the sample surface (bottom)
\cite{Arutyunov R(T) anomaly PRB 1999}.}%
\label{F33 R(T) anomaly current-field}%
\end{center}
\end{figure}

Let us point out that experimentalists frequently use lock-in technique in
order to increase the signal-to-noise ratio. Should a dc component
occasionally occur (e.g., due to improper grounding), experimental
dependencies could become proportional to $dV/dI(T)$ rather than to the
desired $R(T)\equiv V_{ac}/I_{ac}$ resulting in "veird"-looking $R(T)$ curves.
Utilization of rf filters, such as widely used commercial $\pi$-filters, can
also contribute to non-monotonous $R(T)$ transitions due to re-arrangement of
the measuring ac current through the ground loop even at rather low
frequencies $\sim$10 Hz.

Various physical reasons for the resistive transition anomaly in
superconducting nanostructures have been proposed, such as the
fluctuation-governed resistive state \cite{Moschalkov R(T) anomaly
PRB 1994,Moschalkov R(T) anomaly PRB 1997,Arutyunov R(T) anomaly
PRB 1996} nonequilibrium quasiparticle charge imbalance within the
locus of N/S boundary \cite{Kwong R(T) anomaly PRB 1991,Park R(T)
anomaly PRL 1995} or phase-slip centers \cite{Santhanam R(T)
anomaly,Arutyunov R(T) anomaly PRB 1996}. It should be noted that
the resistive transition anomaly has only been reported in
relatively inhomogeneous lift-off fabricated nanostructures with
the width of the $R(T)$ transition strongly exceeding that
predicted by the fluctuation models. It has been shown that the
phenomenon can be qualitatively explained by simple geometric
considerations concerning the shape of the N/S phase boundary of
an inhomogeneous wire in the resistive state \cite{Landau R(T)
anomaly PRB 1997}. The approach has been extended \cite{Arutyunov
R(T) anomaly PRB 1999} demonstrating that the combination of the
two effects provides reasonable agreement with experiments: (i) a
geometry effect of the formation of non-perpendicular (to the wire
axis) N/S boundaries and the corresponding rearrangement of
current across the wire, and (ii) the existence of nonzero and
strongly anisotropic effective resistance of the nonequilibrium
superconducting region close to the N/S interface. Considered
separately, neither of these two contributions could provide
quantitative agreement with the experimental data.

To summarize, it is possible that the resistive transition anomaly originates
from ``dirty physics'' related to inhomogeneity of the finite length lift-off
fabricated wires. This is supported by the fact that no signs of the anomaly
were reported in pure 1D systems \cite{Webb R(T) in Sn whiskers,Tinkham R(T) in Sn whiskers,Meyer RT steps,Arutyunov SOG
whiskers}.

\subsubsection{$T_{C}$ dependence on wire diameter.}

Since early days of low-dimensional superconductivity it is known that the
critical temperature of thin films $T_{C}^{2D}$ and wires $T_{C}^{1D}$\ differ
from the corresponding bulk value $T_{c}^{bulk}$. In indium, aluminum and zink
nanowires $T_{C}$ increases with decreasing the characteristic dimension
\cite{Giordano QPS PRL 1988,Altomare Al nanowire PRL 2006,Zgirski NanoLett
2005,Tian Zn nanowire PRL 2005}. In lead, niobium and MoGe an opposite
tendency has been observed \cite{Dynes QPS PRL 1993,Bezryadin Nb
nanowires,Bezryadin MoGe and Nb wires,BT,Bezryadin MoGe TAPS EPL-2006}. No
noticeable variations of $T_{C}$ have been reported in tin nanowires
\cite{Tian Sn nanowire APL 2003,Tian Sn nanowire PRB 2005,Piraux Sn nanowire
APl 2004,Piraux Sn nanowire NanoSci NanoTech 2005}.

The early models considered size-dependent modification of the
electron-phonon spectrum and found the corresponding variation of
the critical temperature of thin films \cite{Tc(size) theory
McMillan PR 1968,Tc(size) theory Garland PRL 1968,Tc(size) Allen
PR69}. However, later it was noticed that the phonon spectrum
renormalization can significantly contribute to variations of
$T_{C}$ only in ultra-thin structures, while experimentally the
effect is observed also in rather thick films (e.g. up to
$t\sim50$ nm in the case of aluminum), where the phonon spectrum
is almost identical to that of bulk samples. Additionally, it was
found that the critical temperature $T_{C}^{2D}$ of ultra-thin
quench-condensed films depends not only on thickness, but also on
annealing made \textit{in situ } in the measuring chamber at
cryogenic temperatures \cite{Tc(size) experiment Ga Parshin PRB
1996}, indicating that the morphology of the film is also
important. Another significant effect is an interplay between
disorder and electron-electron interactions which can lead to
substantial suppression of $T_{C}$ in reduced dimensions
\cite{Fin}. On the other hand, the so-called shape resonance
effect \cite{Tc(size) theory Shanenko PRB 2006,Shanenko Tc(size)
PRB 2006} can be responsible for the enhancement of $T_{C}$. To
conclude, after years of intensive experimental and theoretical
studies, the origin of the size dependence of the critical
temperature $T_{C}$ remains not fully understood and in many cases
it can be determined by a number of different factors.

\section{Persistent currents in superconducting nanorings}

\subsection{Persistent currents and quantum phase slips}

\begin{figure}
\begin{center}
 \includegraphics[width=8cm]{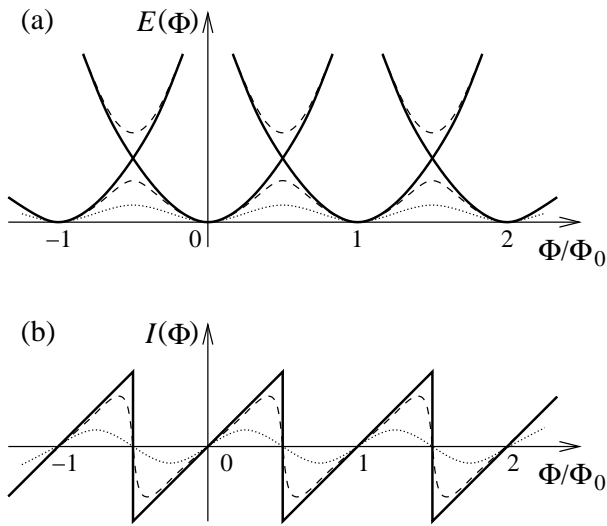}%
\caption{The energy (a) and persistent current (b) for a superconducting
nanoring as functions of magnetic flux without fluctuations, with weak and
with strong QPS effects shown respectively by solid, dashed and dotted lines \cite{MLG}.}%
\label{F34 energy and curent vs flux}%
\end{center}
\end{figure}

It is well known that superconducting rings pierced by external magnetic
flux $\Phi_{x}$ develop circulating persistent currents (PC) which never
vanish. This phenomenon is a fundamental consequence of macroscopic phase
coherence of Cooper pair wave functions. In the case of bulk metallic rings
fluctuations of the phase $\varphi$ of the order parameter can be neglected,
i.e. $\varphi$ can be considered as a purely classical variable. In this case
the total phase difference $\varphi(X)-\varphi(0)$ accumulated along the ring
circumference $X=2\pi R$ is linked to the external flux $\Phi_{x}$ inside the
ring by the well known relation
\begin{equation}
\varphi(X)-\varphi(0)=2\pi p+\phi_{x},
\end{equation}
where $p$ is an integer number, $\phi_{x}=\Phi_{x}/\Phi_{0}$ and, as before,
$\Phi_{0}$ is the superconducting flux quantum. This relation implies the
existence of a phase gradient along the ring which, in turn, means the
presence of discrete set of current and energy states labelled by the number
$p$. At sufficiently low temperatures $T\ll\Delta_{0}$ quasiparticles are
practically irrelevant and the grand partition function of the ring takes the
form
\begin{equation}
{\mathcal{Z}}_{\phi_{x}}=\sum_{p=-\infty}^{\infty}\exp(-E_{p}(\phi
_{x})/T),\label{Z0}%
\end{equation}
where
\begin{equation}
E_{p}(\phi_{x})=\frac{E_{R}}{2}(p+\phi_{x})^{2},\quad
E_{R}=\frac{\pi^2\hbar
^{2}N_0D\Delta_0s}{R}\label{energ}%
\end{equation}
defines flux-dependent energy states of a diffusive
superconducting ring with radius $R$ and cross-section $s$. The
ground state energy $E(\phi_{x})=$min$_{p}E_{p}(\phi_{x})$ is a
periodic function of the flux $\Phi_{x}$ with the period
$\Phi_{0}$. The derivative of the ground state energy with respect
to the flux defines the persistent current
\begin{equation}
I(\Phi_{x})=c\left(  \frac{\partial E(\Phi_{x})}{\partial\Phi_{x}}\right).
\label{defper}%
\end{equation}
Combining eqs. (\ref{energ}) and (\ref{defper}) one finds
\begin{equation}
I=2\pi emv_{s}N_0D\Delta_0s,
\label{evcur}%
\end{equation}
where $m$ is the electron mass and
\begin{equation}
v_{s}(\phi_{x})=\frac{\hbar}{2mR}\mathrm{min}_{p}\left(  p+\frac{\Phi_{x}%
}{\Phi_{0}}\right) \label{st}%
\end{equation}
is the superconducting velocity. Both the current $I$ and the velocity $v_{s}
$ are periodic functions of the magnetic flux $\Phi_{x}$ showing the familiar
sawtooth behavior, see also Fig. \ref{F34 energy and curent vs flux}.%

This picture remains applicable as long as fluctuations of the
superconducting phase $\varphi$ can be neglected, i.e. as long as
the ring remains sufficiently thick. Upon decreasing the thickness
of a superconducting wire $\sqrt{s}$ down to values $\sim10$ nm
range one eventually reaches the new regime in which quantum
fluctuations of the phase gain importance and, as it will be
demonstrated below, essentially modify the low temperature
behavior of superconducting nanorings.

As we have already learned, at low $T$ the most important fluctuations in
superconducting nanowires are quantum phase slips. Each phase slip event
implies transfer of one flux quantum $\Phi_{0}$ through the wire out of or
into the ring and, hence, yields the change of the total flux inside the ring
by exactly the same amount. In other words, each QPS yields a jump between two
neighboring energy states $E_{p}$ (\ref{energ}) (i.e. between two neighboring
parabolas in Fig. 34). As a result of such jumps the flux inside the ring
fluctuates, its average value $\langle\Phi(\Phi_{x})\rangle$ decreases and so
does the persistent current $I$. One can also anticipate that the magnitude of
this effect should increase with the ring perimeter $X$. This is because the
QPS rate $\gamma_{QPS}$ increases linearly with $X$, i.e. the bigger the ring
perimeter the higher the probability for QPS to occur anywhere along the ring.

A quantitative theory of QPS effects in nanorings was proposed by
Matveev \textit{et al.} \cite{MLG}. These authors employed a model
of a closed chain of Josephson junctions in which case QPS can
occur only across such junctions. Here will extend our theory of
QPS (Chapter 5) and analyze the effect of quantum fluctuations on
persistent currents in \textit{uniform} superconducting nanorings.
The key conclusions remain the same for both models.

Before turning to technical details it is instructive to point out a formal
equivalence between the phenomenon discussed here and charging effects in
ultrasmall Josephson junctions (or Cooper pair boxes) \cite{sz}. Indeed, the
energy states of our ring (\ref{energ}) driven by the (normalized) external
magnetic flux $\phi_{x}$ (see Fig. \ref{F34 energy and curent vs flux}) are
fully analogous to such states of a capacitor in a Cooper pair box driven by
the (normalized) gate charge $Q_{x}/2e$. Furthermore, we will demonstrate that
there exists a direct analogy between QPS events changing the flux inside the
ring by $\Phi_{0}$ and tunneling events of single Cooper pairs changing the
capacitor charge by $2e$. This equivalence is reminiscent of the well known
duality between phase and charge representations \cite{sz} and will be
exploited in our consideration.

We start from the case of sufficiently thick superconducting rings where QPS
effects can be neglected. In this case by virtue of Poisson's resummation
theorem (see, e.g., Sec. 3.3.2 in Ref. \cite{sz}) one can identically
transform the ring partition function (\ref{Z0}) to the following expression
\begin{equation}
{\mathcal{Z}}_{\phi_{x}}=\sum_{k=-\infty}^{\infty}\exp(i2\pi k\phi_{x}) \int
d\theta_{0}\int_{\theta_{0}}^{\theta_{0}+2\pi k}{\mathcal{D}}\theta\exp
(-S_{0}[\theta]),\label{Z1}%
\end{equation}
where
\begin{equation}
S_{0}[\theta]=\int_{0}^{\beta}d\tau\frac{1}{2E_{R}}\left( \frac{\partial
\theta}{\partial\tau}\right) ^{2}\label{ss0}%
\end{equation}
and $1/E_{R}$ plays the role of a ''mass'' for a ''particle'' with the
coordinate $\theta$. This coordinate represents an effective angle. It is
formally analogous to the Josephson phase variable in which case $E_{R}$ just
coincides with the charging energy.

Let us now consider thinner rings where QPS effects gain importance. In this
case we should include all transitions between different energy states
(\ref{energ}) labelled by the number $p$. As we already discussed these
transitions are just QPS events with the rate $\gamma_{QPS}$ defined in eq.
(\ref{gammaf}). Let us fix $p=0$ and take into account virtual transitions to
all other energy states and back. Performing summation over all such
contributions we arrive at the series in powers of the rate $\gamma_{QPS}$ (or
fugacity $y$) similar to eq. (\ref{partit}). Assuming that both the ring
perimeter $X$ and its thickness are not too large (the latter condition
restricts the parameter $\mu$) one can neglect weak logarithmic interaction
(\ref{Sint}) between different phase slips. Then spatial QPS coordinates
$x_{n}$ become exact zero modes and can be trivially integrated out. As a
result, we arrive at the following contribution to the partition function
\[
{\tilde Z}(\phi_{x})=\int_{\phi_{x}}^{\phi_{x}}{\mathcal{D}}\phi(\tau)
\sum_{n=0}^{\infty}\frac{1}{2n!}\left( \frac{\gamma_{QPS}}{2}\right) ^{2n}
\int_{0}^{\beta}d\tau_{1}... \int_{0}^{\beta}d\tau_{2n}
\]
\begin{equation}
\times\sum_{\nu_{i}=\pm1} \delta(\dot\phi(\tau) -\dot\phi_{n} (\tau))
\exp\left(-\int_0^\beta d\tau\frac{E_{R}\phi^{2}(\tau)}{2}\right)\label{tZ}%
\end{equation}
where, as before, the summation is carried out over neutral charge
configurations (\ref{nutot}),
\begin{equation}
\phi_{n} (\tau)=\sum_{j=1}^{n}\nu_{j}\Theta(\tau-\tau_{j})
\end{equation}
and $\Theta(\tau)$ is the theta-function. We note that the term in
the exponent describes virtual energy changes occurring due to QPS
events. What remains is to add up similar contributions from all
other parabolas with $p \neq0$. In this way we arrive at the final
expression for the grand partition function of the nanoring
\begin{equation}
{\mathcal{Z}}_{\phi_{x}}=\sum_{p=-\infty}^{\infty}{\tilde Z}(p+\phi
_{x}).\label{Z2}%
\end{equation}
Together with eq. (\ref{tZ}) this result fully accounts for all QPS events in
our system.

In the absence of QPS (i.e. for $\gamma_{QPS} \to0$) eqs. (\ref{Z2}),
(\ref{tZ}) obviously reduce to (\ref{Z0}). For non-zero $\gamma_{QPS}$ one can
rewrite the partition function (\ref{tZ}), (\ref{Z2}) in the equivalent form
of the following path integral
\begin{equation}
{\mathcal{Z}}_{\phi_{x}}=\sum_{k=-\infty}^{\infty}\exp(i2\pi k\phi_{x}) \int
d\theta_{0}\int_{\theta_{0}}^{\theta_{0}+2\pi k}{\mathcal{D}}\theta
\exp(-S[\theta]),\label{Z5}%
\end{equation}
where
\begin{equation}
S[\theta]=S_{0}[\theta] -\gamma_{QPS}\int_{0}^{\beta}d\tau\cos\theta(\tau)
.\label{sss}%
\end{equation}
In order to demonstrate that the partition function (\ref{Z5}), (\ref{sss}) is
identical to that defined in eqs. (\ref{tZ}), (\ref{Z2}) it suffices to
perform the Hubbard-Stratonovich transformation of the kinetic term
(\ref{ss0}) (which amounts to introducing additional path integral over
$\phi(\tau)$), to formally expand $\exp(-S[\theta])$ in powers of
$\gamma_{QPS}$ and then to integrate out the $\theta$-variable in all terms of
these series. After these straightforward steps we arrive back at eqs.
(\ref{tZ}), (\ref{Z2}). We also note that this procedure is described in Sec.
3.3.5 of Ref. \cite{sz}, therefore we can avoid further details here.

Eqs. (\ref{Z5}), (\ref{sss}) define the grand partition function for a quantum
particle in a cosine periodic potential which is, in turn, equivalent to that
for a Josephson junction in the presence of charging effects \cite{sz}. This
partition function can be identically rewritten as (cf. eq. (3.94) in Ref.
\cite{sz})
\begin{equation}
{\mathcal{Z}}_{\phi_{x}}=\sum_{p=-\infty}^{\infty}\exp(-\tilde{E}_{p}(\phi
_{x})/T),\label{Zf}%
\end{equation}
where $\tilde{E}_{p}(\phi_{x})$ are the energy bands of the problem defined by
the solutions of the well-known Mathieu equation. For $\gamma_{QPS}\ll E_{R}$
one has
\begin{equation}
\tilde{E}_{0}(\phi)=\frac{E_{R}}{2\pi^{2}}\arcsin^{2}\left[\left(1-\frac{\pi^{2}%
}{2}\left(\frac{\gamma_{QPS}}{E_{R}}\right)
^{2}\right)\sin(\pi\phi_{x})\right] ,
\end{equation}
i.e. the energy bands remain nearly parabolic
$\tilde{E}_{p}(\phi)\sim E_{p}(\phi)$ except in the vicinity of
the crossing points where gaps open due to level repulsion (see
Fig. \ref{F34 energy and curent vs flux}). In this limit the value
of the gap between the lowest and the first excited energy bands
just coincides with the QPS rate $\delta E_{01}=\gamma_{QPS}$. For
larger $\gamma_{QPS}$ the bandwidth shrinks while the gaps become
bigger. In the limit $\gamma_{QPS}\gg E_{R}$ the lowest band
coincides with
\begin{equation}
\tilde E_{0} (\phi)=
\frac{8}{\sqrt{\pi}}\gamma_{QPS}^{3/4}(E_{R})^{1/4}
e^{-8\sqrt{\gamma_{QPS}/E_{R}}}(1-\cos(2\pi\phi_{x}) ).
\label{grst}%
\end{equation}
The gap between the two lowest bands is $\delta
E_{01}=\sqrt{\gamma _{QPS}E_{R}}$.

These results are sufficient to evaluate PC in superconducting nanorings in
the presence of quantum phase slips. As before, taking the derivative of the
ground state energy with respect to the flux $\Phi_{x}$ and making use of eq.
(\ref{gammaf}) we find that for $\gamma_{QPS}\ll E_{R}$ and outside immediate
vicinity of the points $\phi_{x} =1/2+p$ PC is again defined by eqs.
(\ref{evcur}), (\ref{st}). In the opposite limit $\gamma_{QPS}\gg E_{R}$ we find%
\begin{eqnarray}
I&=&\tilde{I}_{0}\sin(2\pi\phi_{x}),
\nonumber\\
\tilde{I}_{0}&=&\frac{16e}{\sqrt{\pi
\hbar}}\gamma_{QPS}^{3/4}(E_{R})^{1/4}e^{-8\sqrt{\gamma_{QPS}/E_{R}}%
}.\label{esm}%
\end{eqnarray}
We observe that in the latter limit PC is exponentially suppressed
as
\begin{equation}
\tilde I_{0} \propto\exp(-R/R_{c}), \label{lsc}%
\end{equation}
where neglecting a numerical prefactor we estimate
\begin{equation}
R_{c} \sim \xi \exp(S_{\mathrm{core}}
/2).\label{rc}%
\end{equation}
This simple formula sets the size scale beyond which one would
expect PC to be exponentially suppressed in superconducting
nanorings with $\sqrt{s} \lesssim10$ nm. E.g. for $\xi$ of order
100 nm and $S_{\mathrm{core}} \approx 10$  eq. (\ref{rc}) yields
$R_{c}$ of order of a micron.

We would like to emphasize that during our analysis we employed only one
approximation neglecting logarithmic inter-QPS interaction effects. If needed,
such effects can also be included into our consideration and may only lead to
unimportant modifications in our results for rings with very large perimeters
$X$.

The above results demonstrate practically the same qualitative
features as those found within a different approach for the model
of nanorings formed by Josephson chains \cite{MLG}. In particular,
both for granular and for homogeneous rings the dependence of PC
on $\phi_{x}$ gradually changes from sawtooth to sinusoidal as the
ring perimeter $X$ increases. This change is accompanied by
suppression of PC amplitude which eventually becomes an
exponentially decaying function of the ring radius $R$ (\ref{lsc})
in the limit $R>R_{c}$. In other words, eq. (\ref{lsc}) sets the
length scale $R_{c} $ beyond which persistent currents in
superconducting nanorings should be essentially washed out by
quantum fluctuations. In some sense the scale $2\pi R_{c}$ plays
the role of a dephasing length in our problem demonstrating that
zero-point fluctuations can destroy macroscopic phase coherence
down to $T\to 0$ even in superconducting systems.

This conclusion is qualitatively consistent with the results
derived for normal metallic conductors \cite{GZ,GZ1,GZ2,GZ98P} and
single channel rings coupled to dissipative baths
\cite{GZ98P,Paco,Carlos}. In fact, a close similarity between the
partition function (\ref{Z5}), (\ref{sss}) and those studied in
Refs. \cite{Paco,Carlos} exists on a formal level as well: These
problems are described by similar path integrals except in the
case of normal rings coupled to dissipative environments one
should include the corresponding non-local in time term into the
action instead of the last term in eq. (\ref{sss}). At
sufficiently large $X$ these path integrals can also be handled in
exactly the same manner: In both problems the dominating
contribution comes from instantons (kinks) describing quantum
tunneling of the angle variable $\theta$ between different
topological sectors of the problem. In all cases this procedure
yields low temperature PC in the form (\ref{esm}) where one finds
$\tilde{I}_{0}\propto\exp(-(X/L_{\varphi})^{2})$ in the case
of Caldeira-Leggett environments \cite{GZ98P,Paco,Carlos} and $\tilde{I}%
_{0}\propto\exp(-X/L_{\varphi})$ in the case of so-called dirty electron gas
environments \cite{Carlos}.

An important qualitative difference between our present problem and those of
normal nanorings with dissipation lies in the fact that in the latter case
dissipation explicitly \textit{violates time-reversal symmetry} (thus causing
genuine decoherence of electrons), while no such symmetry is violated in eqs.
(\ref{Z5}), (\ref{sss}). Hence, in our case -- in contrast to
\cite{GZ98P,Paco,Carlos} -- exponential suppression of PC (\ref{esm}) can also
be interpreted as a non-trivial coordinate-dependent renormalization effect.
This difference is just the same as that between \textit{dissipativeless}
Cooper pair tunneling (which, just like QPS, opens inter-band gaps in the
spectrum of Josephson junctions) and \textit{dissipative} single electron
tunneling (which opens no such gaps). For more details on the latter subject
we refer the reader to the review \cite{sz}.%

\begin{figure}
\begin{center}
\includegraphics[width=6cm]{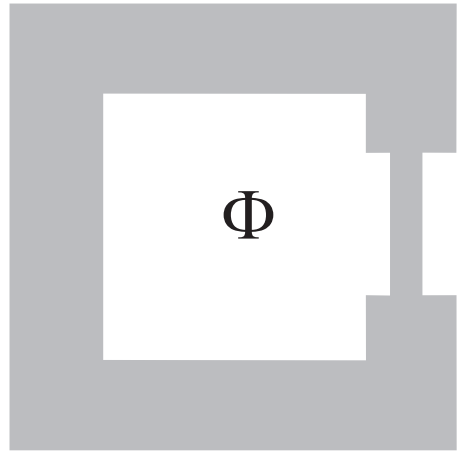}%
\caption{Superconducting ring with a ''quantum phase slip
junction''.}%
\label{F35 ring constriction}%
\end{center}
\end{figure}

Finally we should mention that one can also consider a slightly
modified situation of rings consisting of thicker and thinner
parts, as it is shown in Fig. \ref{F35 ring constriction}.
Assuming that QPS effects are negligible in a thicker part of the
ring and they can only occur in its thinner part (of length $d$)
we arrive at exactly the same results as above in which one should
only replace the ring perimeter $X$ by the length $d$. In
particular, it follows from our analysis that such rings also
exhibit the property of an exact duality to mesoscopic Josephson
junctions if we identify $E_{R}$ with the junction charging energy
$E_{C}$ and the QPS rate $\gamma_{QPS}$ with the Josephson energy
$E_{J}$. For this reason Mooij and Nazarov \cite{MN} suggested to
call systems depicted in Fig. \ref{F35 ring constriction}
\textquotedblright quantum phase slip junctions\textquotedblright\
and argued that any known result on electron transport in circuits
containing Josephson junctions can be exactly mapped onto a dual
result for a QPS junction in a dual circuit. This observation can
be used in metrology, e.g., for practical implementation of the
electric current standard in the above structures. Mooij and
Harmans \cite{MH} proposed to use rings with QPS junctions for
experimental realization of quantum phase slip flux qubits.

It is also worth pointing out that previously a similar exact duality between
phase and charge variables in Josephson junctions was discussed
\cite{AvOd,Z90,sz}, see, e.g., eq. (5.76) in Ref. \cite{sz}. According to this
property the Josephson junction itself can also be a QPS junction if we
interchange the canonically conjugate phase and (quasi)charge variables
$\varphi\to\pi q/e$. In this sense one can identify the angle variable
$\theta$ with the (quasi)charge $\pi q/e$ passing through the QPS junction.

\subsection{Parity effect and persistent currents}

In our previous analysis of QPS effects in superconducting nanorings we
followed the standard technique developed for grand canonical ensembles, i.e.
we implicitly assumed that the total number of electrons in the system $N$ may
fluctuate and the chemical potential $\tilde\mu$ is fixed. Obviously, this
assumption is not correct for rings which are disconnected from any external
circuit. In that case electrons cannot enter or leave the ring and, hence, the
number $N$ is strictly fixed, but the chemical potential $\tilde\mu$, on the
contrary, fluctuates. It turns out that novel effects emerge in this physical
situation. These effects will be discussed below in this section. To a large
extent we will follow the analysis developed in Refs. \cite{SZ,KKZ,Z04,SZ2}.

It is well known that thermodynamic properties of isolated
superconducting systems are sensitive to the parity of the total
number of electrons \cite{AN92,Tuo93,sz94,vDZGT} even though this
number is macroscopically large. This parity effect is a
fundamental property of a superconducting ground state described
by the condensate of Cooper pairs. The number of electrons in the
condensate is necessarily even, hence, for odd $N$ at least one
electron remains unpaired having an extra energy equal to the
superconducting energy gap $\Delta_{0}$. This effect makes
thermodynamic properties of the ground state with even and odd $N$
different. Clear evidence for such parity effect was demonstrated
experimentally in small superconducting islands
\cite{Tuo93,Laf93,RT}.

At the first sight, this parity effect should have little impact on the
supercurrent because of the fundamental uncertainty relation $\delta N
\delta\varphi\gtrsim1$. Should the electron number $N$ be fixed, fluctuations
of the superconducting phase $\varphi$ become large disrupting the
supercurrent. On the other hand, in transport experiments with fluctuations of
$\varphi$ being suppressed the parity effect cannot be observed because of
large fluctuations of $N$.

Consider now isolated superconducting rings pierced by the magnetic flux
$\Phi_{x}$. In accordance with the number-phase uncertainty relation the
\textit{global} superconducting phase of the ring fluctuates strongly in this
case, however these fluctuations are decoupled from the supercurrent and
therefore cannot influence the latter. In this situation the parity effect may
substantially modify PC in superconducting nanorings at sufficiently low
temperatures. In particular, we will show that changing the electron parity
number from even to odd results in spontaneous supercurrent in the ground
state of such rings without any externally applied magnetic flux. In other
words, our fundamental conclusion will be that the \textit{BCS ground state of
a canonical ensemble with odd number of electrons is the state with
spontaneous supercurrent}.

\subsubsection{Parity projection formalism}

In order to systematically investigate interplay between the
parity effect and persistent currents in superconducting nanorings
we will employ the parity projection formalism
\cite{SZ,SZ2,JSA94,GZ294,AN94} which we will briefly outline here.

The grand canonical partition function
\begin{equation}
{\mathcal{Z}}(T,\tilde{\mu})=\mathrm{Tr}e^{-\beta({\mathcal{H}}-\tilde{\mu}N)}%
\end{equation}
is linked to the canonical one $Z(T,N)$ as
\begin{equation}
{\mathcal{Z}}(T,\tilde{\mu})=\sum\limits_{N=0}^{\infty}Z(T,N)\exp
\biggl({\frac{\tilde{\mu}N}{T}}\biggr).\label{6}%
\end{equation}
Here and below ${\mathcal{H}}$ is the system Hamiltonian. Inverting this
relation and defining the canonical partition functions $Z_{e}$ and $Z_{o}$
respectively for even ($N\equiv N_{e}$) and odd ($N\equiv N_{o}$) ensembles,
one gets
\begin{equation}
Z_{e/o}(T)={\frac{1}{2\pi}}\int_{-\pi}^{\pi}due^{-iN_{e/o}u}{\mathcal{Z}%
}_{e/o}(T,iTu),\label{invrel}%
\end{equation}
where
\begin{eqnarray}
{\mathcal{Z}}_{e/o}(T,\tilde{\mu})&=&{\frac{1}{2}}\mathrm{Tr}\left\{
\big[1\pm(-1)^{N}\big]e^{-\beta({\mathcal{H}}-\tilde{\mu}N)}\right\}
\nonumber\\
&=&{\frac{1}{2}}\left(  {\mathcal{Z}}(T,\tilde{\mu})\pm{\mathcal{Z}}
(T,\tilde{\mu}+i\pi T)\right) \label{pm1}%
\end{eqnarray}
are the parity projected grand canonical partition functions. For $N\gg1$ it
is sufficient to evaluate the integral in (\ref{invrel}) within the saddle
point approximation
\begin{equation}
Z_{e/o}(T)\sim e^{-\beta(\Omega_{e/o}-\tilde{\mu}_{e/o}N_{e/o})},\label{14}%
\end{equation}
where ${\Omega}_{e/o}=-T\ln{\mathcal{Z}}_{e/o}(T,\tilde{\mu})$ are the parity
projected thermodynamic potentials,
\[
{\Omega}_{e/o}={\Omega}_{f}-T\ln\bigg[\frac{1}{2}\Big(1\pm e^{-\beta
({\Omega_{b}}-{\Omega_{f}})}\Big)\bigg]
\]
and ${\Omega}_{f/b}=-T\ln\left[  \mathrm{Tr}\left\{  (\pm1)^{\mathcal{N}%
}e^{-\beta({\mathcal{H}}-\tilde{\mu}N)}\right\}  \right]  $. \textquotedblleft
Chemical potentials\textquotedblright\ $\tilde{\mu}_{e/o}$ are defined by the
saddle point condition $N_{e/o}=-\partial\Omega_{e/o}(T,\tilde{\mu}%
_{e/o})/\partial\tilde{\mu}_{e/o}$.

The main advantage of the above formalism is that it allows to express the
canonical partition functions and thermodynamical potentials in terms of the
parity projected grand canonical ones thereby enormously simplifying the whole
calculation. We further note that $\Omega_{f}$ is just the standard grand
canonical thermodynamic potential and $\Omega_{b}$ represents the
corresponding potential linked to the partition function ${\mathcal{Z}%
}(T,\tilde\mu+i\pi T)$. It is easy to see \cite{GZ294} that in
order to recover this function one can evaluate the true grand
canonical partition function ${\mathcal{Z}}(T,\tilde\mu)$, express
the result as a sum over the Fermi Matsubara frequencies
$\omega_{f} =2\pi T(l+1/2)$ and then substitute the Bose Matsubara
frequencies $\omega_{b} =2\pi Tl$ ($l=0,\pm1,...$) instead of
$\omega_{f}$. This procedure automatically yields the correct
expression for ${\mathcal{Z}}(T,\tilde\mu+i\pi T)$ and, hence, for
$\Omega_{b}$.

Having found the thermodynamic potentials for the even and odd ensembles one
can easily determine the equilibrium current $I_{e/o}$. Consider, as before,
isolated superconducting rings pierced by the magnetic flux $\Phi_{x}$. Making
use of the above expressions one finds PC circulating inside the ring:
\begin{equation}
I_{e/o}=I_{f}\pm\frac{I_{b}-I_{f}}{ e^{\beta({\Omega_{b}} - {\Omega_{f}})}
\pm1},\label{Ie/o}%
\end{equation}
where the upper/lower sign corresponds to the even/odd ensemble and we have
defined
\begin{equation}
I_{e/o}= c \left(  \frac{\partial\Omega_{e/o}}{\partial\Phi_{x}} \right)
_{\tilde\mu(\Phi_{x})},\;\;\; I_{f/b}= c \left(  \frac{\partial\Omega_{f/b}%
}{\partial\Phi_{x}} \right)  _{\tilde\mu(\Phi_{x})}.\label{cureo}%
\end{equation}
Eqs. (\ref{Ie/o}), (\ref{cureo}) represent a direct generalization of the
grand canonical formula (\ref{defper}) to canonical ensembles.

\subsubsection{Homogeneous superconducting rings}

Let us first consider homogeneous nanorings with cross section $s$
and perimeter $X=2\pi R$. As before, rings will be assumed
sufficiently thin, $\sqrt{s} < \lambda_{L}$. On the other hand,
below we will neglect QPS effects, i.e. describe superconducting
properties of such rings within the parity projected mean field
BCS theory. As we have already learned, this description is
justified provided the condition $g_{\xi}\gg1$ is satisfied.
Hence, the ring should not be too thin and the total number of
conducting channels should remain large ${\mathcal{N}} \gg1$. In
addition, the perimeter $X$ should not exceed the scale $2\pi
R_{c}$ (\ref{lsc}). Finally, we will neglect the difference
between the mean field values of the BCS order parameter for the
even and odd ensembles \cite{JSA94,GZ294}. This is legitimate
provided the ring volume is large enough, ${\mathcal{V}}=Xs \gg1/N_{0}%
\Delta_{0}$, where, as before, $N_{0}$ is the density of states at the Fermi level.

Evaluating thermodynamic potentials $\Omega_{f/b}$ and expressing
the result in terms of the excitation energies $\varepsilon_{k}$
and the order parameter $\Delta_{0}$ one finds \cite{GZ294}
\begin{align}
\Omega_{f}= \tilde\Omega & -2T \sum_{k} \ln\left( 2 \cosh\frac{\varepsilon
_{k}}{2T} \right) ,\label{Omf}\\
\Omega_{b}= \tilde\Omega & -2T \sum_{k} \ln\left( 2 \sinh\frac{\varepsilon
_{k}}{2T} \right) ,\label{Omb}%
\end{align}
where $\tilde\Omega= |\Delta_{0}|^{2} /\lambda+\mathrm{Tr}\{\hat\xi\}$,
\begin{equation}
\hat\xi= \frac{1}{2m}{\left( -i\hbar\frac{\partial}{\partial
\mbox{\boldmath$r$}}- \frac{e}{c}%
\mbox{\boldmath$A$}(\mbox{\boldmath$r$})\right) }^{2}
+U(\mbox{\boldmath$r$})-\tilde\mu,
\end{equation}
is the single-particle energy operator, $\varepsilon_{k}=\mbox{\boldmath$p$}
\mbox{\boldmath$v$}_{s}+ \sqrt{\xi_{p}^{2}+ \Delta_{0}^{2}}$, where
$\mbox{\boldmath$p$}$ is the quasiparticle momentum, $\xi_{p} =(p^{2}%
-\tilde{\mu}(\Phi_{x})/2m$ and the superconducting velocity $v_{s}$ is defined
in eq. (\ref{st}).

The above equations allow to fully determine PC in superconducting nanorings
with even and odd number of electrons $I_{e/o}$. The parity effect becomes
observable at sufficiently low temperatures \cite{Tuo93} $T<T^{\ast}%
\approx\Delta_{0}\ln(N_{0}{\mathcal{V}}\sqrt{\Delta_{0}T^{\ast}})$.
Here we consider the most interesting limit $T\ll\hbar v_{F}/X$.
From eqs. (\ref{Ie/o})-(\ref{Omb}) we find that in this limit the
current $I_{e}$ exactly coincides with the standard grand
canonical result $I=en_sv_ss$ with $n_{s}\equiv n_{e}$, while the
current $I_{o}$ for the odd ensemble reads \cite{SZ,SZ2,Kang,Yak}
\begin{equation}
I_{o}=en_{o}v_{s}s-e\frac{v_{F}}{X}\mbox{sgn}v_{s}.\label{evenI}%
\end{equation}
Here we introduced the electron density in the case of even/odd total number
of electrons $n_{e/o}=N_{e/o}/{\mathcal{V}}$.

We observe that in the case of odd ensembles there exists an additional term
which modifies the flux dependence of PC and, as we will demonstrate below,
leads to a number of fundamentally important effects. Unfortunately, in the
case of homogeneous rings the difference between PC in even and odd ensembles
turns out to be hardly observable. Indeed, these difference is inversely
proportional to the total number of conducting channels, $(I_{e}-I_{o}%
)/I_{e}\sim1/{\mathcal{N}}$. For this reason the parity effect remains
vanishingly small in generic metallic rings with ${\mathcal{N}}\gtrsim10^{3}$.
On the other hand, for ultra-thin nanorings with ${\mathcal{N}}\lesssim10$ PC
is essentially wiped out due to proliferation of quantum phase slips.
Estimating $g_{\xi}\sim{\mathcal{N}}l/\xi$, we conclude that for $g_{\xi}%
\sim1$ (i.e. when the QPS fugacity is already large and, hence, quantum
suppression of PC becomes very strong) the number of conducting channels yet
remains parametrically large ${\mathcal{N}}\sim\xi/l\gg1$. This estimate
demonstrates that the parity effect on PC is never important in the case of
\textit{homogeneous} superconducting nanorings and appears to be practically
unobservable in such systems.%

\begin{figure}
\begin{center}
 \includegraphics[width=8cm]{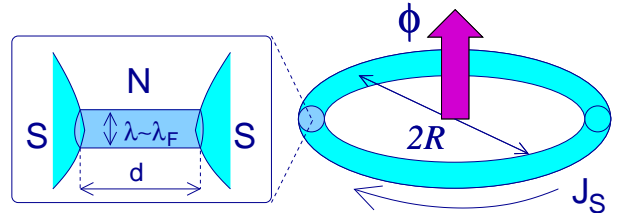}%
\caption{Superconducting ring with embedded SNS junction of length $d$ \cite{SZ}.}%
\label{F36 SNS ring}%
\end{center}
\end{figure}

\subsubsection{SNS rings}

We now turn to the situations in which parity effect gains importance and can
be directly probed in modern experiments. Let us slightly modify our system
and consider a relatively thick superconducting ring with large $g_{\xi}\gg1$
interrupted by a thin wire of length $d$ with only few conducting channels
\begin{equation}
{\mathcal{N}}_{n}\sim1\label{n1}%
\end{equation}
thus forming a weak link inside the superconducting ring, see Fig.
\ref{F36 SNS ring}. Without loss of generality this wire can be considered
normal no matter if it is made of a normal or a superconducting material. In
the latter case quantum fluctuations would fully suppress the order parameter
inside such a wire bringing it into the normal state. In contrast, QPS effects
in superconducting parts of the ring can be neglected thus making the mean
field BCS description applicable. A clear advantage of these structures in
comparison to homogeneous rings is that in the former case the effect of the
electron parity number on PC can be large due to the condition (\ref{n1}).

In order to evaluate PC in such structures we again employ the parity
projection formalism. According to eqs. (\ref{Ie/o}), (\ref{cureo}) we need to
evaluate both the currents $I_{f/b}$ and the difference between the "Fermi"
and \textquotedblright Bose\textquotedblright\ thermodynamic potentials
$\Omega_{b}-\Omega_{f}\equiv\Omega_{bf}$. The currents $I_{f/b}$ can be
conveniently expressed via the phase difference across the weak link
$\varphi\simeq2\pi\phi_{x}$ by means of the general formula \cite{GZ02}
\begin{equation}
I_{f/b}=\frac{2e}{\hbar}\sum_{i=1}^{{\mathcal{N}}_{n}}T\sum_{\omega_{f/b}%
}\frac{\sin\varphi}{\cos\varphi+W_{i}(\omega_{f/b})},\label{fbfb}%
\end{equation}
where the sum runs over conducting channels of the normal wire and the
function $W_{i}(\omega)$ was evaluated in Ref. \cite{GZ02}. The difference of
thermodynamic potentials $\Omega_{bf}$ is defined as a sum of the
contributions from superconducting ($\Omega_{bf}^{(s)}$) and normal
($\Omega_{bf}^{(n)}$) parts of the ring. The former is evaluated with the aid
of eqs. (\ref{Omf}), (\ref{Omb}) which yield the standard result \cite{Tuo93}
\begin{equation}
\beta\Omega_{bf}^{(r)}\simeq N_{0}{\mathcal{V}}\sqrt{\Delta_{0}T}%
e^{-\frac{\Delta_{0}}{T}},
\end{equation}
while the latter is found by integrating $I_{f/b}(\varphi)$ (\ref{fbfb}) over
the phase $\varphi$.

We now consider several important limiting cases. The first limit is that of a
very short normal wire $d\rightarrow0$. This is essentially the limit of a
quantum point contact. In practice this limit is realized for $d\ll\xi$. Even
smaller values of $d$ are required provided the contact transmission is small.
In the limit of a quantum point contact one finds \cite{GZ02}
\begin{equation}
W_{i}(\omega)=(2/{\mathcal{T}}_{i})(1+\hbar^{2}\omega^{2}/\Delta_{0}%
^{2})-1,\label{QPC}%
\end{equation}
where ${\mathcal{T}}_{i}$ define transmissions of the conducting
channels. Substituting this function into the above equations one
arrives at the final result \cite{SZ,SZ2}
\[
I_{e/o}=-\frac{2e}{\hbar}\sum_{i=1}^{N}\frac{\partial\varepsilon_{i}(\varphi
)}{\partial\varphi}\tanh\frac{\varepsilon_{i}(\varphi)}{2T}
\]%
\begin{equation}
\times\left[  1\pm\frac{(\coth\frac{\varepsilon_{i}(\varphi)}{2T})^{2}%
-1}{e^{\beta\Omega_{bf}^{(r)}}\prod\limits_{j=1}^{N}(\coth\frac{\varepsilon
_{j}(\varphi)}{2T})^{2}\pm1}\right]  .\label{result}%
\end{equation}
Here $\varepsilon_{i}(\varphi)=\Delta\sqrt{1-{\mathcal{T}}_{i}\sin^{2}%
(\varphi/2)}$ are Andreev energy levels in a quantum point contact.%

\begin{figure}
\begin{center}
\includegraphics[width=8cm]{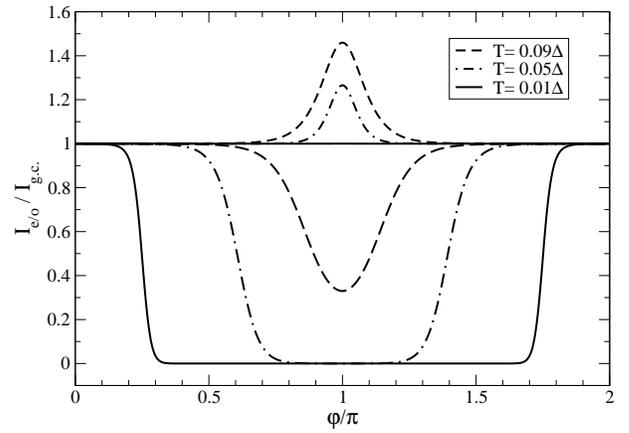}%
\caption{The ratio between canonical and grand canonical values of PC
$I_{e/o}/I_{g.c.}$ (represented by the term in the square brackets in Eq.
(\ref{result}) versus $\varphi$ in a single mode QPC at different temperatures
for even (three upper curves) and odd (three lower curves) ensembles. Here we
have chosen the channel transmission ${\mathcal{T}}=0.99$ \cite{SZ}.}%
\label{F37 PC grand}%
\end{center}
\end{figure}

The first line of eq. (\ref{result}) defines the standard grand
canonical result \cite{KO,KO1} while the term in the square
brackets accounts for the parity effect in our system. For
${\mathcal{N}}_{n}=1$ and at $T=0$ this term reduces to unity for
even ensembles and to zero for odd ones, i.e. PC turns out to be
\textit{totally blocked} in the case of odd number of
electrons \cite{SZ,SZ2,FNJ}.%

\begin{figure}
\begin{center}
\includegraphics[width=8cm]{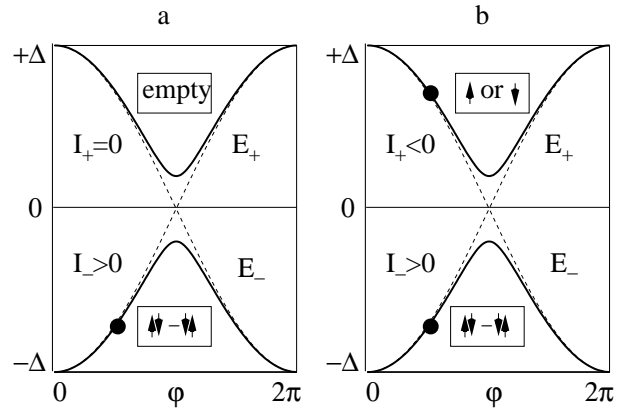}%
\caption{Andreev levels inside a quantum point contact and their occupation at
$T=0$ for even (a) and odd (b) ensembles \cite{SZ}.}%
\label{F38 Andreev levels}%
\end{center}
\end{figure}

The physics of this blocking effect is rather transparent and can
be understood as follows. We first recall that in the limit
$d\rightarrow0$ the Josephson current can be expressed only via
the contributions from discrete Andreev energy states
$E_{\pm}=\pm\epsilon(\varphi)$ as \cite{Furusaki,Furusaki1}
\begin{equation}
I(\varphi)=\frac{2e}{\hbar}\left[  \frac{\partial E_{-}}{\partial\varphi}%
f_{-}(E_{-})+\frac{\partial E_{+}}{\partial\varphi}f_{+}(E_{+})\right]
.\label{jos2}%
\end{equation}
Using the Fermi filling factors for these states
$f_{\pm}(E_{\pm})=[1+\exp (\pm\epsilon(\varphi)/T)]^{-1}$ one
arrives at the standard grand canonical results \cite{KO,KO1}. In
the case of superconducting rings with fixed number of electrons
$N$ these filling factors should be modified. Let us set
$T\rightarrow0$. For even $N$ all electrons are paired occupying
available
states with energies below the Fermi level (see Fig. \ref{F38 Andreev levels}%
a). In this case one has $f_{-}(E_{-})=1$, $f_{+}(E_{+})=0$, the current is
entirely determined by the contribution of the quasiparticle state $E_{-}$ and
eq. (\ref{jos2}) yields the same result as one for the grand canonical
ensemble. By contrast, in the case of odd $N$ one electron always remains
unpaired and occupies the lowest available energy state -- in our case $E_{+}$
-- above the Fermi level. Hence, for odd $N$ one has $f_{\pm}(E_{\pm})=1$
(Fig. \ref{F38 Andreev levels}b), the contributions of the two Andreev states
in eq. (\ref{jos2}) exactly cancel each other, and the current remains zero
for any $\varphi$ or the magnetic flux $\Phi_{x}$. This is just the effect of
parity-induced blocking of PC derived above from formal considerations.

For $T>0$ Eq. (\ref{result}) demonstrates that both for even and especially
for odd $N$ the current-phase relation for QPC may substantially deviate from
that derived for the grand canonical ensemble \cite{KO}, see Fig.
\ref{F37 PC grand}. For even ensembles the supercurrent \textit{increases}
above its grand canonical value. This effect is mainly pronounced for phases
$\varphi$ not very far from $\varphi=\pi$ and -- at sufficiently low $T$ -- it
becomes progressively more important with increasing temperature. On the
contrary, for odd ensembles the supercurrent is always suppressed below its
grand canonical value. This suppression is gradually lifted with increasing
temperature, though at phases $\varphi$ in the vicinity of the point
$\varphi=\pi$ blocking of PC may persist up to sufficiently high $T$. Eq.
(\ref{result}) also shows that in quantum point contacts with several
conducting channels and at $T\rightarrow0$ the current through the most
transparent channel will be blocked by the odd electron. Hence, though
blocking of PC remains incomplete in this case, it may nevertheless be
important also for quantum point contacts with ${\mathcal{N}}_{n}>1$.

Let us now turn to another important limit of superconducting rings containing
a normal wire of length $d>\xi_{0} \sim\hbar v_{F}/\Delta_{0}$. In contrast to
the case $d\rightarrow0$ considered above, the Josephson current in $SNS$
structures cannot anymore be attributed only to the discrete Andreev states
inside a weak link, and an additional contribution from the states in the
continuum should also be taken into account. Furthermore, for any non-zero $d$
there are always more than two discrete Andreev levels in the system.
Accordingly, significant modifications in the physical picture of the parity
effect in such SNS rings can be expected.%

\begin{figure}
\begin{center}
\includegraphics[width=8cm]{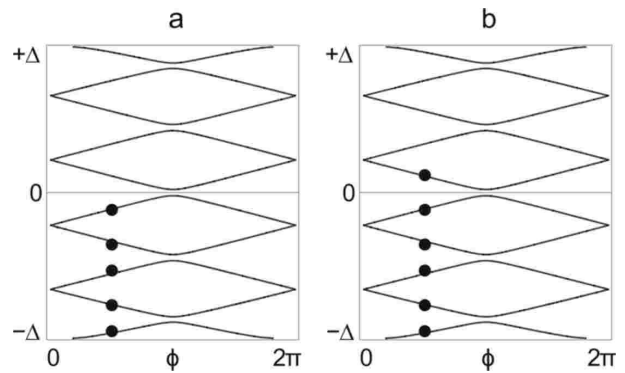}%
\caption{Andreev levels in a single mode $SNS$ junction with $d=6\hbar
v_{F}/\Delta_{0}$ and their occupation at $T=0$ for even (a) and odd (b)
ensembles \cite{SZ}.}%
\label{F39 Andreev SNS}%
\end{center}
\end{figure}

The key difference can be understood already by comparing the typical
structure of discrete Andreev levels in SNS junctions (Fig.
\ref{F39 Andreev SNS}) with that of a quantum point contact (Fig.
\ref{F38 Andreev levels}). As before, in the limit $T\rightarrow0$ all states
below (above) the Fermi level are occupied (empty) provided the total number
of electrons in the system is even (Fig. \ref{F39 Andreev SNS}a). If, on the
other hand, this number is odd the lowest Andreev state above the Fermi energy
is occupied as well (Fig. \ref{F39 Andreev SNS}b) thus providing an additional
contribution to the Josephson current. This contribution, however, cancels
only that of a symmetric Andreev level below the Fermi energy, while the
contributions of all other occupied Andreev levels and of the continuum states
remain uncompensated. Hence, unlike in the case $d\rightarrow0$, in $SNS$
rings one should not anymore expect the effect of PC blocking by the odd
electron but rather some other non-trivial features of the parity effect.%

\begin{figure}
\begin{center}
\includegraphics[width=6cm]{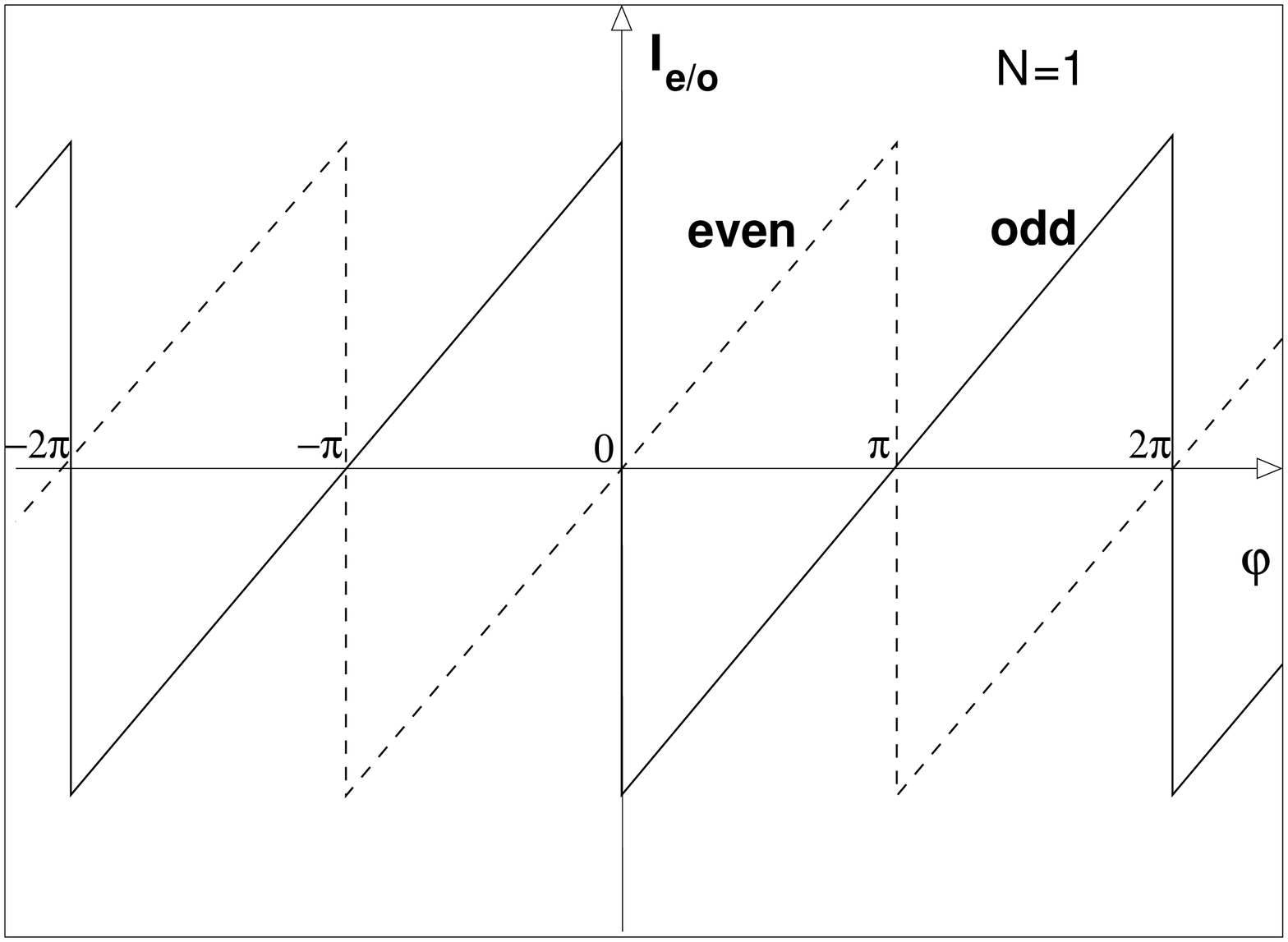}%
\caption{The zero temperature current-phase dependence (\ref{SNS}) for SNS
rings with ${\mathcal{N}}_{n}=1$: $I_{e}(\varphi)$ (dashed) line and
$I_{o}(\varphi)$ (solid) line \cite{SZ}.}%
\label{F40 current phase SNS}%
\end{center}
\end{figure}
This conclusion is fully confirmed by our quantitative analysis
\cite{SZ,SZ2}. Let us restrict our attention to transparent SNS
junctions in which case the function $W_{i}(\omega)\equiv
W(\omega)$ is the same for all transmission channels and reads
\cite{GZ02}
\[
W(\omega)=\left(  \frac{2\hbar^{2}\omega^{2}}{\Delta_{0}^{2}}+1\right)
\cosh\left(  \frac{2\omega d}{v_{F}}\right)
\]%
\begin{equation}
+\frac{2\hbar\omega}{\Delta_{0}}\sqrt{1+\frac{\hbar^{2}\omega^{2}}{\Delta
_{0}^{2}}}\sinh\left(  \frac{2\omega d}{v_{F}}\right)  .
\end{equation}
Substituting this function into (\ref{fbfb}) and repeating the whole
calculation as above, we arrive at the final result \cite{SZ,SZ2} which takes
a particularly simple form in the limit $T\rightarrow0$ and $d\gg\xi_{0} $:
\begin{equation}
I_{e}=\frac{ev_{F}{\mathcal{N}}_{n}}{\pi d}\varphi,\;\;\;\;I_{o}=\frac
{ev_{F}{\mathcal{N}}_{n}}{\pi d}\left(  \varphi-\frac{\pi\mbox{sgn}\varphi
}{{\mathcal{N}}_{n}}\right) \label{SNS}%
\end{equation}
This result applies for $-\pi<\varphi<\pi$ and should be $2\pi$-periodically
continued otherwise. We observe that at $T=0$ the current $I_{e}$ again
coincides with that for the grand canonical ensembles \cite{Kulik,Kulik1},
while in the case of odd ensembles the current-phase relation
is shifted by the value $\pi/N$. This shift has a simple interpretation being
related to the odd electron contribution $(2e/\hbar)\partial E_{1}%
/\partial\varphi$ from the lowest (above the Fermi level) Andreev state
$E_{1}(\varphi)$ inside the SNS junction. As we have expected, this
contribution indeed does not compensate for the current from other electron
states. Rather it provides a possibility for a parity-induced $\pi$-junction
state \cite{Leva} in our system: According to Eq. (\ref{SNS}) for single mode
$SNS$ junctions the \textquotedblleft sawtooth\textquotedblright%
\ current-phase relation will be shifted exactly by $\pi$, see Fig.
\ref{F40 current phase SNS}. More generally, we can talk about a novel $\pi
/N$-junction state, because in the odd case the minimum Josephson energy (zero
current) state is reached at $\varphi=\pm\pi/N$, see Fig.
\ref{F40 current phase SNS}. For any ${\mathcal{N}}_{n}>1$ this is a twofold
degenerate state within the interval $-\pi<\varphi<\pi$. In the particular
case ${\mathcal{N}}_{n}=2$ the current-phase relation $I_{o}(\varphi)$ turns
$\pi$-periodic, see Fig. 41.%

\begin{figure}
\begin{center}
\includegraphics[width=6cm]{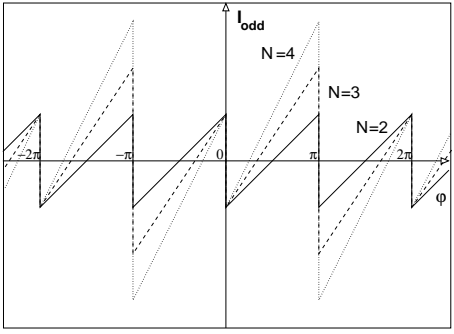}%
\caption{The same as in Fig. \ref{F40 current phase SNS} only for
the odd ensembles (second Eq. (\ref{SNS})) and for
${\mathcal{N}}_{n}=2,3$ and
$4$ \cite{SZ}.}%
\label{F41 current phase SNS odd}%
\end{center}
\end{figure}

Let us recall that the $\pi$-junction state can be realized in $SNS$
structures by driving the electron distribution function in the contact area
out of equilibrium \cite{Volkov,WSZ,Yip}. Here, in contrast,
the situation of a $\pi$- or $\pi/N$-junction is achieved in thermodynamic
equilibrium. Along with this important difference, there also exists a certain
physical similarity between the effects discussed here and in Refs.
\cite{Volkov,WSZ,Yip}: In both cases the electron distribution function in the
weak link deviates substantially from the Fermi function. It is this deviation
which is responsible for the appearance of the $\pi$-junction state in both
physical situations.

Perhaps the most spectacular physical consequence of the parity effect in SNS
rings is the presence of \textit{spontaneous} supercurrents \textit{in the
ground state} of such rings with odd number of electrons. Similarly to the
case of standard $\pi$-junctions \cite{Leva} such spontaneous supercurrents
should flow even in the absence of an externally applied magnetic flux. Unlike
in Ref. \cite{Leva}, however, here the spontaneous current state occurs for
\textit{any} inductance of the ring because of the non-sinusoidal dependence
$I_{o}(\varphi)$.%

\begin{figure}
\begin{center}
\includegraphics[width=6cm]{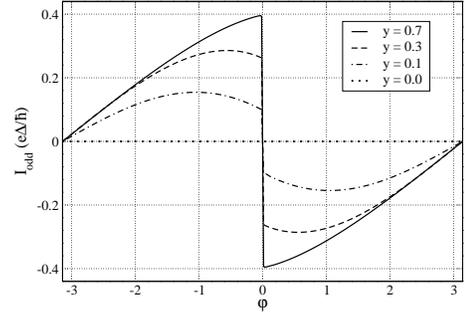}%
\caption{The zero temperature current-phase relation $I_{o}(\varphi)$
($-\pi<\varphi<\pi$) for ${\mathcal{N}}_{n}=1$ and different values of the
parameter $y=d\Delta/\hbar v_{F}$ \cite{SZ}.}%
\label{F42 current-phase}%
\end{center}
\end{figure}

\begin{figure}
\begin{center}
\includegraphics[width=8cm]{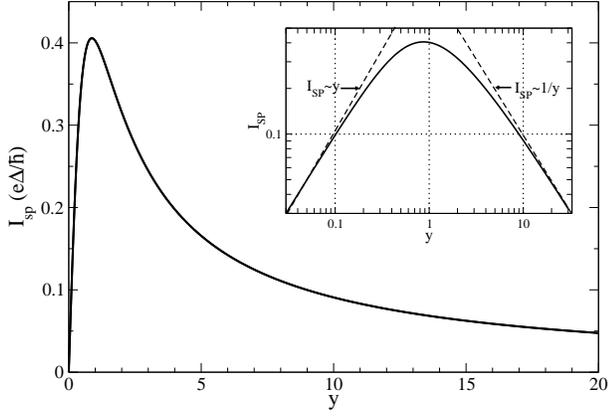}%
\caption{The spontaneous current amplitude $I_{sp}$ as a function of the
parameter $y$ at $T=0$. In the inset, the same function is shown on the
$log-log$ scale. Dashed lines indicate the asymptotic behavior of $I_{sp}(y)$
in the limits of small and large $y$ \cite{SZ}.}%
\label{F43 spontaneous}%
\end{center}
\end{figure}

Consider, for instance, the limit $d\gg\hbar v_{F}/\Delta_{0}$. In
the case of odd number of electrons the ground state energy of an
SNS ring can be written in a simple form
\begin{equation}
E=\frac{\Phi^{2}}{2c{\mathcal{L}}}+\frac{\pi\hbar v_{F}{\mathcal{N}}_{n}}%
{\Phi_{0}^{2}d}\left(  \Phi-\frac{\Phi_{0}\mbox{sgn}\Phi}{2{\mathcal{N}}_{n}%
}\right)  ^{2},\label{E}%
\end{equation}
where the first term is the magnetic energy of the ring (${\mathcal{L}}$ is
the ring inductance) while the second term represents the Josephson energy of
an SNS junction. Minimizing (\ref{E}) with respect to the flux $\Phi$ one
immediately concludes that the ground state of the ring is a twofold
degenerate state with a non-vanishing spontaneous current
\begin{equation}
I=\pm\frac{ev_{F}}{d}\left[  1+\frac{2ev_{F}{\mathcal{N}}_{n}}{d}%
\frac{{\mathcal{L}}}{\Phi_{0}}\right]  ^{-1}\label{Isp}%
\end{equation}
flowing either clockwise or counterclockwise. In the limit of small
inductances ${\mathcal{L}}\rightarrow0$ this current does not vanish and its
amplitude just reduces to that of the odd electron current at $\varphi
\rightarrow0$. One finds \cite{SZ,SZ2}:
\begin{align}
I_{sp}=e\Delta_{0}^{2}d/\hbar^{2}v_{F},\;\;\;\;  & \;\;d\ll\hbar v_{F}%
/\Delta_{0},\label{sp1}\\
I_{sp}=ev_{F}/\pi d,\;\;\;\;  & \;\;d\gg\hbar v_{F}/\Delta_{0}.\label{sp2}%
\end{align}
At $d\sim\hbar v_{F}/\Delta_{0}$ the amplitude of the current $I_{sp}$ can be
evaluated numerically, see Fig. \ref{F43 spontaneous}. One observes that -- in
agreement with eq. (\ref{sp1}) -- $I_{sp}$ increases linearly with $d$ at
small $d$, reaches its maximum value $I_{\mathrm{max}}\sim0.4e\Delta_{0}%
/\hbar$ at $d\sim\xi$ and then decreases with further increase of $d$
approaching the dependence (\ref{sp2}) in the limit of large $d$. For generic
BCS superconductors the magnitude of this maximum current can be estimated as
$I_{\mathrm{max}}\sim10\div100$ nA. These values might be considered as
surprisingly large ones having in mind that this current is associated with
only one Andreev electron state.%

Note that in the above analysis we merely assumed that the normal
wire is sufficiently clean and, on top of that, is in a good
electric contact with superconductors. Since both these
assumptions can be violated in a realistic experiment it is
important to discuss the corresponding modifications of our
results.

Assume first that the transmissions of both NS interfaces are small
$\mathcal{T}_{1,2} \ll1$. In this limit electron transport across the junction
is mainly due to resonant tunneling through discrete energy levels inside the
normal metal. For simplicity we will restrict our analysis to a single channel
junction ${\mathcal{N}}_{n}=1$. The most interesting physical situation is
realized in the limit of short junction $d \ll\hbar v_{F}/\Delta_{0}$. In the
case of a one dimensional metal of length $d$, the level spacing in the
vicinity of the Fermi energy is $\delta\epsilon\sim\hbar v_{F}/d$. Hence, the
condition for the short junction regime can also be represented in the form
$\delta\epsilon\gg\Delta_{0}$. Electron tunneling causes a non-zero linewidth
of the energy levels which is proportional to ${\mathcal{T}}_{1,2}%
\delta\epsilon$. This value is much smaller than $\delta\epsilon$, hence, the
resonances remain well separated. In this situation it suffices to take into
account only the closest level to the Fermi energy inside the normal metal.

As before, making use of eqs. (\ref{fbfb}) combined with the proper expression
for the $W$-function \cite{GZ02} one finds \cite{SZ2}
\begin{equation}
I_{f/b}=\frac{e}{\hbar} T\sum_{\omega_{f/b}} \frac{\Delta_{0} ^{2}%
\mathcal{T}\sin\varphi}{\varepsilon^{2}(\varphi) + \omega_{f/b}^{2} \left(
1+4\mathcal{D}\mathcal{T}/\mathcal{T}_{\mathrm{max}} \right)  },\label{fbfb1}%
\end{equation}
where
\begin{equation}
\mathcal{D}= \left( \frac{\Delta_{0}}{\Gamma}\right) ^{2}\left( 1+\frac{
\omega_{f/b} ^{2}} {\Delta_{0}^{2}}\right) + \frac{\Delta_{0} }{\Gamma}
\sqrt{1+\frac{\omega_{f/b}^{2}} {\Delta_{0}^{2}}},\label{fbfb2}%
\end{equation}
$\varepsilon(\varphi)=\Delta\sqrt{1-\mathcal{T}\sin^{2}(\varphi/2)}$,
$\Gamma=\Gamma_{1}+\Gamma_{2} $, $\Gamma_{1,2}/\hbar=\mathcal{T}_{1,2}v_{F}
/2d$ are the tunneling rates, $\mathcal{T}_{\mathrm{max}}=4 \Gamma_{1}
\Gamma_{2}/ \Gamma^{2}$ and the total transmission probability at the Fermi
energy $\mathcal{T}$ is given by the Breit-Wigner formula
\[
\mathcal{T}=\frac{\Gamma_{1} \Gamma_{2}}{\left( \epsilon_{R} \right)
^{2}+\frac{1}{4} \Gamma^{2}},
\]
where $\epsilon_{R}$ is the energy of a resonant level.

\begin{figure}
\begin{center}
\includegraphics[width=6cm]{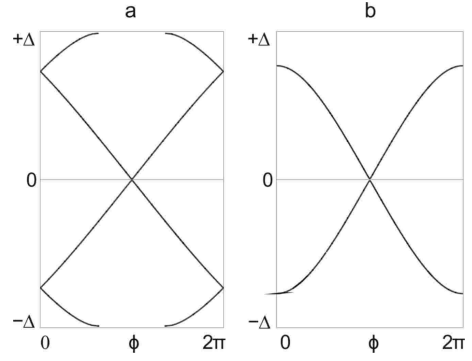}%
\caption{ Andreev levels in a single mode $SNS$ junction with $d=\hbar
v_{F}/\Delta$: (a) $\mathcal{T}_{1,2}=1$ and (b) $\mathcal{T}_{1,2}\ll1$ and
$\Delta/\Gamma=0.5$. In both cases $\mathcal{T}=1$ \cite{SZ2}.}%
\label{F44 Andreev single SNS}%
\end{center}
\end{figure}

It follows from Eqs. (\ref{fbfb1}), (\ref{fbfb2}) that -- although
the transparencies of both barriers are low -- the total
transmission $\mathcal{T}$ and, hence, the Josephson current shows
sharp peaks provided the Fermi energy becomes close to a bound
state inside the junction. On the other hand, eqs. (\ref{fbfb1}),
(\ref{fbfb2}) demonstrate that even in the vicinity of resonances
the behavior of the Josephson current as a function of the phase
difference $\varphi$ and temperature $T$ can substantially deviate
from that for transparent $SNS$ junctions.%

\begin{figure}
\begin{center}
\includegraphics[width=8cm]{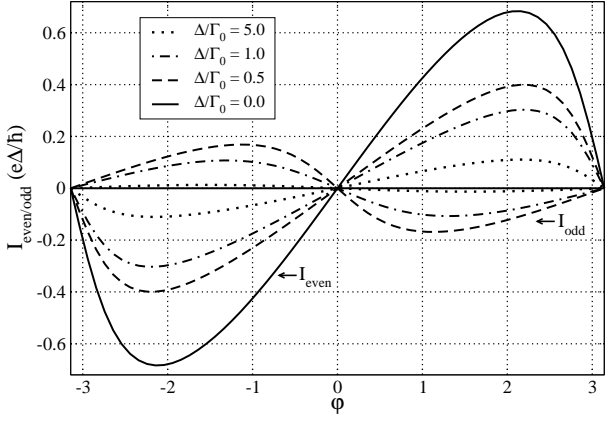}%
\caption{ Zero temperature current-phase relations $I_{e/o}(\varphi)$ for
$\mathcal{T}=0.9$, $\Gamma_{1}=\Gamma_{2}=\Gamma_{0}$ and different values of
the parameter $\Delta_{0}/\Gamma_{0}$ \cite{SZ2}.}%
\label{F45 current phase resonant}%
\end{center}
\end{figure}

In order to understand the physical reasons for such a difference
it is instructive to compare the structure of discrete Andreev
levels for ballistic ($\mathcal{T}_{1,2}=1$) SNS junctions with
that for junctions with weakly transmitting NS interfaces
$\mathcal{T}_{1,2}\ll1$, see Fig. \ref{F44 Andreev single SNS}.
The spectrum of the latter system consists only of a single
non-degenerate state $\varepsilon_{0}(\varphi)$ in the interval
$0<\varepsilon_{0}<\Delta_{0}$ (Fig. \ref{F44 Andreev single
SNS}b). As a result, the behavior of $\varepsilon_{0}(\varphi)$ at
small $\varphi$ is smooth and the derivative of $\varepsilon_{0}$
with respect to $\varphi$ has no jump at $\varphi=0$. In contrast,
in the case of ballistic SNS junctions discrete levels become
split at arbitrary small values of the phase $\varphi$ (Fig.
\ref{F44 Andreev single SNS}a) and the derivative of the lowest
Andreev level with respect to $\varphi$ acquires a jump at
$\varphi=0$. As this feature is absent in resonant SNS junctions
the spontaneous current in the ground state of such systems can
only develop at not very small ring inductances. The results for
PC in SNS rings with resonant transmission are presented in Fig.
\ref{F45 current phase resonant}. They clearly demonstrate
that at sufficiently low temperatures the \textquotedblleft$\pi$%
-junction\textquotedblright\ state should be realized in the case of odd
number of electrons.%

Finally let us turn to the case of a disordered normal wire. The
difference between PC values for odd ($I_{o}$) and even ($I_{e}$)
ensembles is related to the minigap value
$\varepsilon_{g}(\varphi)$ inside the normal metal
\cite{GK,BBS2,Been,Pilgram,Zhou,KKZ}. This relation acquires a
particularly simple form in the limit $T\rightarrow 0$:
\begin{equation}
I_{o}(\Phi)=I_{e}(\Phi)+2e\frac{\partial\varepsilon_{g}(\varphi)}%
{\partial\varphi}.\label{Ioe}%
\end{equation}
The last term in this equation describes the contribution to the current from
the \textquotedblleft odd\textquotedblright\ electron occupying the lowest
available state above the minigap $\varepsilon_{g}(\varphi)$ in the density of
states of the normal metal.

\begin{figure}
\begin{center}
\includegraphics[width=8cm]{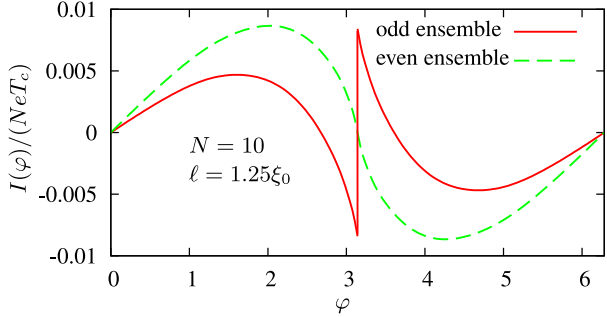}%
\caption{Phase dependence of the Josephson current at $T=0$ for the odd and
even number of electrons in the ring \cite{KKZ}.}%
\label{F46 Jos current odd even}%
\end{center}
\end{figure}

Let us first evaluate PC for the even ensemble $I_{e}$. As before, at $T=0$
this current identically coincides with one calculated for the grand canonical
ensemble. The results for the current-phase relation $I_{e}(\varphi)$ are
displayed in Fig.~\ref{F46 Jos current odd even} at various impurity
concentrations. PC in the odd ensemble $I_{o}$ at $T=0$ can be evaluated from
eq. (\ref{Ioe}). Combining our results for $I_{e}(\varphi) $ with those for
the minigap $\varepsilon_{g}(\varphi)$ derived in Ref. \cite{KKZ} we arrive at
a typical dependence $I_{o}(\varphi)$ displayed in
Fig.~\ref{F46 Jos current odd even}. We observe that at sufficiently large
values of $\varphi<\pi$ the absolute value of the odd electron contribution to
PC $2e\partial\varepsilon_{g}/\partial\varphi$ exceeds the term $I_{e}%
(\varphi)$ and the total current $I_{o}$ changes the sign. This non-trivial
parity-affected current-phase relation is specific for SNS rings with disorder
and it substantially differs from the current-phase relations derived above
for SNS rings with ballistic and resonant transmissions.%

\begin{figure}
\begin{center}
\includegraphics[width=8cm]{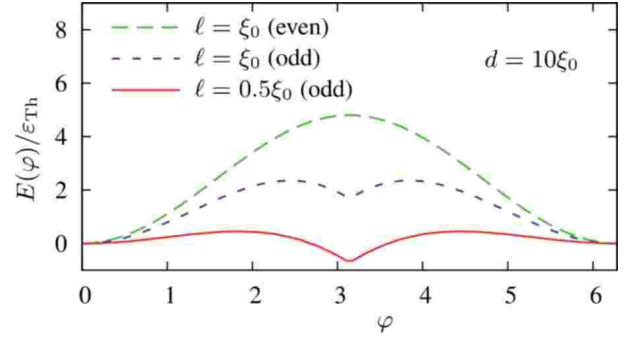}%
\caption{Josephson energy $E(\varphi)$ of an SNS ring as a function of the
phase difference $\varphi$ for the even and odd ensembles. The solid curve
corresponds to a $\pi$-junction state \cite{KKZ}.}%
\label{F47 Omega}%
\end{center}
\end{figure}

At the same time, as in the previous cases, in the odd ensemble
there exists a possibility both for a $\pi$-junction state and for
spontaneous currents in the ground state of the system without any
externally applied magnetic flux. Let us evaluate the ground state
energy of the SNS junction by integrating Eq. (\ref{Ioe}) with
respect to the phase $\varphi$. One finds
\begin{equation}
E_{o}(\varphi)=E_{e}(\varphi)-\varepsilon_{g}(0)+\varepsilon_{g}%
(\varphi),\;\;\;E_{e}(\varphi)=\frac{1}{2e}\int\limits_{0}^{\varphi}%
I_{e}(\varphi)d\varphi,
\end{equation}
where $E_{e/o}(\varphi)$ are the ground state energies of SNS junction for
even and odd number of electrons in the ring. While the energy $E_{e}%
(\varphi)$ is always non-negative and reaches its minimum at $\varphi=0$, in
the odd case the ground state energy $E_{o}(\varphi)$ can become negative
reaching its absolute minimum at $\varphi=\pi$. This physical situation of a
$\pi$-junction is illustrated in Fig. \ref{F47 Omega}.

It is easy to find out under which conditions the $\pi$-junction state becomes
possible. For that purpose it is sufficient to observe that for any impurity
concentration $E_{e}(\pi)=\alpha I_{C}/e$, where $I_{C}$ is the grand
canonical critical current at $T=0$ and the prefactor $\alpha\sim1$ depends on
the form of the current-phase relation. The $\pi$-junction condition
$E_{o}(\pi)<0$ is equivalent to the inequality
\begin{equation}
\varepsilon_{g}(0)>\alpha I_{C}/e.\label{picond}%
\end{equation}
It is obvious from Fig. \ref{F47 Omega}\ that in the many channel limit the
inequality (\ref{picond}) cannot be satisfied for sufficiently large $l$. On
the other hand, for sufficiently short mean free paths $I_{C}\propto l^{2}$
decays faster with decreasing $l$ as compared to the minigap $\varepsilon
_{g}(0)\propto l$, and the $\pi$-junction state becomes possible. In
particular, in the diffusive limit one finds \cite{Dubos} $I_{C}%
\simeq10.82\varepsilon_{\mathrm{Th}}/eR_{N}=1.53e{\mathcal{N}}_{n}v_{F}%
l^{2}/d^{3} $ and $\alpha\simeq1.05$, where $R_{N}$ is the Drude resistance of
a normal metal.%

Combining these results with the expression for the minigap
\cite{BBS2,Zhou}
$\varepsilon_{g}(0)\simeq3.12\varepsilon_{\mathrm{Th}}$, from the
condition (\ref{picond}) we observe that in the odd case the
$\pi$-junction state is realized provided the number of conducting
channels in the junction ${\mathcal{N}}_{n}$ is smaller than
\begin{equation}
{\mathcal{N}}_{n}<0.65d/l.\label{condN}%
\end{equation}
This condition is not very restrictive and it can certainly be
achieved in various experiments. For sufficiently dirty junctions
it allows for a formation of the $\pi$-junction state even in the
many channel limit. The condition (\ref{condN}) can also be
rewritten as $g_{N}<1.73$, where $g_{N}=8{\mathcal{N}}_{n}l/3d$ is
dimensionless conductance of a normal wire.

The condition for the presence of spontaneous currents in the ground state of
SNS rings with an odd number of electrons is established analogously, one
should only take into account an additional energy of the magnetic field
produced by PC circulating inside the ring. The ground state with spontaneous
currents is possible provided the total energy of the ring $E_{\mathrm{tot}%
}(\pi)$ becomes negative, i.e.
\begin{equation}
E_{\mathrm{tot}}(\pi)= 1.8\varepsilon_{\mathrm{Th}} \left[ g_{N}-1.73\right]
+\frac{(\Phi_{0}/2)^{2}}{2\mathcal{L}}< 0.
\end{equation}
This condition is more stringent than that for the $\pi$-junction state, but
can also be satisfied provided ${\mathcal{L}}$ exceeds a certain threshold
value which can roughly be estimated as $\sim0.1 \Phi_{0}^{2}/\varepsilon
_{\mathrm{Th}}$.

We conclude that in the diffusive limit the current-phase relation in the odd
case is entirely different from that in the ballistic case. Also the
restriction on the number of conducting channels ${\mathcal{N}}_{n}$ in the
normal metal (\ref{condN}) is less stringent. This feature of diffusive SNS
rings is rather advantageous for possible experimental observation of the
effects discussed here.

Still, in practice it would be necessary to fabricate SNS rings with few
conducting channels in the normal wire ${\mathcal{N}}_{n}\lesssim10$. This
condition can hardly be met for conventional normal metals. It appears,
therefore, that most promising candidates for practical realization of such
structures are junctions with N-layers formed by carbon nanotubes or organic
molecules. In this respect it is important to point out that observations of
clear signatures of dc Josephson effect in superconducting junctions with a
weak link formed by carbon nanotubes were reported by several experimental
groups \cite{Kas,Delft,VBou,L1,L2,Hels} Also intrinsic superconductivity
in carbon was claimed \cite{HBou,HBou2,Jap}, though at this stage
it appears that more experimental support for these claims would be
desirable. Regardless of this latter issue, SNS
junctions with carbon nanotubes are most likely objects in which it would be
possible to observe the influence of the parity effect on persistent currents
in superconducting nanorings.

\section{Summary}

It is well established that fluctuations play an important role in
structures with reduced dimensionality. In superconducting
materials clear signatures of fluctuation effects exist already
above the critical temperature $T_{C}$
\cite{almt,Maki,Thomson,LV}. Also $T_{C}$ itself can be reduced
due to fluctuation effects. This reduction is particularly
important in disordered 2D films and quasi-1D wires \cite{Fin}. In
this paper we addressed fluctuation effects which occur in
ultra-thin superconducting wires at temperatures below the mean
field BCS critical temperature. Superconducting properties of such
systems have been intensively studied -- both theoretically and
experimentally -- during past years. The key conclusions of these
investigations can be summarized as follows.

Thicker superconducting wires are characterized by very small
Ginzburg numbers $Gi_{1D}\lll$1 and diameters typically $\gtrsim$
100 nm. In such systems the superconducting transition is
broadened due to thermally activated phase slips (TAPS)
\cite{la,mh} which cause non-zero resistance $R(T)$ at
temperatures close enough to the critical temperature $T_{C}-T\ll
T_{C}$. Upon decreasing temperature TAPS events become less likely
and quantum fluctuations of the order parameter take over. This is
the regime of quantum phase slips (QPS) which sets in at
$T\lesssim\Delta_{0}(T)$. As long as the wire is sufficiently
thick and the Ginzburg number remains very small, $Gi_{1D}\lll$1,
QPS events are rare and typically do not lead to any measurable
consequences. Hence, the behavior of thicker wires remains
essentially superconducting outside an immediate vicinity of the
critical temperature. Upon reduction of the wire diameter below
$\sim$50 nm the QPS rate increases drastically. In this regime the
wire resistance $R(T)$ still decreases with temperature but may
remain well in the measurable range down to very low $T$. Provided
the wire diameter is decreased further the dimensionless
conductance of a wire segment of length $\xi$ eventually becomes
of order $g_{\xi}\sim$10 or smaller. In such wires QPS proliferate
causing a sharp \textit{crossover} from a superconducting to a
normal behavior. For generic parameters this crossover is expected
for wire diameters in the 10 nm range \cite{ZGOZ,ZGOZ2,GZ01}. This
crossover was indeed observed in a number of experiments in wires
with thicknesses exactly in this range. Thus, \textit{intrinsic
superconductivity in wires with diameters $\lesssim$ 10 nm is
destroyed by quantum fluctuations of the order parameter at any
temperature down to }$T=0$.

Theoretical analysis of nanowires reveals further interesting
effects, like QPS-binding-unbinding quantum phase transition (QPT)
\cite{ZGOZ} between superconducting and non-superconducting phases
which is predicted to occur as the impedance of a superconducting
wire becomes of order of the quantum resistance unit
$R_{q}\approx$ 6.5 k$\Omega$. For typical parameters this
condition is also achieved for wire diameters in the 10 nm range.
To the best of our knowledge, no clear experimental evidence for
this phase transition exists so far. This can be due to rather
stringent requirements: QPT can only be observed in long nanowires
at sufficiently low temperatures, ideally at $T\to 0$. Interesting
effects may occur also in short nanowires forming weak links
between superconducting electrodes. In many respects such systems
can behave similarly to Josephson nanojunctions and weak links.

Novel effects are also expected in superconducting nanorings.
While rings formed by thicker wires demonstrate the standard
behavior familiar from the bulk samples, the situation changes
drastically as soon as the wire thickness gets reduced down to
$\approx$ 10 nm or below this value. In this case QPS effects
become important leading to strong fluctuations of the magnetic
flux inside the ring. As a result, the amplitude of persistent
current decreases and its flux dependence changes from the
sawtooth-like to a smoother one \cite{MLG}. For such rings even at
$T=0$ persistent current gets exponentially suppressed by quantum
fluctuations provided the ring radius $R$ exceeds the critical
scale $R_{c}\sim \xi\exp(S_{\rm core}/2)$ where $S_{\rm core}\sim
g_{\xi}\sim Gi_{1D}^{-3/2}$ is the action of the QPS core. For
typical wire parameters the length $R_{c}$ can be of order of a
micron or even smaller. This length constitutes another
fundamental scale associated with quantum phase slips.

Yet another important factor which may influence persistent
currents in isolated superconducting nanorings is the electron
parity number. This influence is particularly strong in nanorings
containing a weak link with few conducting modes. Changing the
electron parity number from even to odd may result in
\textit{spontaneous} supercurrent in the ground state of such
rings without any externally applied magnetic flux \cite{SZ}. At
$T=0$ this current is produced by the only unpaired electron which
occupies the lowest available Andreev state. Under certain
conditions this spontaneous supercurrent can reach remarkably
large values up to $\sim e\Delta_{0}/\hbar\sim10\div100$ nA which
can be reliably detected in modern experiments.

\section{Acknowledgements}

We gratefully acknowledge collaboration and/or helpful discussions
with A.A. Bezryadin, G. Blatter, A.V. Galaktionov, M.S. Kalenkov,
C.N. Lau, J.E. Mooij, Yu.V. Nazarov, Y. Oreg, A. van Otterlo, L.
Pryadko, G. Sch\"{o}n, S.V. Sharov, M. Tinkham, D. Vodolazov and
G.T. Zimanyi. This work was supported in part by the European
Community's Framework Programme NMP4-CT-2003-505457 ULTRA-1D
"Experimental and theoretical investigation of electron transport
in ultra-narrow 1-dimensional nanostructures" and by RFBR Grant
06-02-17459.


\appendix

\section{Equilibrium Green-Keldysh functions}

We formally define the sub-blocks $\hat G,\hat F,\hat{\bar G},\hat{\bar F}$ in
the Green-Keldysh matrix (\ref{GGGG}) by means of the following equations:
\begin{align}
\hat G=\frac{1}{i}\left(
\begin{array}
[c]{cc}%
\langle{\mathcal{T}}\hat\psi_{\uparrow}(X_{1})\hat\psi^{\dagger}_{\uparrow
}(X_{2})\rangle & -\langle\hat\psi^{\dagger}_{\uparrow}(X_{2})\hat
\psi_{\uparrow}(X_{1})\rangle\\
\langle\hat\psi_{\uparrow}(X_{1})\hat\psi_{\uparrow}^{\dagger}(X_{2})\rangle &
\langle{\mathcal{T}}^{-1}\hat\psi_{\uparrow}(X_{1})\hat\psi_{\uparrow
}^{\dagger}(X_{2})\rangle
\end{array}
\right) ,\nonumber\\
\hat F=\frac{1}{i}\left(
\begin{array}
[c]{cc}%
\langle{\mathcal{T}}\hat\psi_{\uparrow}(X_{1})\hat\psi_{\downarrow}%
(X_{2})\rangle & -\langle\hat\psi_{\downarrow}(X_{2})\hat\psi_{\uparrow}%
(X_{1})\rangle\\
\langle\hat\psi_{\uparrow}(X_{1})\hat\psi_{\downarrow}(X_{2})\rangle &
\langle{\mathcal{T}}^{-1}\hat\psi_{\uparrow}(X_{1})\hat\psi_{\downarrow}%
(X_{2})\rangle
\end{array}
\right) ,\nonumber\\
\hat{\bar F}=\frac{1}{i}\left(
\begin{array}
[c]{cc}%
\langle{\mathcal{T}}\hat\psi_{\downarrow}^{\dagger}(X_{1})\hat\psi_{\uparrow
}^{\dagger}(X_{2})\rangle & -\langle\hat\psi_{\uparrow}^{\dagger}(X_{2}%
)\hat\psi_{\downarrow}^{\dagger}(X_{1})\rangle\\
\langle\hat\psi_{\downarrow}^{\dagger}(X_{1})\hat\psi_{\uparrow}^{\dagger
}(X_{2})\rangle & \langle{\mathcal{T}}^{-1}\hat\psi_{\downarrow}^{\dagger
}(X_{1})\hat\psi_{\uparrow}^{\dagger}(X_{2})\rangle
\end{array}
\right) ,\nonumber\\
\hat{\bar G}=\frac{1}{i}\left(
\begin{array}
[c]{cc}%
\langle{\mathcal{T}}\hat\psi^{\dagger}_{\downarrow}(X_{1})\hat\psi
_{\downarrow}(X_{2})\rangle & -\langle\hat\psi_{\downarrow}(X_{2})\hat
\psi^{\dagger}_{\downarrow}(X_{1})\rangle\\
\langle\hat\psi^{\dagger}_{\downarrow}(X_{1})\hat\psi_{\downarrow}%
(X_{2})\rangle & \langle{\mathcal{T}}^{-1}\hat\psi^{\dagger}_{\downarrow
}(X_{1})\hat\psi_{\downarrow}(X_{2})\rangle
\end{array}
\right) .\nonumber
\end{align}
Here $\hat\psi_{\alpha},\hat\psi_{\alpha}^{\dagger}$ are the electron
annihilation and creation operators, the symbol ${\mathcal{T}}$ (${\mathcal{T}%
}^{-1}$) stands for the time (anti-time) ordering.

Let us introduce the complete basis of the eigenfunctions $\chi_{n}(\bm{r})$
of the single electron Hamiltonian $H_{0}$:
\begin{align}
H_{0}\chi_{n}(\bm{r})=\xi_{n} \chi_{n}(\bm{r}).
\end{align}
Here $\xi_{n}$ are the electron eneries in the normal state. Defining the
quasiparticle energies in the superconducting state $E_{n}=\sqrt{\xi_{n}%
^{2}+\Delta_{0}^{2}}$ as well as the BCS coherence factors
\begin{align}
u_{n}^{2}=\frac{1}{2}\left( 1+\frac{\xi_{n}}{E_{n}}\right) ,\;\; v_{n}%
^{2}=\frac{1}{2}\left( 1-\frac{\xi_{n}}{E_{n}}\right) ,
\end{align}
we arrive at the explicit expressions for the sub-blocks $\hat G, \hat
F,\hat{\bar G},\hat{\bar F}$:
\begin{align}
i\hat G(t,\bm{r}_{1},\bm{r}_{2}) & =\sum_{n}\chi_{n}(\bm{r}_{1})\chi
_{n}(\bm{r}_{2}) \left[ u_{n}^{2}\hat A_{n}(t) + v_{n}^{2}\hat B_{n}(t)\right]
,\nonumber\\
i\hat{\bar G}(t,\bm{r}_{1},\bm{r}_{2}) & =\sum_{n}\chi_{n}(\bm{r}_{1})\chi
_{n}(\bm{r}_{2}) \left[ v_{n}^{2}\hat A_{n}(t) + u_{n}^{2}\hat B_{n}(t)\right]
,\nonumber\\
i\hat F(t,\bm{r}_{1},\bm{r}_{2}) & =i\hat{\bar F}(t,\bm{r}_{1},\bm{r}_{2}) =
\sum_{n}\chi_{n}(\bm{r}_{1})\chi_{n}(\bm{r}_{2}) u_{n}v_{n}
\nonumber\\ &\times\,
\left[ \hat
A_{n}(t) - \hat B_{n}(t)\right] ,\label{G}%
\end{align}
where
\begin{align}
\hat A_{n}(t) & =e^{-iE_{n}t}\left(
\begin{array}
[c]{cc}%
\theta(t)-f_{n} & -f_{n}\\
1-f_{n} & \theta(-t)-f_{n}%
\end{array}
\right) ,\nonumber\\
\hat B_{n}(t)  & = e^{iE_{n}t}\left(
\begin{array}
[c]{cc}%
f_{n}-\theta(-t) & -(1-f_{n})\\
f_{n} & f_{n}-\theta(t)
\end{array}
\right) .
\end{align}
Here we introduced the occupation probabilities of the quasiparticle states
$f_{n}$. In thermodynamic equilibrium these filling factors take the universal
form
\begin{align}
f_{n}=\frac{1}{1+e^{E_{n}/T}}.
\end{align}

\section{Ward identities}

One can demonstrate that the matrix Green-Keldysh functions satisfy the
following Ward identities%

\begin{align}
i\hat F\hat\beta+ i\hat\beta\hat F  & = \hat G\hat\sigma_{z}\left( -\hat
{\dot\beta}+\frac{i}{2m}\{\nabla,\hat\beta\}\right) \hat F
\nonumber\\ &
+\,\hat F\hat
\sigma_{z}\left( \hat{\dot\beta}+\frac{i}{2m}\{\nabla,\hat\beta\}\right)
\hat{\bar G}\nonumber\\
&  +\,2i\Delta_{0} \hat G\hat\sigma_{z}\hat\beta\hat{\bar G} - 2i\Delta_{0}
\hat F\hat\sigma_{z}\hat\beta\hat F,\nonumber\\
-i\hat{\bar F}\hat\beta- i\hat\beta\hat{\bar F}  & = \hat{\bar F}\hat
\sigma_{z}\left( -\hat{\dot\beta}+\frac{i}{2m}\{\nabla,\hat\beta\}\right) \hat
G
\nonumber\\ &
+\,\hat{\bar G} \hat\sigma_{z}\left( \hat{\dot\beta}+\frac{i}{2m}\{\nabla
,\hat\beta\}\right) \hat{\bar F}\nonumber\\
&  +\,2i\Delta_{0} \hat{\bar F}\hat\sigma_{z}\hat\beta\hat{\bar F} -
2i\Delta_{0} \hat{\bar G}\hat\sigma_{z}\hat\beta\hat G,\label{Ward2}%
\end{align}
\begin{align}
\hat G\hat\beta-\hat\beta\hat G & = i\hat G\hat\sigma_{z}\left( -\hat
{\dot\beta}+\frac{i}{2m}\{\nabla,\hat\beta\}\right) \hat G
\nonumber\\ &
-\,i\hat F\hat
\sigma_{z}\left( \hat{\dot\beta}+\frac{i}{2m}\{\nabla,\hat\beta\}\right)
\hat{\bar F},\nonumber\\
\hat{\bar G}\hat\beta-\hat\beta\hat{\bar G} & = i\hat{\bar F}\hat\sigma
_{z}\left( -\hat{\dot\beta}+\frac{i}{2m}\{\nabla,\hat\beta\}\right) \hat F
\nonumber\\ &
-\,i\hat{\bar F}\hat\sigma_{z}\left( \hat{\dot\beta} +\frac{i}{2m}\{\nabla
,\hat\beta\}\right) \hat{\bar F},\label{Ward1}%
\end{align}
where
\begin{align}
\hat\beta= \left(
\begin{array}
[c]{cc}%
\beta_{F}(t,{\bm r}) & 0\\
0 & \beta_{B}(t,{\bm r})
\end{array}
\right)
\end{align}
is an arbitrary diagonal $2\times2$ matrix.

\section{Kernels}

Explicit expressions for the functions $\chi_{J}$ read
\begin{align}
\chi_{\Delta}(Q) & = -\frac{2N_{0}}{\lambda} +\frac{N_{0}}{\pi}\int d\xi_{1}%
d\xi_{2} \frac{Dq^{2}}{(\xi_{1}-\xi_{2})^{2}+D^{2}q^{4}}
\nonumber\\ &\times\,
\bigg\{ -\left(
1+\frac{\xi_{1}\xi_{2}-\Delta_{0}^{2}}{E_{1}E_{2}}\right)
\frac{(E_{1}+E_{2})(1-f_{1}-f_{2})}{(\omega+i0)^{2}-(E_{1}+E_{2}%
)^{2}}
\nonumber\\ &
+\,\left( 1-\frac{\xi_{1}\xi_{2}-\Delta_{0}^{2}}{E_{1}E_{2}}\right)
\frac{(E_{1}-E_{2})(f_{1}-f_{2})}{(\omega+i0)^{2}-(E_{1}-E_{2})^{2}}
\bigg\},
\label{chiDelta}%
\end{align}
\begin{align}
\chi_{J}(Q) & = -\frac{2e^{2}N_{0}}{\pi}\int d\xi_{1}d\xi_{2} \frac{Dq^{2}%
}{(\xi_{1}-\xi_{2})^{2}+D^{2}q^{4}}\frac{\Delta_{0}^{2}}{E_{1}E_{2}}%
\nonumber\\&  \times\,
\bigg\{ \frac{(E_{1}+E_{2})(1-f_{1}-f_{2})}{(\omega+i0)^{2}-(E_{1}%
+E_{2})^{2}}
\nonumber\\ &
+\,\frac{(E_{1}-E_{2})(f_{1}-f_{2})}{(\omega+i0)^{2}-(E_{1}%
-E_{2})^{2}} \bigg\},\label{chiJ}%
\end{align}
\begin{align}
\chi_{L}(Q) & = -\frac{2m^{2}N_{0}D}{\pi}\int d\xi_{1}d\xi_{2} \frac{(\xi
_{1}-\xi_{2})^{2}}{(\xi_{1}-\xi_{2})^{2}+D^{2}q^{4}}\frac{\Delta_{0}^{2}%
}{E_{1}E_{2}}
\nonumber\\ &\times\,
 \bigg\{ \frac{(E_{1}+E_{2})(1-f_{1}-f_{2})}{(\omega+i0)^{2}-(E_{1}+E_{2})^{2}}
\nonumber\\ &
+\,\frac{(E_{1}-E_{2})(f_{1}-f_{2})}{(\omega+i0)^{2}-(E_{1}-E_{2})^{2}} \bigg\},
\label{chiL}
\end{align}
\begin{align}
\chi_{D}(Q)  & = \frac{e^{2}N_{0} D}{\pi}\int d\xi_{1}d\xi_{2} \frac{1}%
{(\xi_{1}-\xi_{2})^{2}+D^{2}q^{4}}\nonumber\\
& \times\, \bigg\{ \left( 1+\frac{\xi_{1}\xi_{2}+\Delta_{0}^{2}}{E_{1}E_{2}%
}\right) \frac{(E_{1}-E_{2})(f_{1}-f_{2})}{(\omega+i0)^{2}-(E_{1}-E_{2})^{2}%
}\nonumber\\
&  -\,\left( 1-\frac{\xi_{1}\xi_{2}+\Delta_{0}^{2}}{E_{1}E_{2}}\right)
\frac{(E_{1}+E_{2})(1-f_{1}-f_{2})}{(\omega+i0)^{2}-(E_{1}+E_{2})^{2}}
\bigg\}.\label{chiD}%
\end{align}
Here we defined $E_{1,2}=\sqrt{\xi_{1,2}^{2}+\Delta_{0}^{2}}$, $f_{1,2}%
=1/(1+\exp[E_{1,2}/T])$.

In a number of limiting cases the kernels $\chi_{j}$ can be evaluated exactly.
First let us set $T=0$ and consider the limit of small wave vectors $Dq^{2}%
\ll|\omega|,\Delta_{0}$. This regime is relevant, e.g., in the context of
microwave absorption. At $|\omega|<2\Delta_{0}$ we find
\begin{align}
\chi_{\Delta}  & =-2N_{0}\frac{\sqrt{4\Delta_{0}^{2}-\omega^{2}}}{\omega
}\arctan\frac{\omega}{\sqrt{4\Delta_{0}^{2}-\omega^{2}}},\nonumber\\
\chi_{J}  & =\frac{8e^{2}N_{0}\Delta_{0}^{2}}{\omega\sqrt{4\Delta_{0}%
^{2}-\omega^{2}}}\arctan\frac{\omega}{\sqrt{4\Delta_{0}^{2}-\omega^{2}}%
},\nonumber\\
\chi_{L}  & =\frac{2m^{2}\sigma_{D}\Delta_{0}}{e^{2}}K\left(  \frac{\omega
}{2\Delta_{0}}\right)  ,\nonumber\\
\chi_{D}  & =\frac{2\sigma_{D}\Delta_{0}}{\omega^{2}}\left[  K\left(
\frac{\omega}{2\Delta_{0}}\right)  -E\left(  \frac{\omega}{2\Delta_{0}%
}\right)  \right]  ,
\end{align}
while at $|\omega|>2\Delta_{0}$ one finds
\begin{align}
\chi_{\Delta}  & =-2N_{0}\frac{\sqrt{\omega^{2}-4\Delta_{0}^{2}}}{\omega
}\left[  \ln\left(  \frac{\omega}{2\Delta_{0}}+\sqrt{\frac{\omega^{2}}%
{4\Delta_{0}^{2}}-1}\right)  -i\frac{\pi}{2}\right]  ,\nonumber\\
\chi_{J}  & =-\frac{8e^{2}N_{0}\Delta_{0}^{2}}{\omega\sqrt{\omega^{2}%
-4\Delta_{0}^{2}}}\left[  \ln\left(  \frac{\omega}{2\Delta_{0}}+\sqrt
{\frac{\omega^{2}}{4\Delta_{0}^{2}}-1}\right)  -i\frac{\pi}{2}\right]
,\nonumber\\
\chi_{L}  & =\frac{2m^{2}\sigma_{D}\Delta_{0}}{e^{2}}\left[  \frac{2\Delta
_{0}}{\omega}K\left(  \frac{2\Delta_{0}}{\omega}\right)
\right.
\nonumber\\ &
\left.
+\,i\frac{4\Delta_{0}%
}{\omega+2\Delta_{0}}K\left(  \frac{\omega-2\Delta_{0}}{\omega+2\Delta_{0}%
}\right)  \right]  ,
\nonumber\\
\chi_{D}  & =\frac{\sigma_{D}}{\omega}\left[  K\left(  \frac{2\Delta_{0}%
}{\omega}\right)  -E\left(  \frac{2\Delta_{0}}{\omega}\right)  \right]
\nonumber\\
& +\,i\frac{\sigma_{D}}{\omega}\left[  \left(  1+\frac{2\Delta_{0}}{\omega
}\right)  E\left(  \frac{\omega-2\Delta_{0}}{\omega+2\Delta_{0}}\right)
\right.
\nonumber\\ &
\left.
-\,\frac{4\Delta_{0}}{\omega+2\Delta_{0}}K\left(  \frac{\omega-2\Delta_{0}%
}{\omega+2\Delta_{0}}\right)  \right]  .
\end{align}
Here and below $K(x)$ and $E(x)$ are complete elliptic integrals of the first
and second kind respectively. Utilizing
the relation (\ref{Qsigma}) one can verify that the expressions for $\chi_{L}$
and $\chi_{D}$ agree with the well known results \cite{MB}, \cite{AG}.

At temperatures close to $T_{C}$ one can derive analytic expressions for the
kernels in the limit $|\omega|,Dq^{2},\Delta_{0}\ll2\pi T_{C}$. At
$|\omega|<2\Delta_{0}$ these expressions read
\begin{align}
\chi_{\Delta} & = -\frac{7\zeta(3)}{2\pi^{2}}\frac{N_{0}\Delta_{0}^{2}}{T^{2}}
-\frac{N_{0}}{2T}\frac{|\omega|Dq^{2}}{2\Delta_{0}} \left[ \frac{\pi\Delta
_{0}}{|\omega|}+K\left( \frac{\omega}{2\Delta_{0}}\right)  \right. \nonumber\\
&  \left.  - \frac{\omega^{2}+D^{2}q^{4}-4\Delta_{0}^{2}}{\omega^{2}%
+D^{2}q^{4}} \Pi\left( \frac{\omega^{2}}{\omega^{2}+D^{2}q^{4}},\frac{\omega
}{2\Delta_{0}}\right) \right]
\nonumber\\&
+\,i\frac{N_{0}\Delta_{0}}{T}\frac{\omega}{Dq^{2}} \left[  \left(
1-\frac{\omega^{2}}{4\Delta_{0}^{2}}\right)  K\left( \sqrt{1-\frac{\omega^{2}%
}{4\Delta_{0}^{2}}}\right)
\right.
\nonumber\\ &
\left.
+\,\left( \frac{\omega^{2}}{4\Delta_{0}^{2}}%
-1+\frac{D^{2}q^{4}}{\omega^{2}+D^{2}q^{4}}\right)  \right. \nonumber\\
& \times\, \left.  \Pi\left(  \frac{D^{2}q^{4}}{\omega^{2}+D^{2}q^{4}}%
,\sqrt{1-\frac{\omega^{2}}{4\Delta_{0}^{2}}} \right)  \right]
,\label{chiDeltaap1}%
\end{align}
\begin{align}
\chi_{J}  & = \frac{7\zeta(3)}{2\pi^{2}}\frac{e^{2}N_{0}\Delta_{0}^{2}}{T^{2}}
\nonumber\\ &
+\,\frac{e^{2}N_{0}\Delta_{0}}{T}\frac{|\omega|Dq^{2}}{\omega^{2}+D^{2}q^{4}}
\Pi\left( \frac{\omega^{2}}{\omega^{2}+D^{2}q^{4}},\frac{\omega}{2\Delta_{0}%
}\right) \nonumber\\
&  -\,i\frac{e^{2}N_{0}\Delta_{0}}{T}\frac{\omega}{Dq^{2}}\left[  K\left(
\sqrt{1-\frac{\omega^{2}}{4\Delta_{0}^{2}}}\right) -\frac{\omega^{2}}%
{\omega^{2}+D^{2}q^{4}} \right. \nonumber\\
& \times\, \left.  \Pi\left(  \frac{D^{2}q^{4}}{\omega^{2}+D^{2}q^{4}}%
,\sqrt{1-\frac{\omega^{2}}{4\Delta_{0}^{2}}} \right)  \right] ,\label{chiJap1}%
\end{align}
\begin{align}
\chi_{L}  & = \frac{2m^{2}N_{0}D\Delta_{0}^{2}}{T}\left[  \frac{\pi}{2}
+\frac{|\omega|}{2\Delta_{0}} K\left(  \frac{|\omega|}{2\Delta_{0}} \right)
\right.
\nonumber\\ &
\left.
-\,\frac{|\omega|}{2\Delta_{0}} \frac{D^{2}q^{4}}{\omega^{2}+D^{2}q^{4}}
\Pi\left( \frac{\omega^{2}}{\omega^{2}+D^{2}q^{4}},\frac{\omega}{2\Delta_{0}
}\right)  \right]
\nonumber\\
&  -\,i\frac{m^{2}N_{0}D\Delta_{0}}{T}\frac{\omega^{3}}{\omega^{2}+D^{2}q^{4}}
\Pi\left( \frac{D^{2}q^{4}}{\omega^{2}+D^{2}q^{4}},\sqrt{1-\frac{\omega^{2}%
}{4\Delta_{0}^{2}}}\right) .\label{chiLap1}%
\end{align}
Here $\Pi(x,y)$ is the complete elliptic integral of the third kind.
At higher frequencies
$|\omega|>2\Delta_{0}$ we find
\begin{align}
\chi_{\Delta} & = -\frac{7\zeta(3)}{2\pi^{2}}\frac{N_{0}\Delta_{0}^{2}}{T^{2}}
-\frac{N_{0}}{2T}Dq^{2}\left[ \frac{\pi}{2}+K\left( \frac{2\Delta_{0}}{\omega
}\right)
\right.
\nonumber\\ &
\left.
+\left( \frac{4\Delta_{0}^{2}}{\omega^{2}+D^{2}q^{4}}-1\right)
 \Pi\left(  \frac{4\Delta_{0}^{2}}{\omega^{2}+D^{2}q^{4}%
},\frac{2\Delta_{0}}{\omega} \right) \right]
\nonumber\\ &
-\,i\frac{N_{0}}{2T}\frac{\omega
}{|\omega|}\frac{\omega^{2}+D^{2}q^{4}-4\Delta_{0}^{2}}{Dq^{2}}
\nonumber\\
& \times\, \left[  K\left( \sqrt{1-\frac{4\Delta_{0}^{2}}{\omega^{2}}}\right)
-\Pi\left( -\frac{D^{2}q^{4}}{\omega^{2}},\sqrt{1-\frac{4\Delta_{0}^{2}%
}{\omega^{2}}}\right)  \right.
\nonumber\\
&  \left.  -\, \Pi\left( \frac{4\Delta_{0}^{2}-\omega^{2}}{D^{2}q^{4}}%
,\sqrt{1-\frac{4\Delta_{0}^{2}}{\omega^{2}}}\right)  \right]
,\label{chiDeltaap}%
\end{align}
\begin{align}
\chi_{J}  & = \frac{7\zeta(3)}{2\pi^{2}}\frac{e^{2}N_{0}\Delta_{0}^{2}}{T}
\nonumber\\ &
+\,\frac{2e^{2}N_{0}\Delta_{0}^{2}}{T}\frac{Dq^{2}}{\omega^{2}+D^{2}q^{4}}
\Pi\left( \frac{4\Delta_{0}^{2}}{\omega^{2}+D^{2}q^{4}},\frac{2\Delta_{0}%
}{\omega}\right)
\nonumber\\
&  -\, i\frac{2e^{2}N_{0}\Delta_{0}^{2}}{T}\frac{\omega}{|\omega|Dq^{2}}
\left[  K\left( \sqrt{1-\frac{4\Delta_{0}^{2}}{\omega^{2}}}\right)
\right.
\nonumber\\ &
-\Pi\left(
-\frac{D^{2}q^{4}}{\omega^{2}},\sqrt{1-\frac{4\Delta_{0}^{2}}{\omega^{2}}%
}\right)
\nonumber\\
&  \left.  -\, \Pi\left( \frac{4\Delta_{0}^{2}-\omega^{2}}{D^{2}q^{4}}%
,\sqrt{1-\frac{4\Delta_{0}^{2}}{\omega^{2}}}\right)  \right] ,\label{chiJap}%
\end{align}
\begin{align}
\chi_{L} & = \frac{2m^{2}N_{0}D\Delta_{0}^{2}}{T}
\left[  \frac{\pi}{2}+K\left(\frac{2\Delta_{0}}{\omega}\right)
\right.
\nonumber\\ &
\left.
-\,\frac{D^{2}q^{4}}{\omega^{2}+D^{2}q^{4}}
\Pi\left(  \frac{4\Delta_{0}^{2}}{\omega^{2}+D^{2}q^{4}},\frac{2\Delta_{0}}{\omega} \right)  \right]
\nonumber\\
&  +\,i \frac{2m^{2}N_{0}D\Delta_{0}^{2}}{T}\frac{\omega}{|\omega|}
\left[K\left( \sqrt{1-\frac{4\Delta_{0}^{2}}{\omega^{2}}}\right)
\right.
\nonumber\\ &
-\Pi\left(
-\frac{D^{2}q^{4}}{\omega^{2}},\sqrt{1-\frac{4\Delta_{0}^{2}}{\omega^{2}}%
}\right)
\nonumber\\
&  \left.  -\, \Pi\left( \frac{4\Delta_{0}^{2}-\omega^{2}}{D^{2}q^{4}}%
,\sqrt{1-\frac{4\Delta_{0}^{2}}{\omega^{2}}}\right)  \right] .
\label{chiLap}
\end{align}

\begin{widetext}
\section{Relations between the phase and the electromagnetic potentials}

Integrating out the electromagnetic fields in eq. (\ref{a105}) we arrive at
the action (\ref{a116}) where the function ${\mathcal{F}}(\omega,q)$ is
defined by the following general expression
\begin{equation}
{\mathcal{F}}(\omega,q)=\frac{1}{s}\frac{\left(  \frac{\tilde{\chi}_{J}}{4e^{2}}
\omega^{2}+\frac{\tilde{\chi}_{L}}{4m^{2}}q^{2}\right)  \left(  \frac{C}{sL}
+\tilde{\chi}_{D}\left[  C\frac{\omega^{2}}{c^2}+\frac{q^{2}}{L}\right]  \right)
+\frac{\tilde{\chi}_{J}\tilde{\chi}_{L}}{4m^{2}}\left[  C\frac{\omega^{2}}{c^2}
+\frac{q^{2}}{L}\right]  }{\left(  \frac{C}{s}+\tilde{\chi}_{J}+\tilde{\chi
}_{D}q^{2}\right)  \left(  \frac{1}{sL}+\tilde{\chi}_{E}\frac{\omega^{2}}{c^2}+\frac
{e^{2}}{m^{2}c^2}\tilde{\chi}_{L}\right)  -\tilde{\chi}_{D}^{2}\frac{\omega^{2}q^{2}}{c^2}
}.\label{calF}
\end{equation}
The electromagnetic potentials are linked to the fluctuating phase of the
order parameter field via the following saddle point conditions
\begin{align}
V  & =\frac{\tilde{\chi}_{J}\left(  \frac{1}{sL}+
\tilde{\chi}_{D}\frac{\omega^{2}}{c^2}
+\frac{e^{2}}{m^{2}c^2}\tilde{\chi}_{L}\right)  + \frac{e^{2}}{m^{2}c^2}
\tilde{\chi}_{D}\tilde{\chi}_{L}q^{2}}{\left(  \frac{C}{s}+\tilde{\chi}
_{J}+\tilde{\chi}_{D}q^{2}\right)  \left(  \frac{1}{sL}+\tilde{\chi}_{D}
\frac{\omega^{2}}{c^2}+\frac{e^{2}}{m^{2}c^2}\tilde{\chi}_{L}\right)  -\tilde{\chi}_{D}
^{2}\frac{\omega^{2}q^{2}}{c^2}}\left(  \frac{-i\omega}{2e}\varphi\right)  ,\label{vphi}\\
A  & =\frac{\frac{e^{2}}{m^{2}}\tilde{\chi}_{L}\left(  \frac{C}{s}+\tilde
{\chi}_{J}+\tilde{\chi}_{D}q^{2}\right)  +\tilde{\chi}_{D}\tilde{\chi}
_{J}\omega^{2}}{\left(  \frac{C}{s}+\tilde{\chi}_{J}+\tilde{\chi}_{D}
q^{2}\right)  \left(  \frac{1}{sL}+\tilde{\chi}_{D}\frac{\omega^{2}}{c^2}+\frac{e^{2}
}{m^{2}c^2}\tilde{\chi}_{L}\right)  -\tilde{\chi}_{D}^{2}\frac{\omega^{2}q^{2}}{c^2}}\left(
\frac{iq}{2ec}\varphi\right)  .\label{Aphi}
\end{align}
\end{widetext}


\begin{thebibliography}{999}                                                                                              %

\bibitem {Kamerlingh-Onnes}H. Kamerlingh Onnes, Akad. van Wetenschappen
(Amsterdam) 14 (1911) 113.

\bibitem {almt}L.G. Aslamazov, and A.I. Larkin, Fiz. Tverd. Tela 10 (1968)
1140 [Sov. Phys. Solid State 10 (1968) 875].

\bibitem {Maki}K. Maki, Prog. Theor. Phys. 39 (1968) 897.

\bibitem {Thomson}R.S. Thompson, Phys. Rev. B 1 (1970) 327.

\bibitem {LV}A.I. Larkin and A. Varlamov, Theory of Fluctuations in
Superconductors (Clarendon, Oxford, 2005).

\bibitem {HMW}P.C. Hohenberg, Phys. Rev. 158 (1967) 383.

\bibitem {Mermin 1966}N.D. Mermin and H. Wagner, Phys. Rev. Lett. 17 (1966) 1133.

\bibitem {b}V.S. Berezinskii, Sov. Phys. JETP 32 (1971) 493; \textit{ibid.}
34 (1971) 610.

\bibitem {kt}J.M. Kosterlitz and D.J. Thouless, J. Phys. C6 (1973) 1181.

\bibitem {Kosterlilz 1974}J.M. Kosterlitz, J. Phys. C7 (1974) 1046.

\bibitem {Little}W.A. Little, Phys. Rev. 156 (1967) 396.

\bibitem {Meyer IV}J.D. Meyer and G.V. Minnigerode, Phys. Lett. A 38 (1972) 529.

\bibitem {Tidecks}R. Tidecks, Current-Induced Nonequilibrium
Phenomena in Quasi-One-Dimensional Superconductors (Springer, NY,
1990).

\bibitem {ivko}B.I. Ivlev and N.B. Kopnin, Adv. Phys. 80 (1984) 33.

\bibitem {Kopnin book}N. Kopnin, Nonequilibrium superconductivity\textit{\ }%
(Oxford University Press, NY, 2001).

\bibitem {la}J.S. Langer and V. Ambegaokar, Phys. Rev. 164 (1967) 498.

\bibitem {mh}D.E. McCumber and B.I. Halperin, Phys. Rev. B 1 (1970) 1054.

\bibitem {Webb R(T) in Sn whiskers}J.E. Lukens, R.J. Warburton, and W.W. Webb,
Phys. Rev. Lett. 25 (1970) 1180.

\bibitem {Tinkham R(T) in Sn whiskers}R.S. Newbower, M.R. Beasley, and M.
Tinkham, Phys. Rev. B 5 (1972) 864.

\bibitem {Giordano QPS PRL 1988}N. Giordano, Phys. Rev. Lett. 61 (1988) 2137.

\bibitem {sz}G. Sch\"{o}n and A.D. Zaikin, Phys. Rep. 198 (1990) 237.

\bibitem {sm}S. Saito, and Y. Murayama, Phys. Lett. A 135 (1989) 55;
\textit{ibid.} 139 (1989) 85.

\bibitem {Duan}J.-M. Duan, Phys. Rev. Lett. 74 (1995) 5128.

\bibitem {Chang}Y. Chang, Phys. Rev B 54 (1996) 9436.

\bibitem {cl}A.O. Caldeira, and A.J. Leggett, Phys. Rev. Lett. 46 (1981) 211.

\bibitem {cl1}A.O. Caldeira, and A.J. Leggett, Ann. Phys. (N.Y.) 149 (1983) 347.

\bibitem {weiss}U. Weiss, Quantum Dissipative Systems (World Scientific,
Singapore, 2nd Edition, 1999).

\bibitem {ZGOZ}A.D. Zaikin, D.S. Golubev, A. van Otterlo, and G.T. Zimanyi,
Phys. Rev. Lett. {78} (1997) 1552.

\bibitem {ZGOZ2}A.D. Zaikin, D.S. Golubev, A. van Otterlo, and G.T. Zimanyi,
Usp. Fiz. Nauk 168 (1998) 244 [Physics Uspekhi 42 (1998) 226].

\bibitem {GZ01}D.S. Golubev and A.D. Zaikin, Phys. Rev. B 64 (2001) 014504.

\bibitem {ogzb}A. van Otterlo, D.S. Golubev, A.D. Zaikin, and G. Blatter, Eur.
Phys. J. B 10 (1999) 131.

\bibitem {BT}A. Bezryadin, C.N. Lau, and M. Tinkham, Nature 404
(2000) 971.

\bibitem {MLG}K.A. Matveev, A.I. Larkin, and L.I. Glazman, Phys. Rev. Lett. 89
(2002) 096802.

\bibitem {SZ}S.V. Sharov and A.D. Zaikin, Phys. Rev. B 71 (2005) 014518.

\bibitem {MB}D.C. Mattis and J. Bardeen, Phys. Rev. 111 (1958) 412.

\bibitem {AG}A.A. Abrikosov and L.P. Gor'kov, Sov. Phys. JETP 8 (1959) 1090.

\bibitem {LK}A. Levchenko and A. Kamenev, Phys. Rev. B 76 (2007) 094518.

\bibitem {ms}J.E. Mooij and G. Sch\"{o}n, Phys. Rev. Lett. 55 (1985) 114.

\bibitem {ZhLKV}A. Zharov, A. Lopatin, A.E. Koshelev, and V.M. Vinokur, Phys.
Rev. Lett. 98 (2007) 197005.

\bibitem {Langer}J.S. Langer, Phys. Rev. Lett. 21 (1968) 973.

\bibitem{GZ08}D.S. Golubev and A.D. Zaikin, to be published.

\bibitem {ABC}See, e.g., A.I. Vainstein, V.I. Zakharov, V.A. Novikov, and M.A.
Shifman, Usp. Fiz. Nauk 136 (1982) 553 [Sov. Phys. Uspekhi 25
(1982) 195].

\bibitem {Lau MoGe PRL}C.N. Lau, N. Markovic, M. Bockrath, A. Bezryadin, and
M. Tinkham, Phys. Rev. Lett. 87 (2001) 217003.

\bibitem {Bezryadin MoGe review JPCM 2008}A. Bezryadin, J. Phys.: Cond. Mat.
20 (2008) 043202.

\bibitem {Zgirski NanoLett 2005}M. Zgirski, K.P. Riikonen, V. Tuboltsev, and
K. Arutyunov, Nano Lett. 5 (2005) 1029.

\bibitem {Zgirski QPS PRB 2008}M. Zgirski, K.P. Riikonen, V. Tuboltsev, and
K. Arutyunov, Phys. Rev. B 77 (2008) 054508.

\bibitem {Giordano QPS PRL 1989}N. Giordano and E.R. Schuler, Phys. Rev. Lett.
63 (1989) 2417.

\bibitem {Giordano QPS PRB 1991}N. Giordano, Phys. Rev. B 43 (1991) 160.

\bibitem {Giordano Physica B 1994}N. Giordano, Physica B 203 (1994) 460.

\bibitem {Altomare Al nanowire PRL 2006}F. Altomare, A.M. Chang, M.R.
Melloch, Y. Yong, and C.W. Tu, Phys. Rev. Lett. 97 (2006) 017001.

\bibitem {LO}A.I. Larkin and Yu.N. Ovchinnikov, Zh. Eksp. Teor. Fiz. 86 (1984)
719 [Sov. Phys. JETP 59 (1984) 420].

\bibitem {ZP}A.D. Zaikin and S.V. Panyukov, Zh. Eksp. Teor. Fiz. Pis'ma Red.
43 (1986) 518 [JETP Lett. 43 (1986) 670].

\bibitem {Lukens}D.B. Schwartz, B. Sen, C.N. Archie, and J.E. Lukens, Phys.
Rev. Lett. 55 (1985) 1547.

\bibitem {s}A. Schmid, Phys. Rev. Lett. 51 (1983) 1506.

\bibitem {s1}S.A. Bulgadaev, Zh. Eksp. Teor. Fiz. Pis'ma Red. 39 (1984) 264
[JETP Lett. 39 (1984) 315].

\bibitem {s2}F. Guinea, V. Hakim, and A. Muramatsu, Phys. Rev. Lett. 54 (1985) 263.

\bibitem {s3}M.P.A. Fisher and W. Zwerger, Phys. Rev. B 32 (1985) 6190.

\bibitem {Buchler}H.P. Buchler, V.B. Geshkenbein, and G. Blatter, Phys. Rev.
Lett. 92 (2004) 067007.

\bibitem {many}P. Bobbert, R. Fazio, G. Sch\"{o}n, and A.D. Zaikin, Phys. Rev. B
45 (1992) 2294.

\bibitem {FvdZ}R. Fazio and H. van der Zant, Phys. Rept. 355 (2001) 235.

\bibitem {Chbm}S. Chakravarty, Phys. Rev. Lett. 49 (1982) 681.

\bibitem {Chbm1}A.J. Bray and M.A. Moore, Phys. Rev. Lett. 49
(1982) 1545.

\bibitem {PLA1}S.V. Panyukov and A.D. Zaikin, Phys. Lett. A 124 (1987) 325.

\bibitem {Fi87}M.P.A. Fisher, Phys. Rev B 36 (1987) 1917.

\bibitem {Z88}A.D. Zaikin, Physica B 152 (1988) 251.

\bibitem {Ch88}S. Chakravarty, G.L. Ingold, S. Kivelson, and G.T. Zimanyi,
Phys. Rev. B 37 (1988) 3283.

\bibitem {PZ89}S.V. Panyukov and A.D. Zaikin, J. Low Temp. Phys. 75 (1989) 365;
\textit{ibid.} 75 (1989) 389.

\bibitem {Kor}S.E. Korshunov, Europhys. Lett. 9 (1989) 107.

\bibitem {Kor1}S.E. Korshunov, Sov. Phys. JETP 68 (1989) 609.

\bibitem {Zw}W. Zwerger, Europhys. Lett. 9 (1989) 421.

\bibitem {Bobbert}P. Bobbert, R. Fazio, G. Sch\"{o}n, and G.T. Zimanyi, Phys.
Rev. B 41 (1990) 4009.

\bibitem {PLA2}S.V. Panyukov and A.D. Zaikin, Phys. Lett. A 156 (1991) 119.

\bibitem {BD}R.M. Bradley and S. Doniach, Phys. Rev. B 30 (1994) 1138.

\bibitem {Chak}S. Tewari, J. Toner, and S. Chakravarty, Phys. Rev. B 73 (2006) 064503.

\bibitem {Dem}G. Refael, E. Demler, Y. Oreg, and D.S. Fisher, Phys. Rev. B 75
(2007) 014522.

\bibitem {Refnew}G. Refael, E. Demler, and Y. Oreg, arXiv: cond-mat/0803.2515v1
(2008).

\bibitem {Khl}S. Khlebnikov, Phys. Rev. Lett. 93 (2004) 090403.

\bibitem {KhlPr}S. Khlebnikov and L. Pryadko, Phys. Rev. Lett. 95 (2005) 107007.

\bibitem {Sachdev}S. Sachdev, P. Werner, and M. Troyer, Phys. Rev. Lett. 92
(2004) 237003.

\bibitem {HG}F.W.J. Hekking and L.I. Glazman, Phys. Rev. B 55 (1997) 6551.

\bibitem {Pai}G.V. Pai, E. Shimshoni, and N. Andrei, Phys. Rev. B 77 (2008) 104528.

\bibitem {PZ91}S.V. Panyukov and A.D. Zaikin, Phys. Rev. Lett. 67 (1991) 3168.

\bibitem {GZ94}D.S. Golubev and A.D. Zaikin, Phys. Rev. B 50 (1994) 8736.

\bibitem {GZ94-1}D.S. Golubev and A.D. Zaikin, JETP Lett. 63 (1996) 1007.

\bibitem {Naz}Yu.V. Nazarov, Phys. Rev. Lett. 82 (1999) 1245.

\bibitem {GZ00}D.S. Golubev and A.D. Zaikin, Phys. Rev. Lett. 86 (2001) 4887.

\bibitem {GZ04}D.S. Golubev and A.D. Zaikin, Phys. Rev. B 69 (2004) 075318.

\bibitem {GZ04-1}D.S. Golubev and A.D. Zaikin, Phys. Rev. B 70 (2004) 165423.

\bibitem {BN}D.A. Bagrets and Yu.V. Nazarov, Phys. Rev. Lett. 94 (2005) 056801.

\bibitem {Bezryadin MoGe TAPS EPL-2006}A.T. Bollinger, A. Rogachev and A.
Bezryadin, Europhys. Lett. 76 (2006) 505.

\bibitem {Bezryadin condmat07}A.T. Bollinger, R.C. Dinsmore III, A.
Rogachev, and A. Bezryadin, arXiv: 0707.4532v2 (2007).

\bibitem {GZ06}A.V. Galaktionov and A.D. Zaikin, Phys. Rev. B 73 (2006) 184522.

\bibitem {Zgirski inhomogeneity PRB}M. Zgirski and K.Yu. Arutyunov, Phys.
Rev. B 75 (2007) 172509.

\bibitem {Tinkham superconductivity book}M. Tinkham, Introduction to
superconductivity (Mc. Graw-Hill, Inc., 2nd edition, 1996) ISBN
0-07-114782-9.

\bibitem {Superconductor Material}Superconductor Material Science. Metallurgy,
Fabrication and Apllications. Edited by S. Foner and B.
B. Schwartz (Plenum Press, NY, 1981).

\bibitem {Kittel solid state}C. Kittel, Introduction to Solid State Physics
(J. Wiley \& Sons, Inc. 1996) ISBN 0-471-11181-3.

\bibitem {Ashcroft - Mermin}N.W. Ashcroft and N.D. Mermin, Solid State
Physics (Sounders College Publishing, 1976) ISBN 0-03-049346-3.

\bibitem {Bezryadin private}A. Bezryadin, private comminication.

\bibitem {Lau privat}J. Lau, private comminication.

\bibitem {Johansson InO nanowire}A. Johansson, G. Sambandamurthy, D. Shahar,
N. Jacobson, and R. Tenne, Phys. Rev. Lett. 95 (2005) 116804.

\bibitem {Johansson privat}A. Johansson, private comminication.

\bibitem {Tian Zn nanowire  PRL 2005}M. Tian, N. Kumar, S. Xu, J. Wang , J.S.
Kurtz, and M.H.W. Chan, Phys. Rev. Lett. 95 (2005) 076802.

\bibitem {Dynes QPS PRL 1993}F. Sharifi, A.V. Herzog, and R.C. Dynes, Phys.
Rev. Lett. 71 (1993) 428.

\bibitem {Bezryadin Nb nanowires}A. Rogachev and A. Bezryadin, Appl. Phys.
Lett. 83 (2003) 512.

\bibitem {Bezryadin MoGe and Nb wires}A. Rogachev, A.T. Bollinger, and A.
Bezryadin, Phys. Rev. Lett. 94 (2005) 017004.

\bibitem {Tian Sn nanowire APL 2003}M. Tian, J. Wang, J. Snyder,
J. Kurtz, Y. Liu, P. Schiffer, T.E. Mallouk,
and M.H.W. Chanet, Appl. Phys. Lett. 83 (2003) 1620.

\bibitem {Tian Sn nanowire PRB 2005}M. Tian, J. Wang, J.S. Kurtz, Y. Liu, M.
H.W. Chan, T.S. Mayer, and T.E. Mallouk, Phys. Rev. B 71 (2005)
104521.

\bibitem {Piraux Sn nanowire APl 2004}S. Michotte, L. Piraux, F. Boyer, F.R.
Ladan, J.P. Maneval, Appl. Phys. Lett. 85 (2004) 3175.

\bibitem {Piraux Sn nanowire NanoSci NanoTech 2005}L. Piraux, A. Encinas, L.
Vila, S. M\'{a}t\'{e}fi-Tempfli, M. M\'{a}t\'{e}fi-Tempfli, M.
Darques, F. Elhoussine, and S. Michotte, J. Nanosci. Nanotechnol.
5 (2005) 372.

\bibitem {Fin}Y. Oreg and A.M. Finkelstein, Phys. Rev. Lett. 83 (1999) 191.

\bibitem {Shanenko Tc(size) PRB 2006}A.A. Shanenko, M.D. Croitoru, M.
Zgirski, F.M. Peeters, and K. Arutyunov, Phys. Rev. B 74 (2006)
052502.

\bibitem {Givargizov whisker growth}E.I. Givargizov, Highly Anisotropic
Crystals, Series: Materials Science of Minerals and Rocks
(Springer, 1987), ISBN: 978-90-277-2172-73.

\bibitem {Fisher whisker growth}R.M. Fisher, L.S. Darken, and K.G. Carrol, Acta
Metallurgica 2 (1954) 368.

\bibitem {Lutes Sn filaments}O.S. Lutes, Phys. Rev. 105 (1957) 1451.

\bibitem {Gaidukov whiskers}Y.P. Gaidukov, N.P. Danilova, and R.S.
Georgius-Mankarius, Sov. Phys. JETP 66 (1987) 605.

\bibitem {Arutyunov SOG whiskers}K.Yu. Arutyunov, T.V. Ryyn\"{a}nen, J.P.
Pekola, and A.B. Pavolotski, Phys. Rev. B 63 (200) 092506.

\bibitem {Arutyunov patents in nanotechnology}K.Yu. Arutyunov, Recent Patents
on Nanotechnology 1 (2007) 129.

\bibitem {Taylor micricylinder method}G.F. Taylor, Phys. Rev. 23 (1924) 655.

\bibitem {Taylor microcylinder patent}G.F. Taylor, Patent US1793529 (1931).

\bibitem {Ulitovski microcylinder method}A.V. Ulitovsky, Micro-technology in
design of electric devices (Leningrad) 7 (1951) 6.

\bibitem {Ulitovski microcylinder patent1}A.V. Ulitovski and N.M. Avernin,
Method of fabrication of metallic microwire, Patent No161325
(USSR), 19.03.64. Bulletin No.7, 14.

\bibitem {Ulitovski microcylinder patent2}A.V. Ulitovsky, I.M. Maianski, and
A.I. Avramenko, Method of continuous casting of glass coated
microwire. Patent 128427 (USSR), 15.05.60. Bulletin. No.10, 14.

\bibitem {Nikolaeva sub 1 mkm Bi wires}T.E. Huber, A. Nikolaeva, D. Gitsu, L.
Konopko, C.A. Foss Jr., and M.J. Graff, Appl. Phys. Lett. 84
(2004) 1326.

\bibitem {Nikolaeva Bi-Te nanofilaments}A.A. Nikolaeva, D.V. Gitsu, T.E.
Huber, L.A. Konopko, and G. Para, Phys. Stat. Sol. (c) 11 (2004)
2654.

\bibitem {Nikolaeva confinement in Bi nanowires}A. Nikolaeva, D. Gitsu, T.
Huber, and L. Konopko, Physica B 346 (2004) 282.

\bibitem {Arutyunov Sn and In filaments JAP}K.Yu. Arutyunov, N.P. Danilova,
and A.A. Nikolaeva, J. Appl. Phys. 76 (1994) 7139.

\bibitem {Arutyunov Sn and In filaments Physica C}K.Yu. Arutyunov, N.P.
Danilova, and A.A. Nikolaeva, Physica C 235 (1994) 1967.

\bibitem {Santhanam nMR anomaly}P. Santhanam, C.P. Umbach, and C.C. Chi,
Phys. Rev. B 40 (1989) 11392.

\bibitem {Santhanam R(T) anomaly}P. Santhanam, C.C. Chi, S.J. Wind, M.J.
Brady, and J.J. Bucchignano, Phys. Rev. Lett. 66 (199) 2254.

\bibitem {Moschalkov Little-Parks anomaly}H. Vloeberghs, V.V. Moshchalkov, C.
Van Haesendonck, R. Jonckheeere, and Y. Bruenseraede, Phys. Rev.
Lett. 69 (1992) 1268.

\bibitem {Moshchalkov R(T) anomaly}V.V. Moshchalkov  L. Gielen, G.
Neuttiens, C. van Haesendonck, and Y. Bruynseraede, Phys. Rev. B 49
(1994) 15412.

\bibitem {Arutyunov non-locality}K.Y. Arutyunov, J.P. Pekola, A.B.
Pavolotski, and D.A. Presnov, Phys Rev B 64 (2001) 064519.

\bibitem {Bezryadin molecular decoration 1}A. Bezryadin, A. Bollinger, D.
Hopkins, M. Murphey, M. Remeika, and A. Rogachev, Dekker
encyclopedia of nanoscience and nanotechnology, 3761 (2004),
DOI:10.1081/E-ENN 120013540.

\bibitem {Bezryadin molecular decoration 2}D.S. Hopkins, D. Pekker, P.M.
Goldbart, and A. Bezryadin, Science 308 (2005) 1762.

\bibitem {Bezryadin molecular decoration 3}M. Remeika and A. Bezryadin,
Nanotechnology 16 (2005) 1172.

\bibitem {InO film Shahar PRB 1992}D. Shahar and Z. Ovadyahu, Phys. Rev. B 46
(1992) 10917.

\bibitem {Meidan}D. Meidan, Y. Oreg, and G. Refael, Phys. Rev. Lett. 98 (2007) 187001

\bibitem {Meidan2}D. Meidan, Y. Oreg, G. Refael, and R.A. Smith, Physica C
468 (2008) 341.

\bibitem {Graybeal QPS in MoGe PRL 1987}J.M. Graybeal, P.M. Mankiewich, R.
C. Dynes, M.R. Beasley, Phys. Rev. Lett. 59 (1987) 2697.

\bibitem {Graybeal 2D MoGe films  PRB 1984}J.M. Graybeal, M.R. Beasley,
Phys. Rev. B 29 (1984) 4167.

\bibitem {Mooji}A.J. van Run, J. Romijn, and J.E. Mooij, Jpn. J. Appl. Phys.
26 (1987) 1765.

\bibitem {Giordano QPS PRB 1990}N. Giordano, Phys. Rev. B 41 (1990) 6350.

\bibitem {Giordano step technique APL 1980}D.E. Prober, M.D. Feuer, and N.
Giordano, Appl. Phys. Lett. 37 (1980) 94.

\bibitem {Girdano step technique PRL 1979}N. Giordano, W. Gilson, and D.E.
Prober, Phys. Rev. Lett. 43 (1979) 725.

\bibitem {Dynes QPS and nMR PRL 1997}P. Xiong, A.V. Herzog, and R.C. Dynes,
Phys. Rev. Lett. 78 (1997) 927.

\bibitem {Dynes QPS PRL 1996}A. V. Herzog, P. Xiong, F. Sharifi, and R.C.
Dynes, Phys. Rev. Lett. 76 (1996) 668.

\bibitem {Mikko}J.S. Penttila, P.J. Hakonen, M.A. Paalanen, and E.B. Sonin,
Phys. Rev. Lett. 82 (1999) 1004.

\bibitem {GK}A.A. Golubov and M.Yu. Kupriyanov, J. Low Temp. Phys. 70 (1988) 83.

\bibitem {BBS2}W. Belzig, C. Bruder, and G. Sch\"{o}n, Phys. Rev. B 54 (1996) 9443.

\bibitem {Been}K.M. Frahm, P.W. Brouwer, J.A. Melsen, and C.W.J. Beenakker,
Phys. Rev. Lett. 76 (1996) 2981.

\bibitem {Pilgram}S. Pilgram, W. Belzig, and C. Bruder, Phys. Rev. B 62 (2000) 12462.

\bibitem {Zhou}F. Zhou, P. Charlat, B. Spivak, and B. Pannetier, J. Low Temp.
Phys. 110 (1998) 841.

\bibitem {KKZ}M.S. Kalenkov, H. Kloos, and A.D. Zaikin, Phys. Rev. B 74
(2006) 184502.

\bibitem {FNT}A.D. Zaikin and G.F. Zharkov, Fiz. Nizk. Temp. 7 (1981) 375
[Sov. J. Low Temp. Phys. 7 (1981) 181].

\bibitem {WZK}F.K. Wilhelm, A.D. Zaikin, and G. Sch\"{o}n, J. Low Temp. Phys.
106 (1997) 305.

\bibitem {Dubos}P. Dubos, H. Courtois, B. Pannetier, F.K. Wilhelm, A.D.
Zaikin, and G. Sch\"{o}n, Phys. Rev. B 63 (2001) 064502.

\bibitem {Bogomolov capillary wire Sov Phys Sol St 1971}V. Bogomolov, Sov.
Phys. Solid. State. 13 (1971) 815.

\bibitem {Michael In microcylinders JLTP 1974}P. Michael and D.S. McLachlan,
J. Low Temp. Phys. 14 (1974) 607.

\bibitem {Yi Pb electrodeposited nanowires APL 1999}G. Yi and W. Schwarzacher,
Appl. Phys. Lett. 74 (1999) 1746.

\bibitem {Piraux IV in Pb and Sn nanowire PRL 2003}D.Y. Vodolazov,  F.M.
Peeters, L. Piraux , S. M\'{a}t\'{e}fi-Tempfli, and S. Michotte,
Phys. Rev. Lett. 91 (2003) 157001.

\bibitem {Piraux Pb nanowire PRB 2004}S. Michotte, S. Matefi-Tempfli, L.
Piraux, D.Y. Vodolazov, and F. Peeters, Phys. Rev. B 69 (2004)
094512.

\bibitem {Piraux Pb nanowire IV APL 2008}A. Adam, F. de Menten de Horne, L.
Pirauz, and S. Michotte, Appl. Phys. Lett. 92 (2008) 012516.


\bibitem {Piraux multicontact Sn nanowire APL 2007}D. Lucot, F. Pierre, D.
Mailly, K.Yu. Zhang, S. Michotte, F. de Menten de Horne, and L.
Piraux, Appl. Phys. Lett. 91 (2007) 042502.

\bibitem {Piraux Pb and Sn granular nanowires APL 2003}S. Michotte, S.
M\'{a}t\'{e}fi-Tempfli, and L. Piraux, Appl. Phys. Lett. 82 (2003)
4119.

\bibitem {Tian Zn nanowire PRB 2006}M. Tian, N. Kumar, J. Wang, S. Xu,
and M.H.W. Chan, Phys. Rev. B 74 (2006) 014515.

\bibitem {Tian specific heat Zn nanowire PRL 2007}J.S. Kurtz, R.R. Johnson,
M. Tian, N. Kumar, Z. Ma, S. Xu, and M.H.W. Chan, Phys. Rev. Lett.
98 (2007) 247001.

\bibitem {Piraux single nanowire probing  APL 2002}L. Vila, L. Piraux, J.M.
George, and G. Faini, Appl. Phys. Lett. 80 (2002) 3805.

\bibitem {Piraux single wire contact Nanotechnology 2005}S. Fusil, L. Piraux,
S. M\'{a}t\'{e}fi-Tempfli, M. M\'{a}t\'{e}fi-Tempfli, S. Michotte,
C.K. Saul, L.G. Pereira, K. Bouzehouane, V. Cros, C. Deranlot, and
J.-M. George, Nanotechnology 16 (2005) 2936.

\bibitem {Antiproximity theory Clarke PRL 2006}H.C. Fu, A. Siedel, J. Clarke,
and D.H. Lee, Phys. Rev. Lett. 96 (2006) 157005.

\bibitem {Altomare AuPd and Al nanowire fabrication APL 2005}F. Altomare,
A.M. Chang, M.R. Melloch, Y. Hong, and C.W. Tu, Appl. Phys. Lett.
86 (2005) 172501.

\bibitem {Altomare AuPd and Al nanowire fabrication cond-mat 2007}F. Altomare,
A.M. Chang, M.R. Melloch, Y. Hong, and C.W. Tu, arXiv:
cond-mat/04122102v2 (2007).

\bibitem {Altomare Al nanowire cond-mat-2005}F. Altomare, A.M. Chang, M.R.
Melloch, Y. Hong, and C.W. Tu, arXiv: cond-mat/0505772 v1 (2005).

\bibitem {Savolainen IBE method APA 2004}M. Savolainen, V. Tuboltsev, P.
Koppinen, K.-P. Riikonen, and K. Arutyunov, Appl. Phys. A 79
(2004) 1769.

\bibitem {Zgirski IBE method Nanotechnology 2008}M. Zgirski, K.P. Riikonen,
V. Tuboltsev, P. Jalkanen, T.T. Hongisto and K.Yu. Arutyunov,
Nanotechnology 19 (2008) 055301.

\bibitem {Bezr08}M. Sahu, M.-H. Bae, A. Rogachev, D. Pekker, T.-C. Wei,
N. Shah, P.M. Golbart, and A. Bezryadin, arXiv: 0804.2251 (2008).


\bibitem {Magnetic ions in Al Jalkanen SolStComm 2007}P. Jalkanen, V.
Tuboltsev, A. Virtanen, K.Yu. Arutyunov, J. R\"{a}is\"{a}nen, O.
Lebedev, and G. van Tendeloo, Solid St. Commun. 142 (2007) 407.

\bibitem {Bezryadin nMR MoGe and Nb PRL 2006}A. Rogachev, T.-C. Wei, D.
Pekker, A.T. Bollinger, P.M. Golbart, and A. Bezryadin, Phys.
Rev. Lett. 97 (2006) 137001.

\bibitem {Bezryadin nMR theory EPL 2006}T.-C. Wei, D. Pekker, A Rogachev, A.
Bezryadin, and P.M. Golbart, Europhys. Lett. 75 (2006) 943.

\bibitem {Kadin pair breaking JLTP 1978}A.M. Kadin, W.J. Skockpol, and M.
Tinkham, J. Low. Temp. Phys. 33 (1978) 481.

\bibitem {Arutyunov R(T) anomaly PRB 1999}K.Yu.Arutyunov, S.V.Lotkhov,
A.B.Pavolotski, D.A. Presnov, and L. Rinderer, Phys. Rev. B 59
(1999) 6487.

\bibitem {PA}D. Pesin and A. Andreev, Phys. Rev. Lett. 97 (2006) 117001.

\bibitem {nMR Physica B 2001}Y. Terai, T. Yakabe, C. Terakura, T. Terashima,
T. Takamasu, S. Uji, and G. Kido, Physica B 298 (2001) 536.

\bibitem {Arutyunov nMR Physica C 2008}K.Yu. Arutyunov, Physica C 468 (2008) 272.

\bibitem {R(T) steps Webb-Warburton PRL 1968}W.W. Webb and R.J. Warburton,
Phys. Rev. Lett. 20 (1968) 461.

\bibitem {Meyer RT steps}J.D. Meyer, Appl. Phys. 2 (1973) 303.

\bibitem {Kim R(T) anomaly Physica B 1994}J.J. Kim, J.H. Kim, S.J. Lee,
H.J. Lee, K.W. Park, H.J. Shin, and E.H. Lee, Physica B 194-196
(1994) 1035.

\bibitem {Kwong R(T) anomaly PRB 1991}Y.K. Kwong, K. Lin, P.M. Hakonen,
M.S. Isaacson, and J.M. Parpia, Phys. Rev. B 44 (1991) 462.

\bibitem {Park R(T) anomaly PRL 1995}M. Park, M.S. Isaacson, and J.M.
Parpia, Phys. Rev. Lett. 75 (1995) 3740.

\bibitem {Strunk R(T) anomaly PRB 1996}C. Strunk, V. Bruyndoncx, C. Van
Haesendonck, V.V. Moshchalkov, Y. Bruynseraede, B. Burk, C.J.
Chien, and V. Chandrasekhar, Phys. Rev. B 53 (1996) 11332.

\bibitem {Strunk R(T) anomaly Superlat 1996}B. Burk, C.J. Chien, V.
Chandrasekhar, C. Strunk, V. Bruyndoncx, C. Van Haesendonck, V.V.
Moshchalkov, and Y. Bruynseraede, Superlattices and
Microstructures, 20 (1996) 575.

\bibitem {Strunk R(T) anomaly JAP 1998}B. Burk, C. J. Chien, V. Chandrasekhar,
C. Strunk, V. Bruyndoncx, C. Van Haesendonck, and V.V.
Moshchalkov, J. Appl. Phys. 83 (1998) 1549.

\bibitem {Strunk R(T) anomaly PRB 1998}C. Strunk, V. Bruyndoncx, C. Van
Haesendonck, V.V. Moshchalkov, Y. Bruynseraede, C.J. Chien, B.
Burk, and V. Chandrasekhar, Phys. Rev. B 57 (1998) 10854.

\bibitem {Moschalkov R(T) anomaly PRB 1994}V.V. Moshchalkov, L. Gielen, G.
Neuttiens, C. Van Haesendonck, and Y. Bruynseraede, Phys. Rev. B
49 (1994) 15412.

\bibitem {Moschalkov R(T) anomaly PRB 1997}V. Moshchalkov, L. Gielen, G.
Neuttiens, C. Van Haesendonck, and Y. Bruynseraede, Phys. Rev. B
56 (1997) 6352.

\bibitem {Arutyunov R(T) anomaly PRB 1996}K.Yu. Arutyunov, Phys. Rev. B 53
(1996) 12304.

\bibitem {Landau R(T) anomaly PRB 1997}I.L. Landau and L. Rinderer, Phys.
Rev. B 56 (1997) 6348.

\bibitem {Tc(size) theory McMillan PR 1968}W.L. McMillan,  Phys. Rev. 167
(1968) 331.

\bibitem {Tc(size) theory Garland PRL 1968}J.W. Garland, K.H. Bennemann, and
F.M. Mueller, Phys. Rev. Lett. 21 (1968) 1315.

\bibitem {Tc(size) Allen PR69}R.E. Allen and F.W. de Wette, Phys. Rev. 187
(1969) 883.

\bibitem {Tc(size) experiment Ga Parshin PRB 1996}I.A. Parshin, I.L. Landau,
and L. Rinderer, Phys. Rev. B 54 (1996) 1308.

\bibitem {Tc(size) theory Shanenko PRB 2006}A.A. Shanenko and M.D. Croitoru,
Phys. Rev. B 73 (2006) 012510.

\bibitem {GZ}D.S. Golubev and A.D. Zaikin, Phys. Rev. Lett. 81 (1998) 1074.

\bibitem {GZ1}D.S. Golubev and A.D. Zaikin, Phys. Rev. B 59 (1999) 9195.

\bibitem {GZ2}D.S. Golubev and A.D. Zaikin, Physica E 40 (2007) 32.

\bibitem {GZ98P}D.S. Golubev and A.D. Zaikin, Physica B 255 (1998) 164.

\bibitem {Paco}F. Guinea, Phys. Rev. B 65 (2002) 205317.

\bibitem {Carlos}D.S. Golubev, C.P. Herrero, and A.D. Zaikin, Europhys. Lett.
63 (2003) 426.

\bibitem {MN}J.E. Mooij and Yu.V. Nazarov, Nat. Phys. 2 (2006) 169.

\bibitem {MH}J.E. Mooij and C.J.P.M. Harmans, New J. Phys. 7 (2005) 219.

\bibitem {AvOd}D.V. Averin and A.A. Odintsov, Phys. Lett. A 140 (1989) 251.

\bibitem {Z90}A.D. Zaikin, J. Low Temp. Phys. 80 (1990) 223.

\bibitem {Z04}A.D. Zaikin, Fiz. Nizk. Temp. (Kharkov) 30 (2004) 756
[Low Temp. Phys. 30 (2004) 568].

\bibitem {SZ2}S.V. Sharov and A.D. Zaikin, Physica E 29 (2005) 360.

\bibitem {AN92}D.V. Averin and Yu.V. Nazarov, Phys. Rev. Lett. 69 (1992) 1993.

\bibitem {Tuo93}M.T. Tuominen, J.M. Hergenrother, T.S. Tighe, and M. Tinkham,
Phys. Rev. Lett. 69 (1992) 1997.

\bibitem {sz94}G. Sch\"{o}n and A.D. Zaikin, Europhys. Lett. 26 (1994) 695.

\bibitem {vDZGT}J. von Delft, A.D. Zaikin, D.S. Golubev, and W.
Tichy, Phys. Rev. Lett. 77 (1996) 3189.

\bibitem {Laf93}P. Lafarge, P. Joyez, D. Esteve, C. Urbina, and M.H. Devoret,
Phys. Rev. Lett. 70 (1993) 994.

\bibitem {RT}C.T. Black, D.C. Ralph, and M. Tinkham, Phys. Rev. Lett. 76 (1996) 688.

\bibitem {JSA94}B. Janko, A. Smith, and V. Ambegaokar, Phys. Rev. B 50 (1994) 1152.

\bibitem {GZ294}D.S. Golubev and A.D. Zaikin, Phys. Lett. A 195 (1994) 380.

\bibitem {AN94}D.V. Averin and Yu. V. Nazarov, Physica B 203 (1994) 310.

\bibitem {Kang}K. Kang, Europhys. Lett. 51 (2000) 181.

\bibitem {Yak}H.-J. Kwon and V.M. Yakovenko, Phys. Rev. Lett. 89 (2002) 017002.

\bibitem {GZ02}A.V. Galaktionov and A.D. Zaikin, Phys. Rev. B 65 (2002) 184507.

\bibitem {KO}I.O. Kulik and A.N. Omel'yanchuk, Sov. J. Low Temp. Phys. 4
(1978) 142.

\bibitem {KO1}W. Haberkorn, H. Knauer, and J. Richter, Phys. Stat. Solidi (A)
47 (1978) K161.

\bibitem {FNJ}M. Hayashi and H. Ebisawa, Phys. Rev. B 67 (2003) 014524.

\bibitem {Furusaki}A. Furusaki and M. Tsukada, Physica B 165-166 (1990) 967.

\bibitem {Furusaki1}C.W.J. Beenakker and H. van Houten, Phys. Rev. Lett. 66
(1991) 3056.

\bibitem {Kulik}I.O. Kulik, Sov. Phys. JETP 30 (1970) 944.

\bibitem {Kulik1}C. Ishii, Progr. Theor. Phys. 44 (1970) 1525.

\bibitem {Leva}L.N. Bulaevskii, V.V. Kuzii, and A.A. Sobyanin, JETP Lett. 25
(1977) 290.

\bibitem {Volkov}A.F. Volkov, Phys. Rev. Lett. 74 (1995) 4730.

\bibitem {WSZ}F.K. Wilhelm, G. Sch\"{o}n, and A.D. Zaikin, Phys. Rev. Lett. 81
(1998) 1682.

\bibitem {Yip}S.K. Yip, Phys. Rev. 58 (1998) 5803.

\bibitem {Kas}A.Yu. Kasumov, R. Deblock, M. Kociak, B. Reulet, H. Bouchiat,
I.I. Khodos, Yu.B. Gorbatov, V.T. Volkov, C. Journet, and M. Burghard,
Science 281 (1998) 540.

\bibitem {Delft}P. Jarillo-Herrero, J.A. van Dam, and L.P. Kouwenhoven,
Nature 439 (2006) 953.

\bibitem {VBou}J.-P. Cleuziou, W. Wernsdorfer, V. Bouchiat, T. Ondarcuhu, and
M. Monthioux, Nature Nanotechnology 1 (2006) 53.

\bibitem {L1}H.I. Jorgensen, K. Grove-Rasmussen, T. Novotny, K. Flensberg, and
P.E. Lindelof, Phys. Rev. Lett. 96 (2006) 207003.

\bibitem {L2}K. Grove-Rasmussen, H.I. Jorgensen, and P.E. Lindelof, New J.
Phys. 9 (2007) 124.

\bibitem {Hels}T. Tsuneta, L. Lechner, and P.J. Hakonen, Phys. Rev. Lett. 98
(2007) 087002.

\bibitem {HBou}M. Kociak, A.Yu. Kasumov, S. Gueron, B. Reulet, I.I. Khodos,
Yu.B. Gorbatov, V.T. Volkov, L. Vaccarini, and H. Bouchiat,
Phys. Rev. Lett. 86 (2001) 2416.

\bibitem{HBou2}A. Kasumov, M. Kociak, M. Ferrier, R. Deblock, S. Gueron,
 B. Reulet, I. Khodos, O. Stephan, and H. Bouchiat, Phys. Rev. B 68 (2003)
214521.

\bibitem {Jap}I. Takesue, J. Haruyama, N. Kobayashi, S. Chiashi, S. Maruyama,
T. Sugai, and H. Shinohara, Phys. Rev. Lett. 96 (2006) 057001.
















\end{thebibliography}
\end{document}